\documentclass[twoside]{easychithesis}
\usepackage{latexsym}
\usepackage{amsmath,amssymb}
\usepackage{graphicx}
\usepackage{times}
\usepackage{bm}
\allowdisplaybreaks

\DeclareGraphicsExtensions{.eps,.ps}

\newcommand{\newfootnote}[1]{\footnote{#1}}
\newcommand{\beq}{\begin{equation}}
\newcommand{\eeq}{\end{equation}}
\newcommand{\bea}{\begin{eqnarray}}
\newcommand{\eea}{\end{eqnarray}}
\newcommand{\ek}{\epsilon_{\mathbf{k}}}
\newcommand{\ekq}{\epsilon_{\mathbf{k-q}}}
\newcommand{\Ek}{E_{\mathbf{k}}}

\newcommand{\phik}{\varphi_{\mathbf{k}}}
\newcommand{\phikq}{\varphi_{{\mathbf{k}}-{\mathbf{q}}/2}}
\newcommand{\mb}[1]{{\mathbf{#1}}}
\newcommand{\sumk}{\sum_{\mathbf{k}}}

\newcommand{\sumq}{\sum_{\mathbf{q}}}

\newcommand{\Omegaq}{\Omega_{\mathbf{q}}}
\newcommand{\Gammaq}{\Gamma_{\mathbf{q}}}
\newcommand{\uk}{u_{\mathbf{k}}}
\newcommand{\vk}{v_{\mathbf{k}}}
\renewcommand{\text}[1]{{#1}}

\begin{document}

\raggedbottom
\title{Generalization of BCS theory to short coherence length
  superconductors: A BCS--Bose-Einstein crossover scenario}
\author{Qijin Chen} 
\date{August 2000} 
\department{Physics}
\division{Physical Sciences} 
\degree{Doctor of Philosophy} 
\maketitle

\dedication

\dedication
\begin{center}
       {\large {\textit{To my parents}}}
\end{center}

\dedication
\vfill
\centerline{Copyright \copyright\ 2000 by Qijin Chen.}
\centerline{All rights reserved.}
\vspace*{1.7cm}

\topmatter{Abstract}

The microscopic theory of superconductivity by Bardeen, Cooper and
Schrieffer (BCS) is considered one of the most successful theories in
condensed matter physics.  In ordinary metal superconductors the
coherence length $\xi$ is large, and a simple mean field approach, such as
BCS, is thereby justified.  This theory has two important features: the
order parameter and excitation gap are identical, and the formation of pairs
and their Bose condensation take place at the same temperature $T_c$.  It is
now known that BCS theory fails to explain the superconductivity in the
underdoped copper oxide superconductors: the excitation gap $\Delta$ is
finite at $T_c$ and thus distinct from the order parameter
$\Delta_{sc}$. Since these superconductors belong to a large class of
inter-related, generally small $\xi$ materials, this failure has the
potential for widespread impact.

In this thesis, we have extended BCS theory in a natural way to short
coherence length superconductors, based on a BCS--Bose-Einstein condensation
(BEC) crossover scenario. We arrive at a simple physical picture in which
incoherent, finite momentum pairs become progressively more important as the
pairing interaction becomes stronger.  The ensuing distinction between
$\Delta$ and $\Delta_{sc}$ can be associated with these pairs.  These finite
momentum pairs are treated at a mean field level which addresses the pairs
and the single particles on an equal footing.  Within our picture, the
superconducting transition from the fermionic perspective and Bose-Einstein
condensation from the bosonic perspective are just two sides of the same
coin.

In contrast to many other theoretical approaches, our theory is capable of
making quantitative predictions, which can be tested.  This theory was
applied to the cuprates to obtain a phase diagram.  In addition, because
this fitted (with one free parameter) phase diagram represented experiment
quite well, it was possible to quantitatively address derived quantities,
such as the hole concentration and temperature dependences of the in-plane
penetration depth and specific heat.  The mutually compensating
contributions from fermionic quasiparticles and bosonic pair excitations
have provided a natural explanation for the quasi-universal behavior of the
normalized in-plane superfluid density as a function of reduced temperature.
Our bosonic pair excitations also provide an intrinsic mechanism for the
long mysterious linear $T$ terms in the specific heat.  We found new
qualitative effects as well, associated with predicted low $T$ power laws,
which arise from our incoherent pair contributions.  These power laws seem
to be consistent with existing experiments, although more systematic
experimental studies are needed. Finally, we demonstrated that the onset of
superconducting long range order leads to sharp features in the specific
heat at $T_c$, (although the excitation gap is smooth across $T_c$), which
are consistent with experiment.

\topmatter{Acknowledgements}
I would like thank my thesis advisor, Professor Kathryn Levin, for her
support throughout my graduate research. This thesis work was completed
under her careful guidance. Her support has made it possible for me to
present our work at various conferences, and to communicate with the
community of high $T_c$ superconductivity. Both her insights in physics and
her expertise in communicating science have greatly enhanced these
presentations and communications. Her constant enthusiasm for physics has
been an excitement and an encouragement that help me make rapid progress
during the research. Her careful reading and revision have significantly
enhanced this manuscript.

I am very grateful to have Ioan Kosztin as a collaborator and a friend
during my graduate career at the University of Chicago. His support has
greatly facilitated this research, and his insights in physics have been
critical in extending the BCS-BEC crossover theory below $T_c$. The
countless memorable late nights we spent together in the Research Institute
building have been very fruitful.  I am also grateful to Boldisz\'ar Jank\'o
for his  generous help, useful advice, and, especially, for
collaborations during the early stages of this project. In addition, I wish
to thank Ying-Jer Kao for collaborations on the collective mode issues and
for other useful discussions during my graduate study at the University.

I wish to thank G. Deutscher, A.~J. Leggett, G.~F. Mazenko, M.~R. Norman,
B.~R. Patton, N.~E. Phillips, M. Randeria, A.~A.  Varlamov, and P.~B.
Wiegmann for helpful discussions, D.~A.  Bonn, A. Carrington, R.~W.
Giannetta, W.~N. Hardy, S. Kamal, N. Miyakawa, C. Panagopoulos, T. Xiang,
J.~F.  Zasadzinski, and X.~J. Zhou for useful discussions and for sharing
their experimental data with us, J.~R.  Cooper, S. Heim, A. Junod,
I.~O.  Kulik, J.~W.  Loram, and K.~A. Moler for useful communications.

Finally, my gratitude goes to my thesis committee members, Gene Mazenko,
Jeffery Harvey, and Thomas Rosenbaum, for their careful reading of this
manuscript and for their precious time spent on the committee meetings and
the thesis defense. I also wish to thank Woowon Kang for his time on my
first committee meeting.

\tableofcontents

\listoffigures
\addcontentsline{toc}{chapter}{List Of Figures}

\dedication

\mainmatter

\chapter{Introduction and Overview}
\label{Chap_Introduction}

\section{Background: High \textit{T}$_{\bf \lowercase{c}}$ \textit{problem}}

Since its discovery in 1986, high critical temperature ($T_c$)
superconductivity in the cuprates has been a great challenge for scientists.
While people celebrate the miracle that for the first time mankind can
achieve superconductivity at liquid nitrogen temperatures (77K) and thus
make superconductors industrially applicable, they find themselves left with
a puzzle: Why is the superconducting transition temperature so high?  How do
we describe superconductivity in these materials?  And what is the
underlying physics? Although the beautiful microscopic theory by Bardeen,
Cooper, and Schrieffer (BCS) \cite{Schrieffer} has been extremely successful
in explaining superconductivity in ordinary metals, scientists have yet to
find an answer to these questions, after more than a decade of research.

Many theories have been put forward to attack the high $T_c$ puzzle, yet
none has been very successful. Due to the lack of a better theory, for
experimentalists, BCS theory is still by far the most widely applied theory
to interpret experimental data, and to extract physical quantities. In this
work, we will show that BCS theory is a special case of a more general mean
field approach and that this generalization can accommodate quantitatively
various aspects of experimental observations.

\subsection{Overview of BCS theory}

As in all superconductors, electrons are paired in the superconducting
phase. This pairing arises from an (attractive) pairing interaction. In BCS
theory, pairing takes place only between electrons with opposite momenta
($\pm {\bf k}$), and is negligible otherwise.  This can be described by the
following Hamiltonian:
\begin{equation}
H^{BCS}  =  \sum_{{\bf k}\sigma} \epsilon^{\ }_{\bf k}
c^{\dag}_{{\bf k}\sigma} c^{\ }_{{\bf k}\sigma}
+ \sum_{\; \bf k k'} V_{\bf k, k'} 
c^{\dag}_{{\bf k}\uparrow} 
c^{\dag}_{-{\bf k}\downarrow} 
c^{\ }_{-{\bf k'}\downarrow} 
c^{\ }_{{\bf k'}\uparrow}\,,
\label{BCS_H}
\end{equation}
where $c^\dagger_{{\bf k}\sigma}$ creates a particle in the momentum state
${\bf k}$ with spin $\sigma $, and $\epsilon_{\mathbf{k}}$ is the energy
dispersion measured from the chemical potential $\mu$ (we take
$\hbar=k_B=1$). As usual, we assume a separable potential, $V_{\bf k,k'} = g
\varphi^{\ }_{\bf k}\varphi^{\ }_{\bf k'}$, where $g=-|g|$ is the
coupling strength; the momentum dependent function $\varphi^{\ 
  }_{\mathbf{k}}$ will determine the symmetry of the order parameter.  For
ordinary metal superconductors, the pairing interaction originates from the
electron-phonon interaction, and $\varphi^{\ }_{\bf k}=1$ for {\bf k} within
a narrow shell of the Fermi sphere, and zero otherwise. The thickness of the
shell is determined by the Debye frequency of the materials. Note in
Eq.~(\ref{BCS_H}) we have assumed singlet pairing, which is relevant for
both ($s$-wave) normal metal superconductors and the ($d$-wave) cuprate
superconductors.

In the normal state, the quantum expectation value of the pair operator $
c^{\ }_{-{\bf k}} c^{\ }_{{\bf k}}$ (and its complex conjugate) vanishes (we
have suppressed the spin index, following the usual practice). In the
superconducting state, where the electrons are paired into Cooper pairs,
this expectation value does not vanish: this defines the superconducting
order parameter,
\begin{equation}
\Delta_{\mathbf{k}} \equiv -\sum_{\bf k'} V_{\bf k,k'}  
\langle c^{\ }_{-{\bf k'}} c^{\ }_{{\bf k'}}\rangle = \Delta \phik^{\ } \:, 
\qquad
\Delta \equiv |g|\sum_{\bf k} \varphi^{\ }_{\mathbf{k}} 
\langle c^{\ }_{-{\bf k}} c^{\ }_{{\bf k}}\rangle \,.
\label{BCS_OP}
\end{equation}
One key assumption of BCS theory is that the difference between the quantum
operator $c^{\ }_{-{\bf k}} c^{\ }_{{\bf k}}$ and its mean field value,
$\langle c^{\ }_{-{\bf k}} c^{\ }_{{\bf k}}\rangle$ is small, so that the
quadratic term $(c^{\ }_{-{\bf k}} c^{\ }_{{\bf k}} -\langle c^{\ }_{-{\bf
    k}} c^{\ }_{{\bf k}} \rangle)^2$ is negligible. As a consequence of such
a mean field treatment, the Hamiltonian is linearized,
\begin{equation}
  H^{BCS} = \sum_{{\bf k}\sigma} \epsilon^{\ }_{\bf k} c^{\dag}_{{\bf k}\sigma}
  c^{\ }_{{\bf k}\sigma} - \sum_{\bf k} \left( \Delta \varphi^{\ }_{\mathbf{k}}
    c^{\dag}_{{\bf k}} c^{\dag}_{-{\bf k}} + \Delta^* \varphi^{\ }_{\mathbf{k}}
    c^{\ }_{-{\bf k}} c^{\ }_{{\bf k}}\right) -\frac{\Delta^2}{g\;} \,.
\label{BCS_H_Linear}
\end{equation}
This Hamiltonian is not diagonal, and the particles and holes will thus be
mixed via the equations of motion for $c^{\ }_{{\bf k}}$ and
$c^{\dag}_{-{\bf k}}$.  Eq.~(\ref{BCS_H_Linear}) can, however, be
diagonalized via a Bogoliubov transformation
\begin{equation} 
\alpha^{\ }_{\mathbf{k}}=\uk^{\ } c^{\ }_{\bf k} + v^{\ }_{-{\mathbf{k}}}
c^{\dag}_{-{\bf k}} \qquad \alpha_{\mathbf{k}}^{\dag} 
= \uk^{\ } c^{\dag}_{\bf k} + v^{\ }_{-{\mathbf{k}}} c^{\ }_{-{\bf k}} \;,
\label{Bogoliubov_Transform}
\end{equation}
provided 
\begin{equation}
\uk^2 = \frac{1}{2}\left( 1+\frac{\ek}{\Ek}\right), \qquad 
\vk^2 = \frac{1}{2}\left( 1-\frac{\ek}{\Ek}\right)\,,
\end{equation}
where $\Ek = \sqrt{\ek^2 + \Delta^2 \phik^2} $ is the excitation energy of
quasiparticles created by $\alpha^\dag_{\mathbf{k}}$. Note here and
throughout this thesis, we choose $\Delta = \Delta^*$ to be real, which is
always possible in the absence of a supercurrent.

At zero temperature, the ground state wavefunction can be expressed
as
\begin{equation}
\Psi_0 = \prod_{\bf k} (\uk^{\ } + \vk^{\ } c^{\dag}_{{\bf k}} c^{\dag}_{-{\bf
    k}}) |0\rangle \,.
\label{Wavefunction}
\end{equation}
At $T\le T_c$, the diagonalized Hamiltonian takes the form
\begin{equation}
  H^{BCS} = E_0+ \sum_{{\bf k}} \Ek (\alpha^\dag_{\bf k}\alpha^{\ }_{\bf
  k} +\alpha^\dag_{-{\bf k}}\alpha^{\ }_{-{\bf k}})\,,
\label{BCS_H_diagonal}
\end{equation}
where 
\begin{equation}
E_0 %= 2\sumk \ek \vk^2 -2\Delta\sumk \uk \vk -\frac{\Delta^2}{g\;} 
=\sumk (\ek-\Ek) -\frac{\Delta^2}{g\;}\,
\label{BCS_E0}
\end{equation}
is $T$-dependent. The system energy (per unit volume), measured from the
bottom of the band, is thus given by
\begin{equation}
E = \mu n + E_0 +2 \sumk \Ek f(\Ek)\,,
\label{BCS_E}
\end{equation}
where $f(x)$ is the Fermi distribution function, and $n$ is the electron
density.  Note here I consider the general case where the chemical potential
$\mu$ is not pinned at $E_F$. This becomes important as the attractive
interaction $g$ increases.

Now to create a quasiparticle excitation of the ground state $\Psi_0$, it
takes at least an energy $2\Delta$ to break a Cooper pair. Thus the
excitation spectrum is gapped, unlike that in a Fermi liquid.
The value of the excitation gap can be determined via Eq.~(\ref{BCS_OP}).
Substituting Eq.~(\ref{Bogoliubov_Transform}) into Eq.~(\ref{BCS_OP}), one
obtains
\begin{equation}
\Delta = -g \sum \phik^{\ } \uk^{\ } \vk^{\ } [1-2f(\Ek)] 
= -g \sumk \frac{\Delta}{2\Ek}
[1-2f(\Ek)]\phik^2 \,,
\label{BCS_Gap_Eq0}
\end{equation}
or 
\begin{equation}
1+g\sumk \frac{1-2f(\Ek)}{2\Ek} \phik^2 =0 \,.
\label{BCS_Gap_Eq}
\end{equation}

BCS theory predicts that the normalized excitation gap $\Delta(T)/\Delta(0)$
as a function of reduced temperature $T/T_c$ follows a universal curve.
Moreover, the ratio $2\Delta(0)/T_c = 3.52$ is also universal.

\begin{figure}
\centerline{\hskip 0.2in
\includegraphics[bb = 104 78 465 460, width=2.25in, clip]{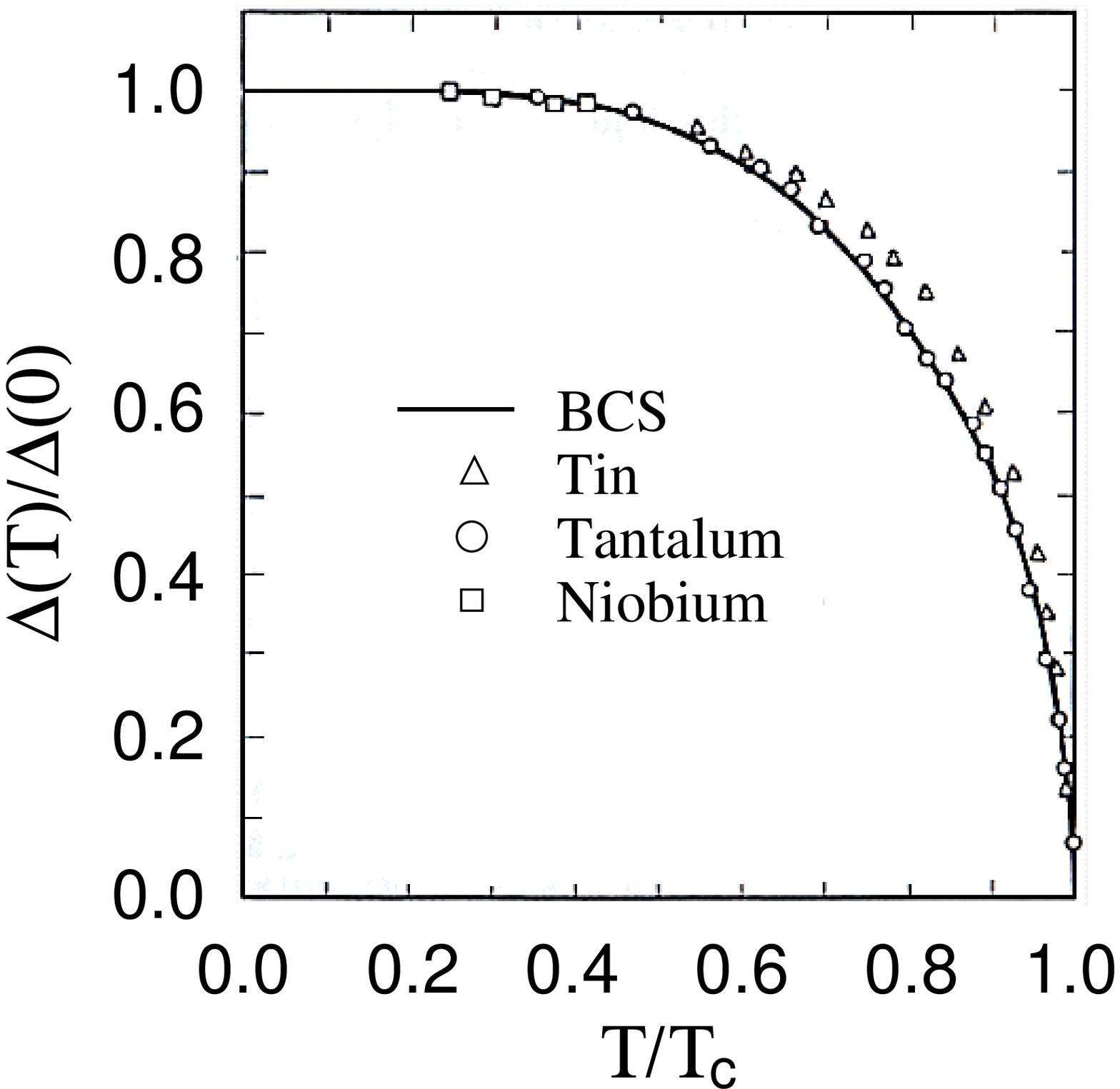}}
\vskip -1.6in
\hskip 1.47in \rotatebox{90}{\includegraphics{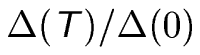}}
\vskip -1.52in
\hskip 1.73in {\footnotesize 1.0}
\vskip 0.225in \hskip 1.73in {\footnotesize 0.8}
\vskip 0.225in \hskip 1.73in {\footnotesize 0.6}
\vskip 0.225in \hskip 1.73in {\footnotesize 0.4}
\vskip 0.225in \hskip 1.73in {\footnotesize 0.2}
\vskip 0.225in \hskip 1.73in {\footnotesize 0.0}
\vskip -0.08in \hskip 1.9in {\footnotesize 0.0}
\hskip 0.225in {\footnotesize 0.2}
\hskip 0.225in {\footnotesize 0.4}
\hskip 0.225in {\footnotesize 0.6}
\hskip 0.225in {\footnotesize 0.8}
\hskip 0.225in {\footnotesize 1.0}
\vskip 0.00in \hskip 2.95in {\small \sffamily 
\textit{T/T}$_{\!\!\mbox{\scriptsize\textit{c}}}$}
\vskip -0.05in
\caption{Comparison between BCS prediction and experimental measurements for
  the excitation gap for ordinary metal superconductors.}
\label{BCS_Gaps}
\end{figure}

Figure \ref{BCS_Gaps} shows the BCS prediction and experimental measurements
for the excitation gap for a variety of metals. It is clear that the
agreement between theory and experiment is remarkable. 

\begin{figure}
  \centerline{\includegraphics[width=2in]{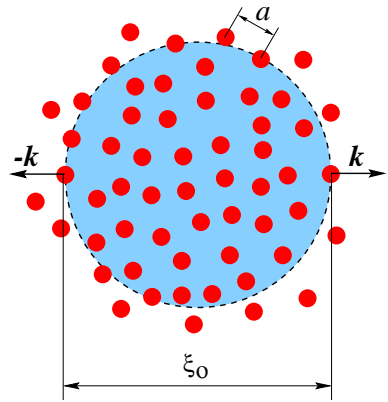}} 
\caption{In BCS theory Cooper pairs highly overlap with each other.}
\label{CooperPairs} 
\end{figure}

This figure also shows that the excitation gap closes at $T_c$. In fact,
\textit{one important feature of BCS theory is that the excitation gap and
  the (magnitude of the) order parameter are identical}. This is a
consequence of the linearization procedure in the mean field treatment of
the Hamiltonian. This procedure may not be valid when the difference $c^{\ 
  }_{-{\bf k}} c^{\ }_{{\bf k}} -\langle c^{\ }_{-{\bf k}} c^{\ }_{{\bf k}}
\rangle$ is not small. Indeed, what makes BCS theory work so well for
normal metal superconductors is that the coherence length $\xi$ (or,
equivalently, the Cooper pair size) for these materials is extremely large.
A typical value is of the order of $10^3 - 10^4$ \AA, which is about
100-1000 times the size of the lattice constant $a$.  In this way the pairs
highly overlap with each other, as schematically shown in
Fig.~\ref{CooperPairs}.  In this case, indeed, $c^{\ }_{-{\bf k}} c^{\ 
  }_{{\bf k}} -\langle c^{\ }_{-{\bf k}} c^{\ }_{{\bf k}} \rangle$ is very
small, and thus mean field theory is a very good approximation.

Physically, a large pair size usually means a weak pairing interaction. In
such a case, Cooper pairs form only at and below $T_c$, and, therefore, the
gap vanishes at $T_c$.

\subsection{Failure of BCS theory in high \textit{T}$_{\bf \lowercase{c}}$
  superconductors}

Unlike in ordinary metal superconductors, where the symmetry factor
associated with the order parameter, $\phik^{\ }$, is isotropic, in the
cuprate superconductors, $\phik^{\ }$ has been experimentally established to
have a $d_{x^2-y^2}$ symmetry; it changes sign across the diagonals of the
two dimensional (2D) Brillouin zone, and, consequently, the gap has four
nodes on the Fermi surface, and four maxima at the $(\pi, 0)$ points.

\begin{figure}
  \centerline{\includegraphics[width=2.5in]{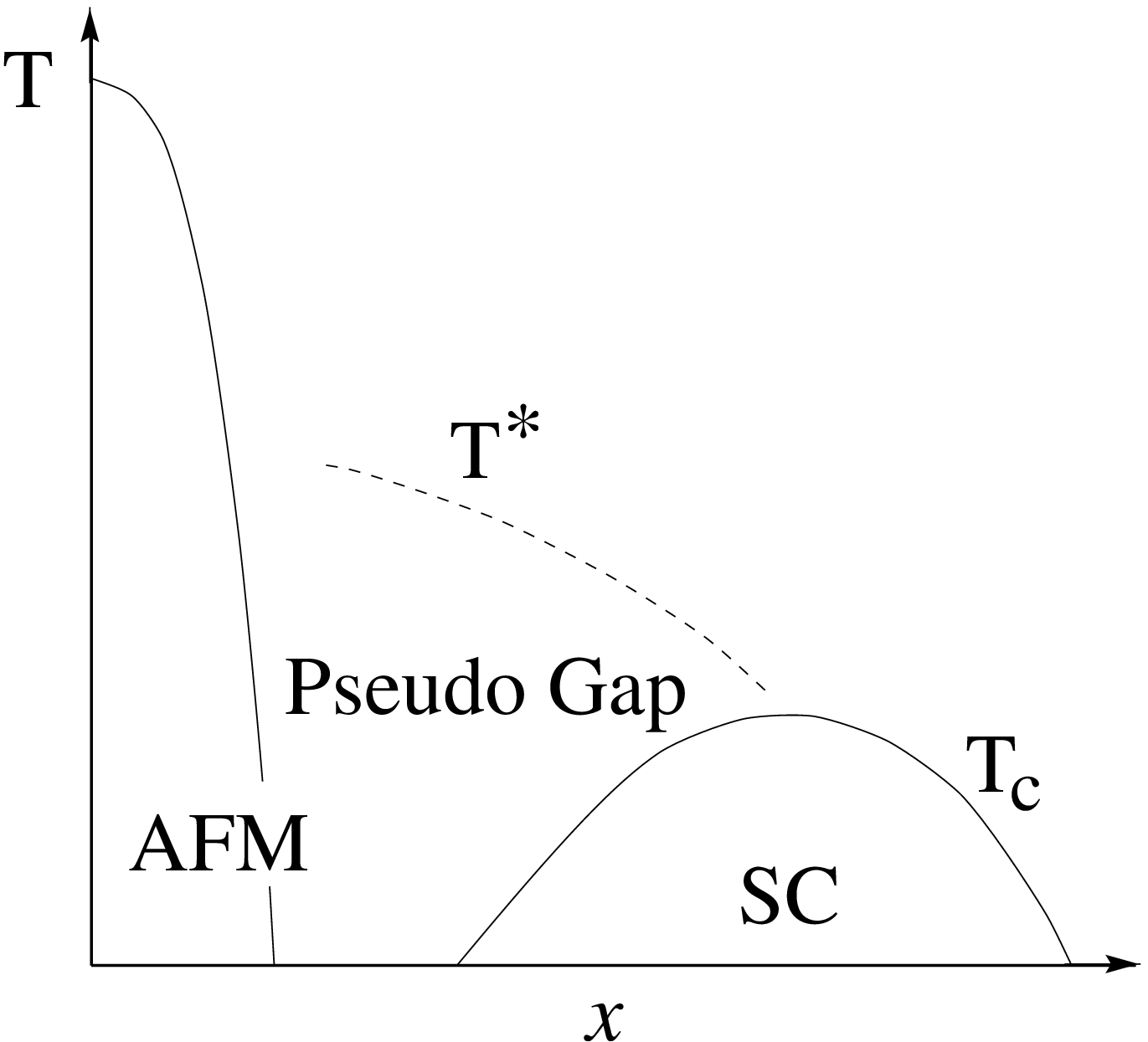}\hskip 1cm
 \includegraphics[width=3in]{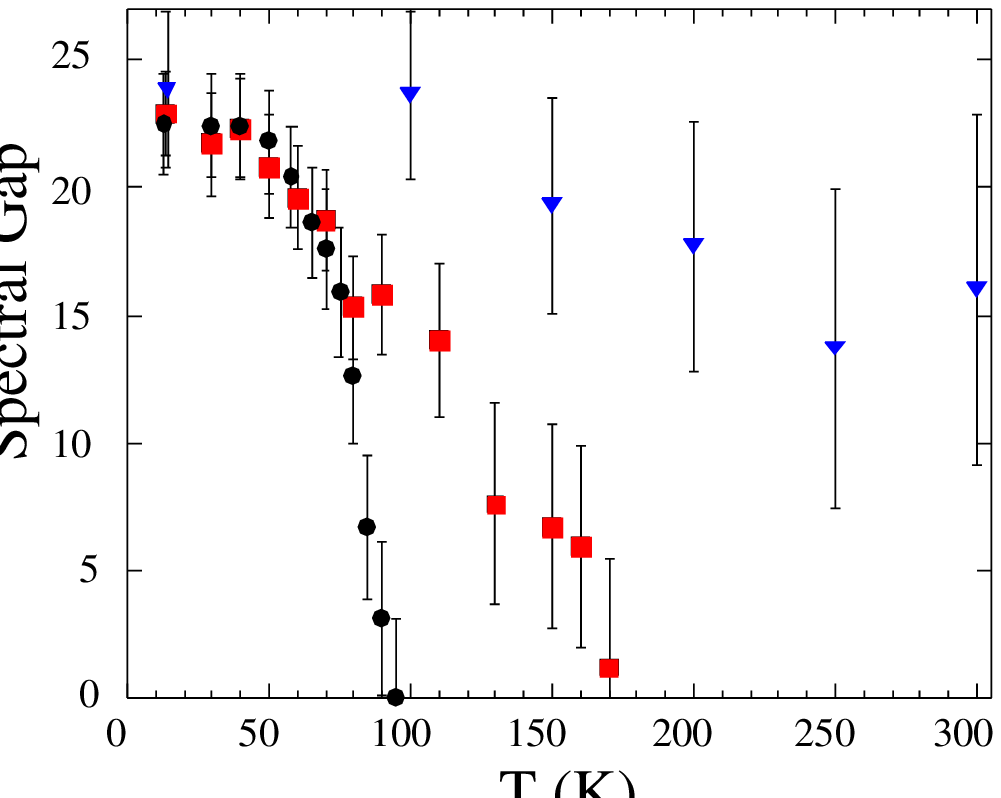}} 
\vskip -2.3in \hskip 1.5in \textbf{\large (a)} \hskip 3.7in \textbf{\large (b)}
\vskip 1.237in \hskip 1.65in $\bullet$
\vskip -0.175in \hskip 1.55in {\tiny $\blacksquare$}
\vskip 0.16in \hskip 1.15in {\scriptsize $\blacktriangledown$}
\vskip 0.3in
\caption[Schematic phase diagram for the cuprates and evidence from ARPES 
measurements for pseudogap in underdoped samples from ARPES.] {(a) Schematic
  phase diagram for the cuprate superconductors (The horizontal axis is the
  doping concentration), and (b) ARPES measurement of the temperature
  dependence of the excitation gap at ($\pi$, 0) in a near-optimal $T_c=87$~K
  sample ({\large $\bullet$}), underdoped 83~K ({\scriptsize
    $\blacksquare$}) and 10~K ($\blacktriangledown$) samples.  The units for
  the gap are meV. (b) is taken from Ref.~\cite{ARPES_ANL}.}
\label{PhaseDiagram_ARPES} 
\end{figure}

An important property of the cuprates, is that, although it works very well
for normal metal superconductors, BCS theory manifestly breaks down when
applied to high $T_c$ superconductors.  We can explore this breakdown by
exploring the schematic phase diagram for the cuprates, shown in
Fig.~\ref{PhaseDiagram_ARPES}(a).  In contrast to ordinary metal
superconductors, the parent compounds of these materials, with precisely one
electron per unit cell, are Mott insulators as well as antiferromagnets at
half filling.  The electronic motion is highly confined within the
copper-oxide planes, with very weak inter-plane coupling.  Therefore, they
are a highly anisotropic three dimensional (3D) or quasi-2D electronic
system with tetragonal or near-tetragonal symmetry.  As the system is doped
with holes, the insulator is converted into a metal as well as superconductor
(at low $T$).  As indicated by Fig.~\ref{PhaseDiagram_ARPES}(a), there exists
an optimal doping concentration, around 0.15 hole per unit cell, which gives
the highest $T_c$ value.  In the underdoped regime a new and important
feature has been observed; this is the unexpected excitation gap (called
pseudogap) above $T_c$. This gap persists up to a much higher crossover
temperature, $T^* > T_c$.

Shown in Fig.~\ref{PhaseDiagram_ARPES}(b) are the angle resolved
photoemission spectroscopy (ARPES) measurements of the excitation gap as a
function of temperature for $\mathrm{Bi_2Sr_2CaCu_2O_{8+\delta}}$ (Bi2212 or
BSCCO) with a variety of doping concentrations.%
\newfootnote{The measured gap values were not very accurate due to the
  limitation of the energy resolution (22-27 meV) of ARPES in 1996, as
  reflected by the large error bars in Fig.~\ref{PhaseDiagram_ARPES}(b). But
  this does not change the qualitative behavior of the excitation gap as a
  function of temperature and of doping concentration.}
Although the gap roughly closes at $T_c$ for the near-optimal sample, it
persists well above $T_c$ for the 83~K underdoped sample.  The more
underdoped sample has a much lower $T_c = 10$ K and a higher zero
temperature gap $\Delta(0)$, but the gap now persists to a much higher
temperature.  The gap evolves continuously across $T_c$. Moreover, it has
been confirmed that the pseudogap above $T_c$ follows the same symmetry
(i.e., $|d_{x^2-y^2}|$) as the gap does below $T_c$.
Figure~\ref{PhaseDiagram_ARPES}(b) should be contrasted with
Fig.~\ref{BCS_Gaps} for the BCS case.

Note that the superconducting order parameter (which we will refer to as
$\Delta_{sc}$ from now on) has to vanish above $T_c$. Thus, without
referring to other experimental quantities, one arrives at the conclusion
that BCS theory breaks down for the cuprates. This breakdown can be
characterized as follows:
\begin{itemize}
\itemsep 0in
\item The excitation gap and the order parameter are distinct, $\Delta \neq
  \Delta_{sc}$;
\item The ratio $2\Delta(0)/T_c$ increases with decreasing doping
  concentration $x$, no longer a universal value.
\end{itemize}

What causes BCS theory to fail?  Some of the most obvious reasons proposed
are that the cuprates all have
\begin{itemize}
\itemsep 0in
\item very short coherence lengths, $10\sim 15$ \AA, only a few times larger
  than the lattice constant; and
\item quasi-two dimensionality.
\end{itemize}
In the case of a short coherence length, the fluctuation $c^{\ }_{-{\bf k}}
c^{\ }_{{\bf k}} -\langle c^{\ }_{-{\bf k}} c^{\ }_{{\bf k}} \rangle$ is no
longer small, and, thus, a mean field theory such as BCS theory should not
work. The low dimensionality further enhances these fluctuations, as is
consistent with the Mermin-Wagner theorem.

\subsection{A successful theory for high \textit{T}$_{\bf c}$ 
  superconductors is yet to come}

Since the discovery of the cuprate superconductors, many theories have been
put forward. Shortly after their discovery, Anderson proposed \cite{RVB} 
a resonating valence bond (RVB) theory. Here the combination of proximity
to a Mott insulating phase and low dimensionality would cause the doped
material to exhibit new behavior, including superconductivity, not
explicable in terms of conventional metal physics. Anderson 
further proposed that as
the system is doped with holes, the antiferromagnetic order would be
destroyed by quantum fluctuations, and the resulting ``spin liquid'' would
contain resonating valence bonds which are electron pairs whose spins are
locked in a singlet configuration. Anderson argued that the valence bonds
resemble the Cooper pairs in BCS theory, and thus could lead to
superconductivity. Although this theory was able to explain qualitatively
the doping dependence of $T_c$ in the underdoped regime \cite{Anderson87},
it is still considered controversial.

Within a closely related RVB framework, the pseudogap phenomena and the
antiferromagnetism of the parent compounds have led other
people \cite{Spin-ChargeSeparation} to postulate a picture based on the
spin-charge separation idea from one dimensional Luttinger liquid physics.
In this picture, the strong antiferromagnetic correlation forces the spins
(spinons) to form singlets, and thus causes a gap in the electron excitation
spectrum above $T_c$.  In this way, the pseudogap is a spin gap. On the
other hand, the charge carriers are holes (holons). These holons form Cooper
pairs below $T_c$.  Thus the gaps below and above $T_c$ have different
origins.  This picture has successfully addressed the doping dependence of
$T_c$ in the underdoped regime \cite{LeeWen}, but it still may be
problematical in the overdoped regime. In addition, it does not seem to
explain why the gap evolves continuously across $T_c$, or why the
superconducting gap in the very underdoped regime is so large, where the
charge carrier concentration is very low. So far, there has been no
microscopic theory which establishes the validity of spin-charge separation
in a 2D system.

In a slightly different vein, Zhang \cite{SO5} proposed an SO(5) model, in
an attempt to unify the two-component superconducting and the three
component antiferromagnetic order parameters.  However, this theory has not
yet addressed the phase diagram of Fig.~\ref{PhaseDiagram_ARPES}.  Emery and
Kivelson proposed a model of a very different flavour \cite{EmeryKivelson}.
They argued that Cooper pairs form at temperature $T^*$, and this leads to
an excitation gap.  However, due to the small size of the phase stiffness,
superconductivity will not set in until phase coherence is established at a
lower temperature $T_c$. In this scenario, the superconductivity is
destroyed purely by phase fluctuations of the order parameter. Some concerns
about this approach are that it is associated with a very large critical
regime in underdoped materials, which does not seem to be consistent with
experiment. Moreover, there is concern that the phase mode will be pushed up
to the plasma frequency by the Coulomb interaction, and therefore become
less important.  Finally, this scenario so far has not been able to make
detailed predictions.

The above theories are distinct from BCS theory. While BCS theory fails in
the underdoped regime, it seems to work reasonably well in the overdoped
regime.  This leads naturally to the idea that one should generalize BCS
theory to accommodate the short coherence lengths of the cuprates rather
than abandon it all together.

The school of thought which we have pursued is the BCS -- Bose-Einstein
(BEC) crossover scenario. This is a natural mean field extension of BCS
which is based on the distinction between the excitation gap and the order
parameter.  In this scenario, pairs form above $T_c$, leading to the
pseudogap ($\Delta$), and then these pairs Bose condense at $T_c$, leading
to superconductivity with order parameter $\Delta_{sc}$.  This general
physical picture did not originate with us. But we have made the important
contribution of taking this approach and extending it below $T_c$ and,
thereby, establishing how the weak coupling BCS superconducting state is
altered as one crosses over towards the BEC limit, where the coupling or
attractive interaction is strong.

\section{Crossover from BCS to Bose-Einstein condensation}

\subsection{Relevance of Bose-Einstein condensation}

Our interest in BCS--BEC crossover theory is motivated by the similarities
between the pseudogap phenomena and BEC physics.  Indeed, Figure
\ref{PhaseDiagram_ARPES}(b) appears to be a very natural extension of
Fig.~\ref{BCS_Gaps}, which incorporates BEC behavior.  In a true Bose
system, as the temperature becomes lower than $T_{BEC}$, the zero momentum
state will become macroscopically occupied, as corresponds to superfluidity.
This phenomenon is well known as Bose-Einstein condensation (BEC).  For a
system of fermions which pair to form ``bosons", in the low density, strong
coupling limit, these fermions will form tightly bound pairs, which do not
overlap with each other, and therefore may be treated as point-like bosons.
These composite bosons will Bose condense at the condensation temperature,
$T_{BEC}$. In a charged system, the material will become superconducting, at
the BEC condensation temperature $T_{BEC}$, which is equivalent to the
superconducting transition temperature $T_c$.

\begin{figure}
\centerline{\large \sffamily BCS \textit{vs} Bose-Einstein Condensation}
\vskip 3eX
\centerline{
\mbox{\parbox{4in}{
\begin{tabular*}{4.in}{c@{\extracolsep\fill}c}
 BCS & BEC \\
&\\
 weak coupling (\protect$g\ll g_c$) & strong coupling
 ($g\gg g_c$) \\
&\\
 large pair size & small pair size \\
  {\bf k}-space pairing & {\bf r}-space pairing \\
&\\
 strongly overlapping  & ideal gas of \\
  Cooper pairs & preformed pairs \\
&\\
 \includegraphics[width=1in, clip]{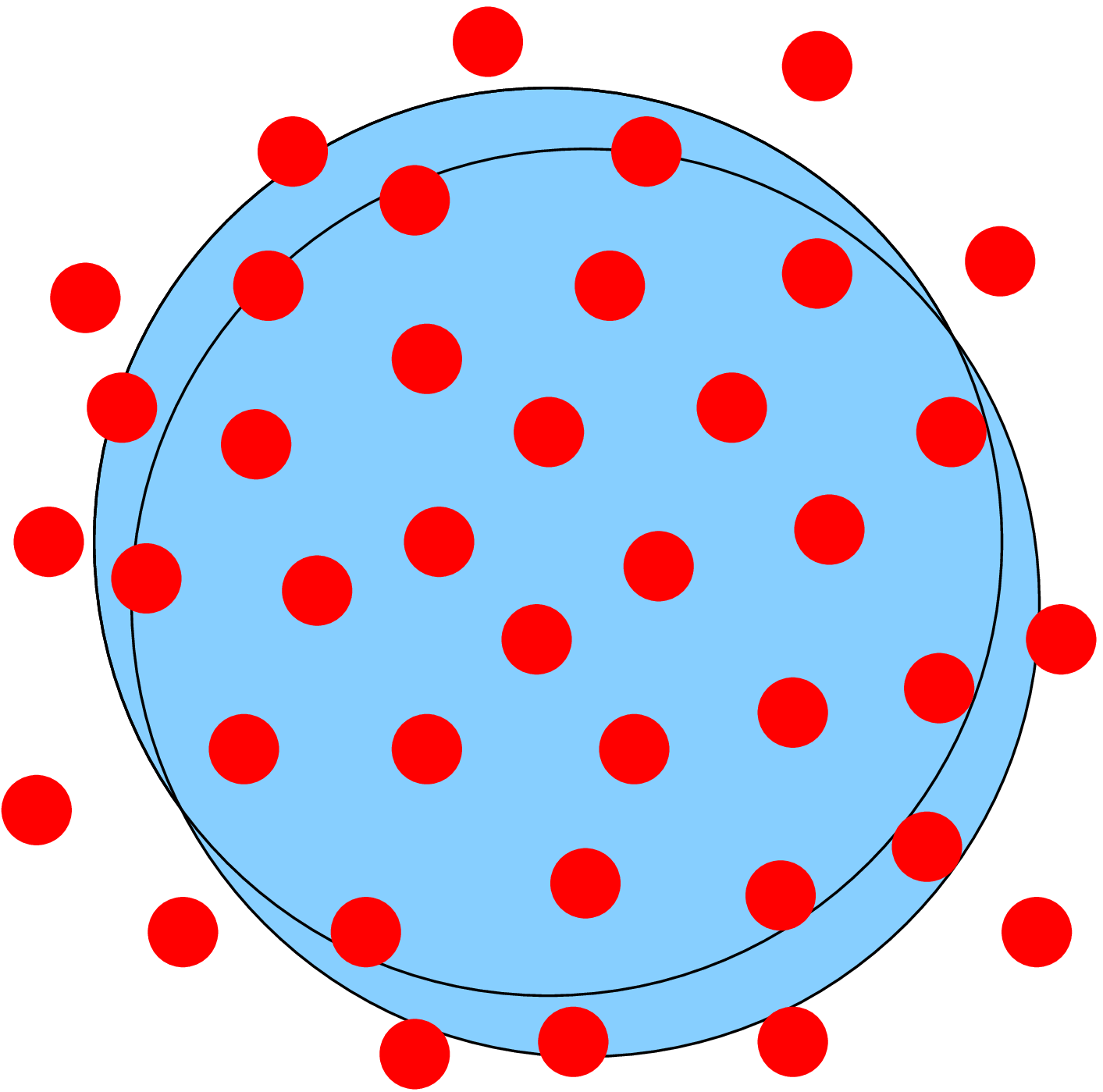} &
\includegraphics[width=1in, clip]{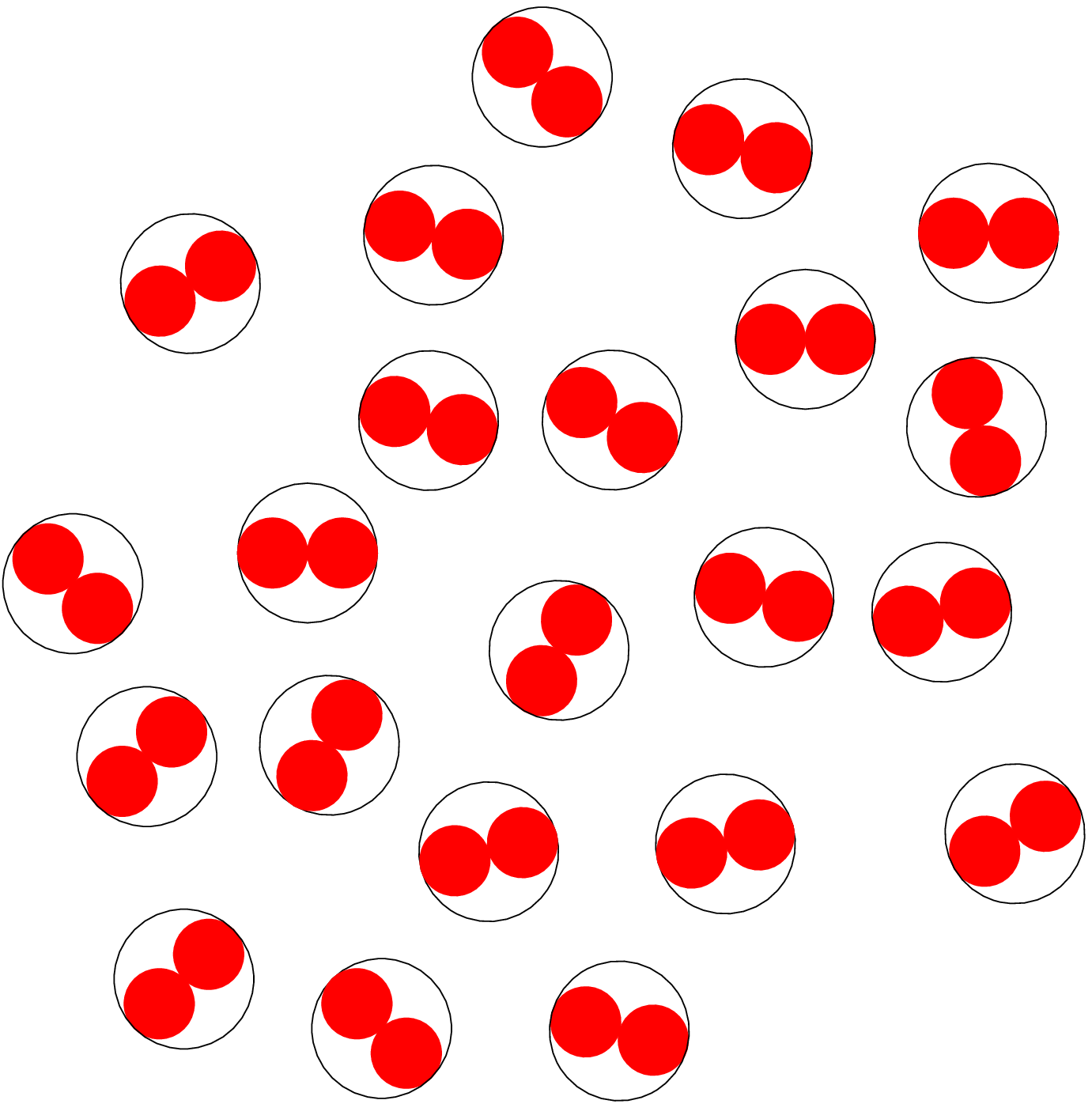} \\ 
&\\
 $T_{pair} = T_c$  & $T_{pair} \gg T_c$ \\ 
\end{tabular*}
}}}
\vskip 6eX
\centerline{Energy to create fermions $\equiv \Delta$}
\vskip 3eX
\centerline{\includegraphics[width=3.5in, clip]{Gaps_Crossover}} 
\vskip 2eX
\caption{Comparison between the weak coupling BCS and strong coupling 
  BEC limits.}
\label{BCS-BEC}
\end{figure}

More generally, superconductivity at weak coupling (associated with the BCS
limit) can be regarded as a special type of Bose-Einstein condensation. Here
the Cooper pairs can be thought of as giant bosons. The suggestion that the
superconducting transition is a Bose condensation of pairs of electrons into
localized bound states (called Schafroth condensation) is one originally
proposed by Schafroth, Blatt, and Butler \cite{Schafroth,Blatt}, before BCS
theory. (Owing to mathematical difficulties associated with what they call
the quasi-chemical equilibrium approach for evaluating the partition
function of the system, they could not carry out calculations.) Unlike in
the strong coupling limit, where the pair formation temperature, $T_{pair}$,
is much higher than the pair condensation temperature, $T_c$, in the BCS
case, \textit{pair formation and pair condensation take place at the same
  temperature}, $T_{pair}=T_c$.  In the weak coupling limit, the Cooper
pairs are highly overlapping, whereas in the strong coupling limit, the pair
size is small. Nevertheless, in both limits, the fermion pairs are bosons,
and the condensed pairs have net momentum zero.

A schematic comparison of the weak coupling BCS and the strong coupling BEC
limits is shown in Fig.~\ref{BCS-BEC}. It can be seen that there are
differences associated with the difference between the pair size (or,
equivalently, the coherence length).  Moreover, these effects lead to
differences associated with the the temperature dependence of the excitation
gap. In the BEC limit the fermionic excitation gap is essentially
temperature independent.

\begin{figure}
\centerline{\includegraphics[width=3.8in]{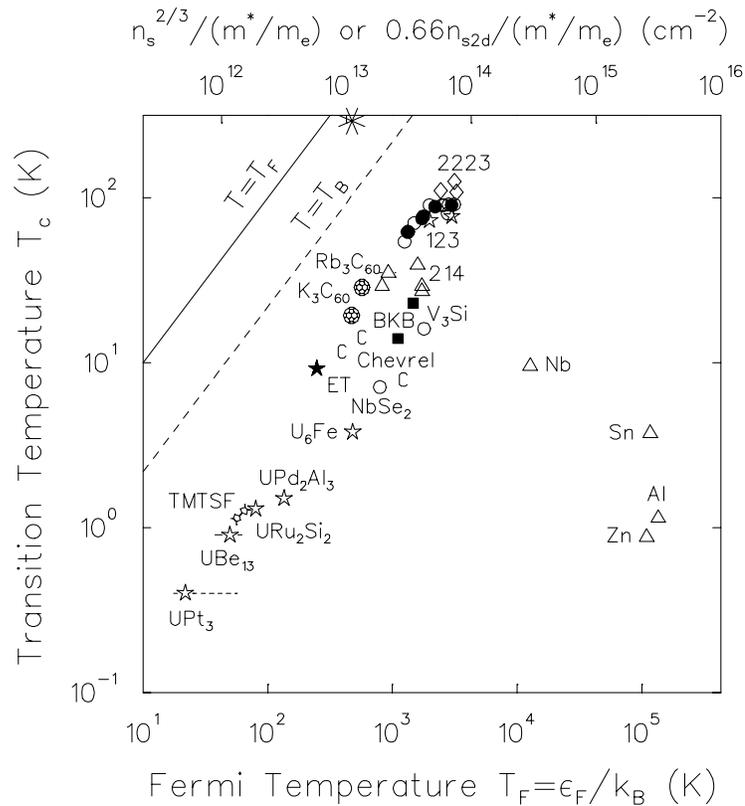}}
\caption[Uemura plot, showing a scaling between $T_c$ and the Fermi energy
or the superfluid density for short coherence length
superconductors.]{Uemura plot. Note the transition temperature scales with
  the Fermi energy or the superfluid density for short coherence length
  superconductors. In comparison, the normal metal superconductors do not
  follow this scaling presumably due to their large coherence length. Taken
  from Ref.~\cite{Uemura}.}
\label{UemuraPlot}
\end{figure}

If one can think of the BCS superconductivity in the weak coupling limit in
terms of BEC, it should be natural to do so for an intermediate coupling
strength. In general, since the pairs already form above $T_c$, one has to
break a pair to excite a single fermion, and therefore there exists a single
fermion excitation gap above $T_c$.  This corresponds in a natural way to
the pseudogap and related phenomena  in the underdoped cuprates shown in
Fig.~\ref{PhaseDiagram_ARPES}.  In addition, the pair size shrinks as the
coupling strength increases, as is consistent with the short coherence
length in the cuprates.  \textit{We may infer from pseudogap phenomena that
  the cuprates are intermediate between the weak coupling BCS and the strong
  coupling BEC limit.} Indeed, the ARPES data shown in
Fig.~\ref{PhaseDiagram_ARPES}(b) indicate that, as the doping concentration
decreases, the excitation gap becomes less sensitive to the temperature,
evolving toward the BEC limit shown in Fig.~\ref{BCS-BEC}.

Further suggestions for the relevance of BEC come from the famous Uemura
plot \cite{Uemura}, as shown in Fig.~\ref{UemuraPlot}. As indicated in this
plot, there exists a universal scaling between $T_c$ and superfluid
stiffness $n_s/m$ (where $n_s$ is the superfluid density) for a variety of
materials, including heavy fermion systems, the cuprate superconductors and
organic superconductors, etc.  That $T_c$ scales with $n_s/m$ seems to be in
line with the expectation from a Bose condensation approach.  Despite the
fact that these materials are very different from each other, they share one
common feature, i.e., short coherence lengths.%
\newfootnote{ In fact, the phase diagram of 2D organic superconductors in the
  $\kappa$-$\mathrm{(BEDT}$-$\mathrm{TTF)_2X}$ family is very similar to
  that of the cuprates, except that the role of doping concentration is
  replaced by pressure \cite{McKenzie}.}
 Note the logarithmic scale in
Fig.~\ref{UemuraPlot}. The ordinary metal superconductors are far off the
scaling behavior, as may be expected, since their extremely large coherence
lengths set them apart.

In summary, given the pseudogap phenomena shown in
Fig.~\ref{PhaseDiagram_ARPES} and the Uemura scaling between $T_c$ and
$n_s/m$ shown in Fig.~\ref{UemuraPlot}, we believe that the cuprates, as
well as other short coherence length superconductors, are intermediate
between the BCS and the BEC limits.

\subsection{Overview of BCS-BEC crossover}

To address the superconducting pairing for arbitrary coupling strength, one
needs to add to the BCS Hamiltonian, Eq.~\ref{BCS_H}, a term which describes
an attractive interaction between pairs of nonzero net momentum:
\begin{equation}
H  =  \sum_{{\bf k}\sigma} \epsilon^{\ }_{\bf k}
c^{\dag}_{{\bf k}\sigma} c^{\ }_{{\bf k}\sigma}
+\sum_{\bf k k' q} V_{\bf k, k'} 
c^{\dag}_{{\bf k}+{\bf q}/2} 
c^{\dag}_{-{\bf k}+{\bf q}/2} 
c^{\ }_{-{\bf k'}+{\bf q}/2} 
c^{\ }_{{\bf k'}+{\bf q}/2} \:.
\label{Hamiltonian}
\end{equation}
Again, here the spin indices in the interaction term have been suppressed.
In this way, we include pairing beyond that of the zero momentum
condensate.

Studies of this Hamiltonian and of the BCS--BEC crossover problem date back
to Eagles \cite{Eagles} early in 1969 when he studied pairing in
superconducting semiconductors.  However, this work was largely overlooked.
In 1980, Leggett \cite{Leggett,Leggett2} gave an explicit implementation of
this crossover at zero temperature, in a variational wavefunction approach.
This work set up a basis for all subsequent studies of the BCS--BEC
crossover theory.  Leggett found that as the coupling $g$ increases, the
system evolves \textit{continuously} from BCS to BEC.  A very important
assumption is that the system can be described by the same BCS form of
wavefunction, $\Psi_0$ (which we will refer to as Leggett ground state).
Moreover, applying a variational approach this ground state is found to be
associated with the same BCS form of gap equation [Eq.~(\ref{BCS_Gap_Eq}),
with $f(\Ek)=0$], provided that the chemical potential is also varied
self-consistently through a fermion number constraint:
\begin{equation}
2\sumk \vk^2 = n\:.
\label{BCS_Number_Eq}
\end{equation}
Leggett pointed out that, in the strong coupling limit, the gap equation
becomes a Schr\"o\-ding\-er equation for ``diatomic molecules'' with a
wavefunction $\psi_{\mathbf{k}}= \uk^* \vk^{\ }= \Delta_{\bf k}/2\Ek$, and
the system is essentially a condensed ideal Bose gas of these molecules.
Meanwhile, the chemical potential $\mu$ becomes a large negative number,
given by half the binding energy of the molecules.

In 1985, Nozi\`eres and Schmitt-Rink (NSR) \cite{NSR} extended this crossover
approach to a calculation for $T_c$, and also discussed for the first time
the effects of lattice band structure. In this pioneering work, NSR
introduced into the number equation the self-energy associated with the
particle-particle ladder diagram (i.e., the scattering $T$ matrix) at the
lowest order. They found that $T_c$, as a function of the coupling strength
$g$, first increases following the BCS calculation, then it starts to
deviate and then decreases after reaching a maximum. Finally it approaches
the Bose condensation temperature $T_{BEC}$ (from above) in the strong
coupling limit. This work suffers from a lack of self-consistency: (1) The
self-energy is not fed back into the $T$-matrix (This is the so-called
``$G_0G_0$ scheme"), as a consequence, (2) the self-energy is not included
in the gap equation.  The latter is derived from a divergence of the $T$
matrix at zero frequency following Thouless \cite{Thouless}. At this level
of theory, there is no excitation or pseudo-gap at $T_c$.

Many papers have been published on the BCS--BEC crossover problem since the
discovery of high $T_c$ superconductors.  Randeria, Duan, and Shieh
\cite{Randeria89,Randeria90} recognized that crossover physics may be
associated with high $T_c$ superconductivity, and studied the cross\-over
problem in 2D at $T=0$ using a variational pairing \textit{Ansatz}.  Their
approach was equivalent to that of NSR.  Also based on this NSR scheme,
Schmitt-Rink, Varma, and Ruckenstein \cite{SVR} found that the 2D Fermi gas
is unstable; bound fermion pairs form at an arbitrarily weak $s$-wave
attraction. Serene \cite{Serene} pointed out that this last result is an
artifact of the truncation of the particle-particle ladder series at the
lowest order as in the original NSR calculations. In an attempt to improve
NSR, Haussman \cite{Haussmann94,Haussmann93} calculated the crossover
behavior of $T_c$ as $g$ increases in a 3D continuous model, using a
conserving approximation for the \textit{T} matrix, in which all single
particle Green's functions are fully dressed by the self-energy. (This is
the so-called ``$GG$ scheme"). He found that $T_{BEC}$ is approached from
below with increasing $g$, as is expected physically.  Within the Haussmann
``GG" scheme, Engelbrecht and coworkers numerically studied the pseudogap
state for $d$-wave superconductors. However, due to technical difficulties,
they were restricted to a 2D lattice model, for which $T_c=0$.  Additional
previous work along these lines has focused on a continuous or jellium
model, at $T=0$ or $T_c$, or on a lattice in strictly 2D.

Jank\'o, Maly and coworkers \cite{Janko,Maly,Maly2} addressed the BCS-BEC
crossover problem at and above $T_c$ in a 3D continuous model, using a third
$T$ matrix approach, the so-called ``$G_0G$ scheme". This scheme is the
basis for the present thesis. These authors have shown that (i) as $T_c$ is
approached from above, the $T$ matrix acquires a sharp resonance. This
resonance finally becomes a divergence at $T_c$. The stronger the
interaction and the closer to $T_c$, the stronger this resonance. (ii) As
$g$ increases, the resonance becomes progressively more important. This
resonance causes a depletion of the fermion density of states at the Fermi
energy, and thereby leads to a pseudogap.

In order to apply the crossover physics to the cuprates and to study the
superconducting state, we will extend this $GG_0$ theoretical approach below
$T_c$, using a realistic anisotropic 3D lattice model.  Detailed
derivations in Chapter 2 will provide support for this theoretical scheme.
%, in the context of the specific ground state wavefunction $\Psi_0$.

\section{Current work --- Generalizing BCS theory to arbitrary coupling strength}

In this thesis, we extend BCS theory to arbitrary coupling strength, by
implementing BCS--BEC crossover physics below $T_c$.  This is the regime in
which (except at $T=0$) there has been virtually no previous work.  We go
beyond BCS by including finite center-of-mass momentum pair excitations
(which we call \textit{pairons}) in the self-energy, and treat the single
particle and two particle propagators on an equal footing. We, then solve
the resulting coupled equations self-consistently, for arbitrary temperature
$T\leq T_c$. Our fundamental equations are derived by truncating the
infinite series of equations of motion at the pair propagator level,
following early work by Kadanoff and Martin \cite{Kadanoff}.
Diagrammatically, these equations can be represented by a $T$ matrix
approximation in a $G_0G$ scheme, where each rung of the particle-particle
ladder contains one bare and one full Green's function.

The importance of the ``$G_0G$ scheme" is that it is consistent with the
Leggett ground state wavefunction for all coupling constants and, moreover
it leads to a gap equation which has the BCS form, given by
Eq.~(\ref{BCS_Gap_Eq}), for
%weak coupling and 
all temperatures below $T_c$. This equation must be supplemented by the
number equation, which is also of the BCS form.  One key result of the
current work is that the pseudogap exists for all non-zero $T$, due to the
existence of finite momentum pair excitations.  In other words, the
pseudogap is present both above and below $T_c$.  A new equation is found,
which must be solved in combination with the excitation gap equation
(\ref{BCS_Gap_Eq}) and the finite $T$ generalization of the fermion number
equation (\ref{BCS_Number_Eq}).  This equation relates the pseudogap
$\Delta_{pg}$ with the density of excited pair states:
\begin{equation}
a_0 ( \Delta ^2 - \Delta_{sc}^2 ) = a_0 \Delta_{pg}^2 = 
\sumq b(\Omegaq) \:. 
\label{Intro_PG_Eq}
\end{equation}
where $b(x)$ is the Bose distribution function, $a_0$ is a constant, and
$\Omegaq$ is the pairon dispersion.  We then have three coupled equations,
in place of the one of BCS theory, and of the two in the Leggett ground
state variational conditions.  In this way we determine $T_c$, as well as
$\Delta$, $\Delta_{sc}$, and $\mu$ for $T\leq T_c$. Note now that the
$\Delta$ in Eq.~(\ref{BCS_Gap_Eq}) is different from the order parameter
$\Delta_{sc}$.

While BCS theory contains only single fermion and zero momentum Cooper
pairs, the new ingredient here is the finite momentum pair excitations.  The
present $T$ matrix formalism corresponds to treating these pairs at a
mean-field level, without including explicitly pair-pair interactions. In
this sense, we have an improved mean field theory, at the non-interacting
pair level.

\begin{figure}
\centerline{\hskip -6mm  \includegraphics[width=4.5in]{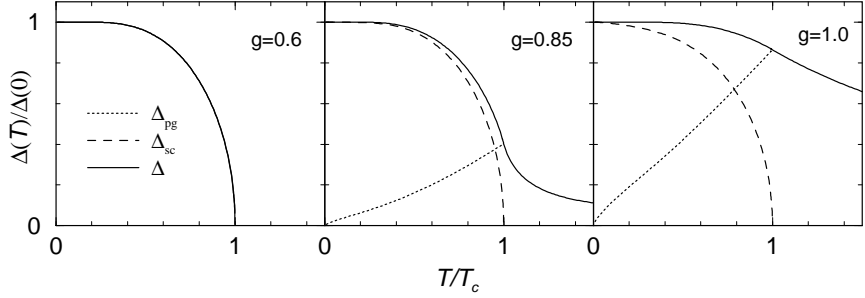}} 
\vskip -0.15in
\caption[Typical evolution of the temperature dependence of the excitation
gap, the order parameter, and the pseudogap with the coupling
strength.]{Typical evolution of the temperature dependence of the excitation
  gap (solid lines), the order parameter (dashed lines) and the pseudogap
  (dotted lines) with coupling strength. The pseudogap increases with $g$.
  The curves are calculated in 3D jellium, with the range of interaction
  $k_0=4k_F$.  Here $g$ is measured in units of $g_c$, the critical coupling
  for two fermions to form a bound state in vacuum.}
\label{Gaps_3D}
\end{figure}

In the weak coupling limit, finite momentum excitations are negligible, and
we recover BCS theory. The gap closes at $T_c$ in this case, and there is no
pseudogap. As the coupling increases, finite momentum excitations become
progressively more important, and the pseudogap develops. In the strong
coupling limit, the system behaves like an ideal Bose gas, and the new
equation (\ref{Intro_PG_Eq}) becomes the boson (i.e., pair) number equation,
and therefore controls the transition temperature $T_c$. The evolution of
the gaps with coupling strength is shown in Fig.~\ref{Gaps_3D}.
As the temperature decreases from $T_c$, the order parameter develops,
whereas the pseudogap decreases. At $T=0$, the pseudogap contribution
vanishes. As a consequence, we recover the Leggett ground state.

\begin{figure}
\centerline{\hskip 0.05in {\large \mbox{$g\ll g_c$}} \hskip 0.8in 
{\large \mbox{$g\sim g_c$}}
    \hskip 0.8in {\large \mbox{$g\gg g_c$}}}
\vskip 0.05in
\centerline{\includegraphics[bb = 0 90 510 247, clip, width=4in]{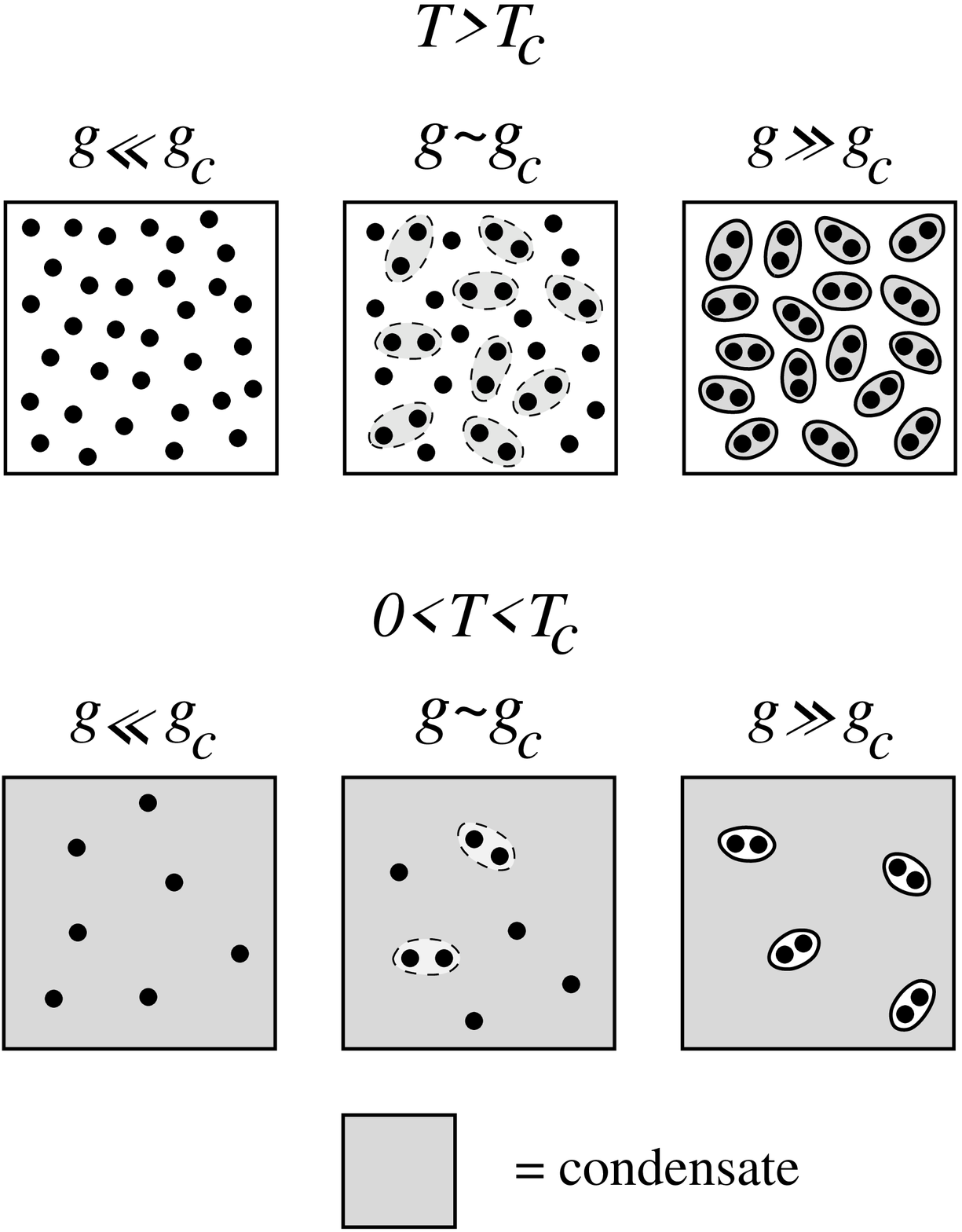}}
\vskip -0.1in
\caption[Evolution of the excitations of the system below $T_c$ as the
coupling strength increases.]{Evolution of the excitations of the system
  below $T_c$ as $g$ increases. The gray background denotes the condensate.}
\label{CrossoverCartoon}
\end{figure}

The physical picture below $T_c$ can be schematically summarized by
Fig.~\ref{CrossoverCartoon}. In the weak coupling limit, the system is
composed of the Cooper pair condensate, and fermionic quasiparticles. In the
strong coupling limit, all fermions form bound pairs, and the system
is composed of a Bose condensate of pairs, and finite momentum pair
excitations. In the intermediate regime, fermionic quasiparticles and
bosonic pair excitations will coexist, besides the condensate.
This is the regime that we think is applicable to the cuprates.

%\vskip 2ex
\clearpage

Some of our key results are as follows:
\begin{enumerate}
\renewcommand{\labelenumi}{(\roman{enumi})}
\itemsep 0in
\item
 At low $T$, pair excitations lead to new power laws in $T$, via the
single particle self-energy. For example, they lead to a new $T^{3/2}$ term in
the superfluid density; in a quasi-2D system, this new power law means a new
linear $T$ term in the specific heat.
\item When applied to $d$-wave superconductors on a lattice (as in the
  cuprates), we find that the superconducting BEC regime is not accessible.
  The cuprates are expected to be intermediate between the BCS and the BEC
  limit, but well within the fermionic regime.
\item We obtain a cuprate phase diagram in (semi-)quantitative agreement
  with experiment.
\item In addition, we predict a universal behavior of the normalized
  superfluid density $n_s(T)/n_s(0)$ as a function of reduced temperature
  $T/T_c$, and a linear $T$ term in the specific heat, in agreement with
  experiment.
\item Finally, we show that the onset of superconducting long range order
  leads to sharp features in the behavior of the specific heat at $T_c$,
  although the excitation gap is smooth across $T_c$.
\end{enumerate}

\vskip 2ex

This thesis is arranged as follows. Next, in Chapter \ref{Chap_Theory}, we
introduce our theoretical formalism, and derive the three equations which
are the basis for all numerical calculations. Then (Chapter \ref{Chap_Tc})
we address the superconducting instability when approached from a
pseudogapped normal state, and study the effects on the behavior of $T_c$ of
low dimensionality, discrete lattice structure, and $d$-wave symmetry of the
pairing interaction. In Chapter 4, we address the superconducting phase,
including the temperature dependence of various gaps, superfluid density,
specific heat, and low $T$ power laws. In Chapter \ref{Chap_CollMode}, we
will address the gauge invariance issue and study the collective modes below
$T_c$.  In the next chapter (Chapter \ref{Chap_Cuprates}), we will apply
this theory to the cuprates, and compare with experiment our predictions for
the phase diagram and a variety of physical quantities.  In Chapter
\ref{Chap_Thermodynamics}, we study thermodynamic signatures of the
transition in the pseudogap regime.  In this context we address the doping
dependence of the tunneling spectrum and the specific heat jump. Finally, in
Chapter \ref{Chap_Conclusion} we will make some concluding remarks, and
discuss some issues not yet addressed.

\chapter{Theoretical Formalism}
\label{Chap_Theory}

In this chapter we derive the three fundamental coupled equations which were
briefly noted in Chapter \ref{Chap_Introduction}. These equations represent
a natural extension of BCS theory, and can be given a simple physical
interpretation.  We begin by briefly describing our physical picture. We
then give a summary of the results of the formalism. Next we go into more of
the details and derive a coupled set of Dyson's equations for the single and
two particle propagators of the system. These coupled equations lead to a
complete set of equations for the excitation gap (or transition
temperature), chemical potential, and the pseudogap, which will then be
deduced from these equations in the subsequent chapters.

\section{Physical picture at arbitrary coupling strength}
\label{Sec_PhysPicture}

BCS theory describes a system of fermions with very weak pairing
interactions. The electrons form pairs only at and below $T_c$. Therefore,
in the normal state, the system is composed of electrons exclusively. Such a
BCS system has a very large coherence length. However, as the coupling, or
attractive interaction increases, the pair size becomes smaller, and
metastable or bound pairs start to form even above $T_c$. There will be an
intermediate temperature regime, in which the single fermions and fermion
pairs coexist. As the temperature is lowered, a condensate of zero momentum
(Cooper) pairs starts to develop at $T_c$. This $T_c$ can be regarded as the
superconducting transition temperature from the fermionic perspective; or
equivalently, it can be treated as a Bose condensation temperature of the
pairs.  When the coupling becomes strong enough, all electrons form small,
tightly bound pairs. In this case, the transition has a completely BEC
nature. Nevertheless, as will be shown in Sec.~\ref{Subsec_Instability} and
Sec.~\ref{Subsec_BelowTc}, the superconducting instability and the BEC
condition are equivalent, as they should be, so that there is a unique and
well defined $T_c$.

\begin{figure}
\centerline{\hskip 0.05in {\large \mbox{$g\ll g_c$}} \hskip 0.65in 
{\large \mbox{$g\sim g_c$}}
    \hskip 0.65in {\large \mbox{$g\gg g_c$}}}
\vskip 0.05in
\centerline{\includegraphics[bb = 0 0 510 150, width=3.5in, clip]{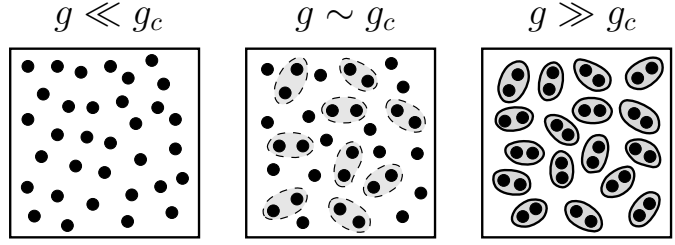}}
\vskip -0.1in
\caption[Schematic picture for BCS--BEC crossover: Evolution of the
excitations in the system with coupling $g$ above $T_c$.]{Schematic picture
  for BCS--BEC crossover: Evolution of the excitations in the system with
  coupling $g$ above $T_c$. Pairs and single particles coexist in the
  intermediate coupling regime.}
\label{CrossoverAboveTc}
\end{figure}

This BCS-BEC crossover is schematically shown in
Fig.~\ref{CrossoverAboveTc}. The short coherence length combined with the
positive (fermionic) chemical potential, suggests that the high $T_c$
superconductors correspond to intermediate coupling strength. Here, we want
to develop a theoretical formalism, to describe what happens across the
transition temperature below $T_c$, with emphasis on the physics of
intermediate coupling.  Based on what we know about the cuprates, we want
our theory to satisfy the following constraints:
\begin{enumerate}
\renewcommand{\labelenumi}{(\roman{enumi})}
\setlength{\itemsep}{-0.3em}
\setlength{\parsep}{0em}
\item It leads to the BCS result in the weak coupling regime;
\item It leads to the ground state given by Leggett \cite{Leggett,Leggett2};
\item In strictly 2D, there is no superconductivity at finite $T$.
\end{enumerate}
We will build our formalism around BCS, and obtain a natural generalization
of BCS through an extended mean field theory.

\subsection*{Overview of the formalism}

In this subsection, we review BCS theory, in terms of a somewhat unfamiliar
but very useful formalism, introduced by Kadanoff and Martin
\cite{Kadanoff}.  This approach is then generalized to the crossover
problem, following the lead of Ref.~\cite{Kadanoff}. For brevity, we
suppress the symmetry factor $\varphi_{\mathbf{k}}$ temporarily, and use a
four vector notation: $K \equiv ({\bf k}, i\omega_n)$, $Q = ({\bf q},
i\Omega_n)$, $\sum_K = T \sum_{{\bf k}, \omega_n}$, etc., where $\omega_n$
and $\Omega_n$ are odd and even Matsubara frequencies, respectively.

BCS theory involves the pair susceptibility ${\displaystyle \chi (Q) =
  \sum_K G (K) G_0 (Q-K)} $, where $G$ is the dressed Green's function
$G^{-1}(K) = G_0 ^{-1}(K) + \Sigma(K) $, which depends on the usual BCS self
energy $ \Sigma(K) = \Sigma_{sc}(K) = -\Delta_{sc}^2 G_0(-K)$.  In this way
the gap equation can be written as
\addtocounter{equation}{-1}
\begin{equation} 
%\refstepcounter{equation}
1 + g \chi (0) = 0,\qquad (T \le T_c)\:.
  \label{eq:gap0} 
\end{equation}  
This equation is equivalent to saying that the particle-particle scattering
matrix (or pair propagator) diverges in the static $Q\rightarrow 0$ limit,
signaling a stable, ordered state in the pairing channel.
At $Q  = 0$, the summand in $\chi$ is the Gor'kov
``$F$" function (up to a multiplicative factor $\Delta_{sc}$) and
\textit{this serves to highlight the central role played in BCS theory by
  the more general quantity} $G(K)G_0(Q-K)$.
Note that (for $ Q \ne 0$), $\chi (Q)$ is \textit{distinct from the pair
  susceptibility of the collective phase mode} which (as we will 
show in Chapter \ref{Chap_CollMode}) enters as
\[
Q_{22}(Q)=- \sum_K \big\{ G (K )[ G (Q-K) + G ( -Q -K)] +2F (K) F
(K-Q)\big\}\:.
\] 
Here each Gor'kov ``$F$'' function introduces effectively one $GG_0$, so
that \textit{the collective mode propagator depends on higher order Green's
  functions than does the gap equation}.  The observations in italics were
first made in Ref.~\cite{Kadanoff} where it was noted that the gap
equation of BCS theory could be rederived by truncating the equations of
motion so that only the one ($G$) and two particle ($t$) propagators
appeared.

The coupled equations \cite{Kadanoff} for $G$ and $t$ can be used to recast
BCS theory in this new language and, more importantly, this formalism can be
readily adapted to include a generalization to the (BCS-like) ground state
at moderate and strong coupling.  $G$ depends on $\Sigma$ which in turn
depends on $t$; we write,
\[
\Sigma (K) = \sum_Q t (Q) G_0(Q-K) \:.
\]
Now, in general, $t$ has two additive contributions (as does the associated
self energy $\Sigma$): from the condensate $t_{sc}(Q) = -\Delta_{sc}^2
\delta (Q)/T$, and from the non-condensed $Q \ne 0$ pairs, with $t_{pg}(Q) =
g/(1 + g \chi (Q))$. Upon expanding $t^{-1}_{pg}(Q)$, it can be easily seen
that the gap equation (\ref{eq:gap0}) implies that effective pair chemical
potential vanishes at $T\leq T_c$, which is the BEC condition. This provides
a reinterpretation of BCS theory, along the lines of ideal Bose gas
condensation.  In the leading order mean field theory $\Sigma = \Sigma_{sc}$
and the BCS gap equation is obtained, as noted above.

More generally, at larger $g$, the above equations hold but we now
include feedback from the finite momentum pairs into Eq.~(\ref{eq:gap0}), via
\[
\Sigma_{pg} (K) = \sum_Q t_{pg}(Q) G_0 (Q- K) \approx G_0 (-K)
\left(\sum_Q t_{pg} (Q)\right) . 
\]
$\sum_Q t_{pg} (Q)\equiv - \Delta^2_{pg}$ defines a new energy scale, called
the ``pseudogap".  Note that under this approximation, $\Sigma_{pg}$ is of
the BCS form, as is the total $\Sigma$.  In this way, one obtains a BCS-like
gap equation (for the fermionic excitation gap), Eq.~(\ref{BCS_Gap_Eq}),
with the order parameter $\Delta_{sc}$ replaced by the total excitation gap
$\Delta=\sqrt{\Delta^2_{sc}+\Delta^2_{pg}}$.  Together with the fermion
number constraint and the pseudogap definition, we have a complete set of
equations. In what follows, we will give a more rigorous derivation of these
equations.

\section{General formalism -- Derivation of the Dyson's Equations}

In 1961, Kadanoff and Martin (KM) \cite{Kadanoff} developed a Green's
function formalism for reformulating BCS theory. This formalism is very
similar to the one first used by Thouless \cite{Thouless} in studying the
superconducting instability. In both KM and Thouless's formalism, the
self-energy which is relevant to superconductivity is assumed to arise from
the particle-particle scattering $T$ matrix. Such an approximation is
usually referred to as a $T$ matrix approximation. The KM formalism was
further extended by Patton \cite{Patton} to study pairing fluctuations near
$T_c$ in a dirty, low dimensional superconductor.
In this context, it should be noted that the $T$ matrix of the present
theory is somewhat analogous to the ``fluctuation propagator" in the early
literature on fluctuation effects in low dimensional superconductors.  While
conventionally this propagator was evaluated at the ``$G_0G_0$ level",
Patton introduced a ``$G_0G$" level approximation, in order to avoid
divergences.
Here we will follow the KM approach, but extend their work slightly to feed
back the effects of finite momentum pair states into the gap equations.
This KM scheme is essentially the ``$G_0G$ approximation", referred to in
the introduction.

\subsection{Truncation of equations of motion}
\label{Sec_Eq_Motion}

We adopt the same notation as KM, and start with a Hamiltonian in real
space with an arbitrary singlet pairing:
\begin{eqnarray}
  H &=& \sum_\alpha \int\!\! d^3 {\mathbf{x}}\; \psi^\dag_\alpha
  ({\mathbf{x}}) \hat{T}({\mathbf{x}}) \psi^{\ }_\alpha ({\mathbf{x}})
  \nonumber\\ 
&&{} + \frac{1}{2} \sum_{\alpha \beta } \int\!\! d^3 {\mathbf{x}} d^3
  {\mathbf{x}}^\prime \; \psi^\dag_\alpha ({\mathbf{x}}) \psi^\dag_\beta
  ({\mathbf{x}}^\prime) V({\mathbf{x}},{\mathbf{x}}^\prime)_{ \alpha, \beta}
  \psi^{\ }_{\beta} ({\mathbf{x}}^\prime) \psi^{\ }_{\alpha}
  ({\mathbf{x}}) \:,
\label{Hamiltonian_RealSpace}
\end{eqnarray}
where $\hat{T}=-\vec{\nabla}^2/2m$ is the kinetic energy operator. The spin
indices are denoted by Greek letters, and we have restricted ourselves
to interactions which do not flip the spin. A singlet pairing will impose
$\alpha = - \beta$. Keeping this in mind, we will neglect the spin indices
in $V({\mathbf{x}},{\mathbf{x}}^\prime)_{ \alpha, \beta}$, for simplicity.
Under this restriction, Eq.~\ref{Hamiltonian_RealSpace} will be equivalent
to Hamiltonian Eq.~\ref{Hamiltonian} in momentum space.

For simplicity, we use notation $ \equiv ({\bf x}, t)$, etc., and use a barred
variable to denote a dummy integration variable. In addition, we define 
\[
V(1-1') \equiv V({\bf x}, {\bf x}^\prime) \delta (t-t^\prime)\:,
\]
\[
\psi(1) = e^{iHt}\psi({\bf x}) e^{-iHt} \:,
\]
and the operator
\[
G^{-1}_0(1) = i \frac{\partial}{\partial t_1} -\hat{T}(1)\:.
\]
The Green's functions for our spin conserving interactions are given by
the standard definition:
\begin{eqnarray}
G(1-1') &=& G^\alpha(1;1')= (-i)\langle T_t \psi_\alpha(1) 
\psi^\dag_\alpha(1')\rangle \:,\nonumber\\
G_2^{\alpha\beta}(12;1'2') &=& (-i)^2\langle T_t\psi_\alpha(1) \psi_\beta(2)
\psi^\dag_\beta(2') \psi^\dag_\alpha(1') \rangle \:,\nonumber\\
G_3^{\alpha\beta\gamma}(123;1'2'3') &=& (-i)^3\langle T_t\psi_\alpha(1) \psi_\beta(2)
\psi_\gamma(3) \psi^\dag_\gamma(3')
\psi^\dag_\beta(2') \psi^\dag_\alpha(1') \rangle \:,
%G_n^{\alpha_1\alpha_2,\ldots,\alpha_n} (12,\ldots,n;1'2',\ldots n')
% =  (-i)^n \langle T_t \psi_{\alpha_1}(1) \psi_{\alpha_2}(2) \ldots
%  \psi_{\alpha_n}(n)  \psi^\dag_{\alpha_n}(n') \ldots
%  \psi^\dag_{\alpha_2}(2')\psi^\dag_{\alpha_1}(1') \rangle \:.
\end{eqnarray}
where $T_t$ is the time ordering operator.

The equation of motion for $\psi$ is given by 
\begin{equation}
i \frac{\partial}{\partial t_1} \psi_\alpha (1) = \hat{T} (1) \psi_\alpha
(1) + \sum_\beta V(1-\bar 1) \psi^\dag_\beta (\bar 1)
\psi_\beta(\bar 1) \psi_\alpha(1)
\nonumber
\end{equation}
and we have the equation of motion for the Green's function $G(1-1')$:
\begin{equation}
G_0^{-1}(1) G(1-1') = \delta(1-1') -i  V(1-\bar 1)
G^{+-}_2(1\bar 1; 1' \bar 1^+) \:,
\label{G_Eq1}
\end{equation}
where $1^+ = \lim_{t'\rightarrow t^+}$. Note here we have imposed singlet
pairing. Stripping the uncorrelated term  $GG$ from $G_2$ 
\[
G^{\alpha\beta}_2 (12;1'2') = G(1-1')G(2-2') + L^{\alpha\beta} (12;1'2') \:, 
\]
and absorbing the Hartree term into $G_0^{-1}$
\begin{equation}
\tilde G_0^{-1} (1) = G_0^{-1}(1) + iV(1-\bar 1) G(\bar 1-\bar 1^+) \:,
\label{G_tilde}
\end{equation}
we can rewrite Eq.~(\ref{G_Eq1}) as
\begin{equation}
\tilde G_0^{-1} (1) G(1-1') = \delta(1-1') -i  V(1-\bar 1)
L^{+-}(1\bar 1; 1' \bar 1^+) \:.
\label{G_Eq2}
\end{equation}
We then have
\begin{equation} 
G(1-1') = \tilde G_0 (1-1') - i \tilde G_0(1-\bar 1) V(\bar 1-\bar 2)
L^{+-} (\bar 1 \bar 2: 1' \bar 2^+) \:.
\label{G_Eq3}
\end{equation}

Similarly, we obtain the equation of motion for $G_2$:
\begin{eqnarray}
G_0^{-1} (1) G_2^{\alpha\beta}(12;1'2') & =& G(2-2')\delta(1-1') -
G(2-1') \delta(1-2')\delta_{\alpha\beta} \nonumber\\
&&{} - i \sum_\gamma V(1-\bar 1)
G_3^{\alpha \beta \gamma} (12\bar 1; 1'2'\bar 1^+) \;,
\end{eqnarray}
where $G_3$ can be decomposed into terms like $GGG$ and $GG_2$, which
correspond to disconnected diagrams, and an irreducible part $L_3$, which
corresponds to connected diagrams:
\begin{eqnarray}
G_3^{\alpha\beta,-\alpha}(12\bar 1; 1'2'\bar 1^+) &=& G(2-2') G_2^{+-} (1\bar
1); 1'\bar 1^+) -\delta_{\alpha\beta}G(2-1')G_2^{+-}(1\bar1;2'\bar 1^+)
\nonumber\\
&&{}  -\delta_{\alpha, -\beta} G(2-\bar 1^+) G_2^{+-}(1\bar 1; 1'2') + G(\bar
1-\bar 1^+) L^{\alpha\beta}(12;1'2') \nonumber\\
&& \hskip -10mm {} +\delta_{\alpha\beta} G(\bar 1-\bar 1^+) G(1-2')G(2-1') -
\delta_{\alpha\beta} G(1-2') L^{+-}(2\bar 1; 1' \bar 1^+) \nonumber\\
&& \hskip -10mm {} -\delta_{\alpha,
  -\beta} G(\bar 1 - 2') L^{+-} (12; 1'\bar 1^+) + G(1-1') L^{-\alpha\beta}
(\bar 1 2; \bar 1^+ 2') \nonumber\\
&& \hskip -10mm {} + \delta_{\alpha, -\beta} G(1-1')G(2-\bar 1^+) G(\bar 1
-2') + L_3^{\alpha\beta,-\alpha}(12\bar 1; 1'2'\bar 1^+) \;.
\end{eqnarray}

\begin{figure}
\hskip 1.5cm \includegraphics[width=5.4in]{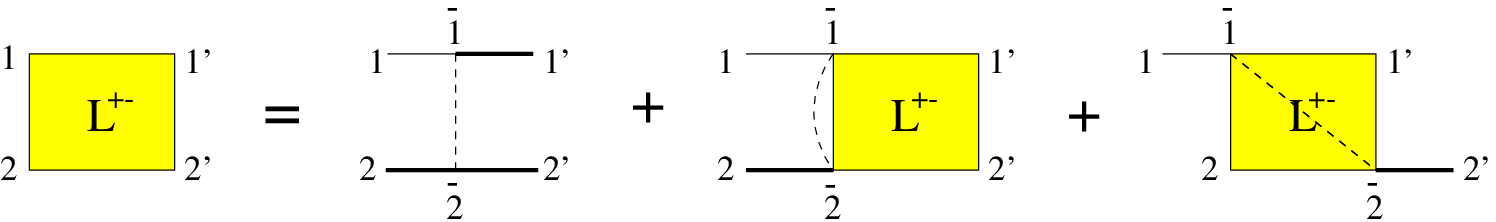}
\vskip 6mm
\hskip 1.75cm \includegraphics[width=5in]{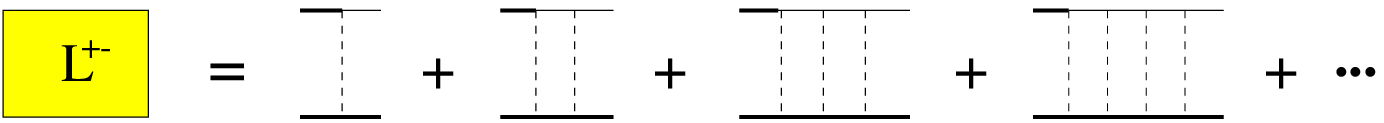}
\vskip -3cm \textbf{(a)} 
\vskip 1.5cm \textbf{(b)}
\vskip 6mm
\caption[Diagrammatic representation of Eqs.~(\ref{L+-_Eq1}) and
(\ref{L+-_Eq2}) for the two particle scattering matrix
$L^{+-}$.]{Diagrammatic representation of (a) Eq.~(\ref{L+-_Eq1}) and (b)
  Eq.~(\ref{L+-_Eq2}).  (b) is generated by reiterating the first two
  diagrams in (a). The thin, thick and dashed lines denote the bare Green's
  function $\tilde G_0$, the full Green's function $G$, and the interaction
  $V$, respectively.}
\label{L+-}
\end{figure}

In principle, one can continue to derive an equation of motion for $G_3$,
and express $G_3$ in terms of $G_4$.  Unfortunately, this is technically
impractical. Here following KM, we truncate the infinite series of equations
of motion at the level of $G_2$, and drop the connected $L_3$ diagrams in
$G_3$.
In this way, we obtain
\begin{eqnarray}
G_3^{+--}(12\bar 1; 1'2'\bar 1^+) &\approx & G(2-2')L^{+-}(1\bar 1; 1'\bar 1^+)
+ G(1-1')G(2-2')G(\bar 1 - \bar 1^+) \nonumber\\
&&{} - G(2-\bar 1^+) G_2^{+-} (1\bar 1; 1'2') + G(\bar 1-\bar
1^+)L^{+-}(12;1'2') \nonumber\\
&&{} - G(\bar 1-2') L^{+-}(12;1'\bar 1^+) \:.
\end{eqnarray}
Note here we have used $L^{--}(12;1'2')=-G(1-2')G(2-1')$ to get rid of
$L^{--}$. Then we have
\begin{eqnarray} 
 G_0^{-1} (1) G_2^{+-}(12;1'2') &=& G(2-2')\delta(1-1') - i
 V(1-\bar 1) G_3^{+--} (12\bar 1; 1'2'\bar 1^+)  \nonumber\\
 &=& G(2-2')\delta(1-1') -i V(1-\bar 1)\left[ G(2-2') L^{+-}(1\bar 1; 1' \bar
  1^+) \right. \nonumber\\
&&{} + G(1-1')G(2-2')G(\bar 1-\bar 1^+) - G(2-\bar 1^+) G_2^{+-}(1\bar 1;
1'2') \nonumber\\
&& \left. {} + G(\bar 1-\bar 1^+) L^{+-}(12;1'2') - G(\bar 1-2') L^{+-}
  (12;1'\bar 1^+) \right] . \nonumber\\
\end{eqnarray}
Using definition Eq.~(\ref{G_tilde}), this can be simplified as
\begin{equation}
\tilde G_0^{-1}(1) L^{+-} (12;1'2') = iV(1-\bar 1) \left[ G(2-\bar 1^+)
  G_2^{+-} (1\bar 1; 1'2') + G(\bar 1 -2')L^{+-}(12;1'\bar 1^+)\right] .
\end{equation}
Then we obtain an integral equation for $L^{+-}$:
\begin{equation}
L^{+-}(12;1'2') = i\tilde G_0(1-\bar 1) V(\bar 1-\bar 2) \left[ G(2-\bar 2)
  G_2^{+-}(\bar 1\bar 2; 1'2') + G(\bar 2-2') L^{+-} (\bar 1 2; 1'\bar
  2^+)\right] .
\label{L+-_Eq1}
\end{equation}
This equation can be represented diagrammatically by Fig.~\ref{L+-}(a). The
first term will produce an infinite series of ladder diagrams. The second
term on the right hand side stands for a non-ladder diagram, in which 
interaction lines cross each other. We will neglect this term, and keep only
ladder diagrams. Finally, we have
\begin{equation}
L^{+-}(12;1'2') = i\tilde G_0(1-\bar 1) V(\bar 1-\bar 2) G(2-\bar 2)
  G_2^{+-}(\bar 1\bar 2; 1'2') \:.
\label{L+-_Eq2}
\end{equation}
Reiteration of this equation will produce the ladder diagram shown in
Fig.~\ref{L+-}(b).

\subsection{Dyson's equations}

In momentum space, these integral equations become algebraic equations. And
one can obtain $L^{+-}$ in terms of $G$ immediately. We define the pair
susceptibility 
\begin{equation}
\chi = -i V \tilde G_0G \:. 
%\qquad \chi^{GG} = -i V GG .
\end{equation}
When each rung of the ladder is not entangled with its neighbors, we obtain
\begin{equation}
L^{+-} = i \frac{GG V \tilde G_0G}{1+\chi} \:.
\label{L+-_Eq3}
\end{equation}
Note this is not always possible for a generic interaction $V$. The
combination $V\tilde G_0G$ in the numerator is not written as $\chi$ because
here $G_0$ and $G$ are external legs.  Now Eq.~(\ref{L+-_Eq3}) and
Eq.~(\ref{G_Eq3}) form a closed set of equations.

The superconducting instability will be given by the divergence of $L^{+-}$,
which signals the formation of stable pairs. Without changing the essential
physics, we make one simplification, by replacing the $GG$ pair in the
numerator with $\tilde G_0 G$. This non-essential approximation
is equivalent to replacing one of the
two full Green's functions in the leftmost rung of the particle-particle
ladders in Fig.~\ref{L+-}(b) with ``bare'' Green's function $\tilde G_0$.

Substituting Eq.~(\ref{L+-_Eq3}) into Eq.~(\ref{G_Eq3}), we have 
\begin{equation}
G = \tilde G_0 + i\tilde G_0 \frac{\chi V}{1+\chi} \tilde G_0 G \:.
\end{equation}
Therefore, we obtain the irreducible self-energy
\begin{equation}
\Sigma = \tilde G^{-1}_0 - G^{-1} = i \frac{\chi V}{1+\chi}\tilde G_0 \:.
\label{SelfEnergy1}
\end{equation}  

\begin{figure}
\centerline{\includegraphics[width=5in]{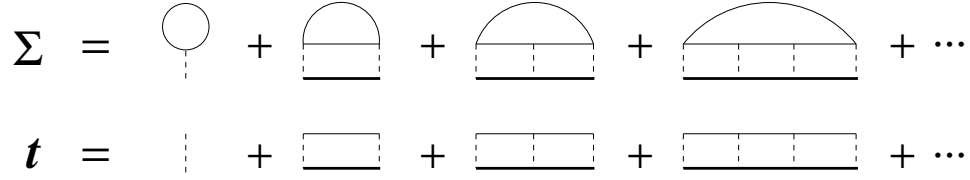}}
\caption{Diagrams for the self-energy and the $T$ matrix.}
\label{Sigma-T}
\end{figure}

To simplify the calculation, we make one more approximation on the
Hartree self-energy. Instead of absorbing the Hartree term in the new
``bare'' Green's function $\tilde G_0$, we will put it back into the
self-energy, Eq.~(\ref{SelfEnergy1}), at the lowest order, namely, $-iVG_0$.
Meanwhile, we will replace $\tilde G_0$ with $G_0$. In most situations, the
Hartree self-energy merely causes a chemical potential shift, or at most, a
slight mass renormalization. Moreover, the Hartree self-energy barely
changes between the normal state and the superconducting state. Therefore,
we do not expect any serious effects of this approximation. With this
approximation, we have
\begin{equation}
\Sigma = G_0^{-1}-G^{-1} = \frac{-iVG_0}{1+\chi} \equiv t G_0 \:,
\qquad
 t = \frac{-iV}{1+\chi} \:.
\end{equation}
The diagrams for the self-energy $\Sigma$ and the $T$ matrix $t$ are given
in Fig.~\ref{Sigma-T}.

We will, from now on, switch to the finite temperature formalism and work in
momentum space. Noting the difference in the Feynman rules and the
definition of the Green's functions between the real time formalism and
finite temperature formalism, the switch can be achieved simply by
multiplying in the above equations each $G$ with $i$, $\Sigma$ with $-i$, 
$G_2$, $L$ or $t$ matrix with -1, and $V$ with $-i$. Then we have
\begin{equation} 
\Sigma = t G_0\:, \qquad t = \frac{V}{1+\chi}\:, 
\qquad \mbox{with} \qquad \chi = VG_0 G \:.  
\end{equation}

Alternatively, we can directly write down the equations based on the
diagrams in Fig.~\ref{Sigma-T}, following proper Feynman rules. For a
separable potential $V_{{\bf k}, {\bf k}^\prime} = g \phik \phik^\prime$, we
can pull $g$ out of $\chi$, so that 
\begin{equation}
\chi(Q) = \sum_K G_0(Q-K) G(K) \phikq^2 \:,
\label{chi_Eq}
\end{equation}
where $K$ and $Q-K$ are the momenta of each single particle line of the
ladder, and $Q$ is the total momentum. Meanwhile, we notice $t(K,Q-K,Q)=
t(Q)\phikq^2$, so that the dependence of the external fermion momentum $K$
(via $\phik$) can be pulled out, and use $t(Q)$ as our new definition for
$T$-matrix. In this way, the new $T$-matrix depends only on the total
momentum $Q$.

Finally, we obtain the Dyson's equations for one and two particle
propagators:
\begin{equation}
\Sigma(K) = G_0^{-1} (K) - G^{-1} (K) = \sum_Q t(Q) G_0 (Q-K) \phikq^2 \:, 
\label{Sigma_Eq}
\end{equation}
and
\begin{equation}
t(Q) = \frac{g}{1+g\chi(Q)} \:.
\label{t_Eq}
\end{equation}
For an arbitrary coupling,  we need to supplement with the fermion number
conservation:
\begin{equation}
n= 2\sum_K G(K) \:.
\label{n_Eq}
\end{equation}
%
%For brevity, here and in what follows, we use a four-vector notation, $K
%\equiv ({\bf k}, i\omega_n)$, $Q = ({\bf q}, i\Omega_n)$, $\sum_K = T
%\sum_{{\bf k}, \omega_n}$, etc., where $\omega_n$ and $\Omega_n$ are odd and
%even Matsubara frequencies, respectively.
%
An essential component of this thesis is to study the solution of
Eqs.~(\ref{Sigma_Eq}-\ref{n_Eq}) for arbitrary coupling strength at $T
\leq T_c$.

Some remarks are in order, before we move on to solve  Dyson's equations.
Note in the particle-particle ladders in Fig.~\ref{Sigma-T}, we have one
single particle Green's function fully dressed while the other is bare. This
is the $G_0G$ scheme of the $T$ matrix approximation. It is different from
the fluctuation exchange (FLEX) scheme \cite{FLEX}, in which both single
particle lines are fully dressed. (This is the $GG$ scheme.) The $GG$ scheme
can be obtained from a thermodynamical dynamical potential $\Omega$ and a
functional $\Phi$ via functional derivatives (see, e.g., Ref.~\cite{Serene}
by Serene).  This can be contrasted with the approximation used by NSR
\cite{NSR}, in which both single particle lines in the ladder are bare, and
hence we refer to this as the $G_0G_0$ scheme.  This is a perturbative
approach, and thus fully dressed Green's functions do not appear in any
Feynman diagram.

While the perturbative $G_0G_0$ scheme has no self-energy feedback in the
$T$ matrix, and thus may not be valid in strong coupling situations, the
symmetric $GG$ scheme is probably overemphasizing feedback effects.
Furthermore, as KM (see Ref.~\cite{Kadanoff}, footnote 13) pointed out, the
resulting Green's functions ``can be rejected in favor of those used in BCS
by means of a variational principle. They can also be rejected experimentally
since they give rise to a $T^2$ specific heat. Finally, the formal
cancellation between the terms in the perturbation series resulting from the
symmetric equation ... can be indicated''.

Our own work focuses on the $G_0G$ scheme, which has a natural derivation via
equations of motion. Due to the $G_0^{-1}$ operator in the equations of
motion, the mix of $G_0$ and $G$ seems to an inescapable consequence.  The
drawback of this approach is that because of this $G_0$ $G$ mix, a
truncation of the series of equations of motion will lead to a theory which
is not $\Phi$ derivable. One cannot write down a simple thermodynamical
potential $\Omega$ and a functional $\Phi$.

\section{Solution to Dyson's equations}
\label{Sec_Dyson_Eq}

\subsection{Superconducting instability condition}
\label{Subsec_Instability}

The superconducting instability is signaled by the divergence of the $T$
matrix at zero momentum and zero frequency. This is usually referred to as
the Thouless criterion \cite{Thouless}. As the $T$ matrix diverges, stable
pairs of zero net momentum can exist, and thus the original Fermi liquid
become unstable. This condition can be expressed as
\begin{equation}
t^{-1}(Q=0) = g^{-1} + \chi(0, T_c) = 0 \:,
\label{Thouless_Criterion1}
\end{equation}
or
\begin{equation}
1+g\sum_K  G_0(-K) G(K)\phik^2 =0 \:.
\label{Gap_Eq}
\end{equation}

In the low frequency, long wavelength limit, one can expand the inverse $T$
matrix into powers of momentum $q$, and frequency $\Omega$:
\begin{equation} 
t^{-1}({\mathbf{q}}, \Omega) = a_1\Omega^2 + a_0 \Omega -\xi^2 q^2
+ \tau_0^\prime +i \Gamma_{{\mathbf{q}}, \Omega}^\prime \;.  
\label{InvT_Expansion}
\end{equation}
Note here the frequency is real in contrast to the discrete Matsubara
frequencies. We have put all imaginary contributions inside
$\Gamma_{{\mathbf{q}}, \Omega}^\prime$.
As shown by Maly \textit{et al.} \cite{Maly2}, this term is very small close
to $T_c$, and vanishes at $Q=0$ and at $T_c$. Thus the Thouless condition
requires $\tau_0^\prime =0$. This is then equivalent to the BEC condition
that the effective chemical potential $\mu_{pair}$ for the pairs vanishes.
Therefore, \textit{the Thouless criterion and the BEC condition are just two
  sides of the same coin.}

\subsection{\textit{T} matrix formalism of BCS theory: Relationship
to the Nambu Gor'kov Green's function scheme}
\label{Subsec_BCS_Tmatrix}

The Nambu-Gor'kov formulation is not suitable for extension to
arbitrary coupling. (In part this is due to the fact that there
are multiple energy gap parameters to be addressed).  Nevertheless,
in the BCS limit, where there is only one energy gap, we can
re-interpret these anomalous Green's functions or
``$F$" functions in terms of the present formalism.
Starting from the linearized BCS Hamiltonian,
Eq.~(\ref{BCS_H_Linear}), the equations of motion for the Green's
function $G$ and the anomalous Green's function $F$ take the following form
in the finite temperature formalism (see Ref.~\cite{Fetter}):
\begin{eqnarray}
G_0^{-1}(1) G(1-1') &=& \delta(1-1') - \Delta(1-\bar 1) F^\dag(\bar 1 -1') \:,
\label{BCS_G_Eq1}
\\
\bar G_0^{-1}(1) F^\dag (1-1') &=& \Delta^*(1-\bar 1) G(\bar 1 -1') \:,
\label{BCS_F_Eq1}
\end{eqnarray}
where the anomalous Green's function is defined as
\begin{equation}
F(1-2) = -\langle T_\tau
\psi_\uparrow(1) \psi_\downarrow(2) \rangle \:,
\label{BCS_F_Def}
\end{equation}
and the order parameter is given by
\begin{equation}
\Delta(1-2) = -V(1-2)F(1-2) \:.
\label{BCS_Delta_Def}
\end{equation}
Note in Eq.~(\ref{BCS_F_Eq1}) $\bar G_0^{-1}(1)$ denotes $G_0^{-1}(1)$ but
with the sign of $\frac{\partial}{\partial \tau}$ and $\vec\nabla$ changed,
satisfying $\bar G_0^{-1}(1) G_0 (1'-1) = \delta(1-1')$, i.e., $\bar G_0
(1-1')= G_0 (1'-1)$.

\begin{figure}
\centerline{\includegraphics[width=4.5in]{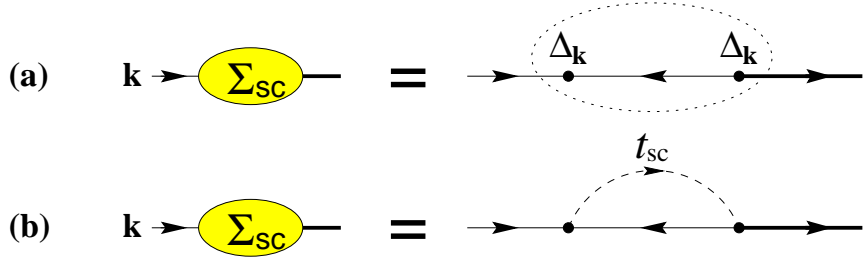}}
\caption[Feynman diagrams for BCS self-energy and its $T$ matrix 
representation.]{Feynman diagrams for (a) BCS self-energy and (b) its $T$
  matrix representation. The dashed line in (b) denotes the singular $T$
  matrix, $t_{sc}$.}
\label{BCS_Sigma}
\end{figure}

From Eq.~(\ref{BCS_F_Eq1}), we have 
\begin{equation}
F^\dag (1-1') = \Delta^*(\bar 1-\bar 2) G_0(\bar 1- 1) G(\bar 2-1') \:.
\label{BCS_F_Eq2}
\end{equation}
Substituting this equation into Eq.~(\ref{BCS_G_Eq1}), we obtain
\begin{equation}
G_0^{-1}(1) G(1-1') = \delta(1-1') - \Delta(1-\bar 1) \Delta^*(\bar 3-\bar
2) G_0(\bar 3- \bar 1) G(\bar 2-1') \:,
\label{BCS_G_Eq2}
\end{equation}
or
\begin{equation} 
G(1-1') = G_0(1-1') - G_0(1-\bar 1) \left[ \Delta(\bar 1 -\bar 4)
  \Delta^*(\bar 3 -\bar 2) G_0(\bar 3-\bar 4) \right] G(\bar 2-1') \:.
\label{BCS_G_Eq3}
\end{equation}
Now we can read off the self-energy $\Sigma$ from the part inside
``[ $\ldots$ ]'':
\begin{equation}
\Sigma(1-2) = -\Delta(1-\bar 1)  G_0(\bar 2-\bar 1) \Delta(\bar 2-2) \:.
\label{BCS_Sigma_Eq1}
\end{equation}
This equation can be represented diagrammatically by
Fig.~\ref{BCS_Sigma}(a). Note that in this diagram, we have contracted the two
points 1 and 2 of $\Delta(1-2)$.  The vertices $\Delta_{\bf k}$ can be easily
deduced from the linearized Hamiltonian, Eq.~(\ref{BCS_H_Linear}).  As the
diagram indicates, the BCS self-energy arises from a process in which one
particle is converted into a hole, and then is converted back into a
particle, due to the presence of the condensate. Equivalently, this
corresponds to the process in which two electrons form a Cooper pair, and
disappear into the condensate, and then this Cooper pair breaks again.

In momentum space, Eq.~(\ref{BCS_Sigma_Eq1}) can be written as
\begin{equation}
\Sigma(K)=-\Delta_{\bf k}^2 G_0(-K) = \frac{\Delta^2\phik^2}
{i\omega + \epsilon_{-\mathbf{k}}} \:,
\label{BCS_Sigma_Eq2}
\end{equation}
where $\Delta_{\bf k} = \Delta\phik$.  
And Eq.~(\ref{BCS_F_Eq2}) becomes
\begin{equation}
F^\dag (K) = \Delta\phik G_0(-K) G(K) \:.
\label{BCS_F_Eq3}
\end{equation}
Meanwhile, the gap equation is given by the self-consistency condition
Eq.~(\ref{BCS_Delta_Def}):
\begin{equation}
\Delta_{\bf k} = -\sum_{K^\prime} V_{{\bf k}, {\bf k}^\prime} F(K') \:,
\label{BCS_Gap_Eq1}
\end{equation}
or
\begin{equation}
1+g\sum_K  G_0(-K) G(K)\phik^2 =0 \:,
\label{BCS_Gap_Eq2}
\end{equation}
which is equivalent to Eq.~(\ref{BCS_Gap_Eq0}). It should be emphasized that
this equation has exactly the same form as Eq.~(\ref{Gap_Eq}).

\textit{The BCS self-energy Eq.~(\ref{BCS_Sigma_Eq2}) can be expressed by
Eq.~(\ref{Sigma_Eq}) in the $T$ matrix formalism}, provided 
\begin{equation}
t(Q) = -\frac{\Delta^2}{T\;} \delta(Q) \:.
\label{BCS_t_Eq}
\end{equation}
This amounts to splitting $L^{+-}$ in Sec.~\ref{Sec_Eq_Motion} into two
anomalous Green's functions:
\begin{equation}
L^{+-}(12;1'2') = F(1-2) F^\dag (1'-2') \:.
\label{L=FF_Eq}
\end{equation}
Indeed, this is what KM did with the $T$ matrix, in order to obtain 
the BCS limit.

In the $T$ matrix formalism, the BCS self-energy can be represented by the
Feynman diagram shown in Fig.~\ref{BCS_Sigma}(b). That $t(Q)$ is
singular is a consequence of the fact that the 
distribution of the ${\mathbf{q}}=0$ Cooper
pairs does not obey ordinary Bose statistics; the zero momentum pair state
is macroscopically occupied. --- \textit{This is Bose-Einstein
  condensation!}

\subsection{Beyond BCS: Effects of a pseudogap at finite $T\leq T_c$}
\label{Subsec_BelowTc}

At finite temperature below $T_c$, the self-energy comes from two
contributions, one from the superconducting condensate (i.e., ${\bf q}=0$
Cooper pairs), and the other from the finite {\bf q} pair excitations
(pairons). For the former contribution, a particle decays into a virtual
hole and a ${\bf q}=0$ Cooper pair and then recombines; for the latter, a
particle decays into a virtual hole and a ${\bf q}\neq 0$ incoherent pair,
and then recombines.  Corresponding to these two processes, it is natural to
decompose the self energy into two additive contributions, one from the
condensate and the other from the finite momentum pairs.  In this way, we
can write the $T$ matrix as
\begin{equation}
t(Q)= t_{sc}(Q) + t_{pg}(Q) \:,
\label{t_Eq1}
\end{equation}
where
\begin{equation}
t_{sc}(Q) = -\frac{\Delta_{sc}^2}{T\;\;} \delta(Q) 
\label{tsc_Eq}
\end{equation}
 and 
\begin{equation}
t_{pg}(Q) = \frac{g}{1+g\chi (Q)} \;,
\label{tpg_Eq1}
\end{equation}
corresponding to the ${\bf q}=0$ and ${\bf q}\neq 0$ terms, respectively. As
will become clear soon, the ${\bf q}\neq 0$ term leads to the pseudogap.
Note here we use the subscript ``sc'' and ``pg'' to distinguish the
superconducting and pseudogap contributions.

Accordingly, we write the self-energy
\begin{equation}
\Sigma(K) = \Sigma_{sc}(K) + \Sigma_{pg}(K) \:,
\label{Sigma_Eq1}
\end{equation}
where
\begin{equation}
\Sigma_{sc}(K) = \sum_Q t_{sc}(Q) G_0(Q-K) \phikq^2 = 
- \Delta_{sc}^2 G_0(-K)\phik^2 
\label{Sigma_sc_Eq}
\end{equation}
and
\begin{equation}
\Sigma_{pg}(K) = \sum_Q  t_{pg}(Q) G_0(Q-K) \phikq^2 \;\;
\end{equation}
are associated with the superconducting and pseudogap contributions,
respectively. 

The Thouless condition requires $1+g\chi(0)=0$ at $T_c$. However, this
condition also applies to all temperatures below $T_c$, for it is simply the
gap equation. This can be motivated as follows. The zero momentum pair state
is macroscopically occupied. The continuity of the pair dispersion away from
zero momentum is equivalent to the presumption that $\mu_{pair} = 0$. In
this way a generalized Thouless condition continues to hold away from $T_c$.
Indeed, this is the BEC condition, which requires that the effective pair
chemical potential $\mu_{pair}$ vanish,
\begin{equation}
\mu_{pair} =0 \:.
\label{BEC_Condition}
\end{equation}
Thus we have the gap equation (\ref{Gap_Eq}) valid for all $T\leq T_c$.

Here we will make one important approximation to make the calculations and
their interpretation easier. Due to the Thouless criterion, $t_{pg}(Q)$ is
highly peaked at $Q=0$, so that the main contribution to $\Sigma_{pg}$ comes
from the vicinity of $Q=0$. Therefore,
\begin{equation}
\Sigma_{pg}(K) \approx \left[ \sum_Q t_{pg}(Q) \right] G_0(-K)\phik^2 
\equiv -\Delta_{pg}^2 G_0(-K)\phik^2  \:,
\label{Sigma_PG_Approx}
\end{equation}
where we have defined a pseudogap parameter $\Delta_{pg}$ via
\begin{equation}
\Delta_{pg}^2 \equiv -\sum_Q t_{pg}(Q) \:.
\label{PG_Def}
\end{equation}

\begin{figure}
\centerline{\includegraphics[bb = 53 0 408 131, clip, width=4in]{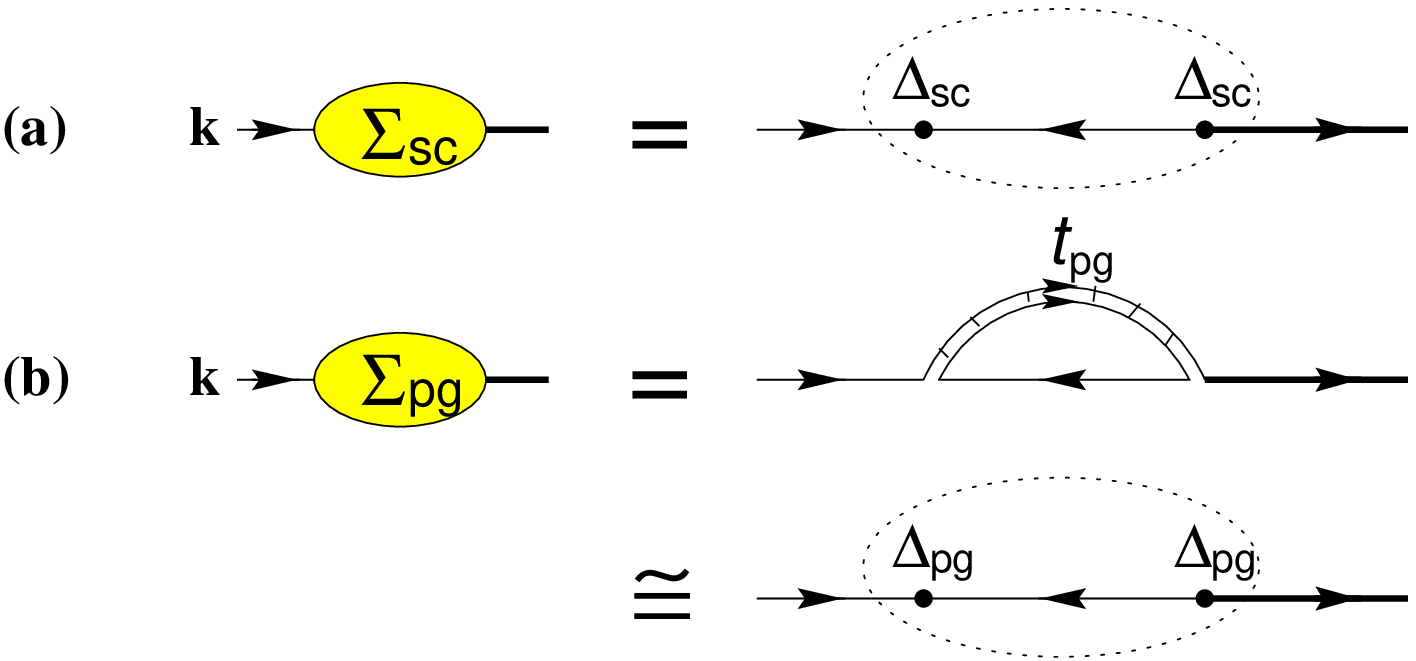}}
\caption{Approximation scheme for the pseudogap self-energy, $\Sigma_{pg}(K)$.}
\label{PG_Approx}
\end{figure}

The approximation made in Eq.~(\ref{Sigma_PG_Approx}) can be
diagrammatically represented by Fig.~\ref{PG_Approx}. It is equivalent to
replacing the finite momentum pairs by a ``pseudo-condensate'' such that
this term gives roughly the same fermionic self energy, $\Sigma_{pg}(K)$.
It is evident that, under this approximation, the pseudogap self-energy
$\Sigma_{pg}(K)$ now has the BCS form. Therefore, we have the total
self-energy
\begin{equation} 
\Sigma(K) = - \Delta^2 G_0(-K)\phik^2 = \frac{\Delta^2 \phik^2}{i\omega +\ek}\:,
\label{Sigma_Eq2}
\end{equation}
where we have used the fact $\ek = \epsilon_{-\mathbf{k}}$. Here $\Delta$
given by
\begin{equation}
\Delta = \sqrt{\Delta_{sc}^2 + \Delta_{pg}^2}
\label{TG_Def}
\end{equation}
corresponds to the total excitation gap. Now we can write down the
full Green's function
\begin{equation}
G(K) = \frac{i\omega +\ek}{(i\omega)^2 - \Ek^2} 
= \frac{\uk^2}{i\omega-\Ek} + \frac{\vk^2}{i\omega+\Ek} \:,
\label{G_Eq}
\end{equation}
where $\Ek = \sqrt{\ek^2 + \Delta^2\phik^2}$ is the quasiparticle dispersion.

Now that the total gap has exactly the BCS form, and that the Thouless
condition, Eq.~(\ref{Gap_Eq}), is formally identical to its BCS counterpart,
Eq.~(\ref{BCS_Gap_Eq2}),
we can immediately write down the \textit{gap
  equation}: 
\begin{equation}
1+g\sumk \frac{1-2f(\Ek)}{2\Ek}\phik^2 =0\;.
\label{Gap_Eq2}
\end{equation}
It should be emphasized that the gap $\Delta$ in this equation is
different from the order parameter $\Delta_{sc}$.
 
Substituting the full $G$, Eq.~(\ref{G_Eq}), into Eq.~(\ref{n_Eq}), we
obtain the \textit{fermion number equation},
\begin{equation}
n = 2\sumk \left[ \vk^2 + \frac{\ek}{\Ek} f(\Ek)\right] \:.
\label{Number_Eq}
\end{equation}

To calculate the pseudogap parameter, we expand the inverse $T$ matrix, as
shown in Eq.~(\ref{InvT_Expansion}). At and below $T_c$, we find that the
Cooper pairs are essentially infinitely long lived, for the imaginary term
$\Gamma^\prime_{{\bf q}, \Omega}$ is negligibly small at small {\bf q} and
small $\Omega$, and vanishes at $\omega={\bf q}=0$.
In the weak coupling limit, the ratio $a_0/a_1$ vanishes.  Thus the pair
dispersion is linear in $q$, as is consistent with particle-hole symmetry in
the BCS limit.  As $g$ increases, $a_0/a_1$ increases, reflecting the
presence of particle-hole asymmetry, and $a_0$ gradually dominates. In this
regime the pair dispersion becomes quadratic in $q$. It should also be
stressed that at small $g$, the residue of the $T$ matrix is vanishingly
small, and thus the contribution of finite {\bf q} pairs is negligible.
Here, we are mostly interested in moderate and strong coupling, where
the contribution from $a_1\Omega^2$ term can be neglected. These issues are
discussed in more detail in Appendix \ref{App_Omegaq}. (In principle, one
can always keep this term, however, it does not lead to significant
quantitative effects in our calculations.)  We may rewrite the $T$ matrix
for finite {\bf q} as
\begin{equation} 
t_{pg}({\mathbf{q}}, \Omega) = \frac{a_0^{-1}}{\Omega - \Omega_{\mathbf{q}}
  + \mu_{pair} + i\Gamma_{{\mathbf{q}}, \Omega}}\:,
\label{t_Expansion}
\end{equation}
where $\mu_{pair}$ is an effective chemical potential for pair excitations.
The pair dispersion $\Omegaq$ can be written as
\begin{eqnarray}
\Omega_{\mathbf{q}} &=& -\frac{1}{a_0}\Big[ \chi({\mathbf{q}}, 0)) -
 \chi({\mathbf{0}}, 0)\Big] \nonumber\\ 
 &=& B q^2 \equiv \frac{q^2} {2
  M^*}  \qquad\qquad\qquad\;\, \mbox{for isotropic systems, or } \nonumber\\
& =& \sum_i B_i q_i^2 \equiv \sum_i \frac{q_i^2} {2
  M_{i}^*}  \qquad\quad \mbox{for anisotropic systems, }
\label{Omegaq}
\end{eqnarray}
which defines the effective pair mass. The expressions for $\Omegaq$ and the
expansion coefficients $a_0$ and $B$ are given in Appendix \ref{App_Omegaq}.
Both of these quantities are essentially a constant as a function of $T$ at
low $T$ or at all $T\leq T_c$ in strong coupling.

%\textit{Qijin this seems repetitive-- can you use this off-set equation
%elsewhere?}
%Once again, Eq.~(\ref{t_Expansion}) shows that the Thouless criterion is
%just the BEC condition:
%%
%\begin{equation}
%\mu_{pair} =0 \:.
%\label{BEC_Condition}
%\end{equation}

Finally, substituting Eq.~(\ref{t_Expansion}) into the pseudogap definition,
Eq.~(\ref{PG_Def}), we obtain an \textit{equation for the pseudogap},
\begin{equation}
a_0\Delta_{pg}^2 = \sum_{\mathbf{q}} b(\Omegaq) \:,
\label{PG_Eq}
\end{equation}
where $b(x)$ is the Bose distribution function. 
From this equation, we see that $\Delta_{pg}^2$ is proportial to the density
of pair excitations.  Indeed, this equation is the boson number equation.
In the BEC regime, where all electrons form pairs, the right hand side must
be equal to $n/2$, the density of pairs, at $T_c$. In the strong coupling,
low density limit with an $s$-wave pairing symmetry, $M^* = 2m$, and this
equation alone completely determines the condensation temperature, $T_c
\approx 0.218 E_F$, as in an ideal Bose gas.

Equations (\ref{Gap_Eq2}), (\ref{Number_Eq}), and (\ref{PG_Eq}) form a
complete set of self-consistent equations.  Setting $\Delta_{sc}=0$, one can
solve for the transition temperature $T_c$, and $\Delta$ and $\mu$ at $T_c$.
Below $T_c$, the first two equations form a complete subset, determining
$\Delta$ and $\mu$ at a given $T$. The third equation indicates how to
decompose $\Delta$ into the pseudogap $\Delta_{pg}$ and superconducting
$\Delta_{sc}$ contributions.

Without any detailed calculations, one can immediately see from the set of
equations the following limiting behavior:
\begin{enumerate}
\renewcommand{\labelenumi}{(\roman{enumi})}
\setlength{\itemsep}{0in}
\setlength{\parsep}{0in}
\item
There exists a pseudogap at $T_c$, for finite coupling strength $g$. 
\item
This pseudogap persists below $T_c$.
\item
At $T=0$, the pseudogap vanishes, i.e., $\Delta_{pg}=0$, via
Eq.~(\ref{PG_Eq}), and $\Delta=\Delta_{sc}$. In this way, we recover the
Leggett ground state. 
\item
In the weak coupling limit, the pair excitations are negligible, and
we recover BCS theory.
\item
In the strong coupling limit, we have essentially an ideal Bose
gas. All pairs are condensed at $T=0$. This is different from the weakly
interacting true Bose liquid. 
\item In 2D, the right hand side of Eq.~(\ref{PG_Eq}) will diverge
  at any finite temperature for the quadratic pair dispersion. This means
  that $T_c =0$ in 2D.
\item Most importantly, $\Delta^2_{pg} \propto T^{3/2}$ at low $T$. This
  will show up in physical quantities as new low $T$ power laws.
\end{enumerate}
It should be obvious that our theory satisfies the requirements we set
forth in Sec.~\ref{Sec_PhysPicture}.

\vskip 4eX

In the following chapters we will solve these coupled equations under various
situations. Since the discussion of most physical quantities, e.g., the
superfluid density, relies on the solution of the gaps, $T_c$, and $\mu$,
the physical consequences of our theory will be deferred until 
after we discuss the solution to the generalized gap equations.

\chapter{\textit{T}\lowercase{$_{\bf c}$} and the Superconducting
instability of the normal state}
\label{Chap_Tc}

In the literature, there have been studies of $T_c$ in the context of the
BCS--BEC crossover approach, for the cases of 3D isotropic continuous
(called ``jellium'') or lattice system and for a strictly 2D lattice.
However, a treatment of the cuprates is somewhat more complicated involving:
quasi-two dimensionality, lattice periodicity, and, finally, the
complexities associated with $d$-wave symmetry in the pairing interaction.
In this Chapter, we solve the set of Eqs.~(\ref{Gap_Eq2}),
(\ref{Number_Eq}), and (\ref{PG_Eq}) for $T_c$, $\mu(T_c)$, and
$\Delta(T_c)$ as a function of $g$, (by setting the order parameter
$\Delta_{sc}=0$), and in this way address the more realistic situations
appropriate to the cuprates.  Our goal is to study the superconducting
instability of the normal state, and the effects on $T_c$ associated with
lattice structure, low dimensionality, and $d$-wave pairing. These studies
are based on previous related work by Maly and coworkers
\cite{Janko,Maly,Maly2} in a 3D jellium model. Our new results are as
follows: (i) Here we find that $T_c$ vanishes logarithmically with the mass
anisotropy ratio as the strict 2D limit is approached, (ii) the chemical
potential $\mu$ smoothly interpolates from $E_F$ to large negative values in
the jellium case, as the coupling varies from weak to strong.  However,
(iii) lattice effects yield a vanishing $T_c$ in the strong coupling limit.
This is associated with a reduction (due to the lattice) in the effective
kinetic energy of the bosons or pairs. (iv) The $d$-wave case on a lattice
is found to be different from the $s$-wave lattice case in one significant
respect: Superconductivity disappears at relatively smaller values of $g$,
as a result of $d$-wave symmetry; the pair size cannot be smaller than a
lattice spacing, and, therefore, the pair mobility is strongly suppressed.
Consequently, the superconducting bosonic regime is never reached for the
$d$-wave case, except for the unphysically dilute limit.

\section{Specifications for various models}
\label{Subsec_Numerics}

Before we do any calculations, in this section we specify in detail the
various dispersion $\ek$ and symmetry factors $\phik$, which characterize
isotropic and anisotropic jellium, $s$-wave and $d$-wave pairing, as well as
discrete lattice structure.  For definiteness, in our quasi-2D calculations
it is assumed that the pairing interaction depends only on the in-plane
momenta.

(i) {\em 3D jellium, s-wave symmetry\/} -- 
As usual, we assume a parabolic dispersion relation,
$\epsilon_{\mathbf{k}}={\mathbf{k}}^2/2m-\mu$, with $\varphi_{\mathbf{k}} =
(1+k^2/k_0^2)^{-1/2}$. The parameter $k_0$ is the inverse range of the
interaction and represents a soft cutoff in momentum space for the
interaction. As will be clear later, $k_0 > k_F$ is assumed in general in
order to access the strong coupling limit.  It is convenient to introduce a
dimensionless scale $g/g_c$ for the coupling constant. Here, following
Ref.~\cite{NSR}, we choose $g_c = -4\pi/mk_0$, which corresponds to
the critical value of the coupling above which bound pairs are formed in
vacuum.

(ii) {\em Quasi-2D jellium, s-wave symmetry}. Here we adopt an anisotropic
energy dispersion
\begin{equation}
  \label{eq:Epsilon-2D-jellium}
  \epsilon_{\mathbf{k}} = 
\frac{{\mathbf{k}}^2_{\parallel}}{2m_\parallel} +
  \frac{k_{\perp}^2}{2m_{\perp}} - \mu\;,
\end{equation}
where $k_\perp$ is restricted to a finite interval ($|k_\perp|\le\pi$) while
$\mathbf{k}_{\parallel}$ is unconstrained.%
\newfootnote{We emphasize that a cut-off for $k_\perp$ is crucial for the quasi
  two-dimensionality of the system; otherwise, by rescaling $k_\perp
  \rightarrow (m_\perp/m_\parallel)^{1/2}\,k_\perp$, the system can be
  transformed into an isotropic 3D jellium model.}
Now $\varphi_{\mathbf{k}} = (1+k^2_\parallel/k_0^2)^{-1/2}$
depends only on the in-plane momentum. 

By tuning the value of the anisotropy ratio $m_{\perp}/m_\parallel$ from one
to infinity, this model can be applied to study effects associated with
continuously varying dimensionality from 3D to 2D.%
\newfootnote{Strictly speaking, a very large cutoff $\Lambda$ for $k_\perp$
  or a very low particle density (i.e., $k_F\ll \Lambda$) is necessary to
  restore the 3D limit if $m_\perp/m_\parallel=1$ is fixed. Alternatively,
  3D can be achieved by letting $m_\perp/m_\parallel \rightarrow 0$ while
  fixing $\Lambda$. This complication does not occur for the lattice case,
  where all momenta are restricted within the first Brillouin zone. }
For convenience, we use the parameter $g_c$ derived for 3D jellium, as a
scale factor for the coupling strength, but call it $g_0$ to avoid
confusion.

The strict 2D limit is approached as $m_{\perp}/m_\parallel \rightarrow
\infty$, and we find $T_c$ vanishes logarithmically with
$m_{\perp}/m_\parallel$. In this limiting case, the right hand side of
Eq.~(\ref{PG_Eq}) vanishes logarithmically at finite $T$, implying that in
2D the superconducting order is not stable against pairing fluctuations
so that $T_c=0$. This result is independent of the details of the dispersion
or order parameter symmetry.  This behavior is consistent with the
Mermin-Wagner theorem.

(iii) {\em Quasi-2D lattice, s- and d-wave symmetry}. On a lattice we adopt
a simple tight-binding model with dispersion
\begin{equation}
  \label{eq:Epsilon-2d-lattice}
  \epsilon_{\mathbf{k}} = 2\,t_\parallel (2-\cos{k_x}-\cos{k_y}) +
  2\;t_{\perp}(1-\cos{k_{\perp}}) - \mu\;,
\end{equation}
where $t_\parallel$ ($t_{\perp}$) is the hopping integral for the in-plane
(out-of-plane) motion. The isotropic 3D lattice is given by setting
$t_\parallel$ =$t_{\perp}$. For $s$-wave pairing, $\varphi_{\mathbf{k}}=1$,
as in the negative $U$ Hubbard model, whereas for $d$-wave pairing,
\begin{equation}
  \varphi_{\mathbf{k}} = \cos{k_x}-\cos{k_y}\:.
  \label{eq:d-wave}
\end{equation}

Equations~(\ref{Gap_Eq2}), (\ref{Number_Eq}), and (\ref{PG_Eq}), together
with the various models for $\varphi_{\mathbf{k}} $ and
$\epsilon_{\mathbf{k}}$, were solved numerically for $\Delta_{pg}$, $\mu$
and $T_c$. The numerically obtained solutions satisfy the appropriate
equations with an accuracy higher than $10^{-7}$.  The momentum summations
were calculated by numerical integration over the whole ${\bf k}$ space for
the jellium case, and over the entire Brillouin zone for the lattice.
However, to facilitate our calculations in the case of the quasi-2D lattice
with a $d$-wave pairing interaction, the momentum integral along the
out-of-plane direction was generally replaced by summation on a lattice with
$N_\perp=16$ sites. For completeness we compared solutions obtained with and
without an approximated low frequency, long wavelength expansion of the pair
dispersion $\Omegaq$ discussed above. Here we found extremely good agreement
between the two different approaches.  In general, we chose the ratios
$m_\perp/m_\parallel =100$ or $t_\perp/t_\parallel=0.01$, although higher
values of the anisotropy were used for illustrative purposes in some cases.

\section{Overview:  \textit{T}\lowercase{$_\mathbf{c}$} and effective mass of 
the pairs}
\label{Sec_overview}

In was pointed out in Ref.~\cite{NSR} in the context of the attractive $U$
Hubbard calculations, that the appropriate description of the strong
coupling limit corresponds to \textit{interacting} bosons on a lattice with
effective hopping integral $t' \approx -2t^2/U$. It, therefore, will
necessarily vanish in the strong coupling limit, as $U \rightarrow \infty$.
In addition to this hopping, there is an effective boson-boson repulsion
which also varies as $V^\prime \approx -2t^2/U$.

This description of an effective boson Hamiltonian can be related to the
present calculations through Eqs.~(\ref{t_Expansion})--(\ref{Omegaq}) which
represent the boson Green's function for such a Hamiltonian and its
parameterization via \textit{only} the pair mass $M^*$.  It is important to
stress that boson-boson interaction effects are not included at the level of
the gap equations, as is consistent with the mean field character of the
ground state wave function.  However, there is a residue of a boson-boson
interaction which enters via $M^*$, which we now discuss.  By solving
Eqs.~(\ref{Gap_Eq2}), (\ref{Number_Eq}), and (\ref{PG_Eq}) self-consistently
and identifying $M^*$ from the effective pair propagator (or $T$ matrix), we
find that our $M^*$ necessarily incorporates Pauli principle
induced pair-pair repulsion. It addition $M^*$ includes pairing symmetry and
density related effects.  Here, an indirect interaction between bosons is
effectively introduced via fermionic self-energy effects which derive from
the interactions between the fermions and bosons. Note, in contrast to
Ref.~\cite{NSR}, in the present work we are not restricted to the bosonic
limit, nor is it essential to consider a periodic lattice. Thus, much of
this language is also relevant to the moderately strong coupling (but still
fermionic) regime, and can even be applied to jellium.

The goal of this section is to establish a natural framework for relating
$M^*$ to $T_c$.  The parameters which enter into $M^*$ via
Eq.~(\ref{Omegaq}) vary according to the length scales in the various
physical models. In the case of jellium, $M^*$ depends in an important way
on the ratio $k_0/k_F$.  For the case of $s$-wave pairing on a lattice,
$M^*$ depends on the inverse lattice constant $\pi/a$ and density $n$.
Finally, for the case of $d$-wave pairing, there is an additional length
scale introduced as a result of the finite spatial extent of the pair. This
enters as if there were an equivalent reduction in $k_0/k_F$ in the
analogous jellium model.  The following factors act to increase $M^*$ or,
alternatively, to reduce the mobility of the pairs: the presence of a
periodic lattice, a spatially extended pairing symmetry (such as $d$-wave)
or, for jellium, small values of the ratio $k_0/k_F \lesssim 0.4$ (i.e.,
high density).

In order to relate $T_c$ to $M^*$, we observe that in an ideal
Bose-Einstein system $T_c$ is inversely proportional to the mass.
Here, this dependence is maintained, in a much more complex theory, as
a consequence of Eqs.~(\ref{PG_Def}) and (\ref{PG_Eq}). Eq.~(\ref{PG_Eq})
is essentially an equation for the number of pairs (bosons), with
renormalized mass $M^*$.  Thus, as we increase $g$ towards the bosonic
regime, it is not surprising that $T_c$ varies inversely with $M^*$.

This leads to our main observations, which apply to moderate and large
$g$, although not necessarily in the strict bosonic regime.  (i) For the
general lattice case, we find that $T_c$ vanishes, either asymptotically
or abruptly, as the coupling increases, in the same way that the inverse
pair mass approaches zero.%
\newfootnote{This is a high density effect at strong coupling. Lattice
  bandstructure effects and $d$-wave pairing serve to enhance this behavior.
  Moreover, it should be noted that the pseudogap $\Delta_{pg}\propto
  (T_c/(m/M^*))^{3/2}$ remains finite as both $T_c$ and $m/M^*$ approach
  zero.}
(ii) For the case of jellium or
low densities on a lattice, both $T_c$ and $M^*$ remain finite and are
inversely proportional. These observations are consistent with, but go
beyond, the physical picture in Ref.~\cite{NSR} that $T_c$ is
expected to be proportional to the pair hopping integral $t'$.  It
should be stressed that in the very weak coupling limit the pair size or
correlation length is large. In this case, the motion of the pairs
becomes highly collective, so that the effective pair mass is very
small.

In the presence of a lattice, the dependence on band filling $n$ is also
important for $M^*$, and thereby, for $T_c$. We find that the bosonic regime
is not accessed for large $n > n_c\approx 0.53$. There are two reasons why
superconductivity abruptly disappears within the fermionic regime. This
occurs primarily (in the language of Ref.~\cite{NSR}) as a consequence of
large indirect pair-pair repulsion, relevant for high electronic densities,
which leads to large $M^*$. In addition, there are effects associated with
the particle-hole symmetry
at half filling.%
\newfootnote{In a negative U Hubbard model, there also exists a competition
  between superconductivity and charge-density wave ordering. This may not
  be relevant for a $d$-wave pairing interaction.}
Precisely at half filling (i.e., the ``filling factor'' $f = 1/2$, or $ 2f=
n = 1$), for the band structure we consider, there is complete particle-hole
symmetry and $\mu$ is pinned at $E_F$.  Similarly, in the vicinity of $n = 1
$, the chemical potential remains near $E_F$ for very large coupling
constants $g$.

By contrast, in the small density (lattice) limit for the $s$-wave case,
($n\approx 0.1$), the indirect pair-pair repulsion is relatively unimportant
in $M^*$ and there is no particle-hole symmetry. In this way the bosonic
regime is readily accessed. Moreover, in this limit we see a precise scaling
of $T_c$ with $1/g$ in the same way as predicted by Ref.~\cite{NSR} (via the
parameter $t' = -2t^2/U$).  Thus in this low density limit superconductivity
disappears asymptotically, rather than abruptly.

The effects of pairing symmetry should also be stressed. Because of the
spatial extent of the $d$-wave function, the pair mobility is strongly
suppressed, and, thus, $M^*$ is relatively larger than for the $s$-wave
case. This lower mobility of $d$-wave pairs leads to the important
result that superconductivity is \textit{always abruptly} (rather than
asymptotically) destroyed with sufficiently large coupling.  Near half
filling we find $\mu$ remains large when $T_c$ vanishes, at large $g$.
As the density $n$ is reduced, away from half filling, $\mu$ decreases
somewhat. It is important to note that the system remains in the
fermionic regime (with positive $\mu$) for all densities down to
$n\approx 0.09$ [See Appendix~\ref{App_n-g} for details].

In all cases discussed thus far, $T_c$ exhibits a non-monotonic dependence
on the coupling constant. It grows exponentially at small $g$, following the
BCS dependence, and shuts off either asymptotically or abruptly at higher
$g$.  One can view this effect as deriving from a competition between
pairing energy scales and effective mass or mobility energy scales.  This
competition is not entirely dissimilar to that found in more conventional
Eliashberg theory where the fermionic renormalized mass and the attractive
interaction compete in such a way as to lead to a saturation in $T_c$ at
large coupling.  However, in the present context, for intermediate and
strong coupling, we are far from the Fermi liquid regime and the effective
mass of the quasi-bound or bound pair is a more appropriate variable.

With this background, it should not be surprising that non-monotonic
behavior will arise, even in situations as simple as in jellium models.
Indeed, in this case we find that for sufficiently long range interactions
or high densities (small $k_0/k_F$) superconductivity disappears abruptly
before the bosonic regime can be reached. (As shown in Fig.~\ref{n-g-S},
$T_c$ is reentrant for certain intermediate densities.)  Even for the case
of short range interactions ($k_0/k_F = 4$), there is a depression in $T_c$
caused by an increase in the pair mass, while still in the fermionic regime.

\begin{figure}
  \centerline{
    \includegraphics[width=3.3in]{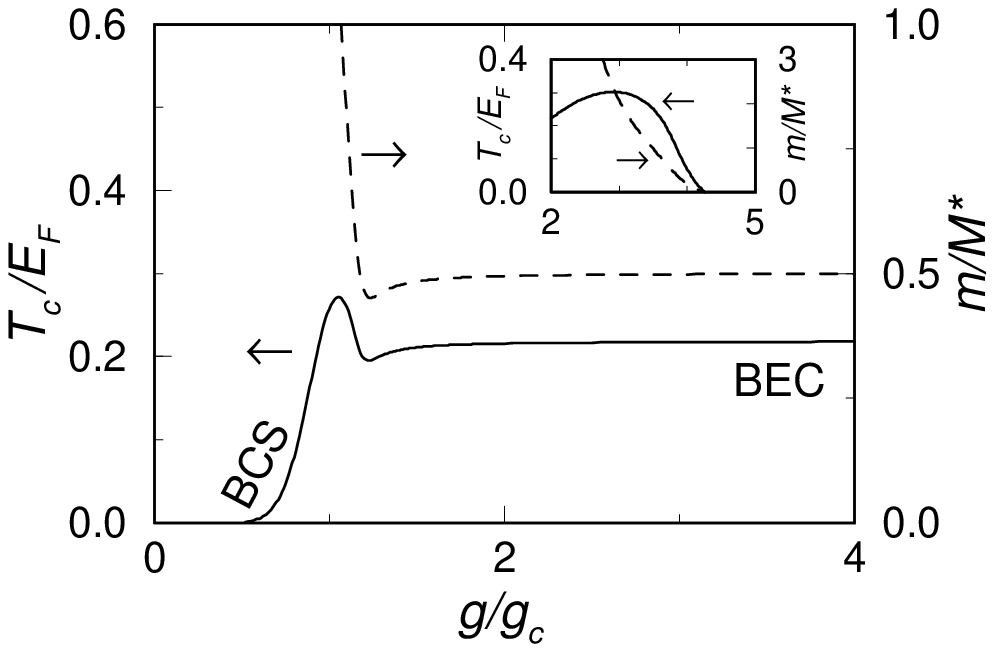}
\includegraphics[width=2.7in]{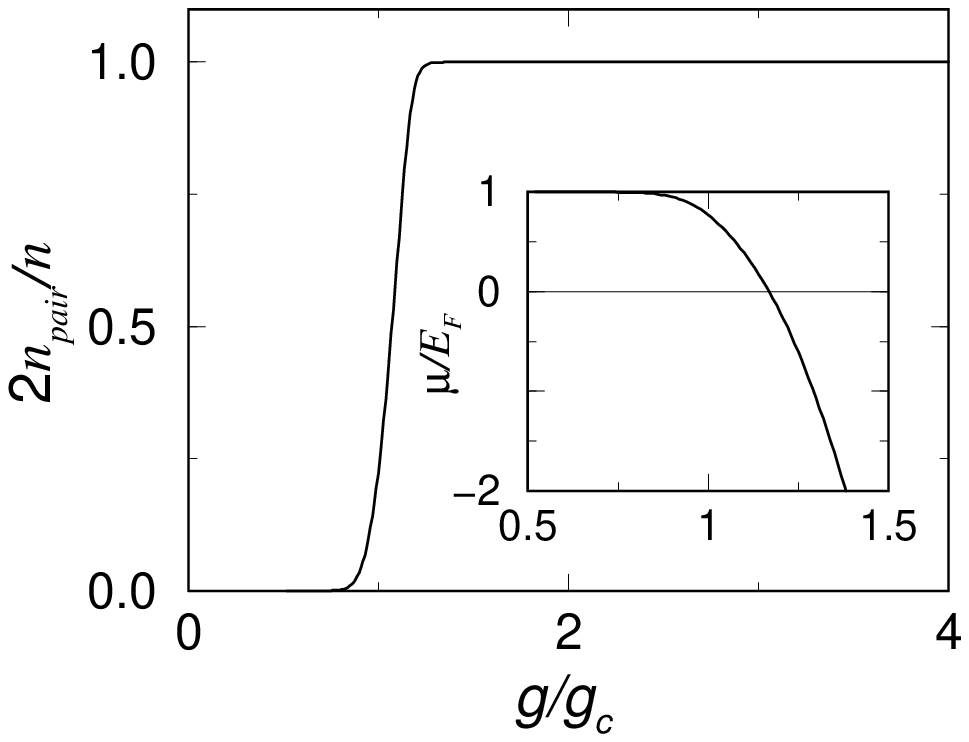}
}
\vskip -2in \hskip 0.65in {\large \textbf(a)} 
\hskip 3.1in {\large \textbf(b)}
\vskip 1.7in
\caption[$T_c$, the inverse pair mass, and the pair density as a function of
coupling constant at low and high densities in 3D jellium. ]{(a) $T_c$ and
  $m/M^*$ as a function of $g/g_c$ in the 3D jellium model with $k_0/k_F=4$
  (main figure) and $k_0/k_F=1/3$ (inset), corresponding to short range (or
  low density), and long range interactions (or high density), respectively.
  Plotted in (b) are the density of pairs (main figure) and the chemical
  potential (inset) as a function of $g/g_c$, corresponding to the main
  figure in (a).  }
\label{Lattice_Fig1}
\end{figure}

In Fig.~\ref{Lattice_Fig1}(a) we plot the calculated $T_c$ for the case of
an isotropic, 3D jellium model with $s$-wave pairing, along with the inverse
pair mass $m/M^*$. 
The density of pairs, $n_{pair}=\sum_\mb{q} b(\Omega_\mb{q})$,
[Eq.~(\ref{PG_Eq})], as well as the chemical potential, are plotted in
Fig.~\ref{Lattice_Fig1}(b). This figure demonstrates that as the system
crosses over into the bosonic regime (i.e., $\mu$ becomes negative),
essentially all electrons form pairs, $n_{pair} = n/2$, so that
Eq.~(\ref{PG_Eq}) becomes a real boson (pair) number equation.  In this
case, Eq.~(\ref{PG_Eq}) alone completely determines the transition
temperature at low densities.
Figure~\ref{Lattice_Fig1} is presented primarily as a base line with which
to compare subsequent plots. The parameter $k_0/ k_F = 4$, is reasonably
large so that the high $g$ asymptote is found to reach the ideal
Bose-Einstein limit ($T_c = T_{BEC}=0.218E_F$) with $M^*=2m$.  The approach
to the high $g$ asymptote is from below, as is expected \cite{Haussmann94}.
This is a result of the decreasing Pauli principle repulsion associated with
increasing $g$, and concomitant reduction in pair size. The non-monotonic
behavior at intermediate $g/g_c \approx 1 $ can be associated with structure
in the effective pair mass, and has been discussed previously from a
different perspective \cite{Maly2}.

In the inset of Fig.~\ref{Lattice_Fig1}(a) are plotted analogous curves for
the case of long range interactions or high densities ($k_0/k_F = 1/3$).
This figure illustrates how superconductivity vanishes abruptly before the
bosonic regime is reached, as a consequence of a diverging pair mass.

\section{Effects of dimensionality}
\label{Sec_Dimensionality}

In this section we illustrate the effects of anisotropy or
dimensionality on $T_c$ (and on $\Delta_{pg}$ and $\mu$ ) within
the context of a jellium dispersion.%
\newfootnote{One may argue that this quasi-2D system is not physical because
  the kinetic energy in the perpendicular direction should have a band
  structure, i.e., $2t_\perp(1-\cos k_\perp$) instead of being simply
  ${\displaystyle \frac{k_\perp^2}{2m_\perp}}$. As shown in
  Sec.~\ref{App_Sec_Q2DJellium}, in this case, ${\displaystyle
    \frac{M^*_\perp}{M^*_\parallel}\propto \left(\frac{m_\perp}{m_\parallel}
    \right)^2}$ (instead of ${\displaystyle \frac{m_\perp}{m_\parallel}}$) ,
  i.e., the pairs see a magnified anisotropy.  This does not affect the
  conclusions here.}
A particularly important check on our theoretical interpolation scheme is to
ascertain that $T_c$ is zero in the strict 2D limit and that $\mu$ varies
continuously from $E_F$ in weak coupling to the large negative values
characteristic of the strong coupling bosonic limit.  The present
calculational scheme should be compared with that of Yamada and
co-workers \cite{Yamada} who included ``mode coupling" or feedback
contributions to $T_c$, but only at the level of the lowest order ``box''
diagram discussed in Ref.~\cite{Janko}. These authors were unable to find a
smooth interpolation between weak and strong coupling, but did successfully
repair the problems \cite{SVR,Serene} associated with the NSR scheme,
which led to negative $\mu$ even in arbitrarily weak coupling.

\begin{figure}
\mbox{\parbox{3in}{
  \centerline{
    \includegraphics[width=3in]{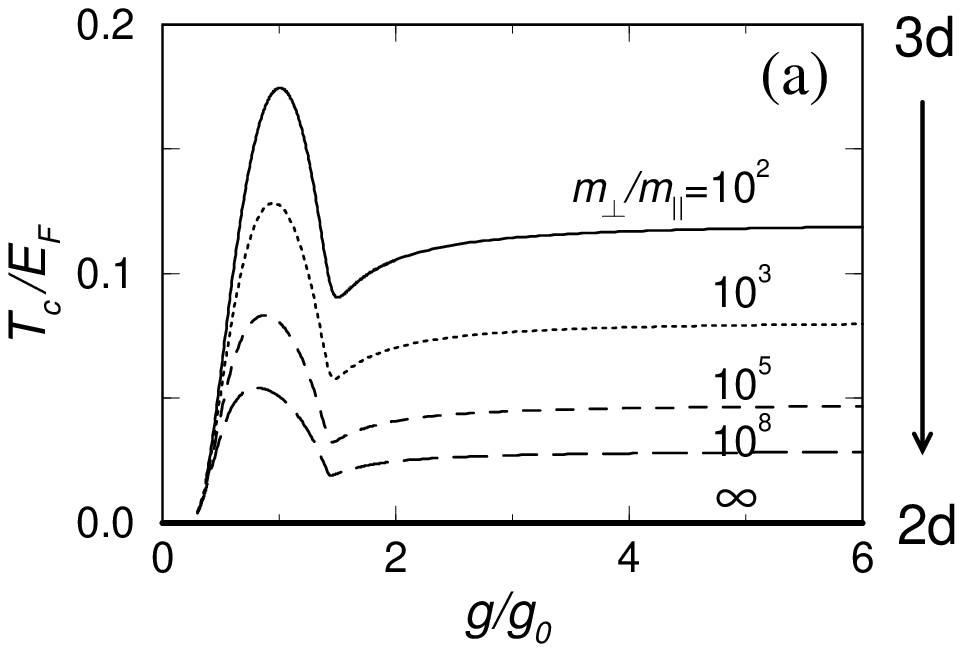} }
  \centerline{\hskip -4mm
    \includegraphics[width=2.82in]{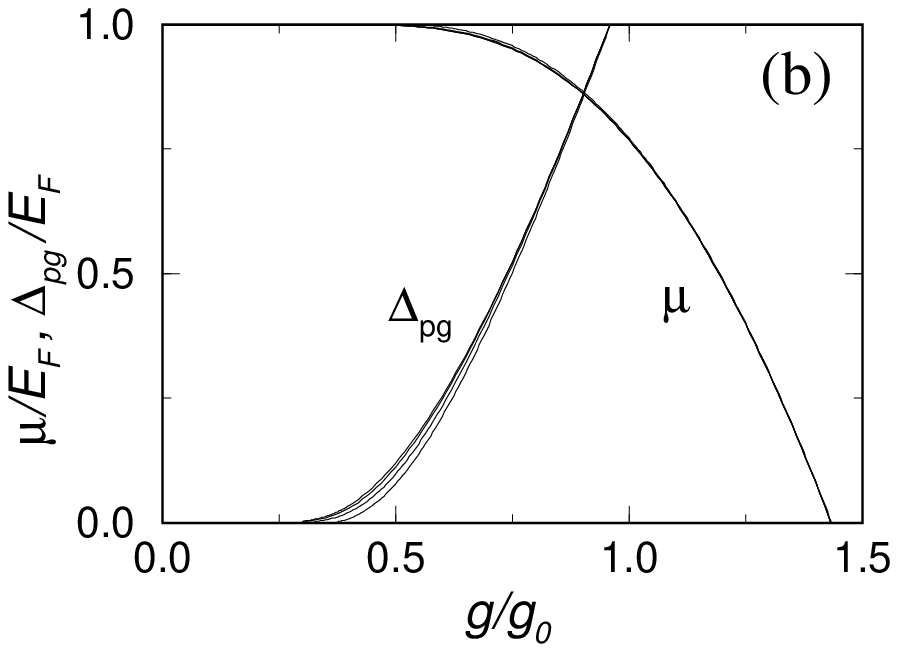} }
}} \hfill
%  \centerline{\hskip -5mm
    \includegraphics[width=2.9in]{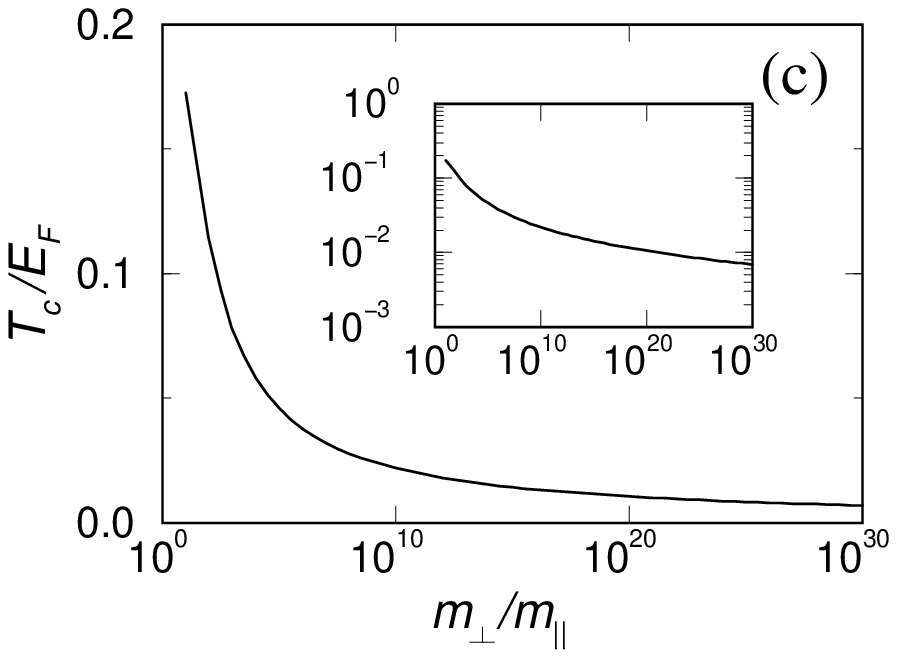} 
%}
\vskip -1.8in
\hfill
\mbox{\parbox{2.9in}{
\caption[Effects of low dimensionality on the crossover behavior of $T_c$,
$\mu$ and $\Delta$.]{Dimensionality crossover in a quasi-2D jellium model.
  (a) $T_c$ as a function of $g$ is seen to vanish for all $g$ as
  $m_\perp/m_\parallel \rightarrow \infty$, while (b) $\mu$ and
  $\Delta_{pg}$ change little.  A continuous variation of $T_c$ versus
  $m_\perp/m_\parallel$ at $g/g_0=4$ is shown in the main portion (semi-log
  plot) and the inset (log-log plot) of (c).  Here $k_0/k_F=4, g_0\equiv
  -4\pi/mk_0$.  }
\label{Lattice_Fig2} 
\vskip 0.1in
}}
\end{figure}

Figures \ref{Lattice_Fig2}(a) and \ref{Lattice_Fig2}(b) show the effect on
$T_c$ and on $\Delta_{pg}$ and $\mu$, respectively, of introducing a
layering or anisotropy into jellium with $s$-wave pairing. The various
curves correspond to different values of the anisotropy ratio
$m_\perp/m_\parallel$. It can be seen from these two figures that $T_c$
approaches zero as the dimensionality approaches 2.  At the same time the
chemical potential $\mu$ interpolates smoothly from the Fermi energy at weak
coupling towards zero at around $g/g_0 = 1.5$ to large negative values (not
shown) at even larger $g$.  The vanishing of the superconducting transition
in strictly 2D was discussed in detail in Sec.~\ref{Subsec_Numerics}.

It should be noted that quasi-two dimensionality will be an important
feature as we begin to incorporate the complexity of $d$-wave pairing.
The essential physical effect introduced by decreasing the dimensionality is
the reduction in energy scales for $T_c$. The chemical potential and
pseudogap amplitude are relatively unaffected by dimensional
crossover effects.%
\newfootnote{The onset coupling $g$ for the pseudogap decreases when the
  dimensionality is reduced, and eventually vanishes in strictly 2D.}
 While $T_c$ rapidly falls off
when anisotropy is first introduced into a 3D system (such as is
plotted in Fig.~\ref{Lattice_Fig1}), the approach to the strict 2D limit is
logarithmic and therefore slow, as can be seen explicitly in
Fig.~\ref{Lattice_Fig2}(c). Thus, in this regime, to get further significant
reductions in $T_c$ associated with a dimensionality reduction
requires extremely large changes in the mass anisotropy.

\section{Effects of a periodic lattice}

The first applications of a BCS Bose-Einstein crossover theory to a periodic
lattice were presented in Ref.~\cite{NSR}. The present approach represents
an extension of the NSR theory in two important ways: we introduce mode
coupling or full self-energy effects which are parameterized by
$\Delta_{pg}$, and which enter via Eq.~(\ref{PG_Def}).  Moreover, the
number equation (\ref{Number_Eq}) is evaluated by including self-energy
effects to all orders.  This is in contrast to the approximate number
equation used in Ref.~\cite{NSR}, which includes only the first order
correction.  In this way we are able to capture the effects which were
qualitatively treated by these authors and which are associated with the
lattice.

\begin{figure}
  \centerline{
    \includegraphics[width=6in]{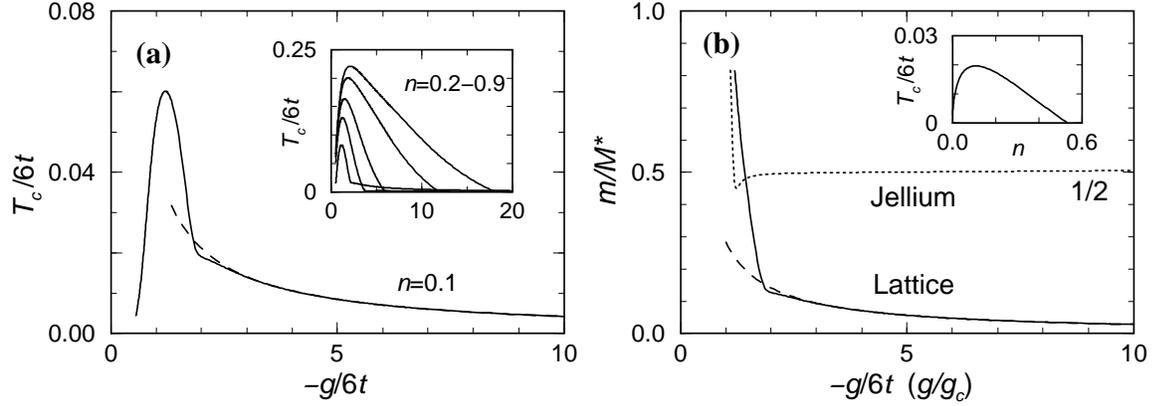} }
\vskip -0.1in
\caption[Lattice effects on the crossover behavior of $T_c$ and the inverse
pair mass with respect to the coupling strength.]{(a) $T_c$ and (b) $m/M^*$
  (solid lines) \textit{vs} $g$ at low filling ($n=0.1$) on a 3D lattice,
  and $T_c$ at larger filling in the inset of (a). A fit to the functional
  form $t^\prime=-2t^2/g$ is plotted (dashed lines) in (a) and (b) with
  adjusted proportionality constants.  For comparison, $m/M^*$ vs $g/g_c$
  for 3D jellium (Fig.~\ref{Lattice_Fig1}) is replotted (dotted line) in
  (b).  From bottom to top, the inset of (a) shows $T_c$ for densities
  $n=0.2, 0.5, 0.7, 0.85$, and 0.9. The inset of (b) shows $T_c$ at $\mu=0$
  as a function of $n$.  }
\label{Lattice_Fig3}
\end{figure}

Figure \ref{Lattice_Fig3}(a) plots the behavior of $T_c$ (solid line) in an
isotropic three-dimensional lattice (with $s$-wave pairing, $\varphi_{\bf
  k}=1$) at a low density $n=0.1$.  The effects of higher electronic filling
are shown in the inset. The low $n$ behavior in the main portion of the
figure can be compared with the jellium calculations of
Fig.~\ref{Lattice_Fig1}.  For small $n$, $T_c$ decreases asymptotically to
zero at high $g$.  For larger $n$, $T_c$ vanishes abruptly before the
bosonic regime ($\mu < 0$) is reached [See inset of
Fig.~\ref{Lattice_Fig3}(b)].  These various effects reflect the analogous
reduction in the effective pair mobility, parameterized by the inverse pair
mass $m/M^*$. To see the correlation with $m/M^*$ in the low density limit,
we plot this quantity in Fig.~\ref{Lattice_Fig3}b, for the lattice as well
as jellium case (where for the latter, $m/M^*\rightarrow 1/2$ at large $g$).
Here the coupling constants are indicated in terms of $g/g_c$ for jellium
and $-g/6t$ for the lattice.  The inflection points at $-g/6t\approx 2$ in
both $T_c$ and $m/M^*$ curves correspond to $\mu=0$, which marks the onset
of the bosonic regime.

Also plotted in both Fig.~\ref{Lattice_Fig3}(a) and \ref{Lattice_Fig3}(b)
(dashed lines) is the effective hopping $t^\prime=-2t^2/g$ for $n=0.1$,
rescaled such that it coincides with $T_c$ and $m/M^*$, respectively, at
high coupling ($-g/6t=30$).  This figure illustrates clearly the effect
first noted by Nozi\`eres and Schmitt-Rink that in the entire bosonic
regime, $T_c$ varies with high precision as $t^\prime$ or equivalently as
$m/M^*$.

Finally, in the inset of Fig.~\ref{Lattice_Fig3}(b), we demonstrate the
limiting value of $n$, above which the bosonic limit can not be accessed.
What is plotted here is the value of $T_c$ at which $\mu$ is zero as a
function of density $n$.  This figure indicates that the bosonic regime can
not be reached for $n>n_c\approx 0.53$.  At densities higher than this, the
pair-pair repulsion increases $M^*$ sufficiently, so that $T_c$ vanishes
abruptly, while $\mu$ is still positive.

\begin{figure}
\centerline{
\includegraphics[width=3in]{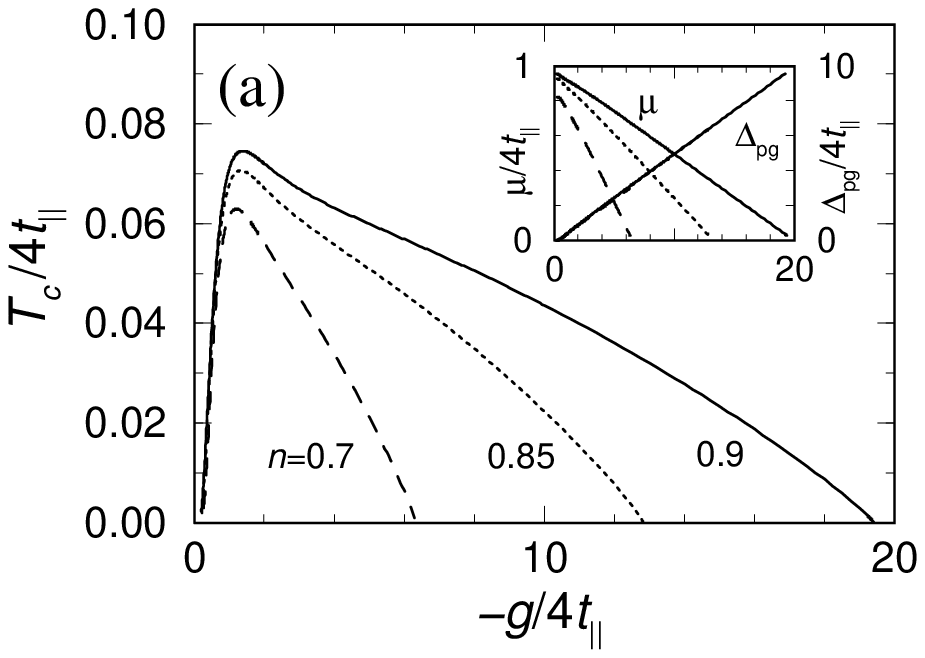}
\includegraphics[width=3in]{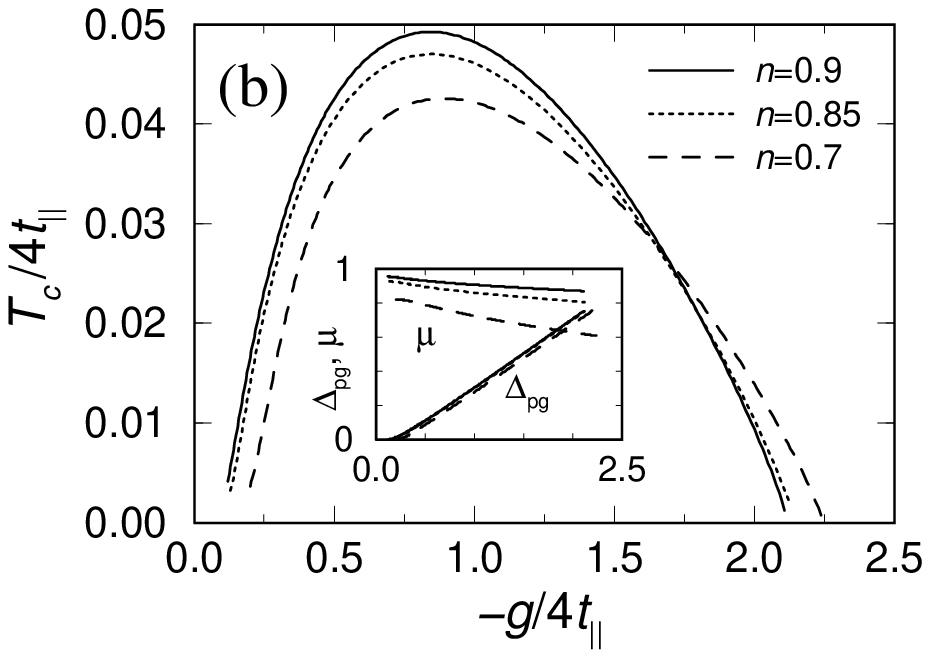}
}
\vskip -0.1in
\caption[Effects of the pairing symmetry on the crossover behavior of $T_c$,
$\mu$ and $\Delta_{pg}$ on a quasi-2D lattice.]{Lattice effects on $T_c$
  (main figure) and $\mu$ and $\Delta_{pg}$ (inset) as a function of $g$ for
  $n$=0.7 (dashed lines), 0.85 (dotted lines), and 0.9 (solid lines) in
  quasi-2D for (a) $s$-wave and (b) $d$-wave pairing symmetries.  Here
  $t_\perp/t_\parallel=0.01$. In (b), $T_c$ vanishes at a much smaller $g$
  than does its $s$-wave counterpart.  }
\label{Lattice_Fig4}
\end{figure}

\section{Effects of \textit{d}-wave symmetry}

We now introduce the effects of a $d$-wave pairing interaction. For the
purposes of comparison we begin by illustrating $T_c$ for the case of
$s$-wave pairing on an anisotropic lattice, shown in
Fig.~\ref{Lattice_Fig4}(a), for three different values (0.7, 0.85, and 0.9)
of the density $n$.%
\newfootnote{We show results for these values of $n$ in order to demonstrate
  the behavior of $T_c$ in over-, optimally, and under-doped cuprates,
  respectively.}
 The inset indicates the behavior of the
pseudogap magnitude and the chemical potential. The plots of $\Delta_{pg}$
for the three different $n$ are essentially unresolvable in the figure.
Note, from the inset, that within small numerical errors $T_c$ and $\mu$
vanish simultaneously.  A comparison of the magnitude of $T_c$ (in the main
figure) with the 3D counterpart shown in the inset of
Fig.~\ref{Lattice_Fig3}(a) illustrates how $T_c$ is suppressed by quasi-two
dimensionality.%
\newfootnote{As in the 3D (s-wave) lattice case, the bosonic regime can be
  accessed at $n < n_c \approx 0.53$ for sufficiently strong coupling.}

In Fig. \ref{Lattice_Fig4}(b), similar plots are presented for the $d$-wave
case.  Here we use the same values of the filling factor as in
Fig.~\ref{Lattice_Fig4}(a), to which Fig.~\ref{Lattice_Fig4}(b) should be
compared. The essential difference between the two figures is the large $g$
behavior.  Lattice effects produce the expected cutoff for $s$-wave pairing.
In the $d$-wave situation this cutoff is at even smaller $g$, and moreover,
corresponds to $\mu\approx E_F$.  Calculations similar to those shown in the
inset of Fig.~\ref{Lattice_Fig3}(b) indicate that superconductivity
disappears while $\mu$ remains positive for all $n$ above the extreme low
density limit (i.e., for $n > n_c \approx 0.09$) (See Appendix \ref{App_n-g}
for details).  This behavior is in contrast to that of the $s$-wave case
where $n_c \approx 0.53$.

In the $d$-wave case, the pair size cannot be made arbitrarily small,
no matter how strong the interaction.  As a result of the extended size
of the pairs, residual repulsive interactions play a more important
role. In this way, the pair mobility is reduced and the pair mass
increased. Thus, as a consequence of the finite pair size, \textit{in
 the d-wave case the system essentially never reaches the
 superconducting bosonic regime}.

\section{Phase diagrams}

In this section we introduce an additional energy scale $T^*$, and in this
way, arrive at plots of characteristic ``phase diagrams" for the crossover
problem. Our focus is on the pseudogap onset, so that attention is
restricted to relatively small and intermediate coupling constants $g$;
consequently, the bosonic regime is not addressed. Here, our calculations of
$T^*$ are based on the solution of Eq.~(\ref{Gap_Eq2}), along with
Eq.~(\ref{Number_Eq}), under the assumption that $\Delta_{pg} = 0$.
This approximation for $T^*$ is consistent with more detailed numerical
work \cite{Maly2} in which this temperature is associated with the onset of a
pair resonance in the $T$ matrix.

\begin{figure}
\mbox{\parbox{3in}{
\includegraphics[width=2.9in]{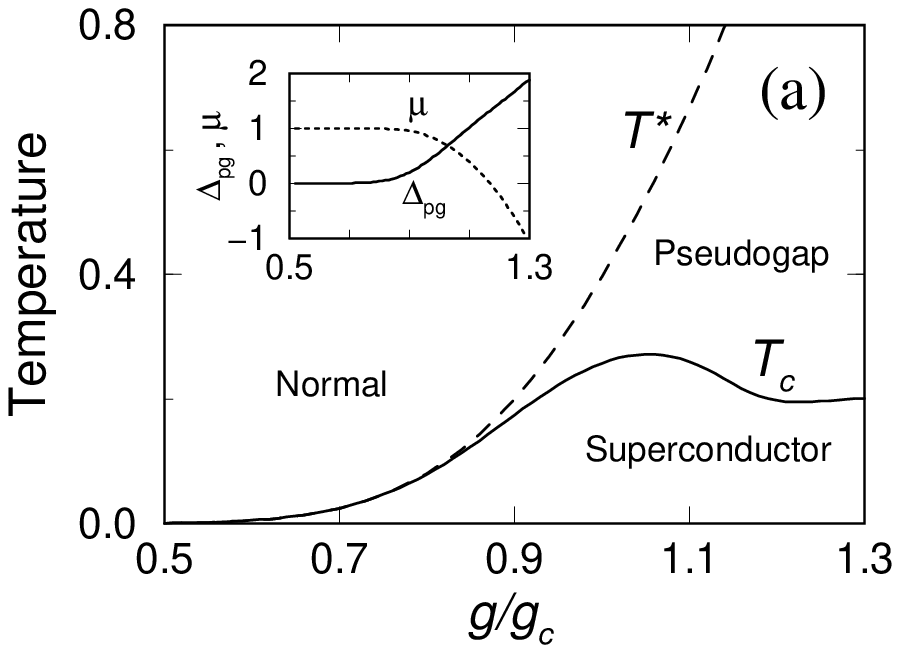}
\includegraphics[width=2.9in]{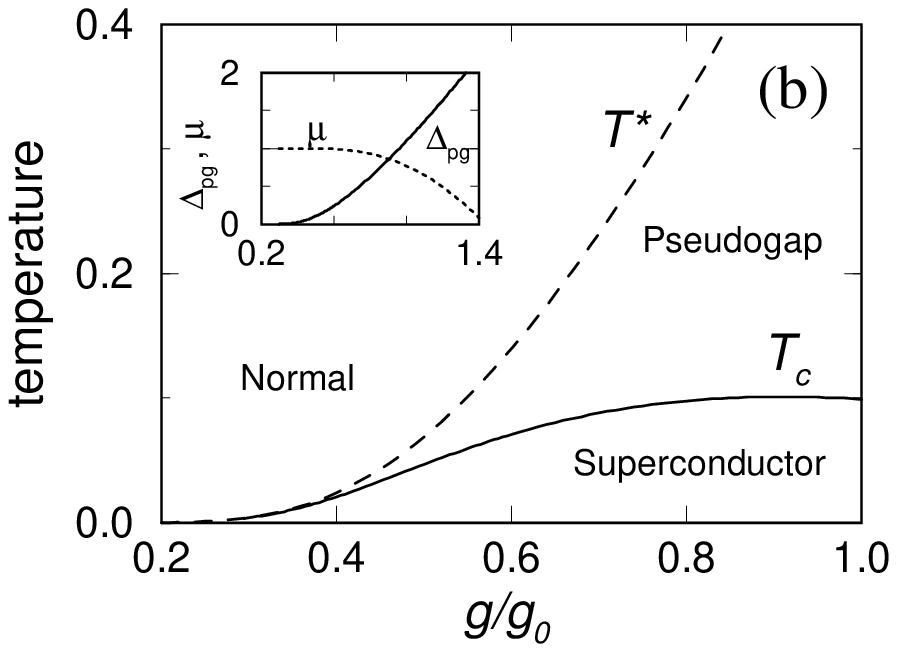}
}}
\includegraphics[width=3in]{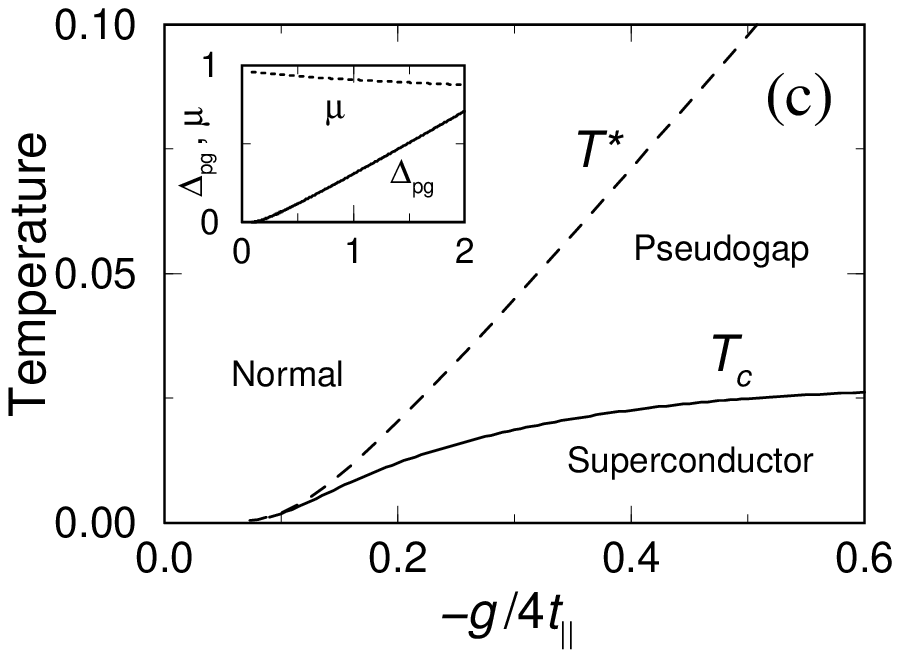}
\vskip -1.4in
\hfill
\mbox{
\parbox{2.8in}{
\caption[Phase diagrams on a temperature -- coupling constant plot for a 3D
and quasi-2D jellium, and a quasi-2D lattice with $d$-wave symmetry.]{Phase
  diagrams of (a) 3D jellium ($k_0/k_F=4$), quasi-2D jellium ($k_0/k_F=4,
  m_\perp/m_\parallel=10^4$), (c) a $d$-wave symmetry on a quasi-2D lattice
  ($t_\perp/t_\parallel=10^{-4}$). Here we take $ n=0.9$.  The same energy
  units are (a), (b) $E_F$ and (c) $4 t_\parallel$.}
\label{Lattice_Fig5}
}}
\end{figure}

In Figs.~\ref{Lattice_Fig5}(a)-(c) our results are consolidated into phase
diagrams for different physical situations. The case of 3D jellium, with
$s$-wave pairing [Fig.~\ref{Lattice_Fig5}(a)] is presented primarily as a
point of comparison.  Figure~\ref{Lattice_Fig5}(b) corresponds to quasi-2D
jellium ($m_\perp/m_\parallel = 10^4$), with $s$-wave pairing and
Fig.~\ref{Lattice_Fig5}(c) to the case of $d$-wave pairing in a quasi-2D
lattice case ($t_\perp/t_\parallel = 10^{-4}$).%
\newfootnote{The phase diagram for the quasi-2D $s$-wave lattice case is
  similar to that of the quasi-2D $d$-wave case for the small coupling
  regime shown in Fig.~\ref{Lattice_Fig5}(c).}
The insets indicate the behavior of $\mu$ and $\Delta_{pg}$.  Comparing
$T^*$ with $T_c$ represents a convenient way of determining the onset of the
pseudogap state.  (For definiteness, we define the onset to correspond to
$T^*=1.1T_c$). It is clear from the first two figures that this occurs for
3D jellium at $g/g_c \approx 0.9$, and for the quasi-2D case at $g/g_0
\approx 0.4$. (Note here $g_0=g_c$.) This observation reinforces the notion
that pseudogap effects are easier to come by in lower-dimensional systems.
Similar behavior is seen in the quasi-2D lattice situation for the $d$-wave
case, although the energy scales on the horizontal and vertical axes reflect
the parameter $t_\parallel$ (rather than $g_c$ and $E_F$).

\vskip 4eX

In summary, in this Chapter, we have applied the theory which was developed
in Chapter~\ref{Chap_Theory} to study the superconducting instability of
the normal state, as a function of $g$.  Our intent was to go beyond the
simple 3D jellium model with $s$-wave pairing. In this way we have
determined the effects of quasi-two dimensionality, of periodic discrete
lattices, and of a $d$-wave pairing interaction. For the $d$-wave case, much
of the discussion presented in this Chapter addressed densities which are
relevant to the cuprate superconductors. A complete study of the $d$-wave
crossover problem at an arbitrary density between 0 and 1, as well as a more
detailed comparison with $s$-wave superconductors, are given in
Appendix~\ref{App_n-g}.

\chapter{Superconducting phase}
\label{Chap_SC_Phase}

The existence of a pseudogap at and above $T_c$ for the underdoped cuprates
is now well established. However, whether the pseudogap or related effects
persist below $T_c$ is still under debate. From the experimental perspective
the relationship (if any) between the pseudo- and superconducting gaps is
unclear.  ARPES \cite{ARPES_ANL,ARPES_Stanford} and other measurements
\cite{Loram93,NMR} on the underdoped cuprates indicate that the normal state
excitation or, equivalently, the pseudogap above $T_c$ evolves smoothly into
the excitation gap in the superconducting phase below $T_c$.  It is unlikely
that a fully developed pseudogap will abruptly disappear as the temperature
falls below $T_c$, but precisely how it connects with the superconducting
\textit{order parameter} is not obvious.  In this Chapter, we will address
the evolution of the various gaps below $T_c$, as functions of both
temperature and coupling strength. Based on the calculations of these gaps,
we will further address the superfluid density and low temperature specific
heat.

We find that for intermediate and strong coupling, as the temperature
decreases the pseudogap $\Delta_{pg}$ persists below $T_c$, and vanishes
eventually as $T^{3/4}$ at $T=0$. In the meantime, the order parameter
$\Delta_{sc}$ develops precisely at $T_c$, and is equal to the excitation
gap $\Delta$ only at $T=0$. At low $T$, while $\Delta$ has an exponential
$T$ dependence, as in BCS, $\Delta_{sc}$ generally has a power law $T$
dependence. We argue that strong pairing correlations exist not only between
a pair of electrons with precisely opposite momenta, but also between a pair
of electrons with a finite net momentum {\bf q}.
The latter correlation makes it possible to create low energy, incoherent
pair excitations.  It is the existence of these finite momentum pairs that
leads to the distinction between the order parameter and the excitation gap.

The central new result of our theory is that physical measurables, in
general, contain contributions from the (usual) fermionic quasiparticles
(via $\Delta$) as well as the bosonic pair states (via $\Delta_{pg}$).
In this Chapter we will demonstrate that pair excitations generally
lead to new low $T$ power laws, e.g, an extra $T^{3/2}$ term in the
superfluid density $n_s$, and a linear in $T$ term in the specific heat
$C_v$ in quasi-2D.

\section{Excitation gap, pseudogap, and superconducting gap}
\label{Sec_Gaps}

To study the temperature evolution of the various gaps, we first solve the
set of coupled equations, Eqs.~(\ref{Gap_Eq2}), (\ref{Number_Eq}), and
(\ref{PG_Eq}), for $T_c$. Then we determine the various gaps and chemical
potential as a function of $T$ below $T_c$ for different situations.
Without loss of generality, in this section, we will work with the simplest
3D jellium model with $s$-wave pairing, as specified in
Sec.~\ref{Subsec_Numerics}.  The gaps for the case of $d$-wave pairing will
be discussed when we address the cuprates in Chapter \ref{Chap_Cuprates}.

\begin{figure}
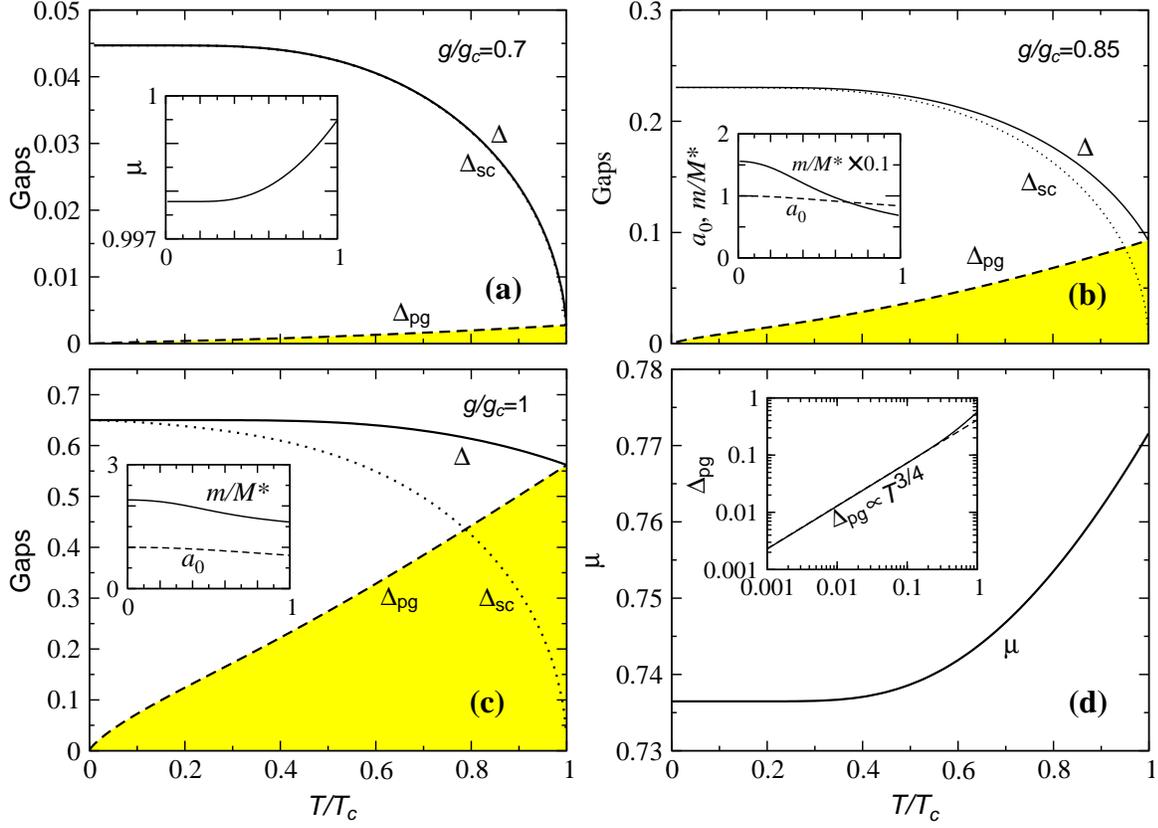

\includegraphics[angle=-90, width=5.98in, clip]{Btk4-g-2}
\centerline{\includegraphics[angle=-90, width=6in, clip]{Btk4g1-2}}
\caption[Temperature dependence of the pseudogap, order parameter and the
excitation gap, as well as the chemical potential and relevant $T$ matrix
expansion coefficients, below $T_c$ for various coupling strength $g$ in 3D
jellium.]{Temperature dependence of various gaps below $T_c$ for (a)
  $g/g_c=0.7$, (b) 0.85, and (c)-(d) 1.0, respectively, in a 3D $s$-wave
  jellium with $k_0 = 4k_F$.  The chemical potential $\mu$, as well as the
  inverse $T$ matrix expansion coefficient $a_0$ (normalized at $T=0$) and
  the inverse pair mass $m/M^*$, are shown as insets, correspondingly.  A
  log-log plot of $\Delta_{pg}$ for $g/g_c=1$ is shown in the inset of (d),
  where the dashed line is a $T^{3/4}$ fit. }
\label{3D_BT_Gaps}
\end{figure}

Shown in Fig.~\ref{3D_BT_Gaps}(a)--(c) are the three self-consistently
determined gaps (superconducting gap $\Delta_{sc}$, pseudogap $\Delta_{pg}$,
and the total excitation gap $\Delta$) calculated at low density,
$k_0=4k_F$, and $g/g_c=0.7$, 0.85, and 1.0 in 3D jellium, for an $s$-wave
pairing.  The cases plotted correspond to weak, intermediate, and strong
pseudogaps. In the weak coupling limit, the pseudogap is expected to vanish,
as in BCS theory.  Indeed, we find that for $g/g_c=0.7$, $\Delta_{pg}$ is
negligible, and $\Delta_{sc}$ and $\Delta$ are essentially the same. As $g$
increases, $\Delta_{pg}$ increases, and the total gap $\Delta$ becomes
independent of $T$ in the strong coupling limit.

From Fig.~\ref{3D_BT_Gaps}(a)--(c), we see that, as the temperature
decreases, the excitation gap $\Delta$ increases, from its value $\Delta =
\Delta_{pg}$ at $T_c$. The order parameter $\Delta_{sc}$ develops at $T_c$,
as expected. It is important to note that the pseudogap persists below
$T_c$. As more and more pairs come into the condensate when $T$ is lowered,
the density of finite momentum pairs is reduced; energetically pairs prefer
to be in the condensate.  Therefore, as $T$ decreases, $\Delta_{pg}$
decreases.  Finally, $\Delta_{pg}$ vanishes at $T=0$, as is consistent with
the fact that these finite \textbf{q} pairs are, indeed, thermal
excitations.

We also show the chemical potential $\mu$ in Fig.~\ref{3D_BT_Gaps}(d) for
$g/g_c=1$, and in the inset of Fig.~\ref{3D_BT_Gaps}(a) for $g/g_c=0.7$.
These plots indicate that $\mu$ does not change much with $T$, as in a Fermi
liquid. In fact, the relative change is the largest in the intermediate
regime ($g\sim g_c$).  As indicated in the inset of
Fig.~\ref{3D_BT_Gaps}(a), in the weak coupling limit, $T_c$ is small, and
$\mu$ is nearly $E_F$ for all $T\leq T_c$; In the strong coupling, low
density limit (not shown here), $\mu$ becomes a large negative number,
$\Delta$ is large, and $T_c$ reaches the upper bound of $T_{BEC}=0.218E_F$,
corresponding to a free Bose gas of fermion pairs.  Here, all $T\leq T_c$
corresponds essentially to the $T\rightarrow 0$ limit. In both the weak and
strong coupling limit, we notice $T_c\ll |\mu|$. In contrast, at $g\sim
g_c$, $T_c$, $\Delta$, and $\mu$ are all comparable. This leads to a
relatively large change in $\mu$ in the intermediate regime, as shown in
Fig.~\ref{3D_BT_Gaps}(d).

From Eq.~(\ref{PG_Eq}), we expect $\Delta_{pg}^2 \propto T^{3/2}$ at low
$T$.  This analytic result relies on the presumption that the coefficient
$a_0$ and the pair dispersion $\Omegaq$ are roughly independent of $T$.  To
support this, in the inset of Fig.~\ref{3D_BT_Gaps}(b) and of
Fig.~\ref{3D_BT_Gaps}(c), $a_0$ is plotted and shown to be roughly a
constant for all $T\leq T_c$. The effective pair mass $M^*$ has a stronger
$T$ dependence, but our presumption is still valid at low $T$.% 
\newfootnote{The leading order corrections for both $a_0$ and $M^*$ are
  quadratic in $T$ in the $s$-wave case, as revealed by detailed numerical
  studies. See Fig.~\ref{Cuprate_Gaps}(e) for a $d$-wave example.}
The $T$ dependence of $M^*$ becomes weaker and weaker as $g$ increases.
Indeed, at low densities, $M^* \rightarrow 2m$ at large $g$ [See
Fig.~\ref{Lattice_Fig1}(a)]. Shown in the inset of Fig.~\ref{3D_BT_Gaps}(d)
is $\Delta_{pg}$ vs $T$ on a log-log plot. The curve fits nearly perfectly
to a $T^{3/4}$ power law (shown as the dashed line) up to $T<0.2 T_c$, i.e.,
$\Delta^2_{pg} \propto T^{3/2}$.

\textit{The bosonic nature of the pair excitations, reflected in
  Eq.~(\ref{PG_Eq}), will lead to extra power law temperature dependences
  for physical quantities which are sensitive to these excitations.} As will
be shown in the next two sections, the superfluid density will acquire an
extra $T^{3/2}$ dependence in 3D and in quasi-2D, and the specific heat will
acquire a $T^{3/2}$ power law in 3D and a linear $T$ dependence in quasi-2D.

Physically, the pseudogap below $T_c$ can be interpreted as
an extra contribution to the excitation gap, reflecting the fact that at
moderate and large $g$, additional energy is needed to overcome the residual
attraction between the fermion components of the excited fermion pairs in
order to produce fermionic-like (Bogoliubov) quasi-particles. In the bosonic
limit, it becomes progressively more difficult to break up these pairs and
the energy $\Delta_{pg}$ increases accordingly.

%We may recapitulate our picture of the pseudogap.  
In general, a fermion is attracted to all other fermions in the system, not
just to the one with opposite momentum and opposite spin. This attraction
makes fermions form pairs with non-zero net momentum, at the cost of a small
amount of energy.  In the BCS limit, because of the light pair mass $M^*$,
the cost of energy increases immediately as the net pair momentum deviates
from zero, and therefore, there is little possibility of forming finite
momentum pairs; all the pairs are frozen in the condensate. As $g$
increases, the pair dispersion becomes softer, and the opportunity to form
finite momentum pairs is greater. It is these pairs that cause the
pseudogap. As a consequence, there will be soft, finite momentum pair
excitations at moderate coupling, in addition to the fermionic
quasiparticles already existing in the weak coupling BCS limit.  However, as
$g$ increases further, it becomes extremely difficult to break up a pair, so
that then the probability of finding fermionic quasiparticles becomes
exponentially small.  In this way, in the strong coupling limit, the system
below $T_c$ will be composed of the condensate and finite momentum pairs
only. This physical picture can be shown schematically in
Fig.~\ref{CrossoverCartoon}.

Note that in our scenario, $\Delta_{pg}$ is temperature dependent below $T_c$
(as well as above $T_c$), and finally vanishes at $T=0$. This is different
from the ``normal state gap'' used phenomenologically by Loram and coworkers
\cite{Loram94}. Upon analysis of data in underdoped cuprates, they
conjectured that the measured excitation gap squared can be expressed as the
sum of the squares of a pseudogap and superconducting order parameter. This
purely {\em phenomenological} analysis leads to a similar
decomposition \cite{Loram97} of the excitation gap, as in
Eq.~(\ref{TG_Def}). However, in contrast to the present work, these authors
presumed that $\Delta_{pg}$ is temperature independent below $T_c$.

\section{Superfluid density}
\label{Sec_Ns}

The pseudogap is an important measure of the distinction between the order
parameter, $\Delta_{sc}$, and the excitation gap, $\Delta$. To establish the
validity of the pseudogap below $T_c$, and to distinguish different possible
theoretical models, one needs experimental tools which measure two of the
three gaps below $T_c$, unambiguously.  ARPES and tunneling probe only the
total excitation gap, $\Delta$. To measure $\Delta_{sc}$, it is necessary to
measure physical quantities which rely crucially on the superconducting
order. To this end, the superfluid density is of interest; it vanishes
identically at and above $T_c$, and can be viewed as an order parameter in
the bosonic regime.

The superfluid density, $n_s$, can be measured via the magnetic London
penetration depth $\lambda$. It takes a finite superfluid density to screen
magnetic flux. In MKSA units, one has
\begin{equation}
  \lambda^{-2}_L = \frac{\mu_0 e^2 n_s}{m} \:,
\label{Lambda_Def_Eq}
\end{equation}
where $\mu_0$ is the magnetic permittivity. For convenience, we will set
$\mu_0 = e=1$. 
Note on a lattice, $n/m$ should be replaced by 
\begin{equation}
\left(\frac{n}{m}\right)_{\alpha\beta} \equiv 2\sumk \frac{\partial^2 \ek}
{\partial k_\alpha \partial k_\beta} n_{\mb{k}} \:, \qquad (\alpha = x,y,z) \:,
\label{Mass_Def_Eq}
\end{equation}
where $n_\mb{k}$ is the fermion density distribution in momentum space. For
the quasi-2D square lattice, the in-plane mass tensor is diagonal, and 
$(n/m)_{xx}= (n/m)_{yy}$. (Since the out-of-plane component will be treated
separately, we can treat $n/m$ as a simple scalar.)
To make our formulae general, we focus on $\lambda^{-2}_L$ instead of
$n_s/m$. Nevertheless, we may use these two quantities interchangeably.

The penetration depth is expressed in terms of the local (static)
electromagnetic response kernel $K(0)$  in linear response
theory \cite{Abrikosov}, 
\begin{equation}
\lambda^{-2}_L=K(0) =  \left( \frac{n}{m}\right) - P_{xx}(0) \:,
\label{Lambda_K0_Eq}
\end{equation}
where $K(0)$ is defined by
\begin{equation} 
J_\alpha(Q) = P_{\alpha\beta}A_\beta(Q) -  \left(\frac{n}
  {m}\right)_{\alpha\beta} A_\beta(Q) = -K_{\alpha\beta}(Q) A_\beta(Q) \:,
\label{K_Def}
\end{equation} 
and the current-current correlation function $P_{\alpha\beta}$ is given by
\begin{equation}
  P_{\alpha\beta}(Q) = -2 \sum_K  \lambda_{\alpha}(K,K+Q)\,G(K+Q)
  \Lambda_{\beta}(K+Q,K)\,G(K) \;.
\label{P_Def_Eq}
\end{equation}
%
%Note here the sign of $P_{\alpha\beta}$ is different from that in
%ref.~\cite{Kosztin}, in order to be consistent with the convention used in
%Sec.~\ref{Sec_Coll_Mode} and ref.~\cite{Kosztin2}. 
The bare vertex is given by
\begin{equation}
\bm{\lambda}(K,K+Q) = \vec{\nabla}_\mb{k}\epsilon_{\mb{k}+\mb{q}/2} =
%\frac{\partial\epsilon_{\mb{k}+\mb{q}/2}} {\partial \mb{k}}=
\frac{1}{m}\left(\mb{k}+\frac{\mb{q}}{2}\right) \:,
\label{lambda_Def}
\end{equation}
where we have used the dispersion for jellium in the last step. The full
vertex $\bm{\Lambda}$ must be deduced in a manner consistent with the
generalized Ward identity, applied here for the uniform static case:
$Q=(\mb{q},0)$, $\mb{q}\rightarrow 0$ \cite{Schrieffer}. The vertex can be
constructed by inserting external legs into the self-energy diagram,
Fig.~\ref{Sigma-T}(a). It is convenient to write
\begin{equation}
\bm{\Lambda} = \bm{\lambda} + \delta\bm{\Lambda}_{pg} +
\delta\bm{\Lambda}_{sc} \:,
\label{Lambda_Eq}
\end{equation}
where the pseudogap contribution $\delta\bm{\Lambda}_{pg}$ to the vertex
correction follows from the Ward identity
\begin{equation} 
\delta\bm{\Lambda}_{pg}(K,K) = \frac{\partial\Sigma_{pg}(K)}{\partial
  \mb{k}} \:,
\label{Lambda_PG_Eq}
\end{equation}
and the superconducting vertex contribution is given by
\begin{equation}
  \delta\bm{\Lambda}_{sc}(K+Q,K) =
  \Delta_{sc}^2\varphi_{\mb{k}}\varphi_{\mb{k}+\mb{q}} G_o(-K-Q) G_o(-K)
  \bm{\lambda}(K+Q,K)\;.
  \label{Lambda_SC_Eq}
\end{equation}
The latter can be obtained by proper vertex insertion to the superconducting
self-energy diagram, Fig.~\ref{BCS_Sigma}, which will lead to the so-called
Maki-Thompson term. A more detailed treatment will be given in
Chapter \ref{Chap_CollMode}, when we address gauge invariance and the
collective modes.

To proceed, we rewrite Eq.~(\ref{Mass_Def_Eq}) by integration by parts,
\begin{eqnarray}
\left(\frac{n}{m}\right) & = & \frac{2}{d} \sum_{\alpha,K} \frac{\partial^2
  \ek}{\partial k_\alpha^2 } G(K) 
 =  - \frac{2}{d} \sum_{\alpha,K} \frac{\partial \ek}{\partial k_\alpha}
\frac{\partial G(K)} {\partial k_\alpha} \nonumber\\
&=& -\frac{2}{d}\sum_{\alpha,K} G^2(K) \frac{\partial \ek} 
{\partial k_\alpha} \left( 
\frac{\partial \ek}{\partial k_\alpha} + \frac{\partial \Sigma_{pg}}
{\partial k_\alpha} + \frac{\partial \Sigma_{sc}}{\partial k_\alpha} \right) 
\:.
\label{n_m_Eq}
\end{eqnarray}
Note here the surface term vanishes in all cases.

Substitute Eqs.~(\ref{P_Def_Eq})--(\ref{n_m_Eq}) and the expression for $G$,
Eq.~(\ref{G_Eq}), into Eq.~(\ref{Lambda_K0_Eq}), after performing the Matsubara
frequency sum, we obtain
\begin{equation} 
\lambda^{-2}_L =
\frac{2}{d}\sum_{\alpha,\mb{k}}\frac{\Delta_{sc}^2}{\Ek^2} \left [
  \frac{1-2f(\Ek)} {2\Ek}+f^\prime(\Ek)\right] \left [
  \left(\frac{\partial\ek}{\partial k_\alpha}\right)^2 \phik^2 -\frac{1}{4}
  \frac{\partial \ek^2}{\partial k_\alpha} \frac{\partial \phik^2} {\partial
    k_\alpha}\right] \:.
\label{Lambda_General_Eq}
\end{equation}
Inserting the appropriate $\ek$ and $\phik$, we can obtain results for 3D
$s$-wave jellium, as well as for a
quasi-2D lattice (in-plane) with $s$- and $d$-wave
pairing. For example, for 3D $s$-wave jellium, we have
\begin{equation}
  n_s=m\lambda^{-2}_L = \frac{2}{3} \sumk
  \frac{\Delta_{sc}^2\phik^2}{\Ek^2}
  \left[\frac{1-2f(\Ek)}{2\Ek} + f'(\Ek)
  \right] \left[\ek(3-\phik^2)+2\mu\right] \;.
\label{Lambda_3D_Eq}
\end{equation}
For a quasi-2D lattice, the $s$-wave in-plane penetration depth, 
(with $\phik=1$), becomes
particularly simple, 
\begin{equation}
  \lambda^{-2}_{ab} =  \sumk
  \frac{\Delta_{sc}^2}{\Ek^2}
  \left[\frac{1-2f(\Ek)}{2\Ek} + f'(\Ek)
  \right] 4t_\parallel^2 (\sin^2 k_x + \sin^2 k_y) \:.
\label{Lambda_2D-S_Eq}
\end{equation}
Finally, for $d$-wave, with $\phik=\cos k_x -\cos k_y$, we find a slightly
more complicated result for the in-plane penetration depth,
\begin{eqnarray}
  \lambda^{-2}_{ab} &=&  \Delta_{sc}^2\sumk
  \frac{\phik^2}{\Ek^2}
  \left[\frac{1-2f(\Ek)}{2\Ek} + f'(\Ek)
  \right]\nonumber\\
&& \times \left[ 4t_\parallel^2 (\sin^2 k_x + \sin^2 k_y)
  -2t_\parallel (\cos k_x + \cos k_y)\ek\right] \:.
\label{Lambda_Dwave_Eq}
\end{eqnarray}

Equation (\ref{Lambda_General_Eq}) is also valid for the out-of-plane
penetration depth, with $d=1$ and $\alpha=z$. For use later, we simply
write it down:
\begin{equation}
\lambda_c^{-2} =  \Delta_{sc}^2\sumk \frac{\phik^2}{\Ek^2}
  \left[\frac{1-2f(\Ek)}{2\Ek} + f'(\Ek)
  \right] (8t_\perp^2 \sin^2 k_z) \:.
\label{Lambda_c_Eq}
\end{equation}

In the absence of a pseudogap, $\Delta_{sc}=\Delta$. Then
Eq.~(\ref{Lambda_General_Eq}) is just the usual BCS formula, as in
Ref.~\cite{Schrieffer}, except that we now allow for a more general $\ek$
and $\phik$. Finally, we can define a relationship between two length scales
\begin{equation}
\lambda_L^{-2} = \frac{\Delta_{sc}^2}{\Delta^2} \lambda_{BCS}^{-2} \:,
\label{Lambda_BCS_Eq}
\end{equation}
where $\lambda_{BCS}^{-2}$ is just $\lambda_L^{-2}$ with the overall prefactor
$\Delta_{sc}^2$ replaced with $\Delta^2$ in
Eq.~(\ref{Lambda_General_Eq}). Obviously, in the pseudogap phase,
$\lambda_{BCS}^{-2}$ does not vanish at $T_c$.

\begin{figure}
\centerline{\includegraphics[angle=-90, width=5.2in, clip]{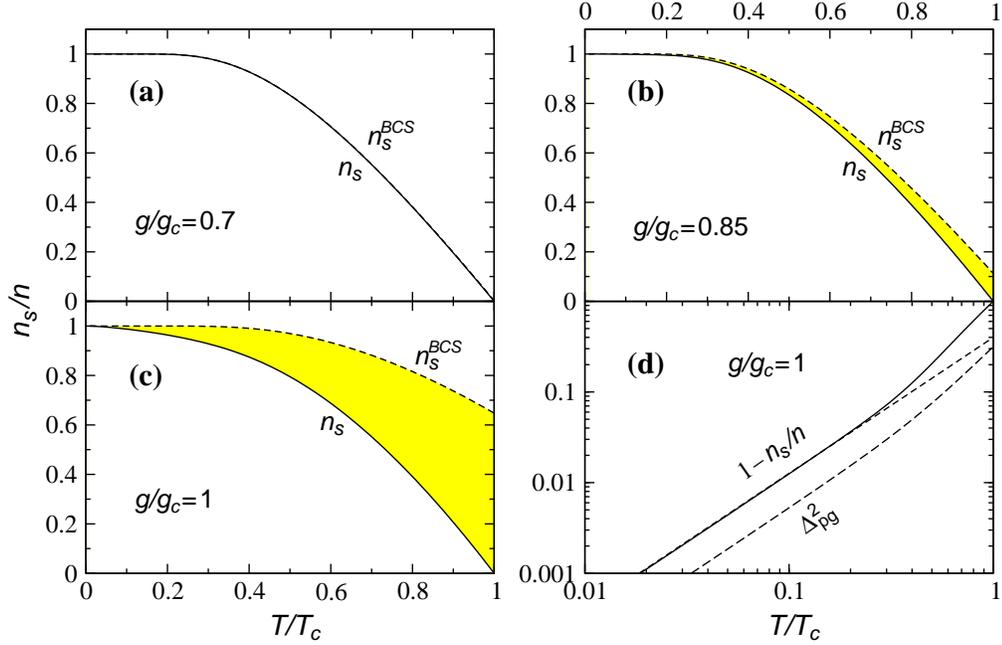}}
\caption[Temperature dependence of the normalized superfluid density for
different coupling strength from weak to moderately strong in 3D
jellium.]{Temperature dependence of the normalized superfluid densities
  $n_s$, as well as comparison with $n_s^{BCS}$, for the same coupling
  strengths as in Fig.~\ref{3D_BT_Gaps}(a)--(d). Shown in (d) are $1-n_s/n$
  and a $T^{3/2}$ power law fit (dashed line) as well as $\Delta_{pg}^2$ as
  a function of $T/T_c$ for $g/g_c=1$ on a log-log scale. $n_s$ and
  $n_s^{BCS}$ are calculated via Eqs.~(\ref{Lambda_General_Eq}) and
  (\ref{Lambda_BCS_Eq}).  The shaded regions help to clarify pseudogap
  effects.}
\label{3D_Ns}
\end{figure}

Shown in Fig.~\ref{3D_Ns} is the temperature dependence of the normalized
superfluid density, $n_s/n$, for 3D $s$-wave jellium, calculated from
Eq.~\ref{Lambda_3D_Eq} for the same three representative values $g/g_c$ as
in Fig.~\ref{3D_BT_Gaps}. For comparison, we also plot $n_s^{BCS}/n$, which
is calculated via $\lambda_{BCS}^{-2}$. In agreement with
Fig.~\ref{3D_BT_Gaps}, for sufficiently weak coupling ($g\leq 0.7$), $n_s$
essentially follows the BCS calculation, the two curves are
indistinguishable. As $g$ increases, the separation between the two curves
becomes evident, particularly in the vicinity of $T_c$.  However, at zero
temperature $n_s = n $, independent of the coupling. This comparison thus
demonstrates how different are these ``pseudogap'' superconductors. Away
from the weak coupling regime, one has to consider the distinction between
the excitation gap $\Delta$ and the order parameter $\Delta_{sc}$.  The
larger the coupling $g$, the larger the difference between
$\lambda_{BCS}^{-2}$ and $\lambda^{-2}_L$.  The superfluid density reflects
most directly the temperature dependence of $\Delta_{sc}$, \textit{not} the
excitation gap. It should be noted that with our calculations of $\lambda$,
\textit{one can uniquely determine the three different gaps below $T_c$}:
Measure $\Delta$ with ARPES or tunneling experiments, then construct
$\lambda_{BCS}^{-2}$. Then measure the penetration depth $\lambda_L$, which
can now be done with fairly good accuracy. Finally one obtains
$\Delta_{sc}^2 = \Delta^2 \lambda^{-2}_L /\lambda_{BCS}^{-2}$ and then
$\Delta_{pg}$.

These bosonic pair excitations will lead to new low $T$ power laws. For an
isotropic $s$-wave superconductor, $\lambda^{-2}_{BCS}$ shows an exponential
behavior at low $T$ due to the isotropic gap in the fermionic excitation
spectrum \cite{Schrieffer}.  With this weak $T$ dependence, the $T$
dependence of $\lambda_L^{-2}$ will be predominantly determined by the
overall prefactor, $\Delta_{sc}^2/\Delta^2$ in Eq.~(\ref{Lambda_BCS_Eq}),
which derives from the bosonic or pair contributions. As expected from
Eq.~(\ref{PG_Eq}), Fig.~\ref{3D_BT_Gaps} shows that $\Delta_{pg}^2$ follows
a $T^{3/2}$ dependence.  Therefore, the temperature dependence of the normal
fluid density $n_n/n\equiv 1-n_s/n$ will also follow a $T^{3/2}$ dependence
in the pseudogap phase at low $T$. Shown in Fig.~\ref{3D_Ns}(d) are $n_n/n$
and $\Delta_{pg}^2$, as well as a $T^{3/2}$ fit (dashed line) on a
logarithmic scale. Obviously, these two quantities have the same $T^{3/2}$
power law.

For $d$-wave pairing, the fermionic quasiparticle density of states is
linear in energy because of the existence of gap nodes on the Fermi surface.
Thus, $\lambda^{-2}_{BCS}$ has a linear $T$ dependence.  The bosonic and
fermionic contributions add so that the the overall low $T$ dependence of
$\lambda^{-2}$ is now given by
\begin{equation}
\frac{\lambda_L(0)^2}{\lambda_L(T)^2} = 1-A\frac{T}{T_c}-B^\prime 
\left(\frac{T}{T_c} \right)^{3/2} \:,
\label{Lambda_D_LowT_Eq}
\end{equation}
where $A$ and $B^\prime$ are expansion coefficients, and explicitly
calculated in Chapter~\ref{Chap_Cuprates}.  In the BCS limit, as expected,
$B^\prime$ vanishes.

%$A=\frac{2\ln 2}{\pi} T_c \lambda_L(0)^2 \frac{v_F}{v_2}$
%This is a quasi-linear
%$T$ dependence, since $T^{3/2}$ power law is very close to linear $T$ 

\begin{figure}
\centerline{\includegraphics[width=4in]{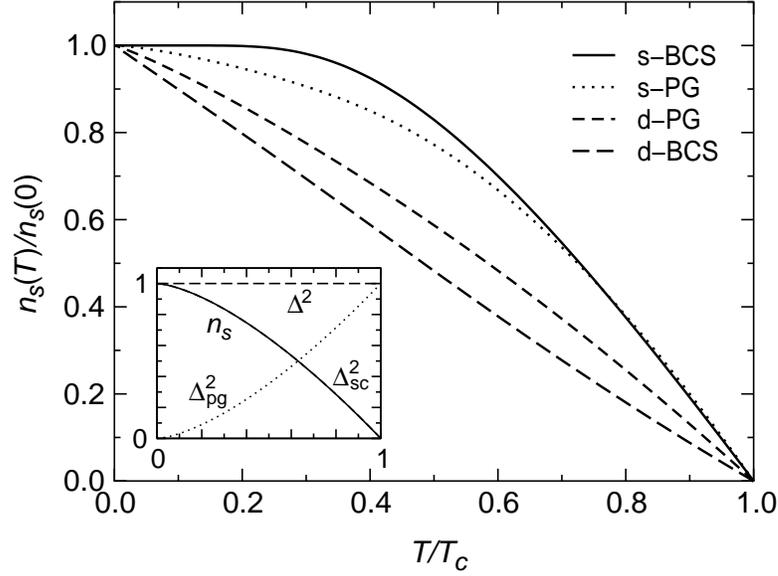}}
\vskip -0.1in
\caption[Temperature dependence of the normalized superfluid density in 
$s$- and $d$-wave, BCS and pseudogap cases on a quasi-2D
lattice.]{Temperature dependence of the normalized superfluid density in
  $s$- and $d$-wave, BCS and pseudogap cases on a quasi-2D lattice with
  $t_\perp/t_\parallel=0.01$. The parameters ($n$, $-g/4t_\parallel$) are:
  for $s$-BCS: (0.5, 0.5); $s$-PG: (0.5, 0.7); $d$-BCS: (0.8, 0.225);
  $d$-PG: (0.92, 0.56). Plotted in the inset are the $T$ dependence of the
  normalized $n_s$, $\Delta_{sc}^2$, $\Delta^2$, and $\Delta_{pg}^2$ for
  $g/g_c=1.5$ and $k_0=4k_F$ in 3D jellium with $s$-wave pairing. The first
  two curves overlap.}
\label{Ns_Comp}
\end{figure}

Plotted in Fig.~\ref{Ns_Comp} is the temperature dependence of the
normalized superfluid density for $s$- and $d$-wave pairing with and without
pseudogap effects.  The fermionic contributions at weak coupling (labeled by
``BCS'') lead to an exponentially flat $s$-wave curve at low $T$, while the
weak coupling (BCS) $d$-wave curve is almost a straight line, going through
the diagonal. In the intermediate coupling regime (labeled by ``PG''), the
pair excitations add a new $T^{3/2}$ power law to both, and bring them
closer. Both the $s$- and $d$-wave curves now have a higher curvature at low
$T$.

The new $T^{3/2}$ power law implies that an extremely careful data analysis
is necessary when one tries to extract information about the order parameter
symmetry from low $T$ power laws. From this figure, we see that, except at
very low $T$, $T^{3/2}$ appears very close to a linear $T$ dependence;
therefore, a pseudogapped $s$-wave curve might be mistaken as $d$-wave.

Both Fig.~\ref{3D_Ns} and Fig.~\ref{Ns_Comp} demonstrate that as $T$
increases, \textit{the superfluid density is destroyed by both
  quasiparticles and pair excitations.} This can be most clearly seen in an
$s$-wave strong coupling case.  Shown in the inset of Fig.~\ref{Ns_Comp} are
the normalized $n_s$, $\Delta^2$, $\Delta_{pg}^2$, and $\Delta_{sc}^2$ in the
BEC regime ($g/g_c=1.5$, $k_0=4k_F$, and $\mu/E_F \approx -3.69$) in 3D
jellium with $s$-wave pairing. Clearly, the $n_s$ and $\Delta_{sc}$ curves
cannot be distinguished. As noted in Chapter \ref{Chap_Theory}, in the BEC
regime, the pseudogap equation (\ref{PG_Eq}) becomes the boson number
equation; $a_0 \Delta^2$, $a_0 \Delta^2_{pg}$, and $a_0 \Delta^2_{sc}$
correspond to $n/2$, $n_n/2$, and $n_s/2$, respectively. Therefore, in the
BEC regime, Eq.~(\ref{Lambda_3D_Eq}) can be simplified as:
\begin{equation}
\frac{n_s}{n\;} = \frac{\Delta_{sc}^2}{\Delta^2} =
1-\left(\frac{T}{T_c}\right)^{3/2} \:, \qquad\qquad \mbox{(in BEC regime)} \:.
\label{Ns_BEC_Eq}
\end{equation}
This equation is familiar from the Bose-Einstein condensation of an ideal
Bose gas. [It is interesting that in the BEC regime, the superfluid density
$n_s$ and the electron density in the condensate are essentially the same.
This is different from the BCS case.]

It should be emphasized that the power index $3/2$, appearing in the
superfluid density, is not very sensitive to the anisotropy in the
dispersion.  However, as the anisotropy ratio $t_\perp/t_\parallel$ becomes
extremely small, this power eventually decreases to values somewhere between
3/2 and 1.
%This insensitivity is not true for the
%specific heat, as we will see below.

\section{Low temperature specific heat}
\label{Sec_Cv}

Finite momentum pairs are expected to show up in thermodynamical properties.
In this section, we explore the consequences of this pseudogap effect on the
specific heat, $C_v$.

Ideally one should calculate the thermodynamical potential for our crossover
problem, from which all thermodynamical properties are determined.
Unfortunately, our approach is based an equation of motion method, which in
general leads to a non-phi derivable theory, and there is no simple
expression for the thermodynamic potential. But nevertheless, we can still
calculate the specific heat based on some reasonable approximations.

In the intermediate coupling regime, the system is composed of the
superconducting condensate and two types of thermal excitations:
quasiparticles and finite center-of-mass momentum pairs (bosons) --- This
leads to a simple three fluid model. Since the phase coherent condensate
does not contribute to the entropy, the entropy comes from the latter two
components. We will now calculate $C_v$ via the entropy.

\begin{figure}
\centerline{\includegraphics[width=2in]{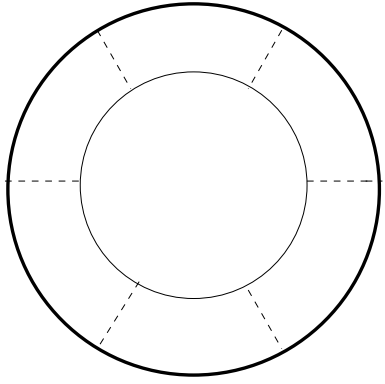}}
\caption[Diagrammatic representation of the pair excitation contribution to
the thermodynamic potential.]{$T$ matrix ring diagram contribution to the
  thermodynamic potential. The thin and thick lines are the bare and full
  Green's functions, respectively.}
\label{G0G_ThermoPotential}
\end{figure}

The approximation we made in Eq.~(\ref{Sigma_PG_Approx}) is equivalent to
treating all the finite momentum pairs as if they were in the condensate (of
course, the distribution of these pairs still obeys a Bose distribution),
solely for the purpose of calculating their contribution to the single
particle self-energy, $\Sigma_{pg}$. As a consequence, the full Green's
function has the BCS form, which depends on the total excitation gap
$\Delta$ instead of the order parameter $\Delta_{sc}$. The contribution to
the thermodynamic potential $\Omega$ from this simplified, BCS-like Green's
function will lead to an entropy given purely by the fermionic quasiparticle
excitations, as in BCS \cite{Schrieffer}:
\begin{equation}
S_{QP}=-2\sumk \Big\{ f(\Ek)\ln f(\Ek) + [1-f(\Ek)] \ln [1-f(\Ek)] \Big\} 
\:.
\label{S_QP_Eq}
\end{equation}
To calculate thermodynamics, this fermionic contribution is not sufficient.
One has to consider the degree of freedom introduced by the pair
excitations, just as one treats the collective mode contribution to the
entropy in Fermi liquid theory. In the first order contribution, we neglect
the possible interactions between pairs. In this way, the pair excitations
will contribute to the entropy as in an ideal Bose gas:
\begin{equation}
  S_{pair} = - \sum_{\mb{q}\ne \mb{0}} \Big\{ b(\Omegaq)\ln b(\Omegaq)
  -[1+b(\Omegaq)] \ln [1+b(\Omegaq)] \Big\} \:.
\label{S_Pair_Eq}
\end{equation}
This pair contribution can be derived from a $T$ matrix ring diagram
contribution to the thermodynamic potential, as shown in
Fig.~\ref{G0G_ThermoPotential}. This corresponds to a term in the
thermodynamical potential:
\begin{equation}
\Omega_{pair} = \sum_Q \ln [1+g\chi(Q)] \:,
\label{ThermoPotential_Eq}
\end{equation}
with $\chi(Q)$ given by Eq.~(\ref{chi_Eq}). In calculations of the
thermodynamical potential the contribution of Eq.~(\ref{ThermoPotential_Eq})
appears (with properly redefined $\chi$) in both the $G_0G_0$ scheme (see
Ref.~\cite{NSR}) and the $GG$ scheme (see Ref.~\cite{Serene}), and we
believe it also appears in the present $G_0G$ scheme. In Appendix
\ref{App_Cv}, we derive Eq.~(\ref{S_Pair_Eq}) from
Eq.~(\ref{ThermoPotential_Eq}), following the Brinkman and Engelsberg
treatment of the paramagnon problem \cite{Brinkman}.

The total entropy is then given by
\begin{equation}
S=S_{QP}+S_{pair} \:.
\label{Entropy_Eq}
\end{equation}
In the weak coupling limit, the quasiparticle contribution $S_{QP}$
dominates. As $g$ increases, the pair contribution $S_{pair}$ becomes more
important. In the strong coupling bosonic regime (which is accessible at low
density with $s$-wave pairing), where the excitation gap becomes too large
for fermionic quasiparticles to be thermally excited, the contribution comes
predominantly from $S_{pair}$, as expected.

It follows that the specific heat is given by
\begin{equation}
C_v = T\frac{\partial S}{\partial T} = 2 \sumk \Ek \frac{\partial
  f(\Ek)}{\partial T} + \sumq \Omegaq \frac{\partial b(\Omegaq)} {\partial
  T} \:.
\label{Cv_Eq}
\end{equation}
Since $C_v$ involves the derivatives ${\displaystyle \frac{\partial
    \mu}{\partial T}}$, ${\displaystyle \frac{\partial \Delta}{\partial
    T}}$, and ${\displaystyle \frac{\partial M^*}{\partial T\;}}$, and these
derivatives are highly non-trivial, in general, we have to calculate $C_v$
numerically from the entropy.
However, we can still obtain the qualitative behavior at low $T$, based on
knowledge from both BCS theory and from calculations involving free bosons.
For $s$-wave pairing, the quasiparticle contribution shows an exponential
behavior at low $T$. For $d$-wave pairing, there are nodes on the Fermi
surface so that the density of states $N(E) \propto E$; therefore, $C_v$
shows a quadratic $T$ dependence. Thus we have
%
%\begin{eqnarray}
%C_{QP} &\propto & (\Delta_0/T)^{1/2} e^{-\Delta_0/T} \:, \qquad \mbox{for
%  \textit{s}-wave}, \nonumber\\
% & \propto& T^2 \:, \qquad \qquad\qquad\qquad \mbox{for \textit{d}-wave}.
%\label{Cv_QP_Eq}
%\end{eqnarray}
%
\begin{equation}
  \renewcommand{\arraystretch}{1.5} C_{QP}\propto \left\{
    \begin{array}{l@{\hspace{1cm}}l} \left({\displaystyle
          \frac{\Delta_0}{T}}\right)^{1/2} \!\!\!
      e^{-\Delta_0/T}\:, & \mbox{for \textit{s}-wave}, \\ T^2\:, & \mbox{for
        \textit{d}-wave}. \end{array} \right.
\end{equation}

As shown in Fig.~\ref{3D_BT_Gaps}, the effective pair mass $M^*$ is
essentially temperature independent at low $T$. Therefore, we may take
$\Omegaq$ as $T$ independent. In this way, we obtain the pair contribution 
\begin{equation}  
C_{pair} \propto \left\{\begin{array}{l@{\hspace{1cm}}l}  T^{3/2} \:, &
    \mbox{for 3D}, \\ T \:, & \mbox{for (quasi-)2D}. \end{array} \right.
\label{Cv_Pair_Eq}
\end{equation}
For quasi-2D, there is a crossover temperature, $T_0$, below which
the system is 3D, but above which the system is essentially 2D. This
crossover temperature depends on the mass anisotropy $t_\perp/t_\parallel$,
and is usually very small in a highly anisotropic system like the
cuprates. For this reason, we will neglect the extremely low $T$ 
(inaccessible) limit.

The low $T$ specific ``coefficient'' $\gamma \equiv C_v/T$ is given by
\begin{equation}
\gamma =\left\{\begin{array}{l@{\hspace{1cm}}l}
 \beta T^{1/2} \:, &
\mbox{for 3D \textit{s}-wave}, \\
\gamma^* \:, &
\mbox{for (quasi-)2D \textit{s}-wave}, \\
 \gamma^* + \alpha T  \:, & \mbox{for (quasi-)2D
  \textit{d}-wave}. \end{array} \right.
\label{Gamma_Eq}
\end{equation}
In the BCS limit, $\beta$ and $\gamma^*$ vanish. 

\begin{figure}
\centerline{\includegraphics[angle=-90, width=6in, clip]{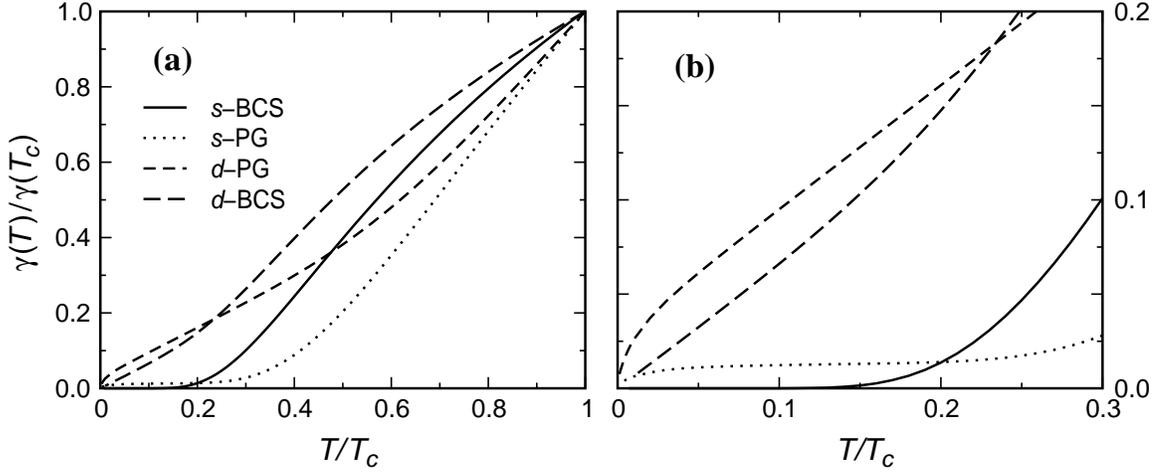}}
\vskip -0.1in
\caption[Temperature dependence (normalized at $T_c$) of the specific
coefficient $\gamma$ for $s$- and $d$-wave pairing with and without
pseudogap.]{(a) Temperature dependence (normalized at $T_c$) of the specific
  coefficient $\gamma$ for $s$- and $d$-wave pairing with and without
  pseudogap. The low $T$ portion is magnified and shown in (b). All other
  parameters are the same as in Fig.~\ref{Ns_Comp}. For the pseudogap cases,
  $\gamma$ has a finite intercept $\gamma^*$ if extrapolated from the data
  between $T/T_c=0.05$ and 0.2.}
\label{Cv_Comp}
\end{figure}

Shown in Fig.~\ref{Cv_Comp}(a) is the temperature dependence of $\gamma(T)$,
normalized at $T_c$, for $s$- and $d$-wave pairing in the BCS and pseudogap
cases. To see the low temperature regime clearly, we magnify the plots, as
shown in Fig.~\ref{Cv_Comp}(b).  In the weak coupling regime (labeled by
``BCS''), the $\gamma(T)$ curves for $s$- and $d$-wave pairing show
exponential and linear $T$ dependences, respectively.  However, in the
intermediate coupling or pseudogap regime (labeled by ``PG''), the
$\gamma(T)$ curves show a finite intercept, $\gamma^*$, on the vertical
axis, when extrapolated from the temperature interval between, say,
0.05$T_c$ and 0.2$T_c$ --- in agreement with Eq.~(\ref{Gamma_Eq}).

To understand why the specific heat is sensitive to the quasi 2D anisotropy,
we note that this occurs because the (anisotropic) pair dispersion $\Omegaq$
appears one more time in the integral for the energy, $E=\sumq \Omegaq
b(\Omegaq)$, whereas in calculating $n_s$, the relevant integral is
$\Delta^2_{pg} \propto \sumq b(\Omegaq)$. It turns out that the latter
integral is far less sensitive to the mass anisotropy ratio used in the
above calculations ($t_\perp/t_\parallel =0.01$), than is the former.

It is worth pointing out that the (spin singlet) pair excitations do not
make their presence explicitly felt in the spin susceptibility or Knight
shift. There are no new low $T$ power laws in these properties or in the
nuclear spin relaxation rate.

\chapter{Gauge invariance and the collective modes}
\label{Chap_CollMode}

\section{Introduction}

In previous chapters, we demonstrated that at finite coupling strength,
there exist soft, finite momentum pair excitations. These excitations are
characteristic of the system in equilibrium, in the absence of the order
parameter oscillations, or collective modes.  It is important to stress that
the pair excitations we have found are different from the collective modes.
As shown in Chapter~\ref{Chap_Theory}, they are determined by a pole of the
mean field $T$ matrix ($t_{pg}$), which is associated with a particular
single-particle self-energy diagram (see Fig.~\ref{Sigma-T} in our case).
Moreover, the pair susceptibility $\chi$, associated with this $T$ matrix,
is in general different from its counterpart in the linear response
collective mode calculations, i.e., the generalized two-particle correlation
function.  This can be seen in a rather straightforward way, even in strict
BCS theory.  The more general two-particle correlation function corresponds
to a $4\times4$ matrix containing multiple correlation functions, e.g., the
density-density, current-current, density-current, and correlations between
the phase and amplitude of the order parameter and between the order
parameter and the current.  The long wavelength expansion of the determinant
of this general matrix determines the collective modes of the
superconducting order parameter.
Within this matrix, the phase modes are associated with a susceptibility
which we call $Q_{22}$, given by Eq.~(\ref{Q22_Eq}), and which we
demonstrate below in Sec.~\ref{sec:PG_EM_exp} is different from pair
susceptibility $\chi$.  This comparison, moreover, serves to underline
the fact that the collective modes introduce effectively higher order
Green's functions (via $F \sim G G_0$) than does the gap equation.  As a
consequence, as we will see in this Chapter, the collective modes contain
some degree of boson-boson interaction.
\textit{The principal effect of the incoherent pair excitations on the
  collective modes is via single particle self-energy and related vertex
  corrections which appear in each of the multiple functions of the
  generalized two-particle correlation function}.

In this Chapter, we will calculate the collective mode spectrum, and show
explicitly that the dispersion of the (soft) phase mode is indeed different
from the pair dispersion. As in BCS, to address the collective modes, one
has to go beyond the level of the gap equation \cite{Kadanoff}, to include
gauge invariance in the problem.  In this way, one sees that the collective
modes introduce higher order Green's functions and thus pair-pair
interactions.  By contrast, as discussed in previous chapters, there are no
pair-pair interactions at the level of the gap equations; these pair-pair
interactions enter at the level of the collective modes. In this context the
bosons are no longer free.
%----------------------------------------------------------------

\section{Electromagnetic response and collective modes of a 
  superconductor: Beyond BCS theory }
\label{sec:gauge}

The purpose of this section is to study the gauge invariant (linear)
response of a superconductor to an external electromagnetic (EM) field, and
obtain the associated collective mode spectrum.  Our discussion is generally
relevant to complex situations such as those appropriate to the BCS-BEC
crossover scenario.  An important ingredient of this discussion is
establishing the role of particle-hole asymmetry.
It should be noted that there are fairly extensive discussions in the
literature on the behavior of collective modes within the $T=0$ crossover
scenario \cite{Belkhir,Micnas92,Cote,Engelbrecht97}.  Here we review a
slightly different formulation \cite{Kulik,Zha} that introduces a matrix
extension of the Kubo formalism of the normal state.  We find that this
approach is more directly amenable to extension to finite $T$, where the
pair fluctuation diagrams need to be incorporated.

The definition of the collective modes of a superconductor must be made with
some precision. We refer to the underlying Goldstone boson of the charged or
uncharged superconductor as the ``AB mode,'' after Anderson
\cite{Anderson58} and Bogoliubov \cite{Bogoliubov}.  This AB mode
appears as a pole structure in the gauge invariant formulation of the
electrodynamic response functions, for example, in the density-density
correlation function. Early work by Prange \cite{Prange} referred to
this as the ``ghost mode'' of the neutral system, since this term is not
directly affected by the long range Coulomb interaction.  By contrast, the
normal modes of the charged or uncharged superconductor, which we shall call
the ``collective modes,'' involve a coupling between the density, phase, and
for the BCS-BEC case, amplitude degrees of freedom.  For these, one needs to
incorporate a many body theoretic treatment of the particle-hole channel as
well.  In crossover theories this channel is not as well characterized as is
the particle-particle channel.

\subsection{Gauge invariant electromagnetic response kernel}
\label{ssec:EM-K}

In the presence of a weak externally applied EM field, with four-vector
potential $A^{\mu} = (\phi, {\bf A})$, the four-current density $J^{\mu} =
(\rho, {\bf J})$ is given by
\begin{equation}
  \label{eq:em1}
  J^{\mu}(Q) = K^{\mu\nu}(Q) A_{\nu}(Q)\;,
\end{equation}
where, $Q\equiv q^{\mu}=(\omega,{\bf q})$ is a four-momentum, and
$K^{\mu\nu}$ is the EM response kernel, which can be written as
\begin{equation}
  \label{eq:em4}
  K^{\mu\nu}(Q) = K_0^{\mu\nu}(Q) + \delta{K}^{\mu\nu}(Q)\;.
\end{equation} 
Here 
\begin{equation}
  \label{eq:em2}
  K_0^{\mu\nu}(\omega,{\bf q}) = P^{\mu\nu}(\omega,{\bf q}) + 
\frac{ne^2}{m}
  g^{\mu\nu}(1-g^{\mu 0})
\end{equation}
is the usual Kubo expression for the electromagnetic response.  We define
the current-current correlation function $P^{\mu\nu}(\tau,{\bf q}) =
-i\theta(\tau)\langle[j^{\mu}(\tau,{\bf q}),j^{\nu}(0,-{\bf q})]\rangle$.
In the above equation, $g^{\mu\nu}$ is the contravariant diagonal metric
tensor, with diagonal elements $(1,-1,-1,-1)$, and $n$, $e$, and $m$ are the
particle density, charge and mass, respectively. In what follows, we will
set $e=1$ for simplicity, as we did in Chapter \ref{Chap_SC_Phase}.

The presence of $\delta{K}^{\mu\nu}$ in Eq.~(\ref{eq:em4}) is due to the
perturbation of the superconducting order parameter by the EM field, i.e.,
to the excitation of the \textit{collective modes} of $\Delta_{sc}$. This
term is required to satisfy charge conservation $q_{\mu}J^{\mu}=0$, which
requires that
\begin{mathletters}
\label{eq:em3}
  \begin{equation}
    \label{eq:em3a}
    q_{\mu}K^{\mu\nu}(Q) = 0\;.
  \end{equation}
Moreover,  gauge invariance yields
\begin{equation}
  \label{eq:em3b}
  K^{\mu\nu}(Q) q_{\nu}  = 0\;,
\end{equation}
\end{mathletters}
Note that, since $K^{\mu\nu}(-Q) = K^{\nu\mu}(Q)$, the two constraints
Eqs.~(\ref{eq:em3}) are in fact equivalent.

The incorporation of gauge invariance into a general microscopic theory may
be implemented in several ways.  Here we do so via a general matrix linear
response approach \cite{Kulik} in which the perturbation of the
condensate is included as additional contributions $\Delta_1+i\Delta_2$ to
the applied external field.  These contributions are self consistently
obtained (by using the gap equation) and then eliminated from the final
expression for $K^{\mu\nu}$.  We now implement this procedure.  Let
$\eta_{1,2}$ denote the change in the expectation value of the pairing field
$\hat{\eta}_{1,2}$ corresponding to $\Delta_{1,2}$.  For the case of an
$s$-wave pairing interaction $g<0$, the self-consistency condition
$\Delta_{1,2}=g\eta_{1,2}/2$ leads to the following equations:
\begin{mathletters}
  \label{eq:em9}
  \begin{equation}
    \label{eq:em9a}
    J^{\mu} = K^{\mu\nu}A_{\nu} = K_0^{\mu\nu}A_{\nu} + R^{\mu
      1}\Delta_1 + R^{\mu 2}\Delta_2\;,
  \end{equation}
  \begin{equation}
    \label{eq:em9b}
    \eta_1 = -\frac{2\Delta_1}{|g|} = R^{1\nu}A_{\nu} + Q_{11}\Delta_1 +
    Q_{12}\Delta_2\;,
  \end{equation}
  \begin{equation}
    \label{eq:em9c}
    \eta_2 = -\frac{2\Delta_2}{|g|} = R^{2\nu}A_{\nu} + Q_{21}\Delta_1 +
    Q_{22}\Delta_2\;,
  \end{equation}
\end{mathletters}
where $R^{\mu i}(\tau,{\bf q})=-i\theta(\tau)\langle[j^{\mu}(\tau,{\bf q}),
\hat{\eta}_i(0,-{\bf q})]\rangle$, with $\mu=0,\ldots,3$, and $i=1,2$; and
$Q_{ij}(\tau,{\bf q})=-i\theta(\tau)\langle[\hat{\eta}_i(\tau,{\bf q}),
\hat{\eta}_j(0,-{\bf q})]\rangle$, with $i,j=1,2$.

Thus far, the important quantities $K_0^{\mu\nu}$, $R^{\mu i }$ and
$Q_{ij}$ are unknowns that contain the details of the appropriate
microscopic model.  We shall return to these later in
Sec.~\ref{sec:PG_EM_exp}. The last two of Eqs.~(\ref{eq:em9}) can be
used to express $\Delta_{1,2}$ in terms of $A_{\nu}$:
\begin{mathletters}
  \label{eq:em10}
  \begin{equation}
    \label{eq:em10a}
    \Delta_1 = -\frac{\tilde{Q}_{22}R^{1\nu} -
      Q_{12}R^{2\nu}}{\tilde{Q}_{11}\tilde{Q}_{22} - Q_{12}Q_{21}}
    A_{\nu}\;,
  \end{equation}
  \begin{equation}
    \label{eq:em10b}
    \Delta_2 = -\frac{\tilde{Q}_{11}R^{2\nu} -
      Q_{21}R^{1\nu}}{\tilde{Q}_{11}\tilde{Q}_{22} - Q_{12}Q_{21}}
    A_{\nu}\;,
  \end{equation}
\end{mathletters}
where $\tilde{Q}_{ii} = 2/|g|+Q_{ii}$, with $i=1,2$. Finally, inserting
Eqs.~(\ref{eq:em10}) into Eq.~(\ref{eq:em9a}) one obtains
\begin{mathletters}
\label{eq:em11}
\begin{equation}
  \label{eq:em11a}
  K^{\mu\nu} = K_0^{\mu\nu} + \delta{K}^{\mu\nu}\;,
\end{equation}
with
\begin{equation}
  \label{eq:em11b}
\delta{K}^{\mu\nu} =
  -\frac{\tilde{Q}_{11}R^{\mu 2}R^{2\nu} +
  \tilde{Q}_{22}R^{\mu 1}R^{1\nu} - Q_{12}R^{\mu 1}R^{2\nu} - 
  Q_{21}R^{\mu 2}R^{1\nu}}{\tilde{Q}_{11}\tilde{Q}_{22} - Q_{12}Q_{21}}. 
\end{equation}
\end{mathletters}

As can be seen from the above rather complicated equation, the
electromagnetic response of a superconductor involves many different
components of the generalized polarizability.  Moreover, in the form of
Eq.~(\ref{eq:em11b}) it is not evident that the results are gauge
invariant. In order to demonstrate gauge invariance and reduce the
number of component polarizabilities, we first rewrite $K^{\mu\nu}$ in a
way which incorporates the effects of the amplitude contributions via a
renormalization of the relevant generalized polarizabilities, i.e.,
\begin{mathletters}
  \label{eq:em14}
  \begin{equation}
    \label{eq:em14a}
    K^{\mu\nu} = {K'}_0^{\mu\nu} + \delta{K'}^{\mu\nu}\;,
  \end{equation}
where
\begin{equation}
  \label{eq:em14b}
  {K'}_0^{\mu\nu} = K_0^{\mu\nu}-\frac{R^{\mu
      1}R^{1\nu}}{\tilde{Q}_{11}}\;
\end{equation}
and
\begin{equation}
  \label{eq:em14c}
  {R'}^{\mu 2} = R^{\mu 2}-\frac{Q_{12}}{\tilde{Q}_{11}}R^{\mu 2}\;, 
\quad
  \tilde{Q}'_{22} = \tilde{Q}_{22} - 
\frac{Q_{12}Q_{21}}{\tilde{Q}_{11}}\;.
\end{equation}
\end{mathletters}
In this way we obtain a simpler expression for $\delta{K'}^{\mu\nu}$:
\begin{equation}
  \quad \delta{K'}^{\mu\nu} = -\frac{{R'}^{\mu
      2}{R'}^{2\nu}}{\tilde{Q'}_{22}}\;.
\label{em:deltaK'}
\end{equation}

We now consider a particular (\textit{a priori} unknown) gauge
${A'}^{\mu}$ in which the current density can be expressed as
$J^{\mu}={K'}_0^{\mu\nu}A'_{\nu}$.  The gauge transformation
\cite{Klemm} that connects the four-potential $A_{\mu}$ in an
arbitrary gauge with $A'_{\mu}$, i.e., $A'_{\mu}=A_{\mu}+i\chi q_{\mu}$,
must satisfy
\begin{equation}
  \label{eq:em5}
  J^{\mu} = K^{\mu\nu}A_{\nu} = {K'}_0^{\mu\nu}\left(A_{\nu}+i\chi
  q_{\nu}\right)\;.
\end{equation}
Now invoking charge conservation, one obtains
\begin{equation}
  \label{eq:em6}
  i\chi = -
\frac{q_{\mu}{K'}_0^{\mu\nu}A_{\nu}}{q_{\mu'}{K'}_0^{\mu'\nu'}q_{
\nu'}}\;, 
\end{equation}
and, therefore,
\begin{equation}
  \label{eq:em15}
  K^{\mu\nu} = {K'}_0^{\mu\nu} - \frac{\left({K'}_0^{\mu\nu'}q_{\nu'}\right)
    \left(q_{\nu''}{K'}_0^{\nu''\nu}\right)}{q_{\mu'}{K'}_0^{\mu'\nu'}q_{
      \nu'}}\;.
\end{equation}
The above equation satisfies two important requirements: it is manifestly
gauge invariant and, moreover, it has been reduced to a form that depends
principally on the four-current-current correlation functions.  (The word
``principally'' appears because in the absence of particle-hole symmetry,
there are effects associated with the order parameter amplitude
contributions that enter via Eq.~(\ref{eq:em14}) and add to the complexity
of the calculations).  Equation~(\ref{eq:em15}) should be directly compared
with Eq.~(\ref{eq:em11b}).  In order for the formulations to be consistent
and to explicitly keep track of the conservation laws (\ref{eq:em3}), the
following identities must be satisfied:
\begin{mathletters}
  \label{eq:em13}
  \begin{equation}
    \label{eq:em13a}
    \left(q_{\mu}{K'}_0^{\mu\nu}\right)\tilde{Q'}_{22} =
    \left(q_{\mu}{R'}^{\mu 2}\right) {R'}^{2\nu}\;, 
  \end{equation}
  \begin{equation}
    \label{eq:em13b}
    \left({K'}_0^{\mu\nu}q_{\nu}\right)\tilde{Q'}_{22} =
    {R'}^{\mu 2}\left({R'}^{2\nu}q_{\nu}\right) \;.
  \end{equation}
\end{mathletters}
These identities may be viewed as ``Ward identities'' for the
superconducting two-particle correlation functions \cite{Zha}.  Any
theory that adds additional self-energy contributions to the BCS scheme
must obey these important equations.  We shall return to this issue in
Sec.~\ref{sec:PG_EM_exp}.

\subsection{The Goldstone boson or AB mode}
\label{ssec:AB_mode}

The EM response kernel [cf.~Eqs.~(\ref{eq:em14})--(\ref{eq:em15})] of a
superconductor contains a pole structure that is related to the underlying
Goldstone boson of the system.  Unlike the phase mode component of the
collective mode spectrum, this AB mode is independent of Coulomb effects
\cite{Prange}. The dispersion of this amplitude renormalized AB mode
is given by
\begin{equation}
  \label{eq:em19}
  q_{\mu}{K'_0}^{\mu\nu}q_{\nu} = 0\;.
\end{equation}
For an isotropic system ${K'}_0^{\alpha\beta} = {K'}_0^{11}
\delta_{\alpha\beta}$, and Eq.~(\ref{eq:em19}) can be rewritten as

\begin{equation}
  \label{eq:em20}
  \omega^2 {K'}_0^{00} + {\bf q}^2 {K'}_0^{11} - 2\omega q_{\alpha}
  {K'}_0^{0\alpha} = 0\;,
\end{equation}
with $\alpha=1,2,3$, and in the last term on the left-hand side of
Eq.~(\ref{eq:em20}) a summation over repeated Greek indices is assumed.
It might seem surprising that from an analysis which incorporates a
complicated matrix linear response approach, the dispersion of the AB
mode ultimately involves only the amplitude renormalized four-current
correlation functions, namely the density-density, current-current, and
density-current correlation functions. This result is, nevertheless, a
consequence of gauge invariance.

At zero temperature ${K'}_0^{0\alpha}$ vanishes, and the sound-like AB mode
has the usual linear dispersion $\omega=\omega_{\bf q}=c|{\bf q}|$ with the
``sound velocity'' given by
\begin{equation}
  \label{eq:em22}
  c^2 = {K'}_0^{11}/{K'}_0^{00} \;.
\end{equation}
The equations in this section represent an important starting point for
our numerical analysis.
 
\subsection{General collective modes}

We may interpret the AB mode as a special type of collective mode which
is associated with $A_{\nu} = 0 $ in Eqs.~(\ref{eq:em10}). This mode
corresponds to free oscillations of $\Delta_{1,2}$ with a dispersion
$\omega= c q$ given by the solution to the equation
\begin{equation}
  \label{eq:em17}
  \text{det}|Q_{ij}| = \tilde{Q}_{11}\tilde{Q}_{22}-Q_{12}Q_{21} = 0\;.
\end{equation}
More generally, according to Eq.~(\ref{eq:em9a}) the collective modes of
the order parameter induce density and current oscillations. In the same
way as the pairing field couples to the mean-field order parameter in
the particle-particle channel, the density operator $\hat{\rho}(Q)$
couples to the mean field $\delta\phi(Q)=V(Q)\delta\rho(Q)$, where
$V(Q)$ is an effective particle-hole interaction that may derive from
the pairing channel or, in a charged superconductor, from the Coulomb
interaction. Here $\delta\rho = \langle\hat{\rho}\rangle-\rho_0$ is the
expectation value of the charge density operator with respect to its
uniform, equilibrium value $\rho_0$.  Within our self-consistent linear
response theory the field $\delta\phi$ must be treated on an equal
footing with $\Delta_{1,2}$ and formally can be incorporated into the
linear response of the system by adding an extra term $K_0^{\mu
  0}\delta\phi$ to the right hand side of Eq.~(\ref{eq:em9a}). The other
two Eqs.~(\ref{eq:em9}) should be treated similarly. Note that, quite
generally, the effect of the ``external field'' $\delta\phi$ amounts to
replacing the scalar potential $A^0=\phi$ by $\bar{A^0} = \bar{\phi} =
\phi+\delta\phi$. In this way one arrives at the following set of three
linear, homogeneous equations for the unknowns $\delta\phi$, $\Delta_1$,
and $\Delta_2$:
\begin{mathletters}
  \label{eq:em23}
  \begin{eqnarray}
    0 &=& R^{10}\delta\phi + \tilde{Q}_{11}\Delta_1 + Q_{12}\Delta_2\;,
    \label{eq:em23a} 
\\ 0 &=& R^{20}\delta\phi + Q_{21}\Delta_1 +
    \tilde{Q}_{22}\Delta_2\;, 
\label{eq:em23b} 
\\ \delta\rho =
    \frac{\delta\phi}{V} &=& K_0^{00}\delta\phi + R^{01}\Delta_1 +
    R^{02}\Delta_2 \;.  
\label{eq:em23c}
  \end{eqnarray}
\end{mathletters}
The dispersion of the collective modes of the system is given by the
condition  that the above Eqs.~(\ref{eq:em23}) have a nontrivial 
solution
\begin{equation}
  \label{eq:em24}
  \left| 
    \begin{array}{ccc}
      Q_{11}+2/|g| & Q_{12} & R^{10} \\
      Q_{21} & Q_{22}+2/|g| & R^{20} \\
      R^{01} & R^{02} & K_0^{00} - 1/V
    \end{array}
  \right| = 0\;.
\end{equation}
In the case of particle-hole symmetry $Q_{12}=Q_{21}=R^{10}=R^{01}=0$
and, the amplitude mode decouples from the phase and density modes; the
latter two are, however, in general coupled.

%-----------------------------------------------------

\section{Effect of pair fluctuations on the electromagnetic response: Some examples}
\label{sec:PG_EM_exp}

Once dressed Green's functions $G$ enter into the calculational schemes, the
collective mode polarizabilities (e.g., $Q_{22}$) and the EM response tensor
$K_0^{\mu\nu}$ must necessarily include vertex corrections dictated by the
form of the self-energy $\Sigma$, which depends on the \textit{T}-matrix
$t$, which, in turn depends on the form of the pair susceptibility
$\chi$.  These vertex corrections are associated with gauge invariance and
with the constraints that are summarized in Eqs.~(\ref{eq:em13}). It can be
seen that these constraints are even more complicated than the Ward
identities of the normal state.  Indeed, it is relatively straightforward to
introduce collective mode effects into the electromagnetic response in a
completely general fashion that is required by gauge invariance.  This issue
was discussed in Sec.~\ref{sec:gauge} as well as extensively in the
literature \cite{Schrieffer,Cote}. The difficulty is in the
implementation.  In this section we begin with a discussion of the $T=0$
behavior where the incoherent pair excitation contributions to the
self-energy corrections and vertex functions vanish. In this section, we
shall keep the symmetry factor $\varphi_{\mathbf{k}}$ explicitly.

\subsection{\textit{T} = 0 behavior of the AB mode  and pair susceptibility}
\label{ssec:T=0_AB}

It is quite useful to first address the zero temperature results since there
it is relatively simple to compare the associated polarizabilities of the AB
mode with that of the pair susceptibility $\chi$.  In the presence of
particle-hole symmetry this collective mode polarizability can be associated
with $Q_{22}$, which was first defined in Eq.~(\ref{eq:em9c}).  In the more
general case (which applies away from the BCS limit) $Q_{22}$ must be
replaced by a combination of phase and amplitude terms so that it is given
by $Q'_{22} = Q_{22} - Q_{12}Q_{21}/\tilde{Q}_{11}$.

We may readily evaluate these contributions in the ground state, where
$\Delta_{sc} = \Delta $. The polarizability $Q_{22}$ is given by 
\begin{eqnarray}
Q_{22}(Q)&=&- \sum_P
 \mbox{\large $[$} G(-P)G(P-Q)+G(P)G(Q-P)  \nonumber\\ 
&&{}+ F^\dagger(P)F^\dagger(P-Q)
+F(P)F(P-Q)\mbox{\large $]$} \varphi_{{\mathbf{p}}-{\mathbf{q}}/2}^2,
\label{Q22_Eq}
\end{eqnarray}
where $G(P)$ and $F^\dag(P)=F(P)$ are given by Eqs.~(\ref{G_Eq}) and
(\ref{BCS_F_Eq2}), respectively, with $\Delta=\Delta_{sc}$ at $T=0$.

Using Eq.~(\ref{BCS_F_Eq2}) and changing variable $P_{\pm} = P\pm Q/2$, one
obtains
\begin{eqnarray}
Q_{22}(Q) &=& -\sum_P \left\{ G(-P_{-}) \left[ G(P_{+}) +
    \Delta_{P_+}\Delta_{P_-} G(-P_+)G_0(P_-)G_0(P_+)\right]
    \right.\nonumber\\
&&{}+ \left. G(-P_+) \left[ G(P_-)+\Delta_{P_+}\Delta_{P_-}
    G(-P_-)G_0(P_+)G_0(P_-)\right] \right\}\varphi_P^2 \nonumber\\
&\equiv& - [A(Q)+A(-Q)] \:,
\end{eqnarray}
where in the first line, we have changed the sign of $P \rightarrow -P$.
Using the Dyson's equation 
\begin{equation}
G_0^{-1}(P) = G^{-1}(P) + \Sigma(P) = G^{-1}(P) - \Delta_P^2 G_0(-P) \:,
\end{equation}
$A(Q)$ can be written as
\begin{eqnarray}
A(Q)  &=& \sum_P G(-P_{-}) \left[ G(P_{+}) +
    \Delta_{P_+}\Delta_{P_-} G(-P_+)G_0(P_-)G_0(P_+)\right] \varphi_P^2
    \nonumber\\
&=& \chi(Q) + \sum_P F(P_-)\left[ F(P_+) - \Delta_{P_-} G(P_+) G_0(-P_-)
\right] \varphi_P^2 \:.
\end{eqnarray}
The second term vanishes if and only if $Q=0$.  At finite $Q$, $A(Q)\ne
\chi(Q)$. In addition, $Q_{22}$ is a symmetrized $A(Q)$.  \textit{As a
  consequence, the phase mode dispersion determined by $\tilde{Q}_{22}=0$ is
  necessarily different from the pair excitation spectrum. And this
  difference is apparent even at the level of BCS theory}.

It is important to stress that at $Q=0$, however, we obtain $A(0) =\chi(0)$.
%
%Now, it can be seen that the pair susceptibility $\chi$ in the pairing
%approximation satisfies
%%
%\begin{equation}
%  \sum_P [G(-P)G(P)+F(P)F(P)]\varphi_{\mathbf{p}}^2  = \sum_P
%  G(P)G_0(-P)\varphi_{\mathbf{p}}^2 = \chi(0)
%\end{equation}
and, moreover, $Q_{12}(0) = Q _{21} (0) = 0 $ so that $Q_{22}^\prime(0) =
Q_{22}(0)=-2\chi(0)$. Therefore, we have
\begin{equation}
\frac{2}{|g|}+Q^\prime_{22}(0)=\frac{2}{|g|}[1+g\chi(0)] = 0 \;.
\end{equation}
\textit{In this way, the AB mode propagator is soft under the same
  conditions which yield a soft pair excitation propagator $t_{pg}(Q)$, and
  these conditions correspond to the gap equation Eq.~(\ref{Gap_Eq2}) at
  $T=0$}.  Moreover, it can be seen that $Q'_{22} ( Q ) = Q'_{22}(-Q)$ so
that, upon expanding around $Q=0$, one has
$\tilde{Q}_{22}(Q)=-\alpha_{22}\Omega^2+\beta_{22}q^2$, $Q_{12}(Q) =
-Q_{21}(Q) = i\Omega\alpha_{12}$, and $2/|g|+Q_{11}(Q)=2/|g|-\alpha_{11}$,
where
\begin{eqnarray}
\alpha_{22}&=&\sum_{\mathbf{k}}
\frac{\varphi_{\mathbf{k}}^2}{4E_{\mathbf{k}}^3} \;,
\nonumber\\
\beta_{22}&=&\frac{1}{d}\sum_{\mathbf{k}}
  \frac{1}{4E_{\mathbf{k}}^3} \left[\varphi_{\mathbf{k}}^2
    (\vec{\nabla} \epsilon_{\mathbf{k}})^2 - \frac{1}{4} (\vec{\nabla}
    \epsilon_{\mathbf{k}}^2)\cdot
    (\vec{\nabla}\varphi_{\mathbf{k}}^2)\right] \;,
\nonumber\\
\alpha_{12}&=&\sum_{\mathbf{k}}
\frac{\epsilon_{\mathbf{k}}}{2E_{\mathbf{k}}^3} \varphi_{\mathbf{k}}^2 \;,
\nonumber\\
\alpha_{11}&=&\sum_{\mathbf{k}}\frac{\epsilon_{\mathbf{k}}^2}
{E_{\mathbf{k}}^3} \varphi_{\mathbf{k}}^2 \;,
\label{eq:expansion}
\end{eqnarray}
where $d$ denotes the dimensionality of the system. Thus, one obtains
\begin{equation}
c^2=\frac{\beta_{22}} 
{\displaystyle \alpha_{22}+\frac{\alpha_{12}^2}{2/|g|-\alpha_{11}}} \:.
\end{equation}
At weak coupling in three dimensions, where one has approximately
particle-hole symmetry, $\alpha_{12}= 0$, the amplitude and the phase modes
decouple. This leads to the well-known result $c=v_F/\sqrt{3}$. More
generally, for arbitrary coupling strength $g$, these equations yield
results equivalent to those in the literature
\cite{Belkhir,Micnas92,Cote,Engelbrecht97}, as well as those derived from
the formalism of Sec.~\ref{ssec:AB_mode}.
Finally, it should be noted that since both Eqs.~(\ref{eq:em15}) and
(\ref{em:deltaK'}) have the same poles, the condition $\tilde{Q}'_{22} (Q) =
0$ yields the same AB mode dispersion as that determined from
Eq.~(\ref{eq:em19}). This is a consequence of gauge invariance.

\subsection{AB mode at finite temperatures}

We now turn to finite temperatures where there is essentially no prior work
on the collective mode behavior in the crossover scenario.  At the level of
BCS theory (and in the Leggett ground state) the extended ``Ward
identities'' of Eqs.~(\ref{eq:em13}) can be explicitly shown to be
satisfied.  Presumably they are also obeyed in the presence of impurities,
as, for example, in the scheme of Ref.~\cite{Kulik}. However, in
general, it is difficult to go beyond these simple cases in computing all
components of the matrix response function.  Fortunately, the calculation of
the AB mode is somewhat simpler. It reduces to a solution of
Eq.~(\ref{eq:em20}), which, \textit{in the presence of particle-hole
  symmetry}, involves a computation of only the electromagnetic response
kernel: the density-density, density-current and current-current correlation
functions.

\begin{figure}
\centerline{\includegraphics[width=3.6in]{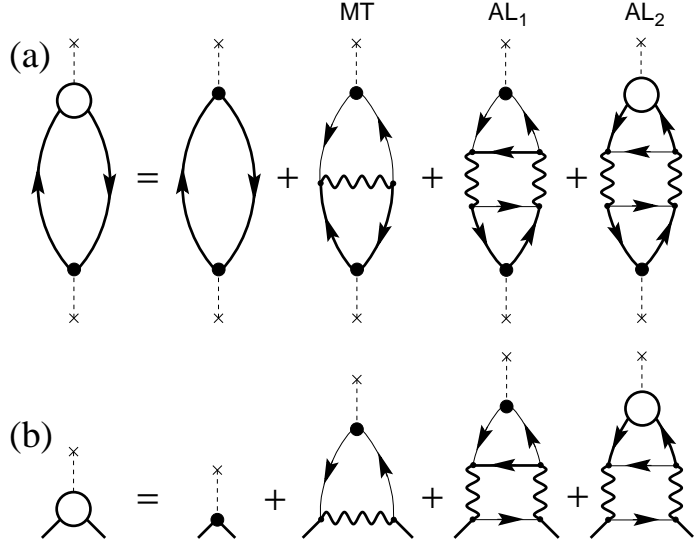}}
  \caption[Diagrammatic representation of the polarization bubble, 
  and the vertex function used to compute the electrodynamic response
  functions.]{Diagrammatic representation of (a) the polarization bubble,
    and (b) the vertex function used to compute the electrodynamic response
    functions. Here the wavy lines represent the pair propagator $t$ and it
    should be noted that the thin and thick lines correspond to $G_0$ and
    $G$ respectively. The total vertex correction is given by the sum of the
    Maki Thompson (MT) and two Aslamazov-Larkin (AL$_1$ and AL$_2$)
    diagrams.}
  \label{fig:diagrams}
\end{figure}

It is the goal of this section to compute these three response functions
within the ``pairing (or $G_0G$) approximation'' to the $T$ matrix.  The
associated \cite{Patton} diagrams are shown in Fig.~\ref{fig:diagrams}.
Because full Green's functions $G$ appear in place of $G_0$ (as indicated by
the heavy lines) these diagrams are related to but different \cite{Patton}
from their counterparts studied by Aslamazov and Larkin and by Maki and
Thompson.  This diagram scheme forms the basis for calculations published by
our group \cite{Kosztin98,Chen99} of the penetration depth within
the BCS-BEC crossover scheme.

Here we make one additional assumption. We treat the amplitude
renormalizations that appear in Eqs.~(\ref{eq:em14}) only approximately,
since these contributions introduce a variety of additional correlation
functions, which must be calculated in a consistent fashion, so as to
satisfy Eqs.~(\ref{eq:em13}).  Because the amplitude mode is gapped, at
least at low $T$, we can approximate these amplitude renormalizations by
their $T=0$ counterparts, which are much simpler to deduce.

The three electromagnetic correlation functions reduce to a calculation
of $P^{\mu\nu}$, which can be written as
\begin{equation}
  P^{\mu\nu}(Q) =2\sum_K \lambda^\mu (K,K-Q) G(K)G(K- Q) \Lambda^\nu
  (K,K-Q)
\label{P_mu_nu}\,,
\end{equation}
where 
\[
\lambda(K,K-Q)=(1, \vec{\nabla}_{\bf k}\epsilon_{{\bf k}-{\bf q}/2})
\]
 and 
\[
\Lambda(K,K-Q)=\lambda(K,K-Q)+\delta
\Lambda_{sc}(K,K-Q) + \delta \Lambda_{pg}(K,K-Q)
\]
are the bare and full vertices, respectively. Note Eqs.~(\ref{lambda_Def})
and (\ref{Lambda_Eq}) in Sec.~\ref{Sec_Ns} are just the spatial components
of these two equations.

To evaluate the vertex function $\Lambda^{\mu}$ we decompose it into a
pseudogap contribution $\Lambda_{pg}$ and a superconducting contribution
$\Lambda_{sc}$. (The latter can be regarded as the Gor'kov $F$ function
contribution, although we do not use that notation here).  The pseudogap
contribution comes from a sum of Maki-Thompson (MT) and Aslamazov-Larkin
(AL$_{1,2}$) diagrams [see Fig.~\ref{fig:diagrams}(b)].  Since these vertex
corrections can be obtained from a proper vertex insertion to the
self-energy, it follows that there is a cancellation between these various
terms that simplifies the algebra.  This cancellation is shown in more
detail in Appendix~\ref{App_Vertex}.
Following the analysis in Appendix~\ref{App_Vertex}, the sum of both $pg$
and $sc$ contributions is given by
\begin{eqnarray}
\label{Vertex}
  \delta \Lambda^\mu (K, K-Q)&\approx& -(\Delta_{sc}^2- \Delta_{pg}^2)
  \varphi_{\mathbf{k}} \varphi_{\mathbf{k-q}}G_0(-K)  G_0(Q-K)
  \lambda^\mu(Q-K,-K)\nonumber\\ 
&&{} - \Delta_{pg}^2 G_0(-K)\frac{\partial
    \varphi^2_{{\mathbf{k}}-{\mathbf{q}}/2}}{\partial k_\mu} \,,
\end{eqnarray}
where use has been made of the fact that $t_{pg}(Q)$ is highly peaked
at $Q=0$, and of the definition of $\Delta_{pg}^2$, Eq.~(\ref{PG_Def}).

The AB mode dispersion involves the sum of three terms that enter into
Eqs.~(\ref{eq:em19}) and (\ref{eq:em20}).  We next substitute
Eqs.~(\ref{Vertex}) into Eq.~(\ref{P_mu_nu}). After performing the Matsubara
frequency summation, and analytically continuing $i\Omega \rightarrow
\Omega+i0^+$, we obtain for small $\Omega$ and $\textbf{q}$
\begin{eqnarray}
  q_\mu K_0^{\mu\nu}q_\nu &=& 
{\mathbf{q}}\cdot \left(\tensor{\frac{\bf n}{\bf m}} + 
\tensor{\bf P}\right) \cdot
{\mathbf{q}} -2 \Omega {\mathbf{q}} \cdot {\mathbf{P}}_0 + 
\Omega^2 P_{00} \nonumber\\
&=& \frac{2}{d}q^2 \sum_{\mathbf{k}}
  \frac{\Delta_{sc}^2}{E_{\mathbf{k}}^2} \left[
    \frac{1-2f(E_{\mathbf{k}})}{2E_{\mathbf{k}}} + f^\prime
    (E_{\mathbf{k}}) \right]\left[\varphi_{\mathbf{k}}^2 (\vec{\nabla}
    \epsilon_{\mathbf{k}})^2 - \frac{1}{4} (\vec{\nabla}
    \epsilon_{\mathbf{k}}^2)\cdot 
(\vec{\nabla}\varphi_{\mathbf{k}}^2)\right] \nonumber\\
&&{} -2\Omega^2 \sum_{\mathbf{k}} \Bigg\{
    \frac{\Delta_{sc}^2\varphi_{\mathbf{k}}^2} {E_{\mathbf{k}}^2} 
\Bigg[ \frac{1-2f(E_{\mathbf{k}})}{2E_{\mathbf{k}}} + f^\prime
      (E_{\mathbf{k}}) \nonumber\\
&&{} - f^\prime (E_{\mathbf{k}}) \frac{\Omega^2
        -({\mathbf{q}}\cdot \vec{\nabla}\epsilon_{\mathbf{k}})^2 - 
\Delta^2 ({\mathbf{q}}\cdot \vec{\nabla} \varphi_{\mathbf{k}})^2} 
{\Omega^2 -({\mathbf{q}}\cdot \vec{\nabla} E_{\mathbf{k}})^2}
    \Bigg]\nonumber\\ 
&&{} +\frac{\Delta_{pg}^2}{4E_{\mathbf{k}}^2} f^\prime 
(E_{\mathbf{k}}) \frac{({\mathbf{q}}\cdot \nabla\epsilon_{\mathbf{k}}^2)
      ({\mathbf{q}}\cdot \vec{\nabla} \varphi_{\mathbf{k}}^2) + 
\Delta^2 ({\mathbf{q}}\cdot \vec{\nabla} \varphi_{\mathbf{k}}^2)^2} 
{\Omega^2 - ({\mathbf{q}}\cdot \vec{\nabla} E_{\mathbf{k}})^2}\Bigg\}\,.
\label{q.P.q}
\end{eqnarray}
Because Eq.~(\ref{q.P.q}) is ill-behaved for long wavelengths and low
frequencies, in order to calculate the AB mode velocity one needs to take
the appropriate limit $\Omega =c q \rightarrow 0$.  By contrast, the
calculation of the London penetration depth first requires setting
$\Omega=0$ (static limit), and then $\textbf{q}\rightarrow \textbf{0}$. The
superfluid density $n_s$ can be calculated from the coefficient of the $q^2$
term in Eq.~(\ref{q.P.q}) [see, also Eq.~(\ref{Q00}) for $Q=0$].  Finally,
the AB mode ``sound'' velocity $c=\Omega/q$, in the absence of the amplitude
renormalization, can be obtained by solving $q_\mu K_0^{\mu\nu}q_\nu =0$.

In the absence of the pseudogap (i.e., when $\Delta_{sc}=\Delta$) the last
term inside $\{\ldots\}$ in Eq.~(\ref{q.P.q}) drops out, and the resulting
analytical expression reduces to the standard BCS result
\cite{Aronov}, which at $T=0$ has the relatively simple form
\begin{equation}
  q_\mu K_0^{\mu\nu}q_\nu = \frac{q^2}{d} \sum_{\mathbf{k}}
  \frac{\Delta_{sc}^2}{E_{\mathbf{k}}^3} \left[\varphi_{\mathbf{k}}^2
    (\vec{\nabla} \epsilon_{\mathbf{k}})^2 - \frac{1}{4} (\vec{\nabla}
    \epsilon_{\mathbf{k}}^2)\cdot
    (\vec{\nabla}\varphi_{\mathbf{k}}^2)\right] -\Omega^2
  \sum_{\mathbf{k}} \frac{\Delta_{sc}^2\varphi_{\mathbf{k}}^2}
  {E_{\mathbf{k}}^3}\,.
\label{qPqatT=0}
\end{equation}
At finite $T$, the AB mode becomes damped, and the real and imaginary parts
of the sound velocity have to be calculated numerically.  Although, the
algebra is somewhat complicated, it can be shown that the AB mode satisfies
$ c \rightarrow 0 $ as $T \rightarrow T_c $, as expected.

To include the amplitude renormalization, using Eq.~(\ref{eq:em14b}), we can
write 
\begin{eqnarray}
 q_\mu {K'}_0^{\mu\nu} q_\nu &=& q_\mu K_0^{\mu\nu}q_\nu-\frac{q_\mu R^{\mu 1}
 R^{1\nu} q_\nu}{\tilde{Q}_{11}}\nonumber\\
& \approx & q_\mu K_0^{\mu\nu}q_\nu - \Omega^2
 \frac{R^{01}R^{10}}{\tilde{Q}_{11}} \;,
\end{eqnarray}
where in the second line, we have used the $T=0$ approximation for the
second term, so that $R^{i1}(0)=0=R^{1i}(0)$ for $i=1, 2, 3$,
$\tilde{Q}_{11}(0) = 2/|g|-\alpha_{11}$, and 
\begin{equation}
R^{10}(0)=R^{01}(0)=-\Delta_{sc}\sum_{\bf k}\frac{\epsilon_{\bf k}} {E_{\bf
    k}^3} \varphi_{\bf k}^2.
\end{equation}
This greatly simplifies the numerical calculations.

It should be noted that the temperature dependence of the amplitude
contribution is always suppressed by the Fermi function $f(E_{\bf k})$. This
amplitude renormalization is negligible at weak coupling strengths, where
the $T$ dependence may be strong near $T_c$ due to the small size of the
gap. On the other hand, when the coupling strength increases, and thus
amplitude effects become more important, the excitation gap becomes large
for all $T\le T_c$, This follows as a result of pseudogap effects.  Hence,
the amplitude contribution is rather insensitive to $T$.  Therefore, it is
reasonable, in both the strong and weak coupling cases, to neglect the $T$
dependence of the amplitude contribution in the numerical analysis.

\section{Numerical results: Zero and finite temperatures}
\label{sec:CM_num_res}

In this section we summarize numerical results obtained for the AB mode
velocity $c$ associated with the electromagnetic response kernel, as
obtained by solving Eqs.~(\ref{eq:em19}) and (\ref{eq:em20}). We also
briefly discuss the behavior for the $T=0$ phase mode velocity $v_{\phi}$
that results from the coupling to density fluctuations, as well [see
Eq.~(\ref{eq:em24})]. The former, which has physical implications for the
behavior of the dielectric constant \cite{Prange,Zha}, is the
more straightforward to compute, because it does not require any new
approximations associated with the effective interactions $V$ in the
particle-hole channel.  The analysis of this section provides information
about the nature of the strong coupling limit, which from the perspective of
the gap equations appears as a ``quasi-ideal'' Bose gas. Here we study this
limit via plots of the infinite $g$ asymptote of the AB mode, called
$c_{\infty}$ and show that in the collective modes there are pair-pair
interactions present (which result from the fact that the collective modes
introduce higher order Green's functions than does the $T$ matrix of the gap
equations).  This analysis also helps to clarify how pair fluctuations
contribute, at finite temperatures, to the collective mode dispersion.  Our
$T=0$ calculations are based on the Leggett ground state which corresponds
to that of the pairing approximation as well. At finite $T$, we numerically
evaluate the AB sound dispersion from Eq.~(\ref{q.P.q}), obtained within the
framework of the pairing approximation.

\begin{figure}
\centerline{\hskip -0.1in \includegraphics[width=6in]{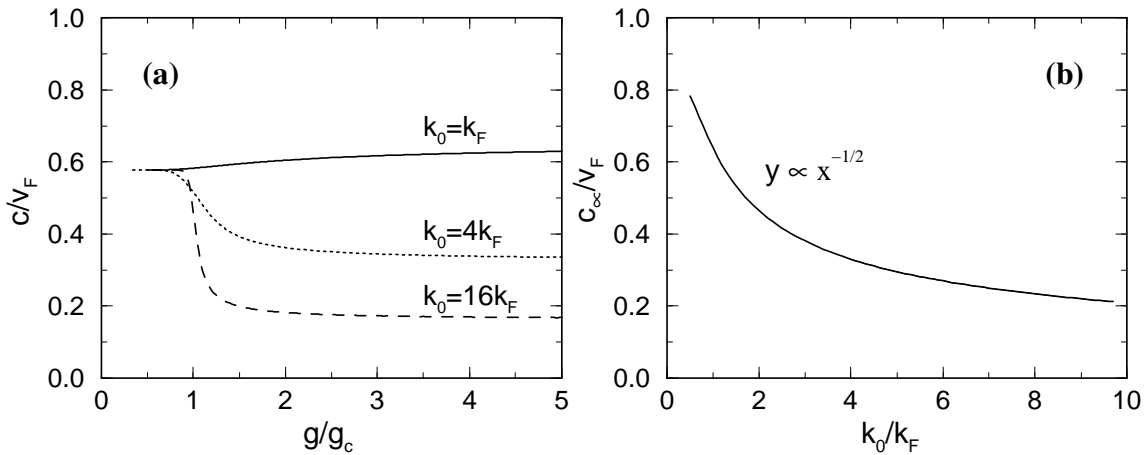}}
\vskip -2.2in 
\hskip 0.68in { \textbf{(a)}} \hskip 4.45in { \textbf{(b)}}
\vskip 1.9in
\caption[AB mode velocity as a function of the coupling strength  
for various densities in 3D jellium, as well as the strong coupling
asymptote as a function of density.]{(a) AB mode velocity $c/v_F$ as a
  function of the coupling strength for various densities characterized by
  $k_0/k_F$ in 3D jellium, as well as (b) the large $g$ asymptote
  $c_{\infty}/v_F$, versus $k_0/k_F$, which varies as $(k_F/k_0)^{1/2}$, as
  expected.  }
\label{CollMode_Fig4}
\end{figure}

In Fig.~\ref{CollMode_Fig4}(a) we plot the zero-temperature value of $c$ as
a function of the coupling strength $g/g_c$ in a 3D jellium model with
$s$-wave pairing at three different electron densities, which are
parameterized via $k_0/k_F$. The most interesting feature of these and
related curves is shown in Fig.~\ref{CollMode_Fig4}(b) where we plot the
asymptotic limit as a function of density or, equivalently, $k_0$.  This
numerically obtained asymptote reflects the \textit{effective} residual
boson-boson interactions in the strong coupling limit of the collective
modes.  This asymptote is close to the value calculated in
Ref.~\cite{Engelbrecht97} whose functional dependence is given by
$c_\infty/v_F\propto\sqrt{k_F/k_0}$ or, equivalently,
$c_\infty\propto\sqrt{n/k_0}$.  Interpreting the physics as if the system
were a true interacting Bose system, one would obtain the effective
interaction $U(0)\approx 3\pi^2/mk_0$, \textit{independent of $g$ in the
  strong coupling limit}.  As expected, these inter-boson interactions come
exclusively from the underlying fermion character of the system, and can be
associated with the repulsion between the fermions due to the Pauli
principle.  All of this is seen most directly
\cite{Haussmann93,SadeMelo,Drechsler} by noting that the behavior displayed
in the inset can be interpreted in terms of the effective scattering length
of the bosons $a_B$, which is found to be twice that of the fermions $a_F$
in the strong coupling limit.  Effects associated with the coupling constant
$g$ are, thus, entirely incorporated into making bosons out of a fermion
pair, and are otherwise invisible.

\begin{figure}
\centerline{ \includegraphics[width=2.97in]{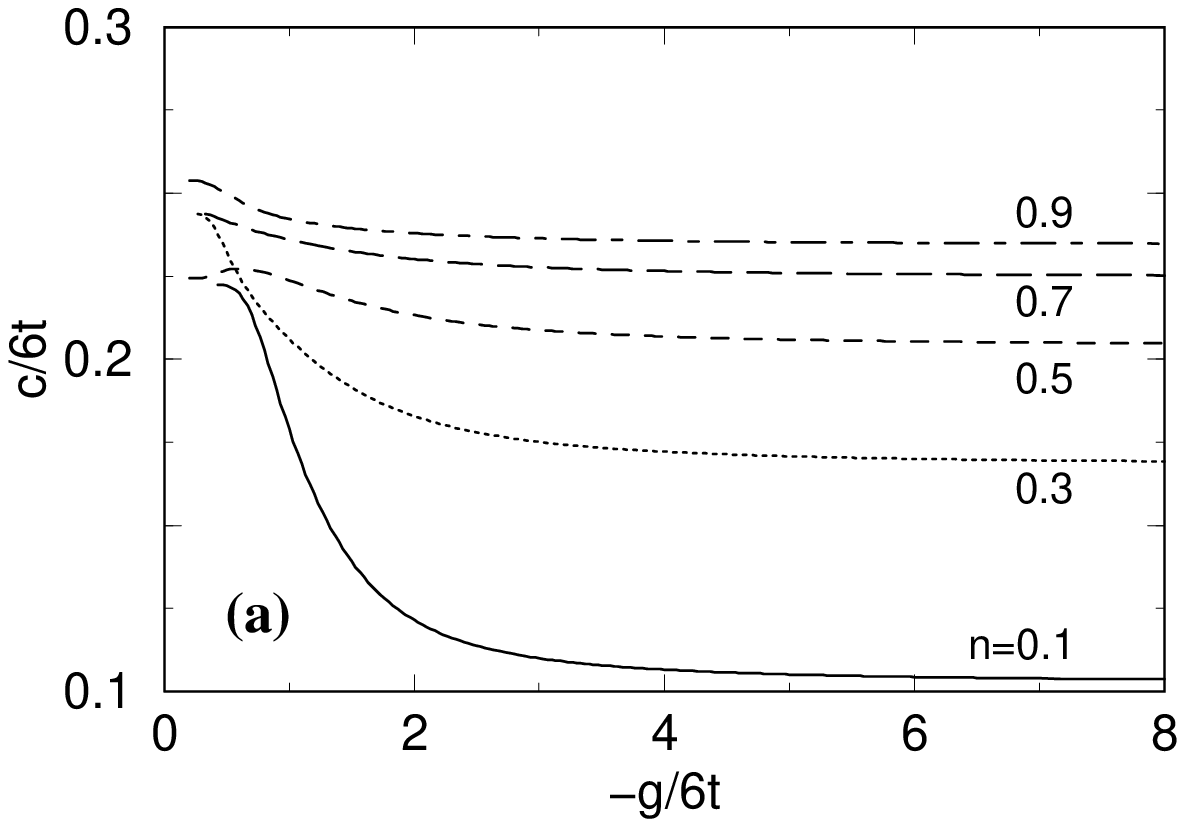}
\includegraphics[width=3.03in]{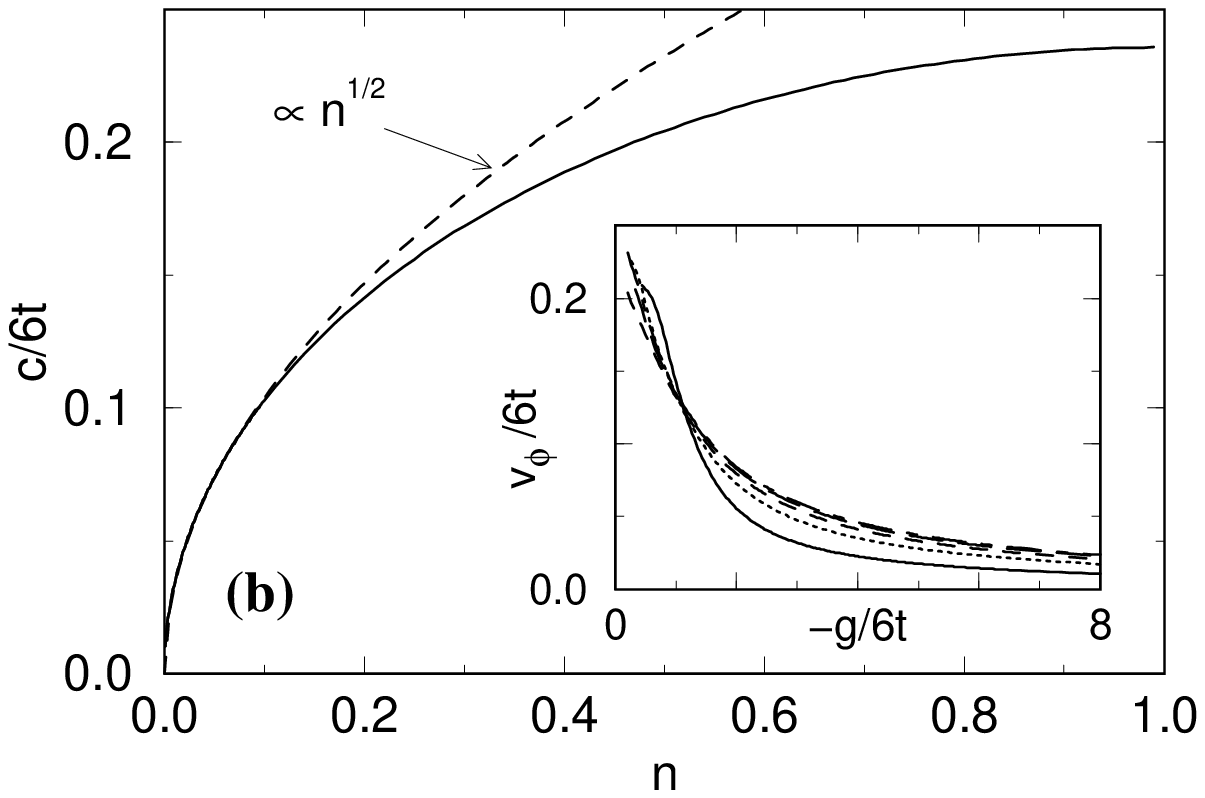}
}
\vskip -0.15in
\caption[Normalized AB mode velocity on a 3D lattice with an $s$-wave 
pairing interaction for various densities as a function of $g$, and the
large $g$ limit for the velocity as a function of density for fixed
$g$.]{(a) Normalized AB mode velocity, $c/6ta$, on a 3D lattice with an
  $s$-wave pairing interaction for various densities as a function of $g$,
  and (b) the large $g$ limit for $c/6ta$ as a function of density $n$ for
  fixed $-g/6t=20$. (Here $6t$ is the half bandwidth). The dashed line in
  (b) shows a fit to the expected low density dependence $n^{1/2}$. Plotted
  in the inset is the velocity of the phase and density coupled collective
  mode $v_{\phi}/ 6ta$ with the particle-hole channel treated at the RPA
  level, for the same $n$ as in (a).  }
\label{CollMode_Fig5}
\end{figure}

The same calculations are repeated in Fig.~\ref{CollMode_Fig5} for a 3D
tight binding lattice band structure with $\varphi_{\bf k}=1$ at $T=0$.
Figure~\ref{CollMode_Fig5}(a) plots the sound velocity for different
densities $n$, as a function of the coupling constant; the behavior of the
large $g$ limit is shown in Fig.~\ref{CollMode_Fig5}(b) as a function of
density for a fixed $g$.  Near half filling, where there is particle-hole
symmetry, the amplitude contributions are irrelevant and the large $g$ limit
for $c$, from Eq.~(\ref{qPqatT=0}), is $c=\sqrt{2}t$, where $t$ is the
hopping integral.  At low $n$ the AB velocity varies as $\sqrt{n}$, which is
consistent with the results shown above for jellium.  In both cases the
behavior again reflects the underlying fermionic character, since it is to
be associated with a Pauli principle induced repulsion between bosons.
Unlike in the jellium case, where $c$ approaches a finite asymptote as $g$
increases, here $c$ vanishes asymptotically due to the increase of the pair
mass associated with lattice effects \cite{NSR,Chen99}. For
completeness, we also show, as an inset in Fig.~\ref{CollMode_Fig5}(b), the
behavior of $v_{\phi}$, where we have used the RPA approximation to
characterize the parameter $V$ in the particle-hole channel. This
approximation is in the spirit of previous work by Belkhir and
Randeria \cite{Belkhir}, although it cannot be readily motivated at
sufficiently large $g$.

\begin{figure}
\centerline{\includegraphics[width=3.5in, clip]{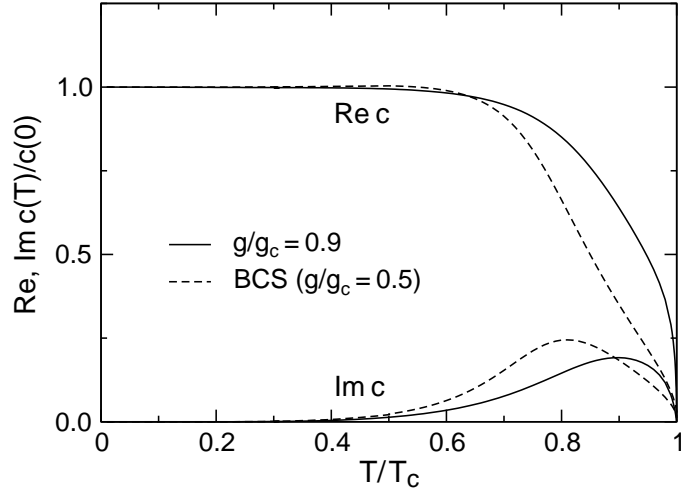}}
\vskip -0.1in
\caption[Temperature dependence of the real and imaginary
parts of the AB mode velocity for moderate coupling and weak coupling BCS in
3D jellium.]{Temperature dependence of the real (Re$\,c$) and imaginary
  (Im$\,c$) parts of the AB mode velocity for moderate coupling (solid
  lines) and weak coupling BCS (dashed lines) in 3D jellium with $k_0=4k_F$.
  The mode is highly damped as $T_c$ is approached.}
\label{CollMode_Fig6}
\end{figure}

Finally, in Fig.~\ref{CollMode_Fig6} we plot the temperature dependence of
the AB mode velocity (both real and imaginary parts), for moderately strong
coupling (solid lines) and the BCS limit (dashed lines).  This figure
suggests that the AB mode velocity reflects the same transition temperature
$T_c$ as is computed via the excited pair propagator or \textit{T}-matrix.
This represents an important self-consistency check on the present
formalism.

\section{Some additional remarks}
\label{sec:conc}

\subsection{Pair excitations \textit{vs} phase fluctuations}

As a result of the distinction between \textit{incoherent} pair
excitations and the \textit{coherent} phase mode of the order parameter, our
theory is different from the phase fluctuation scenario proposed by Emery
and Kivelson \cite{EmeryKivelson}.  In our theory, we deal with a
homogeneous system, in which the order parameter vanishes identically above
$T_c$, and there is no macroscopic occupation of the zero momentum pair
states. In contrast, in the phase fluctuation scenario, the system is
inhomogeneous, consisting of microscopic ``superconducting'' regions, e.g.,
stripes or superconducting grains. In each of these regions, there is a
local (fluctuating) order parameter, and the zero momentum pair state is
macroscopically occupied. Above $T_c$, there is no phase coherence between
these locally condensed regions so that the spatial, as well as temporal,
average of the local order parameter (which defines the order parameter at a macroscopic level) vanishes. 

The pseudogap in both scenarios is associated with the difference between
the excitation gap and the superconducting order parameter. In the present
scenario it is a consequence of incoherent pair excitations. In the
scenario of Emery and Kivelson, this difference is caused by phase
fluctuations.
One can also arrive at a picture (for the $T$-dependence of the excitation
gap and of the order parameter) that is similar to Fig.~\ref{3D_BT_Gaps}
within the phase fluctuation scenario \cite{EmeryKivelson}. What is
different here is the ``tuning parameter,'' which corresponds to the
coupling strength $g$ in the crossover picture, and the phase stiffness
parameter $n_s/m$ in the phase fluctuation picture.

\subsection{Strong coupling limit: Composite \textit{vs} true bosons}

In the strong coupling limit at low density with $s$-wave pairing, our
interpretation of the gap equations leads to the observation that the system
forms essentially an ideal Bose system of tightly bound, non-overlapping
pairs. This quasi-ideal Bose gas behavior is a consequence of the mean field
nature of the ground state wavefunction.  Nevertheless, from the perspective
of the collective modes (which involve effectively higher order
Green's functions), the pairs still have some residual interactions
between each other, owing to the Pauli exclusion between the constituent
fermions. The bosons are composite objects, which are different from true
point-like bosons; for example, they can exchange their constituents through a
scattering process.  Our understanding of composite bosons is still
incomplete although many people have tried to think of this system as a true
Bose liquid. However, \textit{some important distinctions between this
  composite boson system and a true Bose liquid have to be noted.}

A first important distinction can be found, between the composite bosons of
the present theory and ``true'' bosons, at the level of the Leggett ground
state.  From previous work on the BCS-BEC crossover at $T=0$ in the strong
coupling limit, it can be seen that this ground state contains a mix of
quasi-ideal and interacting Bose gas character.  The gap equation is
associated \cite{Leggett} with ``noninteracting diatomic molecules,''
whereas, the collective mode spectrum \cite{Belkhir,Micnas92,Cote} reflects
an effective boson-boson interaction deriving from the Pauli statistics of
the constituent fermions.  This is seen most clearly in jellium models where
the Anderson-Bogoliubov sound velocity remains finite at infinite $g$, with
an asymptote associated with these residual interactions
\cite{Haussmann93,Haussmann94}.

A second important difference between true and the composite bosons
discussed here arises from the fact that this superconducting ground state
corresponds to one in which there is \textit{full condensation} so that, as
in the BCS phase, the condensate fraction $n_0 = n$. By contrast, in a Bose
superfluid there is always a depletion of the condensate at $T=0$, caused by
the existence of an inter-boson repulsion.

As a final important difference, we note that the behavior of the pair
propagator $t(Q)$, which must necessarily be consistent with the $T=0$ gap
and number equations \cite{Leggett}, is highly circumscribed and rather
different from what one might deduce based on the standard model for a Bose
liquid \cite{Fetter}. The fermionic degrees of freedom can never be fully
``integrated out''. The fermionic excitation gap $\Delta$ and the pair
chemical potential $\mu_{pair}=0$ are, moreover, closely inter-related via,
e.g., the gap equation (\ref{Gap_Eq2}).  These effects have no natural
counterpart in the Bose liquid (where the fermionic excitation gap is of no
consequence).

\vskip 4ex

It should finally be noted that we did not consider the Coulomb interaction
in the treatment of the gap-equation-associated $T$ matrix. Coulomb
interactions do, however, affect the collective modes.
Indeed, within the BCS formalism (as well as in the Leggett ground state)
long range Coulomb interactions do not enter in an important way to change
the gap equation structure. If things were otherwise, this would compromise
the self-consistent conditions on $\Delta$ and $\mu$ in the ground state.
Coulomb effects are presumed to be already included in the pairing
interaction.  Indeed, at large $g$, we have seen that essentially all signs
of the two-body ($g$ dependent) fermion-fermion interaction are absent in
the effective boson-boson interaction, which is deduced from the sound mode
velocity.  One might anticipate how Coulomb effects will enter the physics.
For real superconductors, the electron pairs overlap with each other, and
feel the Coulomb interaction predominantly as two nearly independent
constituent fermions, rather than as a rigid point boson.  In this way, the
Coulomb interaction is expected to affect the pair dispersion only
indirectly and rather weakly through the single fermion dispersion.

%In concluding this Chapter, it is worth pointing out one more difference
%between the pair excitations and the order parameter fluctuations. 
%As we see from
%Chapter~\ref{Chap_Theory}, the softness of the pair dispersion is a
%necessary condition for a stable (Cooper) pair to exist in the system. In
%other words, as the temperature decreases from above, the superconductivity
%turns on when and only when the pair dispersion becomes soft (i.e., the $T$
%matrix diverges at $Q=0$, the Thouless criterion). On the contrary, the
%softness of the phase mode tends to destroy the superconductivity. It is for
%this reason that Anderson argued that the Coulomb interaction saved BCS
%theory by pushing the phase mode up to the plasma frequency
%\cite{Anderson58}!
%

%Finally, it is worth pointing out that while the softness of the pair
%dispersion is a necessary condition for establishing the superconducting
%phase, in contrast, a soft phase mode tends to destroy  superconductivity.
%
%

%\include{Chap6}
\chapter{Application to the cuprates}
\label{Chap_Cuprates}

In previous chapters, we have addressed, quite generally, the
superconducting instability of the normal state as well as pseudogap
phenomena for $T\leq T_c$. In this process, we studied the effects of low
dimensionality, lattice band structure, and pairing symmetry.  In addition,
we computed measurable properties below $T_c$ such as the superfluid density
and specific heat.  In this context, we have deduced that there are
important incoherent, finite momentum pair excitations and have seen that
they lead to new low temperature powers in $n_s$ and $C_v$. These power laws
are in addition to the usual contributions arising from the fermionic
quasiparticles.  This series of studies shows how to extend BCS theory (using
the crossover ground state wave function $\Psi_0$) from weak coupling (where
only fermionic quasi-particles are relevant) to moderate coupling (where
both fermionic and bosonic excitations coexist), and finally, to strong
coupling (where the excitations are purely bosonic). Our work represents the
first contribution along these lines to address non-zero temperatures below
$T_c$.

In this Chapter, we will apply these results to the cuprates, and compare
our theoretical predictions with experimental measurements. We begin by
computing a cuprate phase diagram.  Next we study the superfluid density and
make predictions for the quasi-universal behavior of $n_s$ and the $c$-axis
Josephson critical current, as a function of temperature.  We follow with a
study of the low temperature specific heat, and discuss the experimental
evidence for new low $T$ power laws which we predict, in both $n_s$ and
$C_v$.

\section{Cuprate phase diagram}
\label{Sec_CupratePhase}

There is a general consensus \cite{Annett} that the cuprate superconductors
are quasi-2D systems with short coherence lengths $\xi$, having a
$d_{x^2-y^2}$ order parameter symmetry. While the overdoped materials appear
to exhibit Fermi liquid properties, the cuprates become Mott insulators at
half filling, owing to the strong on-site Coulomb repulsion. Both the Mott
insulating state and the generally smooth crossover to Fermi liquid
behavior have to be accommodated in any complete theory of the cuprate phase
diagram.

To address the cuprates, we assume a tight-binding, anisotropic band
structure given by Eq.~(\ref{eq:Epsilon-2d-lattice}), and use the same
$\phik$ given by Eq.~(\ref{eq:d-wave}) in all our calculations.%  
\newfootnote{ It should be noted that experimentalists tend to normalize
  $\phik$ at the $(\pi, 0)$ points in the Brillouin zone, i.e.,
$\phik = \frac{1}{2}(\cos{k_x}-\cos{k_y}) $.
so that when (and only when) the gap values are \textit{involved explicitly in
comparison with experiment} in a given plot, we will multiply the gaps
by a factor of 2. Particularly, we will use $\Delta_0$ to denote the gap
magnitude at $(\pi, 0)$.}
In what follows it should be noted that the coupling strength $g$ enters in
dimensionless form via $g/t_\parallel$, so that one might expect the
characteristic dimensionless coupling strength to increase as the Mott
insulator is approached, since the latter is associated with a decrease in
electronic energy scales, such as $t_\parallel$.  In order to generate
physically realistic values for the various energy scales, we make two
assumptions: (1) We take $g$ as doping-independent (which is not
unreasonable in the absence of any more detailed information about the
pairing mechanism) and (2) incorporate the effect of the Mott transition at
half filling, by introducing a doping concentration ($x$) dependence into
the in-plane hopping matrix elements $t_\parallel$, as would be expected in
the limit of strong on-site Coulomb interactions in a Hubbard
model \cite{Anderson87}. Thus the hopping matrix element is renormalized as
$t_\parallel(x) \approx t_0 (1-n)=t_0x$, where $t_0$ is the matrix element
in the absence of Coulomb effects.  This $x$ dependent energy scale is
consistent with the requirement that the plasma frequency vanish at $x = 0$.

These assumptions now leave us with essentially one free parameter
$-g/4t_0$.  We choose this parameter based on studies in
Appendix \ref{App_n-g}, where we located the cuprates on an $n-g$ phase
diagram, and showed in Fig.~\ref{n-g_Phase} that $0.1< -g/4t_\parallel(x) <
2$. Therefore, there is not much freedom in adjusting this parameter.  We
take $-g/4t_0= 0.047$ to optimize the overall fit of the phase diagram to
experiment. We take $t_\perp/t_\parallel \approx 0.003$,%
\newfootnote{We did not fine-tune the ratio $t_\perp/t_\parallel$ since only
  $T_c$ depends slightly on its value (see Sec.~\ref{Sec_Dimensionality}),
  which is presumably also doping dependent.}
and $t_0 \approx 0.6$~eV, which is reasonably consistent with experimentally
based estimates \cite{t0}. It is worth pointing out that the pseudogap
crossover temperature $T^*$ and the zero temperature excitation gap
$\Delta(0)$ are independent of the choice of $t_\perp/t_\parallel$, as long
as $ t_\perp/t_\parallel \ll 1$. $T_c$ depends on $t_\perp/t_\parallel$ only
logarithmically.

Now we solve the set of equations (\ref{Gap_Eq2}), (\ref{Number_Eq}), and
(\ref{PG_Eq}) for $T_c$ and $\Delta_{pg}(T_c)$ as a function of doping $x$.
We also determine the $T=0$ values for $\Delta(0)$ as a function of $x$
(which derive from the first two equations only).  It should be stressed
that there is essentially an inverse relationship between the $x$-dependence
of $T_c$ and that of $\Delta(0)$ in the underdoped regime.  This can be seen
already in the studies of Chapter 3, where at strong coupling, but still in
the fermionic regime, $T_c$ was suppressed as a result of the growth of the
pseudogap. This comes, in turn, from the fact that, crudely speaking, there
are fewer fermions available to form Cooper pairs, as the pseudogap grows in
strength.

\begin{figure}
\centerline{\includegraphics[width=3.5in]{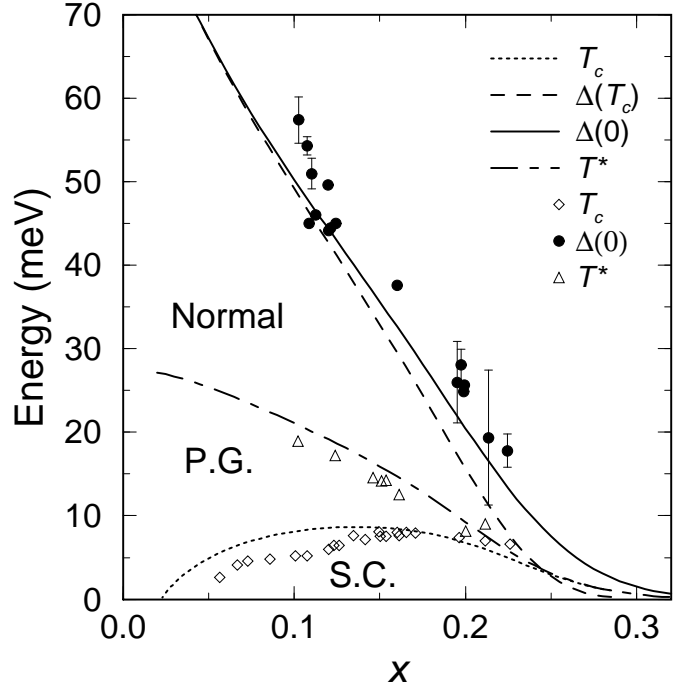}}
\vskip -0.2in
\caption[Cuprate phase diagram, and comparison with experiment.]
{Cuprate phase diagram showing $\Delta (0)$, $T_c$, $\Delta_{pg} (T_c)$, and
  $T^*$, calculated for $-g/4t_0=0.047$, and $t_\perp/t_\parallel =0.003$.
  Shown as symbols are experimental data, taken from: ($\bullet$) Miyakawa
  \textit{et al.} \cite{JohnZ}; ($\diamond$) Rossat-Mignod \textit{et al.}
  \cite{Rossat-Mignod}; ({\scriptsize $\triangle$}) Oda \textit{et al.}
  \cite{Oda}.  The normal, pseudogap, and superconducting phases
  are labeled with ``Normal'', ``P.G.'', and ``S.C.'', respectively. }
\label{Cuprate_Phase}
\end{figure}

As a zeroth order approximation, $T^*$ is roughly given by the mean-field
solution for $T_c$. The results for these quantities are plotted in
Fig.~\ref{Cuprate_Phase}. The relative size of $\Delta_{pg} (T_c)$, compared
to $\Delta (0)$, increases with decreasing $x$.  In the highly overdoped
limit this ratio approaches zero, and the BCS limit is recovered. While $T^*
= T_c$ in the BCS regime, it becomes much larger than $T_c$ in the
underdoped regime. A pseudogap phase exists between $T_c$ and $T^*$, as
observed experimentally. Evidently, the calculated curves are in good
agreement with experimental data taken from the literature, shown as
symbols.  This phase diagram sets the energy scales for use in subsequent
calculations as a function of doping concentration.

Note that, in principle, the anisotropy ratio $t_\perp/t_\parallel$ should
be $x$ dependent; experiment reveals that it is much smaller in the
underdoped regime than in the overdoped regime. If this is taken into
account, $T_c$ will be further suppressed in the underdoped regime, giving 
better agreement for $T_c$ between theory and experiment. However, we choose
a doping independent $t_\perp/t_\parallel$, to keep the physics simpler.

\begin{figure}
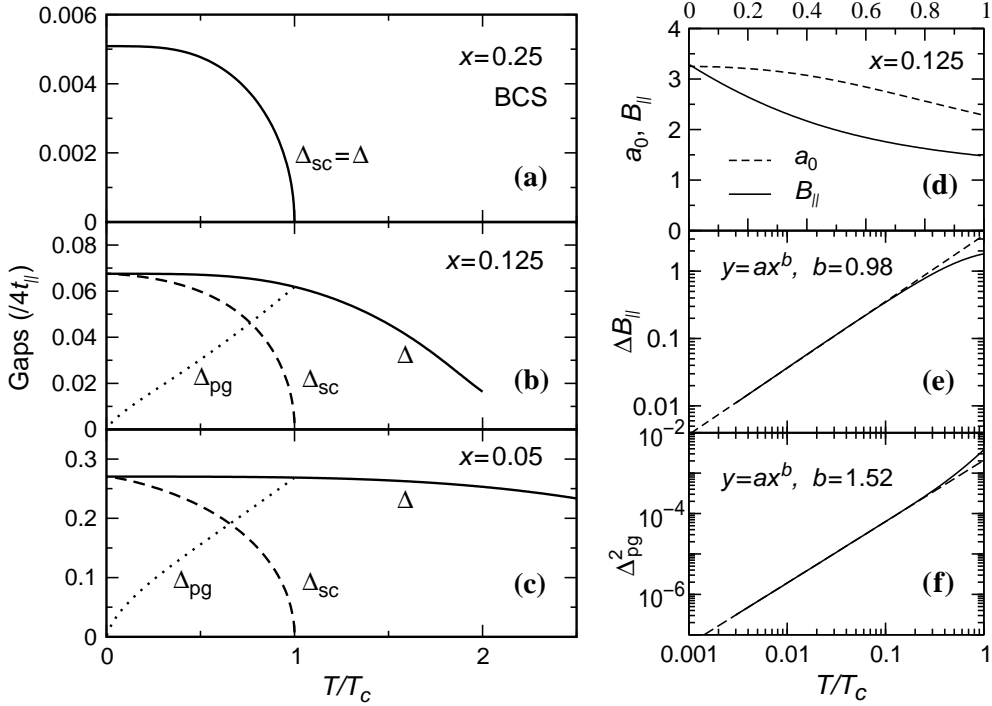

\centerline{\includegraphics[width=3.in, clip]{AtDr100-x-7}\hskip 2ex
\includegraphics[width=2.in, clip]{PMDf.875r100g.36-3.eps}
}
\vskip -0.1in
\caption[Temperature dependence of various gaps for different doping
concentrations.]{Temperature dependence of various gaps for (a) highly
  overdoped, (b) underdoped, and (c) highly underdoped samples. The coupling
  and bandwidth are taken from the theoretical phase diagram,
  Fig.~\ref{Cuprate_Phase}. The gap above $T_c$ is calculated via an
  extrapolation scheme given in Appendix~\ref{App_AboveTc}. Also plotted are
  the temperature dependence of (d) the $T$ matrix expansion coefficients
  $a_0$ and $B_\parallel=1/8t_\parallel M^*_\parallel$, (e) leading order
  correction of $B_\parallel$, and (f) $\Delta^2_{pg}$ on a log-log scale,
  for $x=0.125$.}
\label{Cuprate_Gaps}
\end{figure}

Next we use the different doping concentration dependent energy scales from
this phase diagram, to plot the temperature dependence of the pseudogap
$\Delta_{pg}$, the order parameter $\Delta_{sc}$, and the total excitation
gap $\Delta$ below $T_c$.  Shown in Fig.~\ref{Cuprate_Gaps} are the gaps for
(a) $x=0.25$, (b) $x=0.125$, and (c) $x=0.05$, corresponding to highly
overdoped (BCS), underdoped, and highly underdoped samples, respectively.
We also show the gap \textit{above} $T_c$, which is calculated based on an
extrapolation scheme given in Appendix \ref{App_AboveTc}. In agreement with
experiment, as $x$ decreases, the pseudogap develops, and, thus,
$\Delta_{sc}$ and $\Delta$ separate. In this way, the system evolves from a
BCS-like to a strong pseudogap phase.  Meanwhile, the ratio
$\Delta(T_c)/\Delta(0)$ varies from 0 to 1.  The gap $\Delta$ in (c) is
essentially $T$ independent up to a very high temperature. This can also be
seen from Fig.~\ref{Cuprate_Phase}.  It should be noted that this behavior
is expected as one approaches the BEC regime. See Fig.~\ref{BCS-BEC} , where
this was anticipated.  In comparison with Fig.~\ref{3D_BT_Gaps}, we see
here the decreasing doping level plays the role of an increased coupling
strength $g$, via a shrinking band width. We also show for $x=0.125$ the $T$
matrix expansion coefficients $a_0$ and $B_\parallel$ in
Fig.~\ref{Cuprate_Gaps}(d). Both are roughly a constant at low $T$. In
comparison with the $s$-wave case, the mass $M_\parallel^*$ has a stronger
$T$ dependence. The leading order corrections are quadratic for $a_0$ (as in
$s$-wave) and linear in $T$ for $B_\parallel$. The correction $\Delta
B_\parallel$ is plotted in Fig.~\ref{Cuprate_Gaps}(e). Despite the quasi-two
dimensionality and this slightly stronger $T$ dependence of the pair mass,
$\Delta_{pg}^2$ follows a $T^{3/2}$ law, as shown in
Fig.~\ref{Cuprate_Gaps}(f).
Just as in the $s$-wave case shown in Fig.~\ref{3D_BT_Gaps}, $\Delta_{pg}$
vanishes at $T=0$, owing to the pseudogap equation (\ref{PG_Eq}).%  
\newfootnote{It should, however, be noted that this result is applicable only
  to situations where $T_c$ is finite. It does not apply for $x<x_c\approx
  0.025$, where there is no superconducting phase transition at any finite
  $T$. In this low $x$ regime, the system is always in the normal state, and
  Eq.~(\ref{Gap_Eq2}) cannot be satisfied; nevertheless, Eq.~(\ref{PG_Eq}),
  which parameterizes the normal state self-energy \cite{Janko,Maly}, implies
  that there are pairing fluctuation effects associated with finite
  $\Delta_{pg}$ down to $T=0^+$.}

\section{In-plane penetration depth and \lowercase{\textit{c}}-axis
  Josephson critical current}
\label{Sec_Cuprate_Ns}

\subsection{Quasi-universal behavior of normalized superfluid density vs
  $T/T_c$} 
\label{Subsec_Cuprate_Ns}

Once we have obtained the phase diagram, it is straightforward to calculate
the temperature dependence of the penetration depth as a function of doping
concentration. As discussed in Sec.~\ref{Sec_Ns}, the pseudogap contribution
to the superfluid density introduces a new $T^{3/2}$ power law which may
appear as quasi-linear. We now rewrite Eq.~(\ref{Lambda_D_LowT_Eq}) in a
slightly different form:
\begin{equation}
\frac{\lambda^2_L(0)}{\lambda^2_L(T)} = 1-[A+B(T)]\frac{T}{T_c} \:,
\label{Cuprate_Lambda_LowT_Eq}
\end{equation}
where $B(T)=B^\prime \sqrt{T/T_c}$ depends on $T$ very weakly.

\begin{figure}
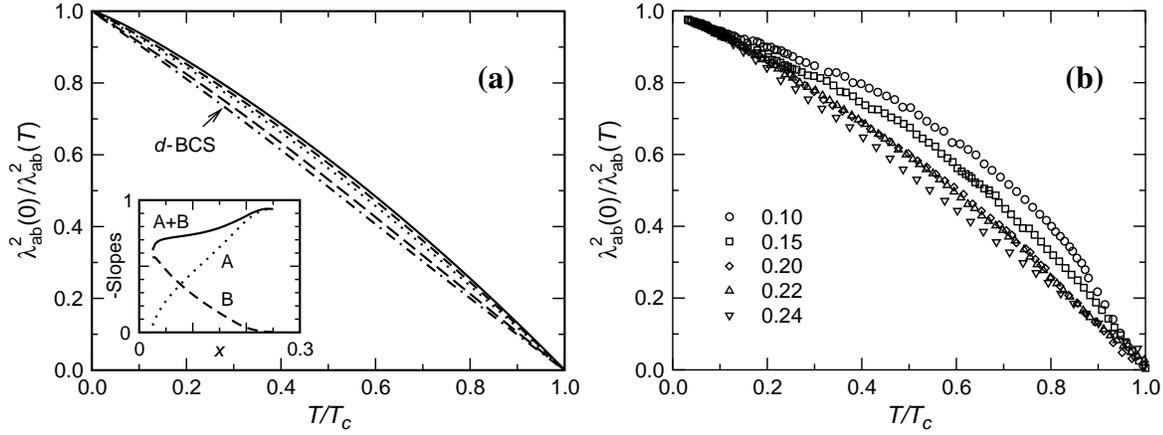

\centerline{\includegraphics[width=3in, clip]{Ns_ab}
\includegraphics[width=3in, clip]{Sr-x-2}}
\vskip -2.in \hskip 2.4in { \textbf{(a)}} 
\hskip 2.8in { \textbf{(b)}}
\vskip 1.72in
\caption[Quasi-universal behavior of the normalized in-plane superfluid 
density as a function of reduced temperature with respect to the doping
concentration.]{(a) Theoretical prediction and (b) experimental measurement
  of the temperature dependence of the ab-plane inverse squared penetration
  depth.  In (a), from bottom to top are plotted for $x=0.25$ (BCS limit,
  dot-dashed line), 0.2 (long-dashed), 0.155 (dotted), 0.125 (dashed) and
  0.05 (solid line). Shown in the inset are (A) the slope given by the low
  temperature expansion assuming a constant $\Delta_{sc}(T)=\Delta (0)$, (B)
  the ratio ${\displaystyle \frac{\Delta_{pg}^2(T)}{\Delta^2
      (0)}\mbox{\Large $/$}\frac{T}{T_c}}$ at $T/T_c=0.2$, and (A+B) the sum
  of two contributions. Experimental data on LSCO shown in (b) are taken
  from Ref.~\protect\cite{Panagopoulos}.}
\label{Cuprate_Ns}
\end{figure}

The normalized in-plane superfluid density, calculated from
Eq.~(\ref{Lambda_Dwave_Eq}), is plotted in Fig.~\ref{Cuprate_Ns}(a) as a
function of $T/T_c$ for several representative values of $x$, ranging from
the highly over- to highly underdoped regimes.  These plots clearly indicate
a ``quasi-universal'' behavior with respect to $x$:
$\lambda^2_L(0)/\lambda_L^2(T)$ \textit{vs.} $T/T_c$ depends only slightly
on $x$. Moreover, the shape of these curves follows closely that of the
weak-coupling BCS theory. The, albeit, small variation with $x$ is
systematic, with the lowest value of $x$ corresponding to the top curve.

This quasi-universal behavior has been confirmed experimentally, as have the
systematic deviations with $x$. Early in 1996, Hardy and
coworkers \cite{Hardy} reported this universality of
the normalized $a$-axis inverse square of the in-plane penetration depth,%
\newfootnote{For the $a$-axis penetration depth, there are no chain effects, in
  YBCO}
$\lambda_a^2(0)/\lambda_a^2(T)$, as a function of $T/T_c$ from underdoped
($T_c=59$~K) to slightly overdoped ($T_c=89$~K) samples in
$\mathrm{YBa_2Cu_3O_{7-\delta}}$ (YBCO) single crystals.% 
\newfootnote{Further evidence comes from the measurements of the ab-plane
  penetration depth on YBCO by the Cambridge group \cite{Cambridge}. These
  authors did not distinguish the $a$- and $b$ axes.}
Recently, Panagopoulos \textit{et al} \cite{Panagopoulos} observed this
quasi-universal behavior of the in-plane penetration depth in
$\mathrm{La}_{2-x}\mathrm{Sr}_x\mathrm{CuO_4}$ (LSCO) polycrystalline
samples with a small, but systematic $x$ dependence of the slope. Their
result, as shown in Fig.~\ref{Cuprate_Ns}(b), are in good agreement with our
predictions.%
\newfootnote{While Hardy \textit{et al.} had the most accurate measurement of
  $\Delta \lambda$ on the best quality YBCO single crystals, they had to
  take the values for $\lambda(0)$ from the literature to calculate the
  superfluid density. It is believed \cite{Xiang} that the values they used
  in Ref.~\cite{Hardy} may be too large for their high purity samples.  In
  addition, since $\lambda(0)$ was unknown in a fully oxygenated (slightly
  overdoped) YBCO sample, they used the value for an optimally doped sample
  instead. As a consequence, the curvature of the superfluid density curves
  were higher than expected. In contrast, Panagopoulos and coworkers
  measured $\lambda(0)$ directly for different doping concentrations.
  Therefore, the slope of the various curves, as well as doping dependence,
  in Fig.~\ref{Cuprate_Ns}(b) may be more reliable. 
%This can be verified
%  from the fact that the curve for the very overdoped sample, $x=0.24$, is
%  very close to the diagonal, as expected from $BCS$ calculations (e.g., the
%  bottom curve in Fig.~\ref{Cuprate_Ns}(a).
  \label{Footnote_Hardy}}

This universal behavior appears surprising at first
sight \cite{LeeWen,Millis} because of the strong $x$ dependence in the ratio
$T_c/\Delta(0)$. To understand the origin of this effect, we plot the
fermionic quasiparticle ($A$) and bosonic pair excitation ($B$)
contributions to the slope, as well as the total slope ($A+B$), in the inset
of Fig.~\ref{Cuprate_Ns}(a). Here $A$ is given by expansion at the gap
nodes:
\begin{eqnarray}
  A &=& \frac{2\ln\!2}{\pi} \frac{v_F}{v_2}\frac{T_c}{(n/m)} =
  \frac{2\ln\!2} {\pi} \frac{T_c}{\Delta_0} \frac{4t_\parallel}{(n/m)} \:,
  \nonumber\\ 
&=& \frac{8 \ln\!2\, }{\pi} \frac{\mu_0 e^2}{d\hbar^2}{t_\parallel
    \lambda^2_{ab}(0)} \frac{k_B T_c}{\Delta_0} \:,
\end{eqnarray}
where we have included $e$, $\mu_0$, $k_B$, and the mean inter-plane distance
$d$ explicitly in the second line for later use. Here $\Delta_0$ is the gap
magnitude at the $(\pi, 0)$ points in the momentum space, i.e., twice
$\Delta(0)$ in our convention. $v_F$ is the Fermi velocity at the nodes,
$v_2$ is the velocity along the tangential direction, ${\displaystyle
  v_2=\frac{\partial \Delta_{\bf k}}{\partial k_2}}$, where ${\displaystyle
  k_2 = \frac{\delta k_x -\delta k_y}{\sqrt{2}}}$. We use a tight binding
band structure and for the $d$-wave $\phik$ we take, ${\displaystyle
  \frac{v_F}{v_2} = \frac{4t_\parallel}{\Delta_0}}$.  As shown in the
figure, $A$ decreases with decreasing $x$, since $T_c/\Delta_0$ decreases.
$B(T)$ is calculated at $T=0.2T_c$, since it vanishes strictly at $T=0$.
The figure shows that $B$ increases as $A$ decreases; the pair excitation
contributions become progressively more important towards the underdoped
regime. This increase of $B$ compensates the decrease of $A$, so that the
total slope $A+B$ is relatively doping independent. This compensation effect
explains the quasi-universality. The destruction of the superconducting
state comes dominantly from pair excitations at small $x$, and
quasiparticles at large $x$. Without the bosonic pair excitations, one would
expect that the slope becomes smaller and smaller with underdoping.

Lee and Wen \cite{LeeWen} proposed an alternative explanation for a strictly
universal behavior of the superfluid density in the \textit{underdoped}
regime, based on a spin-charge separation picture, in which the charge
carrier density is given by $x$, instead of $n = 1-x$.  Their explanation
relies on the assumption that $\Delta_0$ is doping independent (which does
not appear to be confirmed experimentally, as can be seen from the phase
diagram plots of Fig.~\ref{Cuprate_Phase}). In this way they derive the
Uemura scaling \cite{Uemura} result $T_c \propto x$.  According to their
explanation, $\lambda^{-2}_{ab}(T)$ as a function of $T$ should have the
same universality as $\lambda_{ab}^{2}(0)/\lambda_{ab}^{2}(T)$ does as
function of $T/T_c$, since both $\lambda^{-2}_{ab}(0)$ and $T_c$ are
proportional to $x$.  Their result does not seem to be appropriate, in
particular, it does not apply in the overdoped regime.  As will be seen in
Sec.~\ref{Subsec_Ns_Slope}, $\lambda_{ab}^{-2}(T)$ as a function of $T$
appears to have a much stronger doping dependence than they presume.

\subsection{Quasi-universal behavior of \textit{c}-axis Josephson critical 
current}
\label{Subsec_Cuprate_Ic}

\begin{figure}
  \centerline{\includegraphics[width=3.5in, clip]{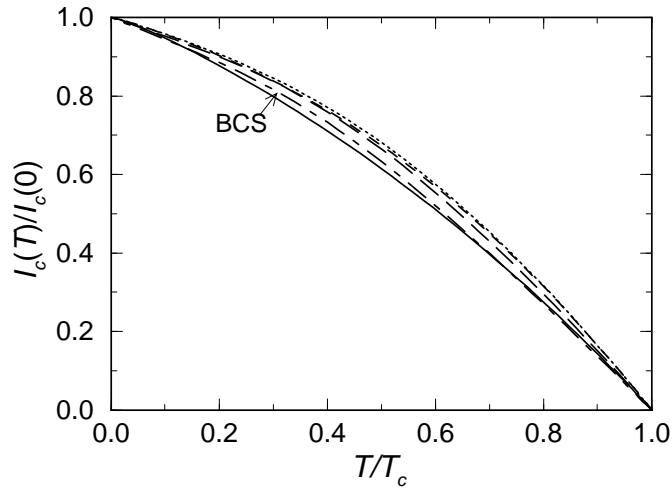}} 
\vskip -0.15in
\caption[Quasi-universal behavior of normalized $c$-axis Josephson critical 
current as a function of reduced temperature $T/T_c$ with respect to doping
concentration.]{Temperature dependence of normalized $c$-axis Josephson
  critical current with doping given by the legends in
  Fig.~\ref{Cuprate_Ns}(a).}
\label{Ic}
\end{figure}

Similar quasi-universal behavior can also be found in the $c$-axis Josephson
critical current, which also depends explicitly on the order parameter as an
overall multiplicative factor.  Following Mahan \cite{Mahan}, we obtain the
general expression for the Josephson critical current
\begin{equation}
I_c = 2e \sum_{\mb{k}\mb{p}} \left|T_{\mb{k}\mb{p}}\right|^2
\frac{\Delta_{sc,\mb{k}} \Delta_{sc,\mb{p}}} {\Ek E_\mb{p}} \left[ 
\frac{1-f(\Ek)-f(E_\mb{p})} {\Ek+E_\mb{p}} + \frac{f(\Ek)-f(E_\mb{p})}
  {\Ek-E_\mb{p}} \right] \:.
\end{equation}
For  $c$-axis Josephson tunneling, the tunneling matrix can presumably be
written as
\begin{equation}
\left|T_{\mb{k}\mb{p}}\right|^2 = \left|T_0\right|^2
\delta_{\mb{k}_\parallel \mb{p}_\parallel} + \left|T_1\right|^2 \:,
\end{equation}
where only the first (coherent) term, which conserves the in-plane momentum,
contributes for a $d$-wave superconductor. As a consequence, we finally
obtain for the cuprates
\begin{eqnarray}
I_c &=& 2e \left|T_0\right|^2 \Delta^2_{sc} \sum_{\mb{k}\mb{p}}
\delta_{\mb{k}_\parallel \mb{p}_\parallel} \frac{\phik \varphi_\mb{p}} {\Ek
  E_\mb{p}} \left[ \frac{1-f(\Ek)-f(E_\mb{p})} {\Ek+E_\mb{p}} +
  \frac{f(\Ek)-f(E_\mb{p})} {\Ek-E_\mb{p}} \right] \nonumber\\  
&\approx& 2e \left|T_0\right|^2 \Delta^2_{sc} \sum_\mb{k}
\frac{\phik^2}{\Ek^2} \left[\frac{1-2f(\Ek)}{2\Ek} + f^\prime(\Ek)\right] \:,
\label{Ic_Eq}
\end{eqnarray}
where in the second line, use has been made of the fact $t_\perp \ll
t_\parallel$. Here we have assumed that the two superconductors are
identical so that $\Delta_{sc}$ is the same for both.  This situation is
relevant to both break junction experiments \cite{JohnZ} and to intrinsic
Josephson tunneling \cite{IJJ} as well. Equation~(\ref{Ic_Eq}), like
Eq.~(\ref{Lambda_Dwave_Eq}), differs from the usual BCS form (as well as
that assumed by Lee and Wen \cite{LeeWen,Millis}) in that the prefactor
$\Delta_{sc}^2$ is no longer the total excitation gap $\Delta^2$.  Comparing
this equation with Eq.~(\ref{Lambda_c_Eq}), one immediately sees, under the
current assumption of the $c$-axis bandstructure, that $ I_c \propto
\lambda_c^{-2} $.  Plotted in Fig.~\ref{Ic} is the temperature dependence of
normalized $c$-axis Josephson critical current with the same doping
concentrations as in Fig.~\ref{Cuprate_Ns}(a). Similar to the superfluid
density, $T_c$ exhibits a quasi-universal behavior, except that now the
small variation of the slope with $x$ is not systematic. This behavior is in
contrast to the strongly $x$ dependent quasiparticle tunneling
characteristics which can be inferred from the temperature dependent
excitation gap plotted in Fig.~\ref{Cuprate_Gaps}. The origin of this
universality is essentially the same as that for $n_s$, deriving from two
mutually compensating contributions.
At this time, there do not appear to be detailed studies of $I_c(T)$ as a
function of $x$, although future measurements will, ultimately, be able to
determine this quantity. In these future experiments the quasiparticle
tunneling characteristics should be simultaneously measured, along with
$I_c(T)$, so that direct comparison can be made to the excitation gap; in
this way, the predictions indicated in Fig.~\ref{Ic} can be tested.
Indications, thus far \cite{JohnZ,Renner}, are that this tunneling
excitation gap coincides rather well with values obtained from photoemission
data (see Fig.~\ref{Cuprate_Phase}).
%\vspace*{1.2in}

%\newpage
\subsection{Search for bosonic pair excitation contributions in penetration depth}
\label{Subsec_Pairon_Search}

At low temperature, the change in the penetration depth, $\Delta \lambda
\equiv \lambda(T) - \lambda(0)$, will acquire the same power law $T$
dependence as the superfluid density, via a Taylor expansion
\begin{equation}
\frac{\Delta\lambda}{\lambda_0} =
\left(\frac{\lambda_0^{2}}{\lambda^2}\right)^{-1/2}-1 \approx \frac{1}{2}
  \Big[A+B(T)\Big] \frac{T}{T_c} \:, 
\end{equation}
where $\lambda_0\equiv \lambda(0)$.  Since the temperature dependence of $B$
is very weak, the experimentally obtained $\Delta\lambda$ will always
appears to be quasi-linear. When there is a large quasiparticle contribution
$A$, it will be hard to separate the fermionic and bosonic contributions to
the the $T$ dependence of $\Delta \lambda$. To make the distinction between
$A$ and $B$ more prominent, we take the temperature derivative,
\begin{equation}
\frac{\mathrm{d} \lambda}{\mathrm{d}T} \approx \frac{1}{2}\frac{\lambda_0}{T_c}
\left[A+\frac{3}{2}B(T)\right] \:. 
\end{equation}
In this way, the $d$-wave quasiparticle contribution is a constant, whereas
the pair excitation contribution still has a $T$ dependence, with a negative
curvature.

\begin{figure}
\centerline{\includegraphics[width=3.5in, clip]{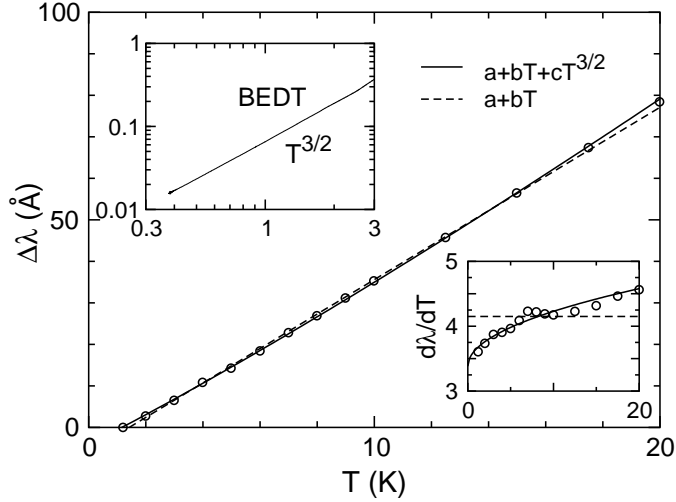}}
\vskip -0.1in
\caption[Comparison of $a$-axis penetration depth data in pure YBCO$_{6.95}$
single crystal, with different theoretical fits corresponding to BCS
$d$-wave and to BCS-BEC predictions.]{Comparison of penetration depth data
  \protect\cite{Hosseini}, $\Delta\lambda$, along $a$-axis, in nominally
  pure YBCO$_{6.95}$ single crystal, with different theoretical fits
  corresponding to BCS $d$-wave (dashed curve) and to BCS-BEC (solid curve)
  predictions. The corresponding derivatives are plotted in the lower inset.
  In the upper inset are experimental data ($\Delta\lambda$ vs $T$) for the
  organic superconductor BEDT from Ref.~\protect\cite{Russ}.}
\label{dLambda_dT}
\end{figure}

We take the experimentally measured $a$-axis penetration depth,
$\Delta\lambda_a$, in optimally doped, high quality, pure YBCO$_{6.95}$ single
crystals from Hardy and coworkers \cite{Hosseini}. Their original data, along
with different theoretical fits corresponding to BCS $d$-wave (dashed curve)
and to BCS-BEC (solid curve) predictions are plotted in
Fig.~\ref{dLambda_dT}. These two fits are essentially indistinguishable, for
the reasons mentioned above. In the lower inset we plot the slopes $d\lambda /
dT $ where the difference between the two sets of curves is more apparent.
Here it is shown that the low temperature downturn of the derivative, seen
to a greater or lesser extent in all $\Delta\lambda(T)$ measurements, fits
our predicted $T^{1/2} + const.$ dependence rather well.  This downturn has
been frequently associated with impurity effects, which yield a linear in
$T$ slope for $\Delta \lambda $ at very low $T$, and in this case, provide a
poorer fit.  While these cuprate experiments were performed on a nearly
optimal sample, the same analysis of an underdoped material yielded
similarly good agreement, but with a $T^{3/2}$ coefficient about a factor of
two larger.  Future more precise and systematic low $T$ experiments on
additional underdoped samples are needed.  

Evidence for the pair excitation contributions also comes from other short
coherence length superconductors.  Plotted in the upper inset of
Fig.~\ref{dLambda_dT} are data \cite{Russ} on the organic superconductor
$\kappa$-(ET)$_2$Cu[N(CN)$_2$]Br (BEDT, $T_c\approx 11$K) which fit a pure
$T^{3/2}$ power law over a wide temperature regime; in contrast to the
cuprates, there is no leading order linear term. These authors have found
this low temperature pure $T^{3/2}$ power law for $\Delta\lambda$ on a
variety of different samples. At present, there seems to be no other
explanation, besides the bosonic pair excitation mechanism presented here,
for this unusual power law at the lowest temperatures. However, this
explanation also implies that these organic superconductors have an order
parameter with $s$-wave symmetry, for which other evidence is somewhat
controversial.

\subsection{Slope of in-plane inverse squared penetration depth: Quantitative
  analysis}
\label{Subsec_Ns_Slope}

In this subsection, we shall calculate \textit{quantitatively} the doping
dependence of $\lambda_0$ and ${\displaystyle \frac{d\lambda^{-2}}{dT\;\;}}$
for a variety of different cuprates, using the experimentally determined
lattice constants.

For $\lambda_0$, we have
\begin{equation}
\lambda_0^{-2} = \frac{e^2\mu_0}{a^2d}\left(\frac{n}{m}\right) \propto x \:,
\end{equation}
where $a$ and $b$ are the in-plane lattice constants, and $d$ is the mean
inter-plane distance. The lattice parameters are:
\begin{itemize}
\item For YBCO: $a\approx b = 3.9$~\AA, $c=11.8$~\AA, $d=5.9$~\AA;
\item For Bi2212: $a=5.4$~\AA, $c=30.8$~\AA, $d=7.7$~\AA;
\item For LSCO: $a=3.79$~\AA, $c=13.2$~\AA, $d=6.6$~\AA.
\end{itemize}
Here the density $n$ is given by the number of electrons per in-plane unit
cell, and $1/m$ implicitly contains a factor $a^2/\hbar^2$. The factor $n/m
\propto x$ in our case is a consequence of the Coulomb-associated reduction
in the bandwidth, whereas the charge carrier density $n=1-x$ changes only
slightly. This is different from alternate scenarios, such as that of Lee
and Wen \cite{LeeWen}, in which the charge carrier density is $x$, with a
doping independent mass $m$.

The temperature derivative is given by
\begin{equation}
\left|\frac{d\lambda^{-2}}{d T\;}\right|=\frac{\lambda_0^{-2}}{T_c\:} 
\left[A+\frac{3}{2} B(T) \right]
%\quad(\; =\frac{2}{\lambda^3}\frac{d\lambda}{dT} \;) 
\:.
\label{Cuprate_Ns_Slope_Eq}
\end{equation}
The first term is associated with the fermionic quasiparticle contribution,
and has a simple, explicit expression
\begin{equation}
\left|\frac{d\lambda^{-2}}{d T\;}\right|_{QP} = \frac{8 \ln\! 2}{\hbar c}
\frac{\alpha k_B}{d} \frac{v_F}{v_2} = \frac{17.65}{d} \frac{v_F}{v_2}\:,
\label{Cuprate_Ns_Slope_QP_Eq}
\end{equation}
where $\alpha =1/137$ is the fine structure constant. Each quantity in the
last equation must be in MKSA units, except for the dimensionless ratio
$v_F/v_2$. Here we write down the full expression for $v_F$ and $v_2$ at the
nodes for a simple tight-binding model with $d$-wave pairing:
\begin{equation} 
v_F = 2\sqrt{2}\, \frac{t_\parallel a}{\hbar} \sin
\frac{k_Fa}{\!\sqrt{2}}\:, \qquad v_2 =
\frac{a}{\hbar}\frac{\Delta_0}{\!\sqrt{2}} \sin \frac{k_Fa}{\!\sqrt{2}}\:,
\qquad \frac{v_F}{v_2} = \frac{4t_\parallel}{\Delta_0} \:.  
\label{Cuprate_Vf_V2_Eq}
\end{equation}
As $x$ increases toward the extreme overdoped limit, $v_F$ increases
proportionally, but $v_2$ (which follows the excitation
gap $\Delta$) must decrease toward zero.
The second term in Eq.~(\ref{Cuprate_Ns_Slope_Eq}) is associated with
bosonic pair excitations. we calculate this term numerically at $T=0.2T_c$,
as we did for the inset of Fig.~\ref{Cuprate_Ns}.

\begin{figure}
\centerline{\includegraphics[angle=-90, width=6in, clip]{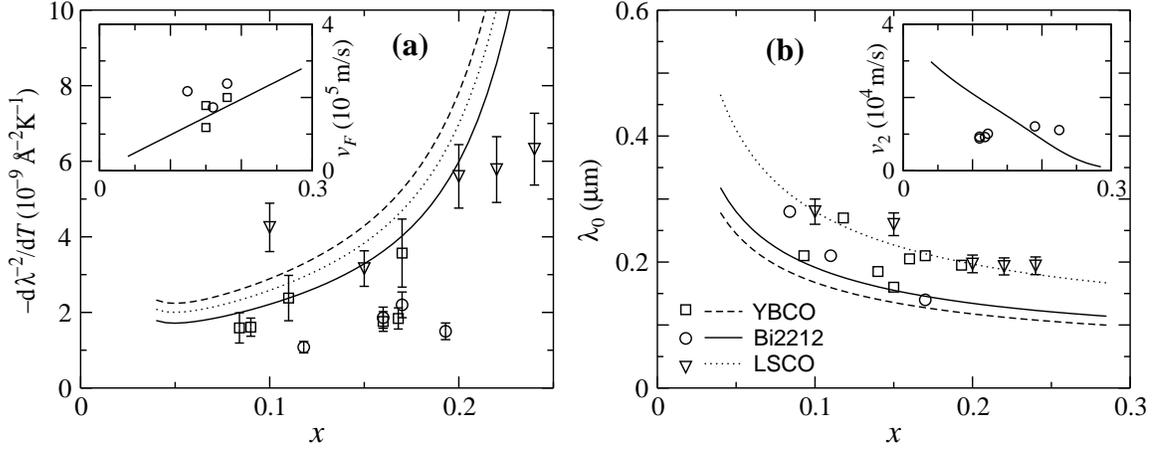}}
\vskip -0.1in
\caption[Quantitative analysis of the doping dependence of the zero
temperature penetration depth, $\lambda_0$, and the temperature derivative
of the inverse squared penetration depth, $|d\lambda^{-2}/dT|$ for various
cuprates.]{Doping dependence of (a) the slope $|d\lambda^{-2}/dT|$ and (b)
  zero temperature penetration depth $\lambda_0$ for various cuprates.
  Theoretical predictions are represented by lines, and experimental data by
  symbols.  Shown in the insets of (a) and (b) are the doping dependence of
  $v_F$ and $v_2$, respectively. The experimental data on $v_F$ are taken
  from ({\large $\circ$}) Refs.~\protect\cite{Valla,Zhou,Waldmann,LeeSF}
  ({\footnotesize $\Box$})\protect\cite{LeeWen,Krishana}, $v_2$ from
  Ref.~\protect\cite{Mesot}, others ({\large $\circ$}, Bi2212)
  Refs.~\protect\cite{Waldmann,LeeSF}; ({\footnotesize $\Box$}, YBCO)
  \protect\cite{Hardy,Cambridge}; ({\footnotesize $\triangledown$}, LSCO)
  \protect\cite{Panagopoulos}.}
\label{Cuprate_Ns_Slope}
\end{figure}

The results for $\lambda_0$ and $d\lambda^{-2}/dT$ for various cuprates are
plotted in Fig.~\ref{Cuprate_Ns_Slope}. The doping dependence of
$v_F$ and $v_2$ are also shown, in the insets of
Fig.~\ref{Cuprate_Ns_Slope}(a) and Fig.~\ref{Cuprate_Ns_Slope}(b),
respectively. Despite the large scatter of the data, our theoretical
predictions are consistent with experiment, for both $\lambda_0$ and
$d\lambda^{-2}/dT$. This agreement is best for LSCO, for which $\lambda_0$
and $\Delta\lambda(T)$ are measured more consistently within
the same group of experiments.%
\newfootnote{The discrepancy in $\Delta\lambda(T)$ for YBCO and Bi2212 may be
  associated with errors in $\lambda_0$, since ${\displaystyle
    \left|\frac{d\lambda^{-2}}{d T}\right|
    =\frac{2}{\lambda^3}\frac{d\lambda}{dT} \:,}$ so that a 10\% error in
  $\lambda_0$ would cause a 30\% error in $d\lambda^{-2}/dT$.}
Unfortunately, it is hard to measure $\lambda_0$ accurately; this quantity
may also be sensitive to the sample quality.%
\newfootnote{For this reason, the Bi2212 data for $d\lambda^{-2}/dT$ from
  Waldman \textit{et al.} \cite{Waldmann} are not as reliable since they had
  difficulty observing the linear $T$ dependence of $\Delta\lambda$
  associated with the nodal quasiparticles. }

As expected from Eqs.~(\ref{Cuprate_Vf_V2_Eq}), the insets of
Fig.~\ref{Cuprate_Ns_Slope} show that the quasiparticle contribution,
${\displaystyle \left| \frac{d\lambda^{-2}}{dT}\right|_{QP} }$ roughly
varies as ${\displaystyle \frac{v_F}{v_2} \sim x^2}$. Without the
contribution from pair excitations, the slope ${\displaystyle \left|
    \frac{d\lambda^{-2}}{dT}\right|}$ would be much smaller in the
underdoped regime, and, therefore, inconsistent with experiment. Bosonic
pair excitations become more and more important with underdoping. Figure
\ref{Cuprate_Ns_Slope} also reveals that the slope $d\lambda^{-2}/dT$ has a
fairly strong doping dependence, particularly in the overdoped regime. This
is in contrast to the quasi-universal behavior of the normalized
$\lambda^{-2}$ vs $T/T_c$.
\vspace*{1cm}

\section{Specific heat at low \textit{T}}
\label{Sec_Cuprate_Cv}

In this section, we apply our three fluid model in the context of specific
heat calculations (based on the discussion in Sec.~\ref{Sec_Cv}) to the
cuprates.  In this way we calculate the coefficients $\alpha$ and $\gamma^*$
in Eq.~(\ref{Gamma_Eq}) as a function of doping.

For $d$-wave superconductors, the fermionic quasiparticle contributes a
quadratic $T$ dependence, $\alpha T^2$, to the specific heat, where $\alpha $
is given by
\begin{equation}
\alpha =\frac{18\zeta(3)}{\pi}\, \frac{k_B^3}{\hbar^2}\, \frac{1}{d}\,
\frac{1}{v_F v_2}\,. 
\label{Cuprate_Cv_Eq}
\end{equation}
The pair excitation contribution, $\gamma^*$,
will be calculated directly from Eq.~(\ref{S_Pair_Eq}).  Here we use our
results for the various gaps and chemical potential below $T_c$ as a
function of both doping and temperature.

\begin{figure}
\centerline{\includegraphics[angle=-90, width=6in, clip]{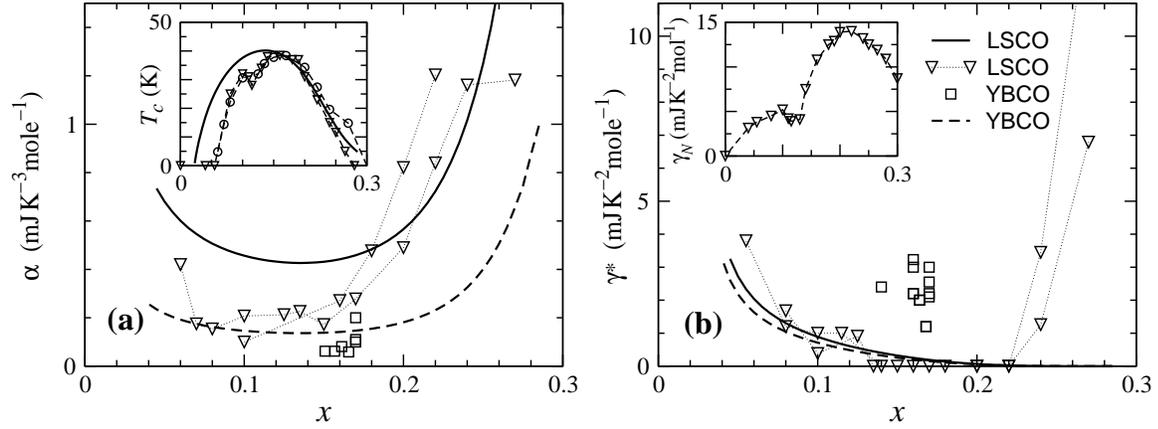}}
\caption[Quantitative ananlysis of the quadratic (associated with
quasiparticles) and the linear (associated with pair excitations) $T$
contributions to the specific heat in various cuprates.]{(a) Quadratic and (b)
  linear $T$ contributions to $C_v$ in various cuprates. $T_c$ is shown in
  the inset of (a). Data are from ({\footnotesize $\triangledown$})
  Refs.~\protect\cite{Loram,Momono} and ({\footnotesize $\Box$})
  Refs.~\protect\cite{Moler,Junod,Wright,Fisher}.  The inset in (b) shows
  normal state result from Ref.~\protect\cite{Momono}. The symbols
  (experiment) and lines (theory) (per mole of formula units) can be
  compared.}
\label{Cuprate_Cv}
\end{figure}

The results for $\alpha$ and $\gamma^*$ are plotted in
Fig.~\ref{Cuprate_Cv}(a) and (b), respectively, for LSCO and YBCO. As can be
seen, the theory is consistent with the experimental data, especially for
LSCO. For YBCO, it has been difficult to see the $\alpha T^2$ term in $C_v$.
Indeed, as shown in Fig.~\ref{Cuprate_Cv}, there is a large discrepancy
between the measurements of different groups, so that the YBCO results may
be less reliable.

There are several important observations from this figure and
Eq.~(\ref{Cuprate_Cv_Eq}) regarding the behavior of $C_v$. (i) As the doping
$x$ decreases, pair excitations become more significant, and, as a
consequence, $\gamma^*$ increases with decreasing $x$.  (ii) On the
\textit{overdoped} side, $\alpha$ increases rapidly with $x$.  This can be
expected from Eq.~(\ref{Cuprate_Cv_Eq}), since $1/v_2 \propto 1/\Delta_0
\propto \gamma_N/T_c$ increases rapidly with overdoping. [Here $\gamma_N$ is
the normal state $\gamma$, which is plotted in the inset of
Fig.~\ref{Cuprate_Cv}(b).] This increase coincides with the rapid drop of
$T_c$, as shown in the inset of Fig.~\ref{Cuprate_Cv}(a).  On simple
physical grounds, a small $T_c$ in the overdoped regime implies that $\gamma
\approx \alpha T$ must increase from 0 to the normal state value $\gamma_N$
within a narrow temperature range, and $\alpha$ has to increase to
accommodate this rapid change.  (iii) In the \textit{underdoped} regime, as
$x$ further decreases, $T_c$ drops rapidly, so that $\alpha \propto
\gamma_N/T_c$ also increases gradually, (although $\gamma_N$ also decreases,
but it does not decrease as fast as $T_c$ does). (iv) Experimentally, as
$T_c$ approaches zero on either side, $\gamma$ should approach $\gamma_N$.
The fact that $\gamma^*$ increases again in the extremely overdoped limit
can be associated with the fact that the low $T$ behavior may be obscured by
the low $T$ upturn in $\gamma(T)$ curve deriving, presumably, from extrinsic
paramagnetic scattering centers, etc.

Our pair excitation mechanism for the widely observed $\gamma^*$ term,
offers a more natural and systematic explanation, than previous claims which
associate this term with extrinsic effects.  The agreement between theory
and experiment [Fig.~\ref{Cuprate_Cv}(b)] seems to be favorable.

\chapter{Thermodynamic signatures of the superconducting transition}
\label{Chap_Thermodynamics}

Both ARPES and tunneling measurements have shown that in a pseudogapped
superconductor, the excitation gap evolves smoothly across the
superconducting phase transition temperature, $T_c$, as plotted in
Fig.~\ref{Cuprate_Gaps}. It, therefore, follows that the superconducting
order parameter and the excitation gap are distinct. Moreover, this
distinction is essential to the more general mean field theory which we have
proposed in this Thesis.  These observations, along with the measured
behavior of the excitation gap, leads to an important question: Will there
be any sign of the phase transition in thermodynamical properties such as
the specific heat and the tunneling spectra (or, equivalently, electronic
density of states) in a pseudogapped superconductor? We know that the answer
must be yes on general physical grounds, since the superconducting and
normal states are distinct phases; thus, we expect thermodynamical
signatures at $T_c$. Moreover, experimentally, it has been established from
specific heat measurements in the cuprates that there is a step
discontinuity or a maxima at $T_c$, depending on the doping level
\cite{Loram,Loram98}. It should be recalled that in BCS theory, the specific
heat jump at $T_c$ results from the opening of the excitation gap. Clearly,
there must be something more subtle going on in the cuprates --- to give
rise to a thermodynamical signature at $T_c$. This is one of the main issues
to be addressed in this Chapter.

In addition, in this Chapter we address the thermodynamic characteristics of
the extrapolated (below $T_c$) ``normal state".  This represents a very
important concept.  The free energy of this extrapolated normal state is
needed in writing down a Landau-Ginzburg expansion for traditional
superconductors.  Here the situation is far more complex and we ask how this
extrapolation should be done when the normal state is not a Fermi liquid?
We pose this question in the context of the condensation energy --- which has
been deduced from specific heat measurements \cite{Loram98}.  So far, the
magnetic field dependent measurements of the normal phase inside a vortex
core \cite{Renner_Vortex} or at a field above $H_{c2}$ \cite{Boebinger} suggest
that for a pseudogapped superconductor the fermionic excitation gap persists
when the superconductivity is destroyed below $T_c$.  These experiments are
consistent with the present physical picture.

The results obtained within the present theoretical scheme show that upon
entering the superconducting phase, the onset of the coherent condensate
leads to dramatic sharpening of the peaks in the electronic spectral
function, which is observable in both tunneling spectra and ARPES
measurements. This sharpening, in conjunction with the temperature
dependence of the excitation gap, will cause a specific heat jump. This
general picture is applied to the cuprates where we show that in the
overdoped regime, this jump (in $C_v$) is mainly attributable to the
temperature dependence of the excitation gap (as in the traditional BCS
case), whereas in the underdoped regime, it is mainly a consequence of the
onset of off-diagonal long range order. This jump becomes smaller towards
underdoping, and behaves more like a $\lambda$ transition in Bose-Einstein
condensation. We also discuss the subtlety of extracting the condensation
energy from specific heat data, and we show that a proper treatment of the
pseudogap in the extrapolated ``normal state'' is important.

In previous chapters, we have used the approximation,
Eq.~(\ref{Sigma_PG_Approx}), in order to simplify the calculations.
Thereby, we defined the pseudogap parameter, $\Delta_{pg}$. Under this
approximation, the pseudogap self-energy $\Sigma_{pg}$ has a BCS-like form,
so that the spectral function is given by two $\delta$-functions at $\pm
\Ek$.  These approximations were justified in the context of the
applications we considered, thus far.  However, in order to study the
thermodynamic behavior at $T_c$ and above $T_c$, in this Chapter, we will
relax this simplifying approximation and allow for lifetime effects in
$\Sigma_{pg}$. This more realistic form incorporates a finite broadening due
to the incoherent nature of the finite center-of-mass momentum pair
excitations.  To avoid extremely heavy and difficult numerical computations,
we will not solve for the broadening and chemical potential, etc.
self-consistently, but rather take the broadening as a phenomenological,
tunable input parameter such that it gives a realistic density of states,
etc. We will use the gaps, $T_c$, and the chemical potential $\mu$ obtained
in chapters \ref{Chap_SC_Phase} and \ref{Chap_Cuprates} in our calculations.

\section{Spectral functions and the density of states}
\label{Sec_Spectral}

We begin with the building block of this chapter, the spectral function
$A(\mb{k}, \omega)$. Both theoretical studies \cite{Maly2} and ARPES data
\cite{Norman} show that above $T_c$, the spectral function can be roughly
approximated by a broadened BCS-like form. Therefore, we can write down
$\Sigma_{pg}$ in a general form,
\begin{equation}
\Sigma_{pg}(\mb{k},\omega) =
\frac{\Delta_{\mb{k},pg}^2}{\omega+\ek+\lambda +i\gamma}+\nu -i\Sigma_0 \:,
\label{SigmaPG_Model_Eq}
\end{equation}
where for given \textbf{k}, $\lambda$ and $\gamma$ are constants, $\nu$ and
$\Sigma_0$ may be a function of $\omega$, and $\nu$ is the real part
contribution introduced by $\Sigma_0$ via the Kramers-Kr\"onig relations.
Since the pseudogap self-energy results from the incoherent finite momentum
pair excitations both above and below $T_c$, this expression will also be
valid in the superconducting state. On the other hand, since the
superconducting self-energy, $\Sigma_{sc}$, results from the phase-coherent
zero-momentum condensate, \textit{there should be no broadening in}
$\Sigma_{sc}$. In this way, we obtain the inverse Green's function
\begin{equation}
G^{-1}(\mb{k},\omega) = \omega-\ek -\nu- \frac{\Delta_{\mb{k},sc}^2} 
{\omega+\ek}
- \frac{\Delta_{\mb{k},pg}^2}{\omega+\ek+\lambda +i\gamma} +i\Sigma_0 \:.
\end{equation}
Thus, the spectral function is given by%
\newfootnote{Note here the spectral function satisfies the normalization
  ${\displaystyle \int_{-\infty}^\infty \frac{\mbox{d}\omega}{2\pi}
    A(\mb{k},\omega) = 1 }$, which follows the convention used in some
  standard textbooks (see, e.g., Refs.~\cite{Fetter} and \cite{Mahan}), but
  is different from that used by some authors in the literature \cite{Norman}.  }
\begin{mathletters}
\label{Spectral_Eq}
\begin{eqnarray}
\label{Spectral_Eq1}
A(\mb{k},\omega)&=&-2\,\mbox{Im}\, G(\mb{k},\omega+i0)\nonumber\\
&=& 2 \frac{(\omega+\ek)\left[(\omega+\ek+\lambda)C-\gamma
    D\right]}{C^2+D^2} \:,
\end{eqnarray}
where 
\begin{eqnarray}
C&=& \gamma \left[\omega^2-\ek^2- \Delta_{\mb{k},sc}^2-\nu
  (\omega+\ek)\right] +
        (\omega+\ek)(\omega+\ek+\lambda)\Sigma_0 \:, \nonumber\\
D&=& (\omega+\ek)\left[\omega^2-\Ek^2-\nu (\omega+\ek)
  -\Sigma_0\gamma\right]  + 
\lambda \left[\omega^2-\ek^2- \Delta_{\mb{k},sc}^2-\nu (\omega+\ek)\right]
 \:. \nonumber
\end{eqnarray}
Above $T_c$, where $\Delta_{sc}=0$, the spectral function can be simplified as
\begin{equation}
A(\mb{k},\omega)=2\frac{(\omega+\ek+\lambda)C^\prime-\gamma
    D^\prime}{{C^\prime}^2+{D^\prime}^2} \:,
%2 \frac{(\omega+\ek+\lambda)\left[ \gamma(\omega-\ek) +
%    \Sigma_0 (\omega+\ek+\lambda) \right] - \gamma \left[ \omega^2 -\Ek^2 +
%    \lambda (\omega-\ek) -\gamma \Sigma_0\right]} {\left[ \omega^2 -\Ek^2 +
%    \lambda (\omega-\ek) -\gamma \Sigma_0\right]^2 + \left
%    [ \gamma(\omega-\ek) + \Sigma_0 (\omega+\ek+\lambda) \right]^2 } \:.
\label{Spectral_Eq2}
\end{equation}
where 
\begin{eqnarray}
C^\prime &=& \gamma(\omega-\ek-\nu) + \Sigma_0 (\omega+\ek+\lambda)\nonumber\\
D^\prime &=& \omega^2 -\Ek^2 -\nu(\omega+\ek) + \lambda (\omega-\ek-\nu) 
-\gamma \Sigma_0 \:. \nonumber
\end{eqnarray}
\end{mathletters}

ARPES data \cite{Norman} show that $\Sigma_0$ is very small for small
$\omega$. However, it can be as large as 300 meV for large $\omega$, and
this is associated with a huge incoherent background, whose origin is still
under debate.  (A proper choice for $\Sigma_0$ can give the peak/dip/hump
features observed in ARPES measurements and in some tunneling data). Here we
are not particularly interested in the high energy features and will take
$\nu=\Sigma_0=0$ to simplify the calculations, without loss of generality.%
\newfootnote{We find that $\Sigma_0$ must decrease with $\omega$ as
  $\omega\rightarrow \pm \infty$, in order to conserve the system
  energy. Otherwise, the energy of the system will diverge. }
In addition, we have no particular reason to support the choice of nonzero
$\lambda$, and hence will also set it to zero. Finally, we obtain a greatly
simplified expression for $A(\mb{k},\omega)$:
\begin{equation}
\renewcommand{\arraystretch}{1.8}
A(\mb{k},\omega) = \left\{ \begin{array}{c@{\hspace{1cm}}l}
{\displaystyle \frac{2\Delta_{\mb{k},pg}^2 \gamma (\omega+\ek)^2}
{(\omega+\ek)^2 (\omega^2-\Ek^2)^2 + \gamma^2
  (\omega^2-\ek^2-\Delta_{\mb{k},sc}^2)^2} \:, } &  (T<T_c), \\
{\displaystyle \frac{2\Delta_{\mb{k}}^2\gamma} {(\omega^2-\Ek^2)^2
+\gamma^2(\omega-\ek)^2} \:,} &  (T>T_c). \end{array} \right.
\label{SpectralF_Eq}
\end{equation}

From Eqs.~(\ref{Spectral_Eq}) and (\ref{SpectralF_Eq}), we see that the
spectral function contains a zero at $\omega=-\ek$ below $T_c$, whereas it
has no zero above $T_c$. This difference is at the heart of the different
thermodynamical behavior of the system above and below $T_c$.

\begin{figure}
\centerline{\includegraphics[width=6in, clip]{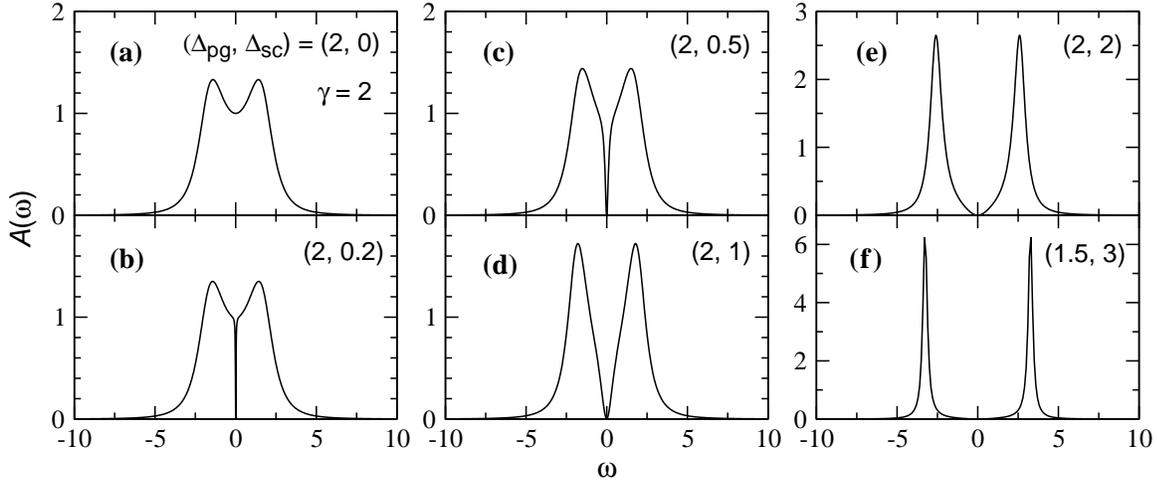}}
\vskip -0.1in
\caption[Effects of the superconducting long range order on the behavior 
of the spectral function at the Fermi level $\ek=0$ as a function of
temperature in a pseudogapped superconductor. ]{Effects of the
  superconducting long range order on behavior of the spectral function at
  the Fermi level $\ek=0$ as a function of temperature in a pseudogapped
  superconductor, in arbitrary units. Here we take $\gamma=2$ for all the
  figures.  The parameters $(\Delta_{pg}, \Delta_{sc})$ are labeled in
  pairs. Figures (a)-(f) correspond to various temperatures decreasing from
  $T=T_c$. A true gap opens up below $T_c$ as a consequence of the onset of
  the condensate.}
\label{SpectralFunction}
\end{figure}

In Fig.~\ref{SpectralFunction}, we plot the spectral function for $\ek=0$
(on the Fermi surface) at different temperatures from slightly above $T_c$
[Fig.~\ref{SpectralFunction}(a)] to temperatures within the
superconducting phase [Fig.~\ref{SpectralFunction}(f)].  Note that in these
plots we do not need to specify the pairing symmetry, but only the magnitude
of the gap along the direction specified by the given wave vector. Since we
have seen (in Fig.~\ref{Gaps_3D}, for example) that varying temperature
corresponds to varying the relative size of $\Delta_{pg}$ and $\Delta_{sc}$,
we chose to vary these energy scale parameters, rather than $T$ directly.
For illustrative purposes, we take $\gamma (T)=\Delta_{pg}(T_c)$. In this
way we ignore any $T$ dependence in $\gamma$ and, thus, single out the long
range order effects associated with $\Delta_{sc}$. Figure
\ref{SpectralFunction}(a) is appropriate for the spectral function at and
slightly above $T_c$. Slightly below $T_c$, a very small condensate
contribution leads to the depletion of the spectral weight at the Fermi
level, as shown in Fig.~\ref{SpectralFunction}(b). As the temperature
continues to decrease, and the superconducting gap increases, the two peaks
in the spectral function become increasingly well separated, as plotted in
Figs.~\ref{SpectralFunction}(c)-(f).  This last panel,
Fig.~\ref{SpectralFunction}(f), is computed with a slightly smaller
$\Delta_{pg}$ and slightly larger $\Delta_{sc}$, as is consistent with the
behavior for $T/T_c \sim 0.7$, seen in, e.g., Fig.~\ref{Cuprate_Gaps}.  Even
at these relatively high temperatures the spectral peaks are quite sharp ---
only slightly broadened relative to their BCS counterparts (where the
spectral function is composed of two $\delta$ functions).  It should be
noted that these narrow peaks derive from long range order, via
$\Delta_{sc}$, and that lifetime effects via $\gamma$ do not lead to
significant peak broadening.  To understand this last point, one needs to
note two facts: (i) A non-zero $\Delta_{sc}$ forces the spectral function to
vanish at the Fermi level $\omega=0$ (for $\ek=0$) so that the two peaks are
necessarily separated; (ii) The imaginary part of the pseudogap self-energy
at the peak location $\Ek$ is given by
\begin{equation}
\gamma^\prime = \gamma \frac{\Delta_{\mb{k},pg}^2}{(\Ek+|\ek|)^2+\gamma^2} =
\gamma \frac{\Delta_{\mb{k},pg}^2}{\Delta_\mb{k}^2+\gamma^2} \:,
\label{Gamma'_Eq}
\end{equation} 
so that the effective peak width, determined by $\gamma^\prime$, decreases
rapidly as $T$ decreases. [Here in the second equation above, we have used
$\ek=0$ which is relevant to Fig.~\ref{SpectralFunction}].  It should be
emphasized that below $T_c$, the spectral functions in
Eqs.~(\ref{Spectral_Eq}) and (\ref{SpectralF_Eq}), are very different from
those obtained using a broadened BCS form; there is no true gap for the
latter, in contrast to the present case.

\begin{figure}
\centerline{\includegraphics[width=6in, clip]{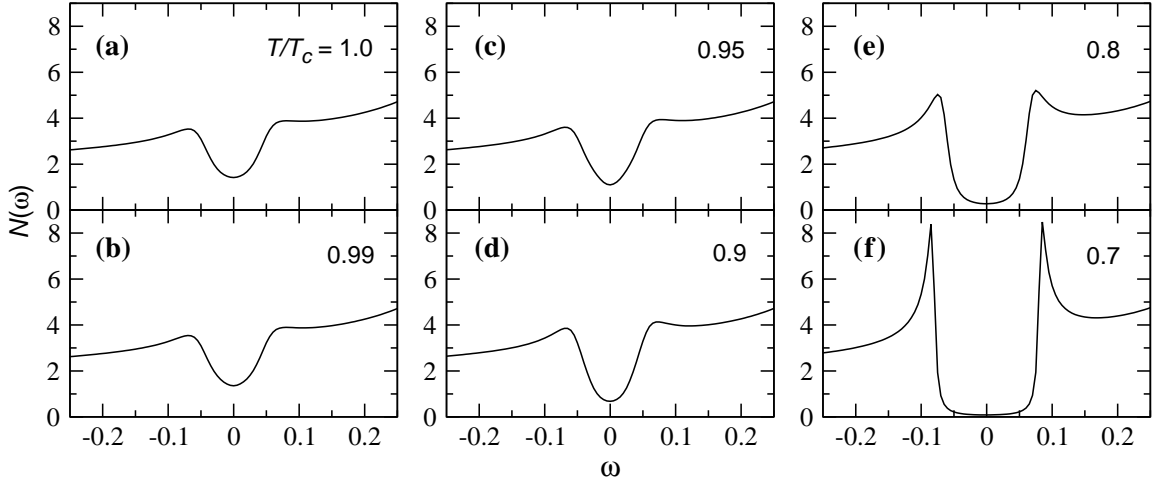}}
\vskip -0.1in
\caption[Effects of superconducting long range order on the behavior 
of the density of states as a function of temperature in a pseudogapped
$s$-wave superconductor. ]{Effects of superconducting long range order on
  the behavior of the density of states as a function of temperature in a
  pseudogapped $s$-wave superconductor with $n=0.5$. Here we take the
  parameters $\Delta_{pg}, \Delta_{sc}$, and $\gamma$ same as those used in
  Fig.~\ref{SpectralFunction}, but measured in terms of the unit
  $4t_\parallel$, the half band width. Figures (a)-(f) correspond to various
  temperatures decreasing from $T=T_c$, as labeled in the figure. At
  $T/T_c\sim 0.7$, as shown in (f), the density of states is close to that
  of strict BCS theory.}
\label{DOS}
\end{figure}

The density of states, which is given by 
\begin{equation}
N(\omega)=\sumk A(\mb{k},\omega) \:,
\end{equation}
can be easily obtained from the spectral function. It should be stressed
that the peak sharpening effects discussed above, associated with
superconducting long range order, will also be reflected in the density
of states. For illustrative purposes, in this section, we present our
results for the case of $s$-wave pairing.  In Fig.~\ref{DOS}(a)-(f), we
plot the density of states for a quasi-2D $s$-wave superconductor.  As
in Fig.~\ref{SpectralFunction}, we emphasize here the effects associated
with long range order, via $\Delta_{sc}$, taking the same energy scales
as used in Fig.~\ref{SpectralFunction}.  Due to the contributions of the
states with $\ek\neq 0$, the narrow dips in
Fig.~\ref{SpectralFunction}(b)-(c) do not show up here. However, as is
evident, the density of states within the gap region decreases quickly,
as the superconducting condensate develops. In comparison with the much
weaker temperature dependence of the density of states above $T_c$, it
can be inferred from the figure that there will be something like a step
discontinuity in the derivative of the density of states with respect to
$T$, $\mbox{d}N(\omega)/\mbox{d}T$.

\section{Specific heat}
\label{Sec_Thermo_Cv}

The rapid decrease of the density of states with decreasing $T$, in the
vicinity of $T_c$, will be reflected in the behavior of the specific heat,
$C_v$. These effects will, thereby, lead to a thermodynamical signature of
the phase transition. To calculate $C_v$ in terms of the more general
spectral function (which now contains lifetime effects) requires an
approach different from that used to obtain Eq.~(\ref{S_QP_Eq}).  Here, we
derive $C_v$ directly from the energy of the system via $C_v=dE/dT$, where
the energy $E$ is calculated using a direct spectral integration of the
energy distribution, as given in Ref.~\cite{Fetter}:
\begin{eqnarray}
E &=& 2T\sum_{\mb{k},n} \frac{1}{2}(i\omega_n + \ek^0 +
\mu)G(\mb{k},i\omega_n) \nonumber\\
&=& \sumk \int_{-\infty}^\infty \frac{\mbox{d}\omega}{2\pi}
(\omega+\ek+2\mu) A(\mb{k},\omega)f(\omega) \:,
\label{EnergyInt_Eq}
\end{eqnarray}
where $\ek^0=\ek+\mu$ is the dispersion measured with respect to the bottom
of the band, which is, therefore, temperature independent.  It is not
difficult to show that for the BCS spectral function,
$A(\mb{k},\omega)=2\pi[\uk^2\delta(\omega-\Ek)+\vk^2\delta(\omega+\Ek)]$,
this equation reduces to Eq.~(\ref{BCS_E}). To determine the various
contributions to the specific heat, we rewrite Eq.~(\ref{EnergyInt_Eq}) as
\begin{equation}
E = \int_{-\infty}^\infty \frac{\mbox{d}\omega}{2\pi}
[(\omega+\mu) N(\omega)+K(\omega)]f(\omega) \:,
\end{equation}
where we have defined $K(\omega)\equiv \sumk \ek^0 A(\mb{k},\omega)$, which
can be regarded as the contribution associated with the
kinetic energy of the system.  In this way, we obtain
\begin{eqnarray} 
C_v &=& \int_{-\infty}^\infty \frac{\mbox{d}\omega}{2\pi} \left\{
  \frac{\partial \mu}{\partial T} N(\omega) f(\omega)
  -\left[(\omega+\mu)N(\omega)+K(\omega)\right]\frac{\omega} {T}
  f^\prime(\omega) \right.\nonumber\\ 
&&{} + \left.
  \left[(\omega+\mu)\frac{\partial N(\omega)}{\partial T} + \frac{\partial
      K(\omega)}{\partial T}\right]f(\omega) \right\} \:.  
\end{eqnarray}
The two terms on the right hand side, appearing in the first line of the
equation, are present in the general Fermi liquid case, whereas the term on
the second line derives
from the temperature dependence of the density of states.%
\newfootnote{Strictly speaking, in a Fermi liquid, the density of states is
  also $T$-dependent, via the chemical potential in $A(\mb{k},\omega)$.
  However, this $T$-dependence is usually negligible.}
In a Fermi liquid at low $T$, where $f^\prime(\omega)\sim -\delta(\omega)$
and $N(\omega)\approx N(0)$, the first two terms lead to a contribution to
$C_v/T$ which is proportional to $N(\omega)$.  However, when $N(\omega)$ is
a function of $T$, the second term will be significant and sometimes
dominant.  \textit{In this case, $C_v/T$ no longer reflects the density of
  states.} Indeed, it is this term that gives rise to the specific heat
discontinuity at $T_c$ in BCS theory.  Similarly, for the case of a
pseudogap superconductor, this non-Fermi liquid term also leads to a step discontinuity.

\begin{figure}
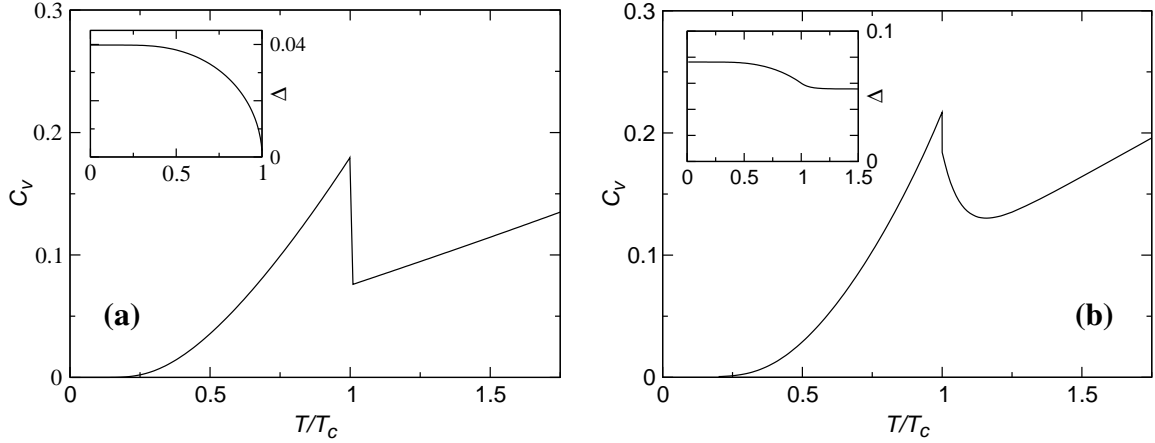

\centerline{\includegraphics[width=2.9in,clip]{CvSf_5g_5BCS-2}\hfill
\includegraphics[width=2.9in,clip]{CvSf_5r100g_6}}
\vskip -.8in
\hskip 0.5in {\textbf{(a)}} \hskip 4.85in {\textbf{(b)}}
\vskip .5in
\caption[Comparison of the temperature dependence of the specific heat in
the weak coupling BCS case and moderate coupling pseudogap case.]{Comparison
  of the temperature dependence of the specific heat in (a) the weak
  coupling BCS case and (b) moderate coupling pseudogap case. Shown here are
  quasi-2D $s$-wave results, at $n=0.5$, $-g/4t_\parallel = 0.5$ and 0.6,
  respectively. The $T$ dependence of the gap is shown as 
  insets. The $C_v$ jump at $T_c$ can be accounted for by the discontinuity
  of $\mbox{d}N(\omega)/\mbox{d}T$ in both cases.}
\label{Cv_S}
\end{figure}

In Fig.~\ref{Cv_S} we plot the temperature dependence of $C_v$ in both (a)
the weak coupling BCS case and (b) the moderate coupling pseudogap case with
$s$-wave pairing. We choose the broadening $\gamma(T)=T$ for the second of
these calculations, although our results are not particularly sensitive to
the form of this parameter, provided it is non-zero.  We also indicate in
the insets, the respective temperature dependent excitation gaps, which have
been assumed in producing the figure.  It can be seen here that for the
pseudogap case, we have presumed there is some small feature in the
excitation gap at $T_c$, which was not evident in our more naive
calculations (see Appendix \ref{App_AboveTc}) which ignored lifetime effects
of the finite momentum pairs.

In the BCS case as shown in Fig.~\ref{Cv_S}(a), the specific heat jump
arises exclusively from the discontinuity of the derivative of this
excitation gap, via
\begin{equation}
 \Delta C_v = -N(0) \frac{\mbox{d}\Delta^2} {\mbox{d}T\;}\:.
\label{Cv_Jump}
\end{equation}
In contrast, in the pseudogap case, the gap $\Delta$ and its derivative
$\mbox{d}\Delta/\mbox{d}T$ are presumed to be continuous across $T_c$ as
shown in Fig.~\ref{Cv_S}(b). In this case, this jump arises from the
discontinuity of $\mbox{d}N(\omega)/\mbox{d}T$ which results from the onset
of the superconducting order. Above $T_c$, $\mbox{d}N(\omega)/\mbox{d}T$
decreases and in this way $C_v$ also decreases. It should be emphasized that
the temperature dependence of $C_v$ at $T\gtrsim T_c$ is dominated by
$\mbox{d}N(\omega)/\mbox{d}T$, instead of $N(\omega)$ itself. In fact the
contributions of these two functions are in opposite directions as a
function of temperature. This observation implies that one cannot extract
the size of the excitation gap from the specific heat without taking proper
account of the $T$ dependence of the density of states. At higher $T$ away
from $T_c$, however, $\mbox{d}N(\omega)/\mbox{d}T$ decreases gradually to
zero, and in this way, $C_v$ is controlled by $N(\omega)$.  The shape of
the anomaly in Fig.~\ref{Cv_S}(b) is more representative of a
$\lambda$-like behavior, although there is a precise step function
discontinuity just at $T_c$. The curvature above, but in the vicinity of
$T_c$, can be associated with the subtle feature at $T_c$ which appears in
the excitation gap.%
\newfootnote{One may ask what happens to the pairon $T^{3/2}$ contribution to
  the specific heat in the present model. To answer this question, we note
  that $C_v$ calculated through the derivative of the energy of the system,
  Eq.~(\ref{EnergyInt_Eq}), in principle, contains all possible
  contributions. However, here we model the pseudogap self-energy by a
  parameter $\gamma$ and have suppressed the pair dispersion. This is not
   sufficient to capture the pairon contributions properly. In fact,
  this finite $\gamma$ modeling is only meant for high $T$, whereas at low
  $T$ when both the quasiparticles and the pair excitations are very long
  lived, the formalism we developed in Sec.~\ref{Sec_Cv} should be used,
  instead.  }

In summary, we have seen above that Eq.~(\ref{Cv_Jump}) is only valid in the
weak coupling BCS limit; it necessarily breaks down in the pseudogap phase.
\textit{We conclude that} $\mbox{d}N(\omega)/\mbox{d}T$ \textit{is a more
  appropriate quantity for describing the specific heat jump at $T_c$ than
  is} $-\mbox{d}\Delta^2/\mbox{d}T$.  This result is implicitly used to
derive the simpler form of Eq.~(\ref{Cv_Jump}).  The important temperature
dependence in the temperature derivative of the density of states comes, in
turn, from the onset of long range order via $\Delta_{sc}$. Comparing
Fig.~\ref{Cv_S}(a) and (b), one can see that the gradual decrease of
$\mbox{d}N(\omega)/\mbox{d}T$ in the pseudogap case leads to a more graduate
decrease in $C_v$ above the transition, so that $C_v$ looks more like what
would be expected in a $\lambda$ transition, as distinct from a BCS step
discontinuity.

\section{Application to the cuprates}
\label{Sec_Appl_Cuprates}

The results obtained in Sec.~\ref{Sec_Spectral} and Sec.~\ref{Sec_Thermo_Cv}
are generally valid for both $s$- and $d$-wave cases, and can be readily
applied to the $d$-wave cuprates.  In this subsection we test the physical
picture and the results obtained above, by studying the tunneling spectra and the
specific heat behavior in the cuprates, as a function of doping and of temperature.

\subsection{Tunneling spectra}
\label{Subsec_Tunneling}

We first turn to calculations of tunneling spectra for
which there is substantial experimental data. These experiments were among the
first to provide information about the excitation gap --- which measurements
seem to be consistent with ARPES data.

Here we treat the hole concentration dependence of the electronic energy
scales as in Chapter \ref{Chap_Cuprates}. However, in order to compare with
tunneling spectra, in this subsection, we use a slightly more realistic band
structure which includes a next-nearest neighbor hopping term, characterized
by $t^\prime$, in the band dispersion, $\ek$. This is discussed in more
detail in Appendix \ref{App_Omegaq}. ARPES measurements indicate that both
the under- and optimally doped cuprates have a hole-like Fermi
surface \cite{Ding}, which can be roughly modeled by choosing
$t^\prime/t\approx 0.4$.  For this choice of $t^\prime$, the van Hove
singularities appear inside the filled lower half band. Because these
effects were not particularly relevant in our previous studies, this
complication was ignored until this Chapter.

For a given density of states $N(\omega)$, the quasiparticle tunneling current
across a super-conducting-insulator-normal (SIN) junction can be readily calculated \cite{Mahan},
\begin{equation}
I_{SIN} = 2eN_0 T^2_0 \int_{-\infty}^\infty \frac{\mbox{d}\omega}{2\pi}
N(\omega)  \left[ f(\omega-eV)-f(\omega)\right] \:,
\end{equation}
where we have assumed a constant density of states, $N_0$, for the normal
metal, and taken $T_0$ as the isotropic tunneling matrix element. At low $T$, one
obtains 
\begin{equation}
\left(\frac{\mbox{d}I}{\mbox{d}V}\right)_{SIN} \approx
\frac{e^2N_0T_0^2}{\pi} N(eV) \:.
\end{equation}
It can be seen that at low $T$, the tunneling spectra and the density of states are
equivalent, up to a multiplicative constant coefficient. However, at high
$T$ comparable to $T_c$, the tunneling spectra reflect a thermally
broadened density of states.

\begin{figure}
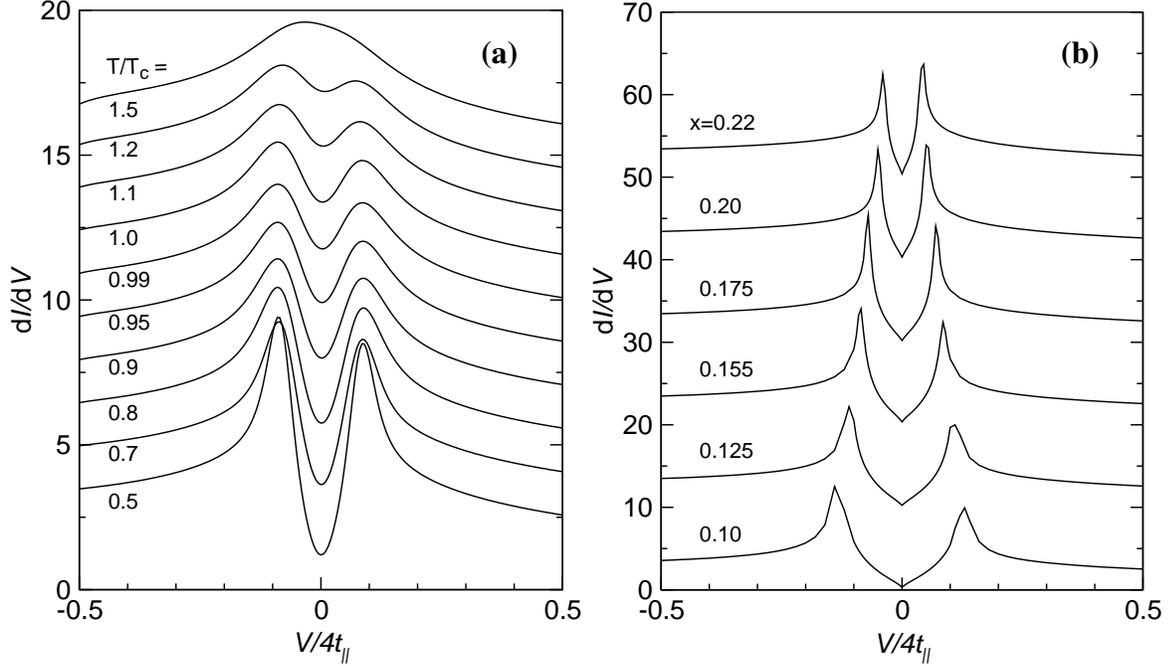

\centerline{\includegraphics[width=3in,clip]{SINDf_845t1_4-2a}
\includegraphics[width=3in,clip]{DOSDt1_4T_5-x-2}}
\vskip -3.35in 
\hskip 2.45in {\textbf{(a)} \hskip 2.8in \textbf{(b)}}
\vskip 3.07in
\caption[Temperature and doping dependence of tunneling spectra across an
SIN junction.]{(a) Temperature and (b) doping dependence of tunneling
  spectra across an SIN junction. Shown in (a) are the $\mbox{d}I/\mbox{d}V$
  characteristics calculated for optimal doping at various temperatures from
  above to below $T_c$.  Shown in (b) are tunneling spectra at low $T$
  (around $0.2T_c$) for various doping $x$. The units for
  $\mbox{d}I/\mbox{d}V$ are $e^2N_0T_0^2/4t_\parallel$. For clarity, the
  curves in (a) and (b) are vertically offset by 1.5 and 10, respectively.}
\label{SIN}
\end{figure}

In Fig.~\ref{SIN}(a), we plot the SIN tunneling spectra, calculated for
optimal doping at temperatures varying from above to below $T_c$. The van
Hove singularity introduces a broad maximum in the spectra at high
temperatures, as seen for the top curve in Fig.~\ref{SIN}(a). [It should be
noted that the density of states contains (pseudo)gap like features which
lead to two peaks even at this relatively high temperature].  The figure
shows that even for this optimal sample, as is consistent with the phase
diagram of Fig.~\ref{SpectralFunction}, [although not so evident in the
first generation experiments summarized in
Fig.~\ref{PhaseDiagram_ARPES}(b)], there is a pseudogap at $T_c$. Because the
optimal sample is reasonably well within the pseudogap regime, it is
reasonable to presume, as in our calculations, that the excitation gap
increases only slightly with
decreasing temperature.%
\newfootnote{Thermal broadening of the spectral peaks tends to push the peaks apart
  and may lead to a seemingly increasing gap with
  $T$ above $T_c$ in some cases. }
In Fig.~\ref{SIN}(b), we show the calculated tunneling spectra at low $T$
for variable doping concentration.%
\newfootnote{These doping dependent measurements are usually taken at 4.2~K}
As expected from the phase diagram (Fig.~\ref{Cuprate_Phase}), which
consolidates both tunneling and ARPES data, the excitation gap seen in
Fig.~\ref{SIN}(b) increases with underdoping.  Indeed, the behavior of the
tunneling spectra shown in both plots is similar to what is observed
experimentally by Renner and coworkers \cite{Renner} and by
Miyakawa \textit{et al} \cite{JohnZ}.  It should be noted that there is no
direct feature in the tunneling curves associated with $\Delta_{sc}$. There
are, however, indirect effects arising from the onset of long range order,
which lead to the sharpening of the density of states peaks.  Note that we
have ignored (as previously) other self-energy contributions (e.g., arising
from the particle-hole channel) which would enter into the parameter
$\Sigma_0$ in Eq.~(\ref{SigmaPG_Model_Eq}). Finally, we point out that the
asymmetry of the high energy background in tunneling data is a natural
consequence of a non-zero $t^\prime$.

\subsection{Specific heat}
\label{Subsec_Cv}

We, next, discuss the behavior of the specific heat in the cuprates as a
function of temperature and hole concentration. Here, too, there is a
substantial amount of experimental data, although primarily from one
experimental group \cite{Loram98}.  In the present calculations there are two
important parameters which we treat phenomenologically, rather than by
iteratively solving our set of three coupled integral equations. They are
the total gap $\Delta$ and the pair lifetime $\gamma$.  In the analyses of
previous chapters the latter played no important role, so that $\Delta$
could be reasonably accurately calculated by ignoring lifetime effects.  For
$C_v$ this is no longer the case and $\gamma$, particularly, must be chosen
with some care.  As is consistent with lifetime estimates in the literature,
we take $\gamma \propto T^3$ below $T_c$ and linear in $T$ above $T_c$.%
\newfootnote{Recalling that $\Delta_{pg}^2\propto T^{3/2}$ at low $T$ and is
  roughly linear in $T$, this choice of $\gamma$ will give rise to a
  temperature dependence of the effective quasiparticle peak width
  $\gamma^\prime \propto T^{4-4.5}$ for $T\leq T_c$, via
  Eq.~(\ref{Gamma'_Eq}). This is consistent with the experimentally observed
  $T^4 - T^5$ dependence of the quasiparticle scattering rate \cite{Hardy2}.
  Above $T_c$, the $T$-dependence of $\Delta_{pg}$ is weak, so that
  $\gamma^\prime \propto T$, in agreement with ARPES data \cite{Valla}.}
For the doping dependence, we assume that $\gamma$ varies inversely with
$\Delta$ since when the gap is large, the available quasiparticle scattering
decreases.  With this reasonable assumption, along with the continuity of
$\gamma$ at $T_c$, we obtain a simple form:
\begin{equation}
\gamma = \left\{ \begin{array}{l@{\hspace{1cm}}l}
a T^3/T_c\Delta \:, & (T< T_c),\\
a TT_c/\Delta \:, & (T> T_c). \end{array} \right.
\label{Gamma_Model_Eq}
\end{equation}
Here, the coefficient $a$ is of the order of, or less than, unity.  
%Besides $\gamma$, another important choice to make is the temperature
%dependence of $\Delta$ above $T_c$, since we cannot calculate $\Delta$ with
%finite broadening as well as the chemical potential self-consistently. We
%allow the gap above $T_c$ to decrease exponentially towards a high $T$
%value, but we impose the continuity of the gap and its derivative with
%respect to $T$ at $T_c$. The gap decreases within certain fraction (fixed
%for all $x$) of $T^*-T_c$, and the gap is filled in at high $T$ by the large
%$\gamma$. The gaps derived this way seems to be consistent with the neutron
%scattering experiment \cite{Dai} and some tunneling spectra, though it is
%slightly different from our simple zero $\gamma$ extrapolation above $T_c$
%as shown in Appendix \ref{App_AboveTc}. 
Our choice for the excitation gaps is shown in Fig.~\ref{Cv-Gaps}. Here, at
this higher level of approximation we expect some, albeit weak, structure to
be introduced at $T_c$ (since $\gamma$ varies around $T_c$). This provides a
small correction to the first order solutions shown in
Fig.~\ref{Cuprate_Gaps}, and discussed in Appendix \ref{App_AboveTc}.

%It should be noted that both a finite $\gamma$ and the right $T$-dependence
%of $\Delta$ are important. Without a finite $\gamma$, the thermodynamical
%behavior would be BCS-like, and there would be no step discontinuity in
%$C_v$.  On the other hand, if we use the gaps (above $T_c$) calculated using
%the extrapolation scheme in Appendix \ref{App_AboveTc},
%$\mbox{d}\Delta/\mbox{d}T$ will increases $C_v$ above $T_c$, so that $C_v$
%may keep growing until it passes a maximum around $T^*$. In short, we do not
%care much about the excitations in deriving $T_c$, whereas we have to when
%the excitations become the subject of study. To keep the same amount of
%density of states depletion, it is expected that the gaps obtained with a
%finite life time broadening should be larger.
%
\begin{figure}
  \centerline{\includegraphics[width=6in,clip]{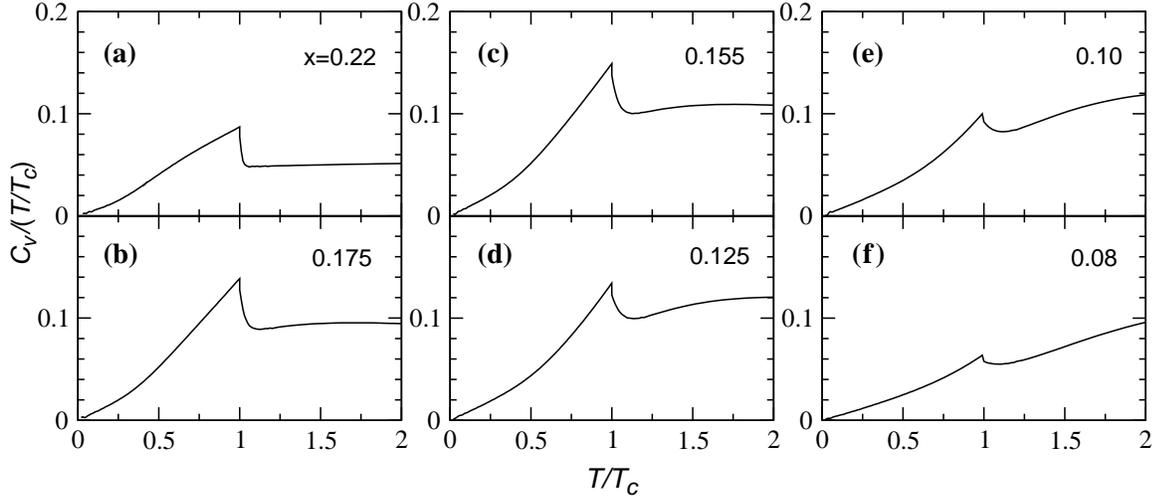}}
\vskip -0.1in
\caption[Temperature dependence of the specific heat for various doping
concentrations.]{Temperature dependence of the specific heat for various
  doping concentrations, calculated with $a=1/4$ in Eq.~\ref{Gamma_Model_Eq}.}
\label{Cv-x}
\end{figure}

The results for $C_v/T$ are plotted in Fig.~\ref{Cv-x}(a)-(f), from over- to
underdoped systems.  As shown in these plots, the behavior of $C_v$ is
BCS-like in the overdoped regime. As the system passes from optimal doping
towards underdoping, the behavior is more representative of a $\lambda$-like
anomaly. This latter shaped curve is similar to what was seen in the
$s$-wave case discussed in Sec.~\ref{Sec_Thermo_Cv} and can be traced to
the very weak feature in the temperature dependence of $\Delta$ at $T_c$.
All these trends seem to be qualitatively consistent with experimental data
\cite{Loram,Loram98}. As can be seen from the figures, in the underdoped
regime at high $T$, there is a maximum in $C_v/T$, around $T^*$.  Note the
difference at low $T$ between the $d$-wave results here and the $s$-wave
behavior of Fig.~\ref{Cv_S}.  Here, the $d$-wave nodes lead to a larger
quasiparticle specific heat at low $T$.

The experimentally observed $\lambda$-like anomaly of $C_v$ at $T_c$ has
been interpreted previously as evidence for a Bose condensation description.
Here, in contrast, we see that within our generalized mean field theory,
this anomaly naturally arises from the temperature dependence of the
fermionic excitation gap which has some structure at, but persists above
$T_c$. (Thus, this is a property of superconductors which have a well
established pseudogap!) That the experimental data, (which, except at
extremely reduced hole concentrations), show a reasonably sharp
($\lambda$-like) structure at $T_c$ --- seems to reinforce the general theme
of this Chapter and Thesis --- that corrections to BCS theory may be
reasonably accounted for by an improved mean field theory, rather than by,
say, including fluctuation effects.

\begin{figure}
  \centerline{\includegraphics[width=3.in,clip]{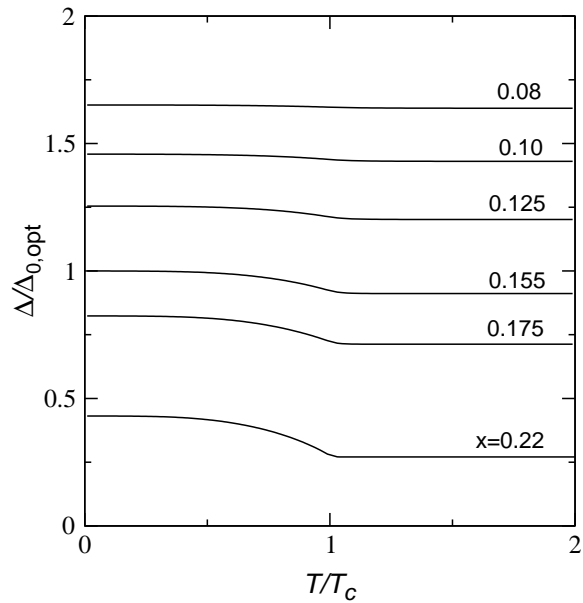}}
\vskip -0.1in
\caption[Temperature dependence of the excitation gaps for various doping
concentrations used for calculations in Fig.~\ref{Cv-x}.]{Temperature
  dependence of the excitation gaps for various doping concentrations used
  for calculations in Fig.~\ref{Cv-x}. The units for the gaps are the zero
  $T$ gap at optimal doping.}
\label{Cv-Gaps}
\end{figure}

\section{Low \textit{T} extrapolation of the pseudogapped normal state}
\label{Sec_Extrapolation}

One important consequence of specific heat measurements is that 
they, in principle, may provide information about
the condensation energy, which is the energy gained upon transforming from
the normal to the superconducting states.  What is required, however, is to make an
extrapolation of the normal state down to low $T<T_c$. This low
$T$ extrapolated (``normal") state has played an important role in traditional
superconductors, where the condensation energy can be precisely determined
by integrating the difference between the entropy of the superconducting
state and that of the extrapolated normal state with respect to $T$. In this
final section, we discuss the thermodynamical variables associated with the
extrapolated normal state in pseudogap superconductors.

As in any phase transition, one has to compare the free energies of the
disordered and ordered phases in order to establish their relative
thermodynamical stability. This extrapolated normal state is at the basis of
the free energy functional of Landau-Ginzburg theory.  Moreover,
considerable attention has been paid recently to the related condensation
energy in the context of determining the pairing ``mechanism" in the high
temperature superconductors \cite{Demler,Norman2}.  However, in order to
address the mechanism at a microscopic level, one needs to extract the
condensation energy accurately along with determining the behavior of the
various extrapolated polarizabilities (e.g.,magnetic and electric) below
$T_c$ .  In this way, one has to address the nature of the extrapolated
normal state and, more specifically, the question of whether this normal
state is a Fermi liquid --- at temperatures below $T_c$, as in BCS theory.
The answer we provide is unequivocally no, on all but the overdoped side of
the cuprate phase diagram.  There are several simple ways to support this
answer using experimental thermodynamical data, which compare quite
favorably with the calculations of this Chapter. We discuss here two ways,
based on the specific heat --- or entropy, and on the tunneling density of
states.

\begin{figure}
  \centerline{\includegraphics[width=5.5in,clip]{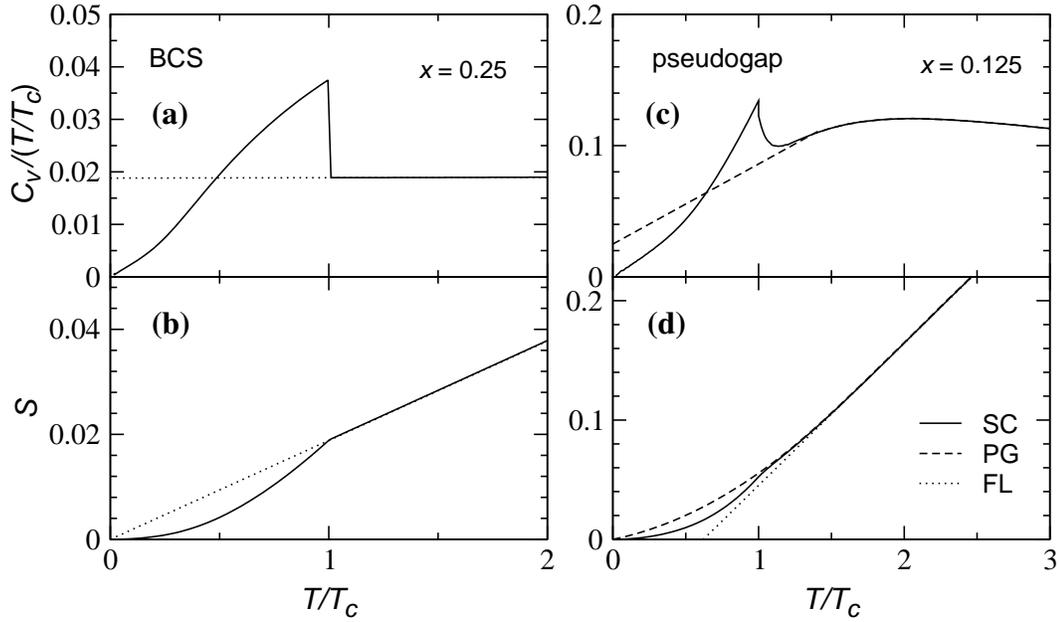}} \vskip -2.8in
  \hskip 1in {\textbf{(a)}} \hskip 2.35in {\textbf{(c)}} \vskip
  0.9in \hskip 1in {\textbf{(b)}} \hskip 2.35in {\textbf{(d)}}
  \vskip 1.4in
\caption[Comparison of the extrapolated normal state below $T_c$ in BCS and
pseudogap superconductors.]{Comparison of the extrapolated normal state
  below $T_c$ in (a)-(b) BCS and (c)-(d) pseudogap superconductors. Shown
  are the extrapolations for $C_v/T$ and the entropy $S$ in the upper and
  lower panels, respectively.}
\label{Cv_Extrapolation}
\end{figure}

In Fig.~\ref{Cv_Extrapolation} we plot the calculated $C_v/T$ and entropy
$S$ for (a)-(b) the BCS case, and compare it with the counterparts obtained
for the pseudogap superconductor in (c) and (d). The dotted lines represent
the Fermi liquid extrapolation. Figures \ref{Cv_Extrapolation}(a) and (b)
show that the Fermi liquid extrapolation is sensible for the BCS case ---
$C_v/T$ is a constant, and $S$ is a straight line going through the origin,
yielding the consistent result: $S=0$ at $T=0$.  In contrast, for the
pseudogap case, the Fermi liquid extrapolation is unphysical, approaching a
negative entropy at low $T$.  Finally, the dashed lines in
Fig.~\ref{Cv_Extrapolation}(c)-(d) show a fairly sensible extrapolated
normal state. The extrapolated $C_v/T$ in Fig.~\ref{Cv_Extrapolation}(c) is
consistent with conservation of entropy, $S=\int_0^T C_v/T \mbox{d}T$,
i.e., the area under the $C_v/T$ curves.  The ``normal" state entropy is
obtained via integrating the extrapolated $C_v/T$, leading to a vanishing
entropy at $T=0$.  Indeed, this is in fact the empirical procedure followed
experimentally \cite{Loram98}, and Figs.~\ref{Cv_Extrapolation}(c)-(d) are
similar to what is observed experimentally \cite{Loram98} in underdoped
cuprates. The ``normal" state $C_v/T$ and $S$ turn out to be linear and
quadratic in $T$, respectively.  A rough estimate of the condensation energy
can be obtained from the integral area between the solid line (for the
superconducting state) and the dashed line (for the extrapolated normal
state) in Fig.~\ref{Cv_Extrapolation}(d).
It can be noted that, while a more precise measure of the condensation
energy may be obtained by computing the field dependent Gibbs free energy,
this is more complicated to implement both theoretically and experimentally.

Finally, it should be stressed that the behavior of the extrapolated $C_v/T$
indicates that \textit{there exists a pseudogap below $T_c$}. This is the
same conclusion as reached in previous empirical studies of the specific
heat \cite{Loram98}.

\begin{figure}
\centerline{\includegraphics[width=6in,clip]{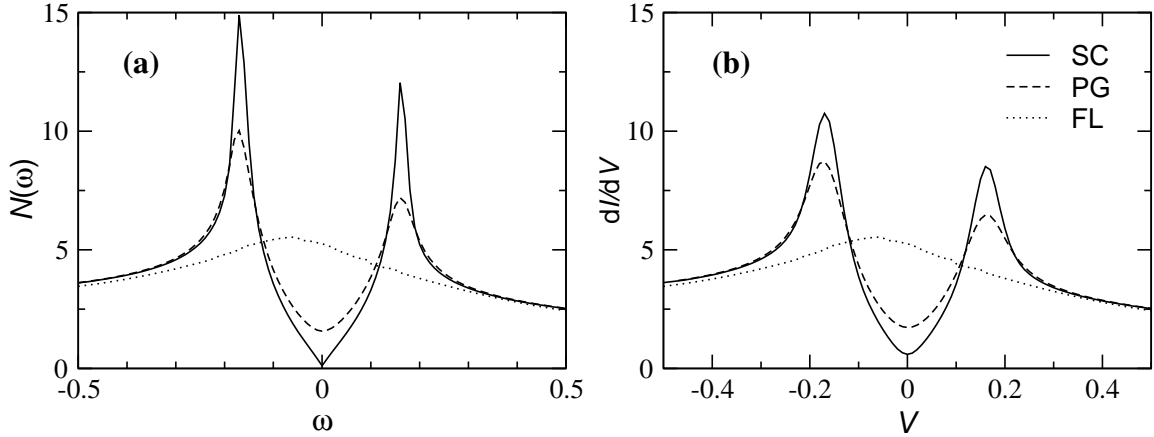}}
\vskip -2.1in 
\hskip 0.6in \textbf{(a)} \hskip 2.85in \textbf{(b)}
\vskip 1.85in
\caption[Density of states and SIN tunneling $\mbox{d}I/\mbox{d}V$ 
characteristics of an extrapolated pseudogap ``normal state'' below
$T_c$.]{(a) Density of states and (b) SIN tunneling $\mbox{d}I/\mbox{d}V$ 
  characteristics of an extrapolated pseudogap ``normal state'' below
  $T_c$. For comparison, also plotted are Fermi liquid extrapolations.}
\label{DOS_PG}
\end{figure}

The pseudogap also appears in the extrapolated normal state density of
states below $T_c$. In Fig.~\ref{DOS_PG}, we show the three analogous curves
for the density of states as well as the tunneling spectra, calculated at
$T=0.5T_c$. The solid curves correspond to the superconducting phase, the
dotted curves to the Fermi liquid extrapolation and the dashed curves to the
calculated normal state of the pseudogap phase. For the latter we assume
that the full excitation gap exists without superconducting long range
order. There is some experimental support for our extrapolated normal state
(or PG) curves which can be obtained from vortex core spectroscopy STM
measurements \cite{Renner_Vortex}.  Thus, \textit{we deduce from tunneling
  measurements that some manifestation of a pseudogap exists below $T_c$.}

These two studies provide reasonably strong support for the present
theoretical picture.

\vskip 1cm

In concluding this Chapter, we note that the onset of the superconducting
long range order at $T_c$ leads to a rapid sharpening of the quasiparticle
spectral peaks, and to a discontinuity of the derivative of the density of
states across $T_c$.  As a consequence, the superconducting phase transition
manifests itself as a step discontinuity or a $\lambda$-like anomaly in
$C_v$ at $T_c$, depending on the doping concentration. We have further shown
that the pseudogap exists in extrapolated thermodynamical properties below
$T_c$, and must be properly accounted for in a proper analysis of the
condensation energy and the ``pairing mechanism" or related calculated
polarizabilities below $T_c$.

\chapter{Concluding remarks}
\label{Chap_Conclusion}

The work outlined in this thesis is viewed as having an impact which goes
beyond understanding high $T_c$ superconductors.  Even in the days before
BCS, there was a longstanding interest in Bose-Einstein condensation and its
relation to superconductivity. These ideas led to the picture of Schafroth
condensation, which did not receive much attention following the BCS
proposal. Today, with the resurgence of interest in BEC in the alkali gases,
it is even more important to understand the relationship between a BCS and
BEC description of superfluidity or superconductivity. Moreover, with the
discovery of more and more ``exotic" superconductors, which have short
coherence lengths, interest has again focused on the connections between BCS
and BEC.  Indeed, BEC-like or pseudogap behavior has been found (for
example, in the Knight shift) in 2D organic (short $\xi$) superconductors.
%For example, the
%temperature dependence of the specific heat at the transition in the
%underdoped cuprates is more like a $\lambda$ transition, characteristic of
%BEC, rather than a BCS jump. 
In this context, the Uemura plot of Fig.~\ref{UemuraPlot} is particularly
important (not only because it suggests a BEC-like scaling between $T_c$ and
the superfluid density but) because it suggests that the cuprates are not
particularly special when compared with other short $\xi$ superconductors,
including organic, heavy fermion superconductors, etc.  This leads naturally
to the conclusion that the most likely reason for the failure of BCS theory
in these materials is the small size of $\xi$. A radical departure from BCS
theory may not be warranted.  At the very least, whatever complicated
physics needs to be additionally introduced, as a \textit{baseline}, it
seems essential to address the fundamental non-BCS-mean-field effects
associated with short $\xi$.

Beyond these general observations one may ask what is the strongest support
for applying our theoretical BCS-BEC based scheme to the cuprates.  After
all, there are alternatives in the literature and these seem to fall into
two main categories. (i) The excitations of the superconducting state are
purely fermionic, and more or less describable by $d$-wave BCS theory,
except for the introduction of Fermi liquid renormalizations.  This school
of thought is called ``Fermi liquid based superconductivity".  (ii) The
excitations of the superconducting state are essentially bosonic.  This
leads to the ``phase fluctuation" approaches, which are based in large part
on the small values of the phase stiffness $n_s/m$ in the very underdoped
regime. Ours is a third alternative in which both fermionic and bosonic
excitations coexist (although the bosonic excitations here are \textit{not}
phase fluctuations).

In establishing support for the present viewpoint, it is important to note
that there are many other features of the cuprates which suggest BCS theory
is not strictly applicable. Impurity effects are found to be highly local
(unlike the behavior predicted by the Abrikosov-Gor'kov generalization of
BCS).  This locality is presumed to arise from short $\xi$.  Moreover, there
are unexpectedly localized states in the vortex cores, which are also
presumed to be associated with short $\xi$. Even more interesting, perhaps,
is that the normal state \textit{below} $T_c$ in a vortex core is found to
have a pseudogap, much as we would expect; it is not a Fermi liquid normal
state, as in BCS theory. In addition there are unusual anomalies in the
thermal conductivity and an unexpectedly small non-linear Meissner effect.

\section{Summary}

In this thesis, we have extended BCS theory in a natural way to short
coherence length superconductors, based on a simple physical picture in
which incoherent, finite momentum pairs become increasingly important as the
pairing interaction becomes stronger. We went beyond the BCS mean field
treatment of the single particles and the superconducting condensate, by
including these finite momentum pairs at a mean field level, and we,
thereby, treated the pairs and the single particles on an equal footing.
Within our picture, the superconducting transition from the fermionic
perspective and Bose-Einstein condensation from the bosonic perspective are
just two sides of the same coin.

In this context, we have found a third equation which we call the pseudogap
equation. This equation must be solved in conjunction with the excitation
gap equation and the fermion number equation.  The single particle number
equation characterizes the single particles, the pseudogap equation, which
is essentially a pair number equation, characterizes the pairs. The single
particles and the pairs are interrelated via the gap equation.  On one hand,
this gap equation ensures the softness of the pair dispersion; on the other
hand, it controls the gapping of the fermionic or single particle
dispersion. In contrast to BCS, the gap equation now contains new
ingredients associated with the finite momentum pairs, in addition to the
zero momentum Cooper pair condensate. The incoherent pairs lead to the
pseudogap above $T_c$ and to the very important distinction between the
excitation gap and the superconducting order parameter below $T_c$, at
intermediate and strong coupling.

In contrast to most other theoretical approaches, our theory is capable of
making quantitative predictions, which can be tested.  This theory was
applied to the cuprates to obtain a phase diagram.  (We believe this may be
the first time that the cuprate phase diagram has been calculated
quantitatively within any single theory). In addition, because this fitted
(with one free parameter) phase diagram represented experiment quite well,
it was possible to quantitatively address derived quantities, such as the
hole concentration and temperature dependences of the in-plane penetration
depth and specific heat.  The mutually compensating contributions from
fermionic quasiparticles and bosonic pair excitations have provided a
natural explanation for the unusual quasi-universal behavior of the
normalized in-plane superfluid density as a function of reduced temperature.
Our bosonic pair excitations also demonstrate an intrinsic origin for the
long mysterious linear $T$ term in the specific heat.  We found new
qualitative effects as well, associated with predicted low $T$ power laws,
which arise from incoherent pair contributions.  These power laws seem to be
consistent with existing experiments, although more systematic experimental
studies are needed. Finally, we showed that the onset of superconducting
long range order leads to sharp features in specific heat at $T_c$, in both
over- and underdoped regime. In addition, we demonstrated that the pseudogap
exists in the extrapolated normal state below $T_c$, and must be properly
accounted for in extracting the condensation energy from specific heat data.

\section{Remarks}

%I THOUGHT RECITING THE VARIOUS PARAMETERS WAS TOO TECHNICAL FOR SUMMARY.
%In contrast to other theoretical approaches, our theory is capable of making
%quantitative predictions, which can be tested. Moreover, we do not have many
%free tuning parameters. Strictly speaking, we have three parameters, with
%\textit{only one free}: (i) The band parameter $t_0$ has to be consistent
%with experiment; (ii) The mass anisotropy ratio $t_\perp/t_\parallel$
%affects $T_c$ (and only $T_c$) only logarithmically, and has to be
%consistent consistent with experiment; (iii) The free parameter $g$ has to
%be chosen to optimize the phase diagram so that all $\Delta_0$, $T^*$, as
%well as $T_c$, are in agreement with experimental data. Once these three
%parameters are chosen, all other physical quantities can be uniquely
%determined.

The BCS form of the gap equation found in our theory is crucially related to
the Leggett ground state and the $G_0G$ scheme within the $T$ matrix
approximation.  Unlike a diagrammatic $GG$ scheme, the present $G_0G$ scheme
derives from a truncation of the equations of motion at the pair level.  Our
approach should be viewed as a generalized mean field theory. This is the
essence of the $T$ matrix approximation. Incoherent pairs are treated at the
mean field level, no explicit pair-pair interactions are involved. To
include inter-pair interactions, one presumably needs to truncate the
equations of motion at a higher order.  In line with this picture, at the
$T$ matrix level, the pair dispersion is always quadratic, (except in the
weak coupling limit when exact particle-hole symmetry is assumed). This
dispersion is clearly different from the order parameter collective modes,
which effectively include inter-pair interactions through higher order
diagrams. The pairs should be thought of as ``single'' objects, rather than
collective excitations.

One natural consequence of the $T$ matrix approximation is the quasi-ideal
Bose gas nature of the system in the strong coupling limit. The pairs
(bosons) are essentially free. There is a full condensation at $T=0$,
consistent with the Leggett ground state. This is different from a
``true'' interacting Bose system. This is required by the BCS form of the
gap equation. Had the Cooper pairs been interacting with each other, one
would presumably not get the BCS gap equation. Requiring the strong coupling
limit to be a true Bose liquid would be equivalent to abandoning the Leggett
ground state equations, at least in the
large $g$ limit.

\section{Speculations}

Would one find a true Bose liquid in the strong coupling limit if one
included higher order Green's functions? We have tried to go beyond the $T$
matrix approximation by truncating the equations of motion at $G_4$, in the
hope of retaining explicitly pair-pair interaction diagrams. Unfortunately,
as pointed out by Kadanoff and Martin, it is technically extremely difficult
to implement this procedure.  We did find terms corresponding to pair-pair
interactions but the set of coupled equations is much more complicated than
would be the case for a true Bose system.  In this procedure the one
particle i.e, fermionic Green's functions also enter since the starting
Hamiltonian involves only inter-fermion interactions.  We are, thus, not
sure whether one would get a true boson like dispersion for the incoherent
pairs in the strong coupling limit if one could carry out these complex
calculations.

An alternative physically intuitive approach to the strong coupling limit
would be to treat the pairs as fundamental objects, which might virtually
decay into two fermions.  (This is analogous to the $\pi$ meson problem, in
which each meson is composed of a quark-anti-quark pair.) In this case, one
would tend to dress the two electrons within a pair in the same way. The
bosons would interact only via virtual processes. It might be possible to
find a true Bose liquid behavior of the system. It should be stressed that
this procedure would amount to a modification of the Leggett ground state.
In addition, the two constraining equations of the Leggett formalism --- the
fermionic chemical potential and excitation gap --- would presumably no longer
be meaningful.  We believe these issues are interesting on general
theoretical grounds, but that this limit is not appropriate when the system
is still in the fermionic regime, as seems to be the case for all known
superconductors.

\section{Future directions}

It is clear that in our formalism, the ratio $T_c/\Delta$ is not given by
the BCS universal constant.  Rather $T_c$ is controlled by the pseudogap in
the intermediate coupling regime.  This may explain the absence of clear
isotope effects. Because the pseudogap is enhanced by lower dimensionality,
a higher mass anisotropy will suppress $T_c$. This may also be relevant to
pressure effects on $T_c$.  $T_c$ has been found to increase with pressure
in most cuprate superconductors. As the pressure increases, the interlayer
distance decreases, and, thus, the interlayer hopping matrix element may
increase substantially; in this way $T_c$ is enhanced.

There are obvious extensions of the present work to include inhomogeneity
and magnetic field effects. A simple speculation regarding the latter would
be that a magnetic field is effective in destroying phase coherence, but
does \textit{not} affect pair formation, so that $T^*$ and $\Delta$ are less
strongly field dependent.  If a pseudogapped superconductor is driven normal
by a large field, it may still exhibit an excitation gap, so that it is not
in a metallic phase.  Said, alternatively, without phase coherence, pairs
can still contribute to the excitation gap, but not to the
superconductivity.  Experiments which involve inhomogeneity are localized
impurity effects, vortex core states, Andreev reflection, etc. An extension
along these lines will be very important.

There are many other experiments we need to address. For example, the
unexpectedly weak non-linear Meissner effect, the unexpectedly strong field
dependence of $C_v$ around $T_c$, the electrical (ac and dc) conductivity,
above $T_c$, and the unusual anomalies in the thermal conductivity recently
reported by Ong \textit{et al} \cite{Ong00}. To address some of these
experiments, it will be necessary to extend the current formalism above
$T_c$ to make it more practical and less computational, as was achieved here
below $T_c$.

\appendix

\chapter{Expression for pair dispersion, $\Omega_{\bf \lowercase{q}}$}
\label{App_Omegaq}

In the long wavelength, low frequency limit, one can expand the inverse $T$
matrix as:%
\footnote{Note here that the $\xi$ which appears in the $q^2$ term,
  following the conventional notation for a Ginzburg-Laudau expansion, is
  different from the coherence length we talk about in this Thesis.}
\begin{equation} 
t^{-1}_{pg}({\mathbf{q}}, \Omega) = a_1\Omega^2 + a_0 \Omega -\xi^2 q^2
+ \tau_0^\prime +i \Gamma_{{\mathbf{q}}, \Omega}^\prime\;.  
\label{Omega_q:exp}
\end{equation}
Here the imaginary terms, which are included in $\Gamma_{{\bf
    q},\Omega}^\prime$, are negligible below $T_c$.
The linear contribution (\textbf{q}) is absent due to the inversion
symmetry (${\bf q} \leftrightarrow - {\bf q}$) of the system.

In the weak coupling limit, the ratio $a_0/a_1$ is vanishingly small; when
the system has exact particle-hole symmetry (e.g., a 2D tight binding band
at half filling with a nearest neighbor hopping), $a_0$ vanishes. In this
case the dispersion determined via
\begin{equation}
t^{-1}_{pg}({\mathbf{q}}, \Omega) = 0
\end{equation}
is linear in $q$, $\Omega_{\bf q} \sim cq$. In the absence of particle-hole
symmetry, as $g$ increases, $a_0/a_1$ increases, thus $a_0 \Omega$ gradually
dominates and we find the important result: $\Omega_{\bf q} \sim Bq^2$. For
any finite $g$ and arbitrarily small $q$, the dispersion is always
quadratic, at the lowest energies.

We are interested in the moderate and strong coupling cases, where we can
drop the $a_1 \Omega^2$ term in Eq.~(\ref{Omega_q:exp}), and hence we have
\begin{equation} 
t_{pg}({\mathbf{q}}, \Omega) = \frac{a_0^{-1}}{\Omega - \Omega_{\mathbf{q}}
  + \mu_{pair} + i\Gamma_{{\mathbf{q}}, \Omega}}\;,
\label{Omega_q:t}
\end{equation}
where 
\begin{equation}
\Omega_{\mathbf{q}} = B q^2 = \frac{\xi^2}{a_0} q^2 \equiv \frac{q^2} {2
  M^*} 
\end{equation}
is quadratic. This defines the effective pair mass, $M^*$. Below $T_c$, we
have $\mu_{pair}=0$, as a consequence of the gap equation, $t^{-1}(Q=0) =
0$, as is consistent with ideal Bose gas condensation.

For an anisotropic 3D or a quasi-2D case, this becomes
\begin{equation}
\Omega_{\mathbf{q}} = B_\parallel q_\parallel^2 + B_\perp q_\perp^2 
 \equiv \frac{q_\parallel^2} {2 M_{\parallel}^*} 
+ \frac{q_\perp^2} {2 M_{\perp}^*} 
\end{equation}

The inverse $T$ matrix can be written explicitly:
\begin{equation}
t^{-1}_{\mb{q},i\Omega_n} = g^{-1} + \sum_\mb{k} \left[ \frac{1-f(\Ek)-f(\ekq)}
  {\Ek + \ekq - i\Omega_n} u_\mb{k}^2 - \frac{f(\Ek) - f(\ekq)} {\Ek - \ekq 
  + i\Omega_n} v_\mb{k}^2 \right] \phikq^2 \;,
\end{equation}
where $\Omega_n = 2n\pi T$ is Matsubara frequency for a boson field. After
analytical continuation, $i\Omega_n \rightarrow \Omega+i0^+$, one obtains 
\begin{equation}
t^{-1}_{\mb{q},\Omega+i0^+} = g^{-1} + \sum_\mb{k} \left[
  \frac{1-f(\Ek)-f(\ekq)} {\Ek + \ekq - \Omega-i0^+} u_\mb{k}^2 -
  \frac{f(\Ek) - f(\ekq)} {\Ek - \ekq + \Omega+i0^+} v_\mb{k}^2 \right]
\phikq^2\;.  
\label{Omega_q:t-exp}
\end{equation}
From Eq.~(\ref{Omega_q:t}), we have
\begin{eqnarray} 
a_0 & = & \left. \frac{\partial}{\partial \Omega}
  t^{-1}_{\mb{q},\Omega}\right|_{q=0, \Omega=0}\nonumber\\ 
    & = & \sum_\mb{k}
 \left[ \frac{1-f(\Ek)-f(\ekq)} {(\Ek + \ekq)^2} u_\mb{k}^2 + 
   \frac{f(\Ek) - f(\ekq)} {(\Ek - \ekq)^2} v_\mb{k}^2 \right]\phik^2
 \nonumber\\
    & = & \frac{1}{2\Delta^2}\sum_\mb{k} \bigg[ [1-2f(\ek)] -
  \frac{\ek}{\Ek} [1-2f(\Ek)]\bigg] \nonumber\\
   &  = & \frac{1}{2\Delta^2} \bigg[ n- 2\sumk f(\ek)\bigg] \;,
\label{Omega_q:a0}
\end{eqnarray}
where use has been made of the number equation (\ref{Number_Eq}) in the last
step in Eq.~(\ref{Omega_q:a0}). The imaginary part of the pair dispersion is
given by
\begin{eqnarray}
\Gamma_{\mb{q},\Omega} &=& \frac{\pi}{a_0} \sumk \mbox{\Large $\{$}
 \left[1-f(\Ek)-f(\ekq)\right] \uk^2 \delta (\Ek+\ekq-\Omega)
  \nonumber\\ 
&&{} +  \left[f(\Ek)-f(\ekq)\right] \vk^2
  \delta(\Ek-\ekq+\Omega)\mbox{\Large $\}$} \phikq^2 \:.
\label{Omega_q:Gamma_q}
\end{eqnarray}
While it clearly depends on $\Omega$ in general, however,
$\Gamma_{\mb{q},\Omega}$ is relevant only at $\Omega=\Omegaq$.  For $s$-wave
pairing, $\Gamma_{\mb{q},\Omegaq}$ vanishes identically when
$-(\Ek-\ekq)_{min} < \Omegaq < (\Ek+\ekq)_{min}$. For $d$-wave pairing,
detailed numerical calculations show that $\Gamma_{\mb{q},\Omegaq}$ is
always much smaller than $\Omegaq$ for small \textbf{q}, $T\leq T_c$. In
both cases, the pair dispersion is well defined for small $({\bf q},
\Omega)$ at $T\leq T_c$, and is given by
\begin{eqnarray}
\Omega_\mb{q} &=& - \frac{1}{a_0} t^{-1}_{\mb{q},\Omega=0} \nonumber\\
 & = & -\frac{1}{a_0} \left\{ g^{-1} + \sum_\mb{k} \left[
  \frac{1-f(\Ek)-f(\ekq)} {\Ek + \ekq} u_\mb{k}^2 -
  \frac{f(\Ek) - f(\ekq)} {\Ek - \ekq} v_\mb{k}^2 \right]
\phikq^2\right\} \nonumber\\
 & = & -\frac{1}{a_0} \left\{ \sum_\mb{k} \left[
  \frac{1-f(\Ek)-f(\ekq)} {\Ek + \ekq} u_\mb{k}^2 -
  \frac{f(\Ek) - f(\ekq)} {\Ek - \ekq} v_\mb{k}^2 \right]
\phikq^2 \right. \nonumber\\
&&{}  -\left. \frac{1-2f(\Ek)}{2\Ek}\phik^2 \right\} \nonumber\\
 & = & \sum_i B_i q_i^2 + \ldots  \;, \qquad\qquad (i=1,2,3).
\end{eqnarray}
Here we have used the gap equation (\ref{Gap_Eq2}) in the 4th line.
We have verified numerically that this leading order expansion of $\Omegaq$
is sufficient, so that higher order terms in the expansion can be dropped.
\begin{eqnarray}
-a_0 \frac{\partial\Omegaq}{\partial \mb{q}\;} & = &
\sumk \left[ -\frac{\uk^2}{\Ek+\ekq} 
  + \frac{\vk^2}{\Ek-\ekq}\right] f^\prime(\ekq)  
 \phikq^2 \vec{\nabla}_\mb{q} \ekq  \nonumber\\
 &&{} - \left[ \frac{1-f(\Ek)-f(\ekq)} {(\Ek + \ekq)^2} u_\mb{k}^2 + 
   \frac{f(\Ek) - f(\ekq)} {(\Ek - \ekq)^2} v_\mb{k}^2 \right]
 \phikq^2 \vec{\nabla}_\mb{q} \ekq \nonumber\\
&&{} + \left[ \frac{1-f(\Ek)-f(\ekq)} {\Ek + \ekq} u_\mb{k}^2 - 
   \frac{f(\Ek) - f(\ekq)} {\Ek - \ekq} v_\mb{k}^2 \right]
\vec{\nabla}_\mb{q} \phikq^2 \nonumber\\
&\equiv & (1) + (2) + (3)\;,
\end{eqnarray}
where $f'(x)=-f(x)[1-f(x)]/T$. Thus,
\begin{equation}
-a_0 \left.\frac{\partial^2\Omegaq}{\partial \mb{q}^2} 
 \right|_{\mb{q}=\mb{0}}  = (\mbox{I}) + (\mbox{II}) + (\mbox{III}), 
\end{equation}
where
\begin{mathletapp}
\begin{eqnarray}
(\mbox{I}) &=& \left. -a_0 \frac{\partial^2\Omegaq}{\partial \mb{q}^2} 
 \right|_{\mb{q}=\mb{0}}  : \mbox{1st term} \nonumber\\
 & = & \left. \sumk \left[ \frac{\uk^2}{(\Ek+\ekq)^2} 
  + \frac{\vk^2}{(\Ek-\ekq)^2} \right] f^\prime(\ekq)  
 \phikq^2 (\vec{\nabla}_\mb{q} \ekq)^2 \right|_{\mb{q}=\mb{0}} 
 \nonumber\\
 & = & \sumk \left[ \frac{\uk^2}{(\Ek+\ek)^2} 
  + \frac{\vk^2}{(\Ek-\ek)^2} \right] f^\prime(\ek)  
 \phik^2 (\vec{\nabla}_\mb{k} \ek)^2 \nonumber\\
 & = & \frac{1}{\Delta^2} \sumk f^\prime(\ek) (\vec{\nabla}_\mb{k} \ek)^2 \;,
%\end{eqnarray}
%
%\begin{eqnarray}
\\
\nonumber\\
(\mbox{II}) &=&  \left. - a_0 \frac{\partial^2\Omegaq}{\partial \mb{q}^2}
\right|_{\mb{q}=\mb{0}} : \mbox{2nd term}\; \nonumber\\
& =&  \sumk \left\{ \left[ \frac{\uk^2}{(\Ek+\ekq)^2} 
  + \frac{\vk^2}{(\Ek-\ekq)^2} \right] f^\prime(\ekq)  
 \phikq^2 (\vec{\nabla}_\mb{q} \ekq)^2 \right. \nonumber\\
 &&{} +2 \left[
  \frac{1-f(\Ek)-f(\ekq)} {(\Ek + \ekq)^3} u_\mb{k}^2 -
  \frac{f(\Ek) - f(\ekq)} {(\Ek - \ekq)^3} v_\mb{k}^2 \right]
\phikq^2  (\vec{\nabla}_\mb{q} \ekq)^2 \nonumber\\
 && {}  - \left[
  \frac{1-f(\Ek)-f(\ekq)} {(\Ek + \ekq)^2} u_\mb{k}^2 +
  \frac{f(\Ek) - f(\ekq)} {(\Ek - \ekq)^2} v_\mb{k}^2 \right] \nonumber\\
 &&{}   \times \left.
  \left[ \phikq^2 \vec{\nabla}^2_\mb{q} \ekq + (\vec{\nabla}_\mb{q} \ekq)
    \cdot (\vec{\nabla}_\mb{q} \phikq^2)\right] \right\} 
  \nonumber \\
 & = & \frac{1}{\Delta^2} \sumk f^\prime(\ek) (\vec{\nabla}_\mb{k} \ek)^2
 \nonumber \\
 &&{} + \frac{1}{\Delta^4} \sumk \Ek \left[ \left( 1+\frac{\ek^2}{\Ek^2}
   \right) [1-2f(\Ek)] - 2 \frac{\ek}{\Ek}[1-2f(\ek)] \right] \phik^{-2}  
   (\vec{\nabla}_\mb{k} \ek)^2 \nonumber \\
 &&{} - \frac{1}{2\Delta^2} \sumk \left[ [1-2f(\ek)] - \frac{\ek}{\Ek} 
   [1-2f(\Ek)] \right] \phik^{-2} \nonumber \\
 &&{} \times \left[ \phik^2 \vec{\nabla}^2_\mb{k} \ek 
   + \frac{1}{2}(\vec{\nabla}_\mb{k} \ek)
    \cdot (\vec{\nabla}_\mb{k} \phik^2) \right] \;,
%\end{eqnarray}
%
%\begin{eqnarray}
\\
\nonumber\\
(\mbox{III}) &=& \left. - a_0 \frac{\partial^2\Omegaq}{\partial \mb{q}^2} 
 \right|_{\mb{q}=\mb{0}} : \mbox{3rd term} \nonumber\\
&=& \sumk \left\{ \left[
  - \frac{1-f(\Ek)-f(\ekq)} {(\Ek + \ekq)^2} u_\mb{k}^2 -
  \frac{f(\Ek) - f(\ekq)} {(\Ek - \ekq)^2} v_\mb{k}^2 \right]
\right. \nonumber\\
&&{} \times 
(\vec{\nabla}_\mb{q} \ekq) \cdot (\vec{\nabla}_\mb{q} \phikq^2)
 \nonumber\\
&&{} +\left. \left[
  \frac{1-f(\Ek)-f(\ekq)} {\Ek + \ekq} u_\mb{k}^2 -
  \frac{f(\Ek) - f(\ekq)} {\Ek - \ekq} v_\mb{k}^2 \right]
  \vec{\nabla}^2_\mb{q} \phikq^2 \right\}\nonumber\\
&=& -\frac{1}{4\Delta^2} \sumk \left[ [1-2f(\ek)] - \frac{\ek}{\Ek} 
   [1-2f(\Ek)] \right] \phik^{-2} (\vec{\nabla}_\mb{k} \ek)
    \cdot (\vec{\nabla}_\mb{k} \phik^2) \nonumber\\
 &&{} + \frac{1}{4} \sumk \frac{1-2f(\Ek)}{2\Ek} \vec{\nabla}^2_\mb{k} \phik^2 
\;.
\end{eqnarray}
\end{mathletapp}
Finally, one has
\begin{eqnarray}
\left. \frac{\partial^2\Omegaq}{\partial \mb{q}^2}\right|_{\mb{q}=\mb{0}}
 & = & -\frac{1}{a_0\Delta^2} \sumk \left\{ \left[2f^\prime(\ek) 
  + \frac{\Ek}{\Delta^2\phik^2} \right.\right. \nonumber\\
&&{}  \times\left.  \left[ \left( 1+\frac{\ek^2}{\Ek^2}
   \right) [1-2f(\Ek)] - 2 \frac{\ek}{\Ek}[1-2f(\ek)] \right] \right]
 (\vec{\nabla}_\mb{k} \ek)^2 \nonumber \\
 &&{} - \frac{1}{2} \left[ [1-2f(\ek)] - \frac{\ek}{\Ek} 
   [1-2f(\Ek)] \right] \left[ \vec{\nabla}^2_\mb{k} \ek 
   + \phik^{-2} (\vec{\nabla}_\mb{k} \ek)
    \cdot (\vec{\nabla}_\mb{k} \phik^2) \right]
\nonumber\\
&&{} + \left. \frac{\Delta^2}{4} \frac{1-2f(\Ek)}{2\Ek} \vec{\nabla}^2_\mb{k} 
\phik^2 \right\} \;,
\label{Omega_q:expression}
\end{eqnarray}
and
\begin{eqnarray}
B&=&\frac{1}{6}\left. \frac{\partial^2\Omegaq}{\partial \mb{q}^2} 
 \right|_{\mb{q}=\mb{0}} \qquad \hskip 1.6in \mbox{for isotropic 3D, or} \nonumber\\
B_\parallel &=& \frac{1}{4}\left. \frac{\partial^2\Omegaq}{\partial
    \mb{q}_\parallel^2} \right|_{\mb{q}=\mb{0}}, \quad
B_\perp = \frac{1}{2}\left. \frac{\partial^2\Omegaq}{\partial
    \mb{q}_\perp^2} \right|_{\mb{q}=\mb{0}}, \qquad \mbox{for quasi-2D}.
\end{eqnarray}

We will list below the results for different physical
situations.

\section{3D \textit{s}-wave jellium}

The fermion dispersion and the symmetry factor are given by
\begin{mathletapp}
\begin{equation}
\ek = \frac{k^2}{2m}-\mu, \qquad 
\phik^2 = \frac{1}{ 1+\left(\frac{k}{k_0}\right)^2}\;.
\end{equation}
\[
(\vec{\nabla}_\mb{k}\ek)^2 = \frac{k^2}{m^2} = \frac{2}{m} (\ek +\mu) \;, 
\qquad
\vec{\nabla}_\mb{k}^2\ek = \frac{3}{m}\;,
\]
\begin{equation}
\vec{\nabla}_\mb{k}^2\phik^2 = \frac{2}{k_0^2}\phik^4 (1-4\phik^2), \qquad
(\vec{\nabla}_\mb{k}\ek)\cdot (\vec{\nabla}_\mb{k} \phik^2)
 = -\frac{2}{m}\phik^2 (1-\phik^2)
\label{3Djellium:derivatives}
\end{equation}
\end{mathletapp}
Substituting Eqs.~(\ref{3Djellium:derivatives}) into
Eqs.~(\ref{Omega_q:expression}), one obtains
\begin{eqnarray}
B &=& -\frac{1}{6a_0\Delta^2}\sumk \left\{ \frac{4}{m}(\ek
  +\mu)f^\prime(\ek) \right. \nonumber\\
&&{} +\frac{2}{m} \frac{\Ek (\ek+\mu)}{\Delta^2 \phik^2} 
\left[ \left( 1+\frac{\ek^2}{\Ek^2}
   \right) [1-2f(\Ek)] - 2 \frac{\ek}{\Ek}[1-2f(\ek)] \right] \nonumber\\
&&{} -\frac{1}{2m} \left[ [1-2f(\ek)] - \frac{\ek}{\Ek} 
   [1-2f(\Ek)] \right] (1+2\phik^2) \nonumber\\
&& {} + \left. \frac{\Delta^2}{2k_0^2} \frac{1-2f(\Ek)}{2\Ek} \phik^4(1-4\phik^2) 
\right\} \;,
\end{eqnarray} 
where $a_0$ is given by Eq.~(\ref{Omega_q:a0})

\section{Quasi-2D \textit{s}-wave jellium}
\label{App_Sec_Q2DJellium}

As in 3D jellium, the fermion dispersion and symmetry factor are given by
\begin{mathletapp}
\begin{equation}
\ek = \frac{k_\parallel^2}{2m_\parallel}+\frac{k_\perp^2}{2m_\perp}-\mu, 
\qquad 
\phik^2 = \frac{1}{ 1+\left(\frac{k_\parallel}{k_0}\right)^2} \qquad m_\perp
\gg m_\parallel \quad -\pi < k_\perp \leq \pi \;.
\end{equation}
\[
(\vec{\nabla}_\mb{k}\ek)^2 = \frac{k_\parallel^2}{m_\parallel^2} \oplus 
\frac{k_\perp^2}{m_\perp^2} = \frac{2\;}{m_\parallel}\epsilon_F \left
  ( \frac{k^2_\parallel}{k^2_F} \oplus  
\frac{m_\parallel^2}{m_\perp^2}\frac{k^2_\perp}{k^2_F}\right) \;,
\qquad
\vec{\nabla}_\mb{k}^2\ek = \frac{2\;}{m_\parallel}  
  \oplus \frac{1}{m_\perp}\;,
\]
\begin{equation}
\vec{\nabla}_\mb{k}^2\phik^2 = \frac{4}{k_0^2}\phik^4 (1-2\phik^2) \oplus 0, 
\qquad
(\vec{\nabla}_\mb{k}\ek)\cdot (\vec{\nabla}_\mb{k} \phik^2)
 = -\frac{2\;}{m_\parallel}\phik^2 (1-\phik^2) \oplus 0\;.
\label{Q2Djellium:derivatives}
\end{equation}
\end{mathletapp}
The ``Fermi energy'' $\epsilon_F = k_F^2/2m_\parallel$ is defined in a
strictly 2D case, i.e., $n = \int^{k_F}_0 2d^2 k/(2\pi)^2$. Note here I have
used direct sum ``$\oplus$'' to indicate that the terms will enter
$B_\parallel $ and $B_\perp$ separately.  Substituting
Eqs.~(\ref{3Djellium:derivatives}) into Eqs.~(\ref{Omega_q:expression}), one
obtains
\begin{mathletapp}
\begin{eqnarray}
B_\parallel &=& -\frac{1}{4a_0\Delta^2}\sumk \Bigg\{ \frac{4}{m_\parallel} 
\epsilon_F f^\prime(\ek)\frac{k^2_\parallel}{k^2_F}  \nonumber\\
&&{} +\frac{2}{m_\parallel} \frac{\Ek \epsilon_F}{\Delta^2 \phik^2} 
\left[ \left( 1+\frac{\ek^2}{\Ek^2}
   \right) [1-2f(\Ek)] - 2 \frac{\ek}{\Ek}[1-2f(\ek)] \right]
\frac{k^2_\parallel}{k^2_F} \nonumber\\
&&{} -\frac{1}{m_\parallel} \left[ [1-2f(\ek)] - \frac{\ek}{\Ek} 
   [1-2f(\Ek)] \right] \phik^2 \nonumber\\
&&{}   + \frac{\Delta^2}{k_0^2} \frac{1-2f(\Ek)}{2\Ek} \phik^4(1-2\phik^2) 
\Bigg\} \;,
%\end{eqnarray} 
%\begin{eqnarray}
\nonumber\\
\\
B_\perp &=& -\frac{1}{2a_0\Delta^2 m_\parallel}\sumk \Bigg\{ 
4\epsilon_F f^\prime(\ek)\frac{k^2_\perp}{k^2_F} \frac{m_\parallel^2}
{m_\perp^2}   \nonumber\\
&&{} + 2 \frac{\Ek \epsilon_F}{\Delta^2 \phik^2} 
\left[ \left( 1+\frac{\ek^2}{\Ek^2}
   \right) [1-2f(\Ek)] - 2 \frac{\ek}{\Ek}[1-2f(\ek)] \right]
\frac{k^2_\perp}{k^2_F} \frac{m_\parallel^2} {m_\perp^2} \nonumber\\
&& {} -\frac{1}{2} \left[ [1-2f(\ek)] - \frac{\ek}{\Ek} 
   [1-2f(\Ek)] \right] \frac{m_\parallel} {m_\perp} 
\Bigg\}  \nonumber\\
& \sim & \frac{1}{2a_0\Delta^2 m_\parallel}\sumk 
\left[[1-2f(\ek)] - \frac{\ek}{\Ek} 
   [1-2f(\Ek)] \right] \frac{m_\parallel} {m_\perp} \nonumber\\
& = & \frac{1}{2m_\perp} \qquad \qquad \mbox{as}\qquad
 \frac{m_\parallel}{m_\perp}
\longrightarrow 0\;.
\end{eqnarray} 
\end{mathletapp}
Here, use has been made of Eq.~(\ref{Omega_q:a0}) in the last line.

If we assume that the
3rd (out-of-plane) dimension corresponds to a periodic lattice so that 
\begin{mathletapp}
\begin{equation}
\ek = \frac{k_\parallel^2}{2m_\parallel} + 
\frac{1}{m_\perp}(1-\cos k_\perp)-\mu, 
\end{equation}
and thus
\begin{equation}
(\vec{\nabla}_\mb{k}\ek)^2 = \frac{k_\parallel^2}{m_\parallel^2} \oplus 
\frac{\sin^2 k_\perp}{m_\perp^2}
\qquad
\vec{\nabla}_\mb{k}^2\ek = \frac{2\;}{m_\parallel}  
  \oplus \frac{1}{m_\perp} \cos k_\perp \;,
\end{equation}
\end{mathletapp}
then
\begin{mathletapp}
\begin{eqnarray}
B_\parallel &=& -\frac{1}{4a_0\Delta^2}\sumk \Bigg\{ 2
 f^\prime(\ek)\frac{k^2_\parallel}{m_\parallel^2}  \nonumber\\
&&{} + \frac{\Ek}{\Delta^2 \phik^2} 
\left[ \left( 1+\frac{\ek^2}{\Ek^2}
   \right) [1-2f(\Ek)] - 2 \frac{\ek}{\Ek}[1-2f(\ek)] \right]
\frac{k^2_\parallel}{m_\parallel^2} \nonumber\\
&&{} -\frac{1}{m_\parallel} \left[ [1-2f(\ek)] - \frac{\ek}{\Ek} 
   [1-2f(\Ek)] \right] \phik^2 \nonumber\\
&&{} +  \frac{\Delta^2}{k_0^2} \frac{1-2f(\Ek)}{2\Ek} \phik^4 (1-2\phik^2) 
\Bigg\} \;,
%\end{eqnarray} 
%\begin{eqnarray}
\nonumber\\
\\
B_\perp &=& -\frac{1}{2a_0\Delta^2 m_\parallel}\sumk \Bigg\{ 
\frac{2\;}{m_\parallel} f^\prime(\ek) \frac{m_\parallel^2}
{m_\perp^2} \sin^2 k_\perp  \nonumber\\
&&{} + \frac{1\;}{m_\parallel} \frac{\Ek}{\Delta^2 \phik^2} 
\left[ \left( 1+\frac{\ek^2}{\Ek^2}
   \right) [1-2f(\Ek)] - 2 \frac{\ek}{\Ek}[1-2f(\ek)] \right]
\frac{m_\parallel^2} {m_\perp^2} \sin^2 k_\perp \nonumber\\
&& {} -\frac{1}{2} \left[ [1-2f(\ek)] - \frac{\ek}{\Ek} 
   [1-2f(\Ek)] \right] \frac{m_\parallel} {m_\perp} \cos k_\perp
\Bigg\}  \nonumber\\
& \sim & \frac{1}{m_\parallel} \left( \frac{m_\parallel}{m_\perp} \right)^2
\qquad \qquad \mbox{as}\qquad \frac{m_\parallel}{m_\perp} 
\longrightarrow 0\;.
\label{Omega_q:Q2Dlattice}
\end{eqnarray} 
\end{mathletapp}
Eq.~(\ref{Omega_q:Q2Dlattice}) indicates that the anisotropy mass ratio
$m_\parallel/m_\perp$ of the single particle dispersion is magnified in the
pair dispersion.

\section{Quasi-2D lattice: \textit{s}- and \textit{d}-wave}

We assume a tight binding model for the electron dispersion and include next
nearest neighbor hopping terms:
\begin{mathletapp}
\begin{equation}
\ek = 2t (2-\cos k_x - \cos k_y)-2t^\prime (1-\cos k_x \cos k_y) -2t_z
(1-\cos k_z) -\mu
\end{equation}
\begin{eqnarray}
\vec{\nabla}^2_\mb{k} \ek &=& 2t (\cos k_x + \cos k_y) 
- 4t^\prime \cos k_x \cos k_y \oplus 2t_z \cos k_z
\nonumber\\
\nonumber\\
(\vec{\nabla}_\mb{k} \ek)^2 &=&  4t^2 (\sin^2 k_x + \sin^2  k_y) 
+ 4{t^\prime}^2 (\cos k_x - \cos k_y)^2 \nonumber\\
&&{} +4t^\prime \left[ -2t(\cos k_x + \cos k_y) + 2t^\prime \cos k_x \cos k_y
\right] (1-\cos k_x \cos k_y) \nonumber\\ 
&&{} \oplus\;  4t_z^2 \sin^2 k_z
\end{eqnarray}
\end{mathletapp}

\subsection{Quasi-2D lattice: \textit{s}-wave}

\begin{equation}
\phik=1, \quad \vec{\nabla}^2_\mb{k} \phik^2 = (\vec{\nabla}_\mb{k} \ek)
\cdot (\vec{\nabla}_\mb{k}\phik^2) =0 \;.
\end{equation}
\begin{mathletapp}
\begin{eqnarray}
B_\parallel &=& -\frac{1}{4a_0\Delta^2}\sumk \left\{ \left[ 2
 f^\prime(\ek)  + \frac{\Ek}{\Delta^2} 
\left[ \left( 1+\frac{\ek^2}{\Ek^2}
   \right) [1-2f(\Ek)] - 2 \frac{\ek}{\Ek}[1-2f(\ek)] \right] \right]  \right.
\nonumber\\
&&{} \times \left[4t^2 (\sin^2 k_x + \sin^2  k_y) 
+ 4{t^\prime}^2 (\cos k_x - \cos k_y)^2 \right. \nonumber\\
&&{} + \left. 4t^\prime \left[ -2t(\cos k_x + \cos k_y) 
 + 2t^\prime \cos k_x \cos k_y
\right] (1-\cos k_x \cos k_y)\right] \nonumber\\ 
&&{} -\left. \frac{1}{2} \left[ [1-2f(\ek)] - \frac{\ek}{\Ek} 
   [1-2f(\Ek)] \right] \left[ 2t (\cos k_x + \cos k_y) 
- 4t^\prime \cos k_x \cos k_y \right] \right\}
%\end{eqnarray} 
%\begin{eqnarray}
\nonumber\\
\\
\nonumber\\ 
B_\perp &=& -\frac{1}{2a_0\Delta^2}\sumk \left\{ \left[
2 f^\prime(\ek) + \frac{\Ek}{\Delta^2} 
\left[ \left( 1+\frac{\ek^2}{\Ek^2}
   \right) [1-2f(\Ek)] - 2 \frac{\ek}{\Ek}[1-2f(\ek)] \right] \right]
\right. \nonumber\\
&&{} \times 4t_z^2 \sin^2 k_z 
 \left. -\frac{1}{2} \left[ [1-2f(\ek)] - \frac{\ek}{\Ek} 
   [1-2f(\Ek)] \right] 2t_z \cos k_z
\right\}  
\end{eqnarray} 
\end{mathletapp}
When $t^\prime=0$, $B_\parallel$ becomes much simpler:
\begin{eqnarray}
B_\parallel &=& -\frac{1}{4a_0\Delta^2}\sumk \left\{ \left[ 2
 f^\prime(\ek)  + \frac{\Ek}{\Delta^2} 
\left[ \left( 1+\frac{\ek^2}{\Ek^2}
   \right) [1-2f(\Ek)] - 2 \frac{\ek}{\Ek}[1-2f(\ek)] \right]\right]   \right.
\nonumber\\
&&{} \times \left[4t^2 (\sin^2 k_x + \sin^2  k_y) 
 \right] \nonumber\\
&&{}- \left. \frac{1}{2} \left[ [1-2f(\ek)] - \frac{\ek}{\Ek} 
   [1-2f(\Ek)] \right] \left[ 2t (\cos k_x + \cos k_y) \right] \right\}
\end{eqnarray} 
Setting $t_z=t$ and $t^\prime=0$, one obtains $B=B_\parallel =B_\perp$ for a
3D isotropic lattice with $s$-wave pairing $\varphi_{\bf k}=1$ within the
nearest neighbor approximation.

\subsection{Quasi-2D lattice: \textit{d}-wave}

\begin{mathletapp}
\begin{equation}
\phik=\cos k_x - \cos k_y , \quad \vec{\nabla}^2_\mb{k} \phik^2 = 
%4(1-\cos^2 k_x -\cos^2 k_y + \cos k_x \cos k_y)
-2(\cos 2k_x + \cos 2k_y) + 4\cos k_x \cos k_y
\end{equation}
\begin{equation}
(\vec{\nabla}_\mb{k} \ek)
\cdot (\vec{\nabla}_\mb{k}\phik^2) = \left[ 4t(\cos k_x + \cos k_y)
  -4t^\prime (1+\cos k_x \cos k_y)\right] \phik^2 \;.
\end{equation}
\end{mathletapp}
\begin{mathletapp}
\begin{eqnarray}
B_\parallel &=& -\frac{1}{4a_0\Delta^2}\sumk \left\{ 
\left[ 2 f^\prime(\ek)  + \frac{\Ek}{\Delta^2 \phik^2} 
\left[ \left( 1+\frac{\ek^2}{\Ek^2}
   \right) [1-2f(\Ek)] - 2 \frac{\ek}{\Ek}[1-2f(\ek)] \right] \right]  \right.
\nonumber\\
&&{} \times \left[4t^2 (\sin^2 k_x + \sin^2  k_y) 
+ 4{t^\prime}^2 \phik^2 \right. \nonumber\\
&&{}  +\left. 4t^\prime \left[ -2t(\cos k_x + \cos k_y) 
 + 2t^\prime \cos k_x \cos k_y
\right] (1-\cos k_x \cos k_y)\right] \nonumber\\ 
&&{} -\frac{1}{2} \left[ [1-2f(\ek)] - \frac{\ek}{\Ek} 
   [1-2f(\Ek)] \right]\nonumber\\
&&{} \times \left[ 6t (\cos k_x + \cos k_y) 
- 4t^\prime (1+2\cos k_x \cos k_y) \right] \nonumber\\
&&{} + \left. \Delta^2 \frac{1-2f(\Ek)}{2\Ek} 
\left[ -\frac{1}{2} (\cos 2k_x + \cos 2k_y) + \cos k_x \cos k_y\right]
\right\} \;,
%\end{eqnarray} 
%\begin{eqnarray}
\label{Omega_q:d-wave_t1}
\\
\nonumber\\
B_\perp &=& -\frac{1}{2a_0\Delta^2}\sumk \left\{ \left[
2 f^\prime(\ek) + \frac{\Ek}{\Delta^2\phik^2} 
\left[ \left( 1+\frac{\ek^2}{\Ek^2}
   \right) [1-2f(\Ek)] - 2 \frac{\ek}{\Ek}[1-2f(\ek)] \right] \right]
\right. \nonumber\\
&&{} \times 4t_z^2 \sin^2 k_z 
 \left. -\frac{1}{2} \left[ [1-2f(\ek)] - \frac{\ek}{\Ek} 
   [1-2f(\Ek)] \right] 2t_z \cos k_z
\right\}  
\end{eqnarray} 
\end{mathletapp}
When $t^\prime=0$, Eq.~(\ref{Omega_q:d-wave_t1}) becomes
\begin{eqnarray}
B_\parallel &=& -\frac{1}{4a_0\Delta^2}\sumk \left\{ \left[ 2
 f^\prime(\ek)  + \frac{\Ek}{\Delta^2} 
\left[ \left( 1+\frac{\ek^2}{\Ek^2}
   \right) [1-2f(\Ek)] - 2 \frac{\ek}{\Ek}[1-2f(\ek)] \right]\right]   \right.
\nonumber\\
&&{} \times \left[4t^2 (\sin^2 k_x + \sin^2  k_y) 
 \right] \nonumber\\
&&{} -\frac{1}{2} \left[ [1-2f(\ek)] - \frac{\ek}{\Ek} 
   [1-2f(\Ek)] \right] \left[ 6t (\cos k_x + \cos k_y) \right]
\nonumber\\
&&{} + \left. \Delta^2 \frac{1-2f(\Ek)}{2\Ek} 
\left[ -\frac{1}{2} (\cos 2k_x + \cos 2k_y) + \cos k_x \cos k_y\right]
 \right\}
\end{eqnarray}

\section{Weak coupling limit}

In this section, we explore in more detail the significance of pair
excitations in the weak coupling limit.  We expand Eq.~(\ref{Omega_q:t-exp}) to
$\Omega^2$, so that
\begin{eqnarray}
a_1 &= & \left. \frac{1}{2} \frac{\partial^2}{\partial \Omega^2}
  t^{-1}_{\mb{q},\Omega}\right|_{q=0, \Omega=0} \nonumber\\
&=&  \left[
  \frac{1-f(\Ek)-f(\ek)} {(\Ek + \ek)^3} u_\mb{k}^2 -
  \frac{f(\Ek) - f(\ek)} {(\Ek - \ek)^3} v_\mb{k}^2 \right]
\phik^2 \nonumber\\
&=&\frac{1}{2\Delta^4} \sumk \Ek \left[ \left( 1+\frac{\ek^2}{\Ek^2}
   \right) [1-2f(\Ek)] - 2 \frac{\ek}{\Ek}[1-2f(\ek)] \right] \phik^{-2}  
\label{Omega_q:a1}
\end{eqnarray}
The main contribution to this integral comes from the neighborhood of
$\epsilon_{\bf k}=0$.  To illustrate this, take for example a simple 3D
isotropic $s$-wave case at $T=0$. When
$|\epsilon_{\bf k}| \gg \Delta$, one has
\begin{equation}
a_1 =  \frac{1}{2\Delta^4} \sumk \frac{(\Ek-|\ek|)^2}{\Ek\phik^2}
\approx \sumk \frac{\phik^2}{8\Ek^3} = \frac{N(0)}{4\Delta^2}\;,
\label{Omega_q:a1_approx}
\end{equation} 
where $N(0)$ is the density of states for one spin orientation. Meanwhile,
under the same approximation, one has
\begin{equation}
a_0 \approx \sumk \frac{\phik^2}{4\Ek^2}\,\mbox{sgn}(\ek) 
\quad \propto \Delta^{-1}\;.
\end{equation}
Therefore, $a_0/a_1 \propto \Delta$, and is vanishingly small as
$g$ approaches zero. In the presence of particle-hole symmetry, $a_0$
vanishes identically.
Similarly, one can obtain approximations for $\xi^2 = a_0 B$ from
Eq.~(\ref{Omega_q:expression}). Now take $\varphi_{\bf k}=1$, and
approximate $(\vec{\nabla}\epsilon_{\bf k})^2$ by $v_F^2$, one obtains
\begin{equation} 
\xi^2 = \frac{1}{6}\left[\frac{3}{2\Delta^2}N(0)v_F^2 + a_0
  \frac{3}{m}\right] \approx
\frac{N(0)}{4\Delta^2}v_F^2 = a_1 v_F^2 \;.
\label{Omega_q:xi2_exp}
\end{equation}
Finally, 
\begin{equation}
c = \sqrt{\xi^2/a_1} = v_F \:, \qquad\qquad (\mbox{for isotropic 3D})\;.
\end{equation}
For quasi-2D, this velocity will become $c=\sqrt{\frac{3}{2}} v_F$.
 
It should be stressed that the residue of the pair propagator vanishes as
$1/\sqrt{a_1 \xi^2} \propto \Delta^{2}$, and therefore does not lead to a
significant contribution to the pseudogap at small $g$, so that one recovers
BCS theory in the weak coupling limit.

An additional important observation should be made.  Note here the velocity
for the pair dispersion is different from that for the collective phase mode
of the superconducting order parameter, given by $v_F/\sqrt{3}$ (in 3D).%
%
%\footnote{In fact, when $\Delta$ is extremely small in the weak coupling
%  limit, in particularly around $T_c$, the $T$ matrix expansion,
%  Eq.~(\ref{Omega_q:exp}), is no longer a good approximation. In this case,
%  the pseudogap effect is completely negligible.}

\section{Effects of the $\bm{\Omega^2}$ term of the inverse \textit{T} matrix}

The two roots of the inverse $T$ matrix are 
\begin{equation}
\Omegaq = \frac{-a_0\pm \sqrt{a_0^2 +4a_1\xi^2 q^2}}{2a_1}\;.
\end{equation}
The contribution to pseudogap comes from the positive root:
\begin{equation}
\Omegaq =\frac{1}{2}\left[ \sqrt{\left(\frac{a_0}{a_1}\right)^2 + 4 \left
      ( \frac{a_0}{a_1}\right) \Omegaq^0} -\frac{a_0}{a_1}\right] \;, 
\end{equation}
where $\Omega^0_{\bf q} = Bq^2 $ is the dispersion in the absence of
$a_1$. In the limit $a_1/a_0 \rightarrow 0$, $\Omega_{\bf q} \rightarrow
\Omega^0_{\bf q}$. 
The residue of the $T$ matrix at this pole is given by
\begin{equation}
\frac{1}{\sqrt{a_0^2+4a_1 \xi^2 q^2}} = \frac{1}{a_0} \frac{1}
{\sqrt{\displaystyle 1+4\left(\frac{a_1}{a_0}\right)\Omegaq^0}}\;.
\end{equation}
Note $\Omega_{\bf q}$ is quadratic in $q$ in the long wavelength limit. 
Now the pseudogap equation is modified slightly:
\begin{equation}
a_0 \Delta_{pg}^2 = \sumq \left[ 1+ 4 \left( \frac{a_1}{a_0}\right)
  \Omegaq^0 \right]^{-1/2} b(\Omegaq)\;.
\end{equation}
The ratio $a_0/a_1$ can be easily obtained from Eq.~(\ref{Omega_q:a0}) and
Eq.~(\ref{Omega_q:a1}). 

Since $a_1$ is considerable only at weak coupling, where the pseudogap is
weak, we do not anticipate a strong effect of $a_1$ in calculations. Indeed,
numerical studies show that $T_c$ is almost unaffected, whereas
$\Delta_{pg}$ is modified (enhanced) only at very weak couplings.

\begin{figure}
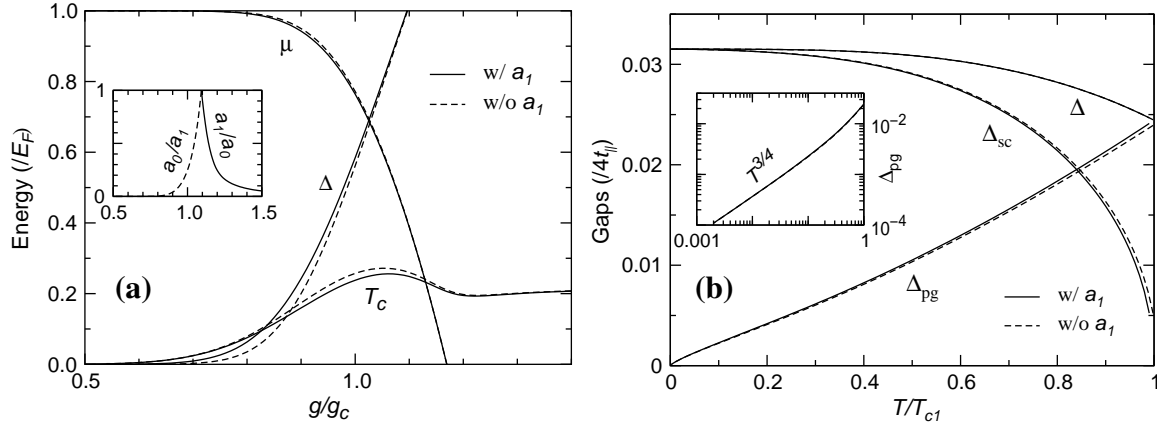

\centerline{\includegraphics[width=2.95in, clip]{3Da1k4-2}\hfill
\includegraphics[width=2.96in, clip]{Da1f_825r100g_26-2}}
\vskip -.85in \hskip 0.55in \textbf{(a)} \hskip 2.8in \textbf{(b)}
\vskip 0.55in
\caption[Effects of the $a_1\Omega^2$ term in the inverse $T$ matrix
expansion on the solutions for various gaps as well as $T_c$ as a function of
$g$ and $T$.]{Effects of the $a_1\Omega^2$ term in the $T$ matrix expansion
  on the solutions for various gaps as well as $T_c$ as a function of $g$ and
  $T$.  The solid line are calculated with the $a_1\Omega^2$ term, whereas
  the dashed ones without this term.  Plotted in (a) are the solutions for
  $T_c$, $\Delta(T_c)$, and $\mu(T_c)$ vs $g$ for $k_0=4k_F$ in 3D jellium
  with $s$-wave pairing.  The ratio $a_0/a_1$ is shown in the set.  Plotted
  in (b) are the three gaps below $T_c$ for the overdoped cuprates with
  doping $x=0.175$, coupling $-g/4t_\parallel=0.26$, and anisotropy
  $t_\perp/t_\parallel=0.01$. Here $T_{c1}/4t_\parallel=0.0210$ and
  $T_{c2}/4t_\parallel=0.0212$ correspond to the solutions obtained with and
  without the $a_1\Omega^2$ term, respectively. The inset shows that the low
  $T$ power law of $\Delta_{pg}\propto T^{3/4}$ is unaffected.}
\label{Effect_a1}
\end{figure}

Among various situations, the effect of the $a_1 \Omega^2$ term is most
prominent in 3D jellium. Shown in Fig.~\ref{Effect_a1}(a) is a comparison
between the solutions for $T_c$, $\mu(T_c)$, and $\Delta(T_c)$ calculated
with and without $a_1 \Omega^2$ as a function of $g$ in a 3D jellium model
with $s$-wave pairing. Even in this case, the difference is very small.  The
values of $T_c$ and $\Delta$ are only slightly suppressed when calculated
with $a_1$.  Similar plots for quasi-2D with $d$-wave pairing shows that the
curves calculated in both ways coincide.

Most importantly, the $a_1 \Omega^2$ term does not modify the low $T$ power
laws in any appreciable way.  In Fig.~\ref{Effect_a1}(b), we plot the gap
values as a function of $T$ below $T_c$ for the overdoped cuprates,
$x=0.175$.  Evidently, the two sets of curves almost completely overlap
except for the tiny (about 1\%) difference in $T_c$. Both the main figure
and the inset show that the $T^{3/4}$ dependence of $\Delta_{pg}$ is
unaffected. As the pseudogap develops with underdoping (or equivalently,
with increasing $-g/4t_\parallel$), the ratio $a_1/a_0$ becomes smaller,
[see, e.g., the inset of Fig.~\ref{Effect_a1}(a)], and, therefore, the
effect of $a_1 \Omega^2$ becomes even weaker. We conclude that the $a_1
\Omega^2$ term in the inverse $T$ matrix expansion can be safely dropped in
all calculations.

\chapter{BCS--BEC crossover for a quasi-2D \textit{d}-wave
superconductor at arbitrary density}
\label{App_n-g}

%\section{Where do the cuprates fit? --- \lowercase{\textit{n}-\textit{g}}
%  phase diagram}
%\label{Sec_2D}

In Chapter~\ref{Chap_Tc}, we studied the effects of low dimensionality,
lattice band structure and pairing symmetry. For the $d$-wave case, we only
presented the situation where the electron density was near half filling, as
is relevant to the cuprates.  In this Appendix, we extend this study to the
whole density-coupling phase diagram, and will locate from a more general
perspective the region appropriate to high $T_c$ superconductors. In this
context, we also study the difference between $d$ and the $s$-wave pairing
symmetries.

\section{Bosonic \textit{d}-wave superconductors: Extreme low density limit}

\begin{figure}
\centerline{\includegraphics{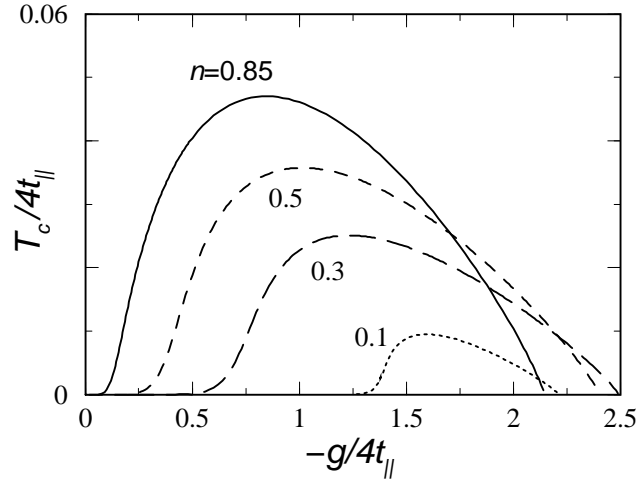}}
\vskip -0.1in
\caption[Crossover behavior of $T_c$ with respect to $g$ for different electron
densities on a quasi-2D lattice with $d$-wave pairing.]{$T_c$ as a function
  of $g$ for various density $n$ on a quasi-2D lattice with $d$-wave
  pairing. Unlike in $s$-wave case, there exists an threshold coupling,
  below which $T_c$ essentially vanishes. $T_c$ survives only a small range
  of $g$.}
\label{DwaveThreshold}
\end{figure}

\begin{figure}
  \centerline{\hskip -2mm \includegraphics[width=6in]{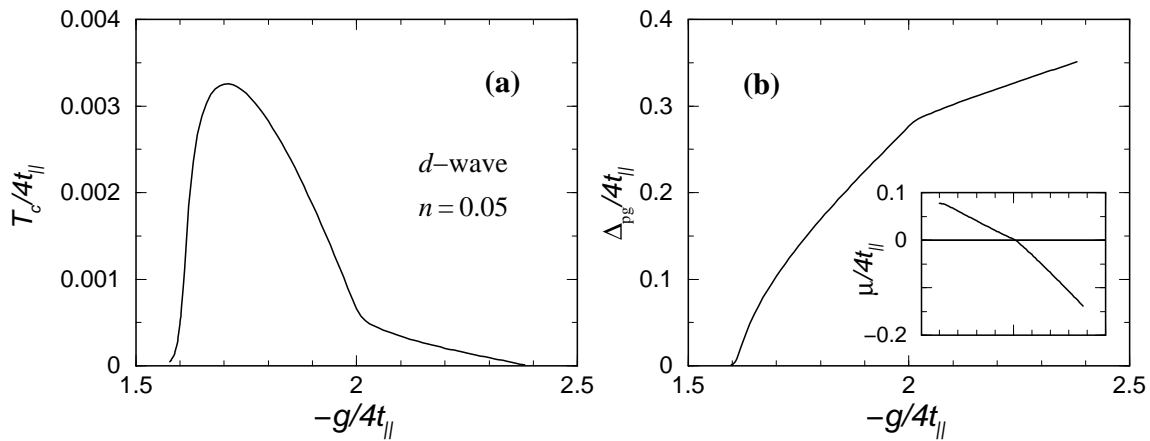}} 
\vskip -0.1in
\caption[Crossover behavior of $T_c$, $\Delta_{pg}(T_c)$, and $\mu(T_c)$ 
unphysically low density $n=0.05$ on a quasi-2D lattice with $d$-wave
pairing symmetry.]{(a) $T_c$, (b) $\Delta_{pg}(T_c)$, and $\mu(T_c)$ (inset)
  behavior as a function of $g$ at very low density, $n=0.05$.  Bosonic
  superconducting regime is accessible for this low $n$. There is a kink in
  all three curves. $t_\perp/t_\parallel=0.01$.}
\label{DwaveBosonic} 
\end{figure}

In contrast to the case of $s$-wave pairing, for a quasi-2D lattice with
$d$-wave pairing, there exists a rather large threshold coupling, $g_{th}$,
below which $T_c$ is essentially zero. This threshold increases with
decreasing electron density. This can be shown in Fig.~\ref{DwaveThreshold},
for various densities from high to low.  From the inset
of Fig.~\ref{Lattice_Fig3}(a), one can see that the threshold for the
$s$-wave case is very small, since the curve follows the BCS exponential
dependence; moreover, it is roughly the same for various $n$. By contrast,
for $d$-wave pairing, Fig.~\ref{DwaveThreshold} shows that this threshold is
very large, and is strongly density dependent.  In addition,
Fig.~\ref{DwaveThreshold} also shows that $T_c$ survives for only a small
range of $g$, in agreement with Fig.~\ref{Lattice_Fig4}(b). Quite generally,
$T_c$ vanishes before $\mu$ becomes negative.

An important consequence of these observations is that the bosonic
superconducting regime on a $d$-wave lattice is accessible only at extremely
low and unphysical density.  In this limit, the inter-particle distance is
much larger than the pair size at sufficiently strong coupling, and
therefore, the internal structure of the pair wavefunction becomes less
important. Shown in Fig.~\ref{DwaveBosonic} are the behavior of $T_c$,
$\Delta_{pg}$, and $\mu(T_c)$ with increasing $g$.  The threshold coupling
is even larger than that shown in Fig.~\ref{DwaveThreshold} because of lower
density. The bosonic superconductor regime is accessible for roughly
$-g/4t_\parallel > 2 $; thereafter, $T_c$ continues to gradually decline.
In contrast to $s$-wave pairing, it is interesting to notice that there is a
kink in all three curves, as the chemical potential $\mu$ changes sign.  As
will be shown below, a similar kink would also appear at high densities if
the bosonic superconducting regime could be accessed.

\begin{figure}
\centerline{\includegraphics[width=4in]{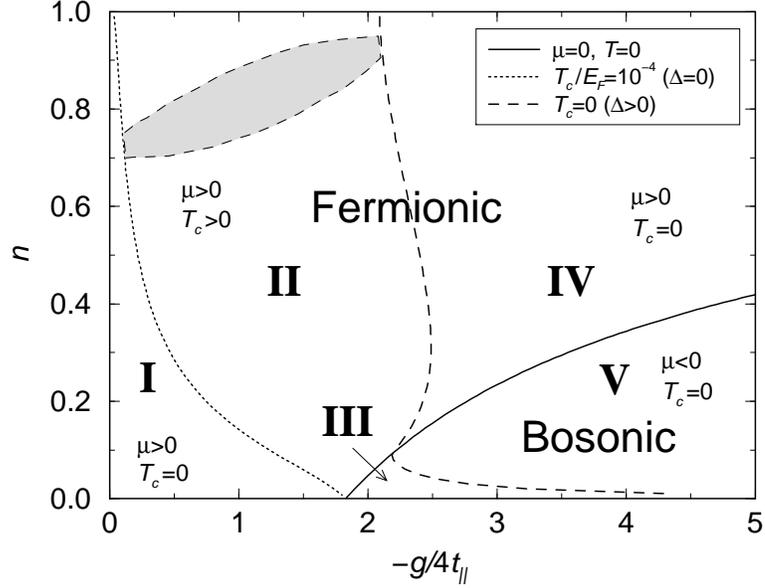}}
\vskip -0.1in
\caption[Phase diagram of a quasi-2D $d$-wave superconductor on a
density--coupling plot.]{$n-g$ phase diagram for a quasi-2D $d$-wave
  superconductor. There are five physically distinct regions. Bosonic
  superconducting regime exists only in region III, the unphysically low
  density limit. }
\label{n-g_Phase}
\end{figure}

\section{\lowercase{\textit{n-g}} phase diagram}

Although the bosonic superconducting regime does not exist for moderate and
high densities for $d$-wave pairing, one can still use BCS mean-field theory
to solve for the chemical potential at $T=0$, as a function of $g$ and $n$,
treating the system as if it were superconducting.  In this way, we can find
where the chemical potential changes sign and associate the sign change with
the bosonic regime.  We now obtain a phase diagram on a $n$-$g$ plot for a
quasi-2D superconductor, as shown in Fig.~\ref{n-g_Phase}. This phase
diagram is composed of five physically distinct regions. The boundaries
which separate different regions are given by the threshold coupling
$g_{th}(n)$ [dotted line], $g_{max}(n)$ where $T_c$ vanishes in the
pseudogap regime [dashed line], and where the chemical potential changes
sign, $\mu(n,g)=0$ [Solid line]. While the former two separate the
superconducting regions from the non-superconducting ones, the latter
separate the bosonic from fermionic regions. In region I, the coupling
strength $g$ is below the threshold $g_{th}$, (or, alternatively, $n$ is too
low), so that $T_c$ is essentially zero.  This is a fermionic,
non-superconducting region.  Region II is a fermionic, superconducting
region; Region III a bosonic, superconducting region; Region IV and V are
non-superconducting, fermionic and bosonic, respectively. Note here, roughly
speaking, this phase diagram is independent of the anisotropy ratio
$t_\perp/t_\parallel$ except when it is extremely small.

\begin{figure}
\centerline{\includegraphics[width=4in]{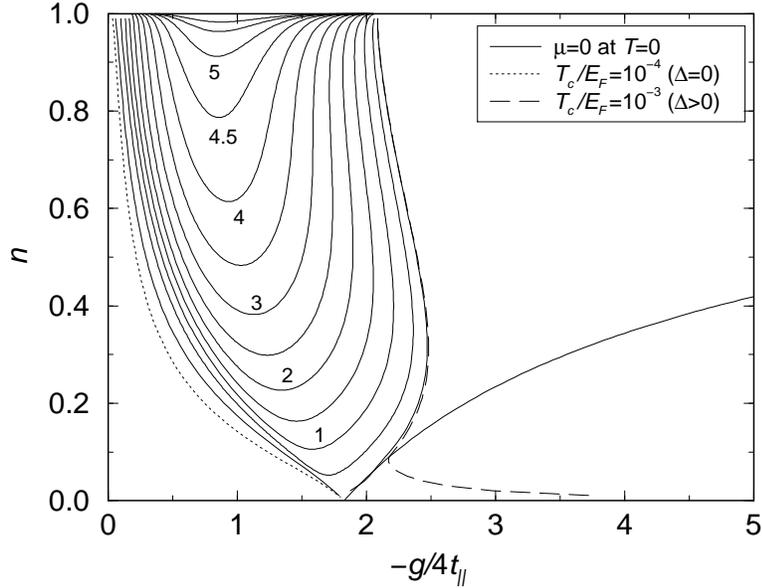}}
\vskip -0.1in
\caption[Contour plot of $T_c$ on the $n-g$ phase diagram of a quasi-2D 
$d$-wave superconductor.]{Contour plot of $T_c$ as a function of $n$ and $g$
  for a quasi-2D $d$-wave superconductor, shown in Fig.~\ref{n-g_Phase}. The
  $T_c$ value is labeled in unit of $10^{-2} E_F$, calculated for
  $t_\perp/t_\parallel =0.01$.}
\label{n-g_Contour}
\end{figure}

The evolution of $T_c$ with $n$ and $g$ can be plotted in a contour plot, as
shown in Fig.~\ref{n-g_Contour} for $t_\perp/t_\parallel= 10^{-2}$. As can
be seen, as a function of $g$ for given $n$, there is a maximum $T_c$, as
also shown in Fig.~\ref{DwaveThreshold}. On the other hand, for a given
$g$, roughly speaking, a higher $n$ corresponds to a higher $T_c$.

Now we are ready to find out where the cuprates fit into this phase diagram.
For the cuprates, $T_c$ is nonzero only when the doping concentration
$x=1-n$ falls between about 0.05 and 0.3. In addition, there is an optimal
$T_c$ around $x=0.15$, i.e., $n=0.85$.
%On the other hand, there have been
%experimental evidence showing that the coupling strength $g$ is larger in
%the underdoped regime than in the overdoped regime. 
Therefore, we conclude that the cuprate superconductors correspond to the
shaded area in Fig.~\ref{n-g_Phase}.

Determining the parameter range for high $T_c$ superconductors is very
important. In the literature (see, e.g, ref.~\cite{Engelbrecht97}), the
coupling constant is often taken to be very large. As we see here, this
larger value corresponds to the non-superconducting region IV, and,
therefore, is not an appropriate choice. In addition, we conclude that the
cuprates are far from the bosonic regime. As shown in the inset of
Fig.~\ref{Lattice_Fig4}(b), the chemical potential is only slightly lower
relative to $E_F$. Therefore, \textit{the pairs in a $d$-wave superconductor
  above $T_c$ are not bound pairs, they are sharp resonant states.} Thus, we
are not in what is frequently referred to as the pre-formed pair limit.

\section{Comparison with \textit{s}-wave superconductors}

\begin{figure}
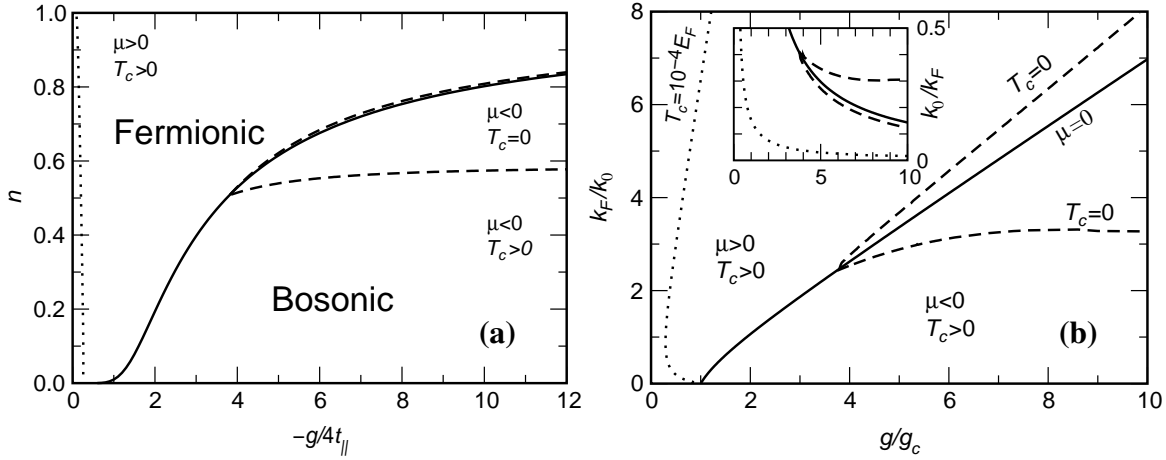

\centerline{
\includegraphics[width=3in]{Swave-Phase}
\includegraphics[width=3in, clip]{3Dk0g-Phase}
}
\vskip -0.8in
\hskip 2.45in {\textbf{(a)}} \hskip 2.8in {\textbf{(b)}}
\vskip 0.52in
\caption[$n-g$ phase diagram of an $s$-wave superconductor on a quasi-2D
lattice, as well as in an isotropic 3D jellium.]{(a) $n-g$ phase diagram for
  an $s$-wave superconductor on a quasi-2D lattice, and (b) $k_F/k_0 - g$
  phase diagram in an isotropic 3D jellium model with $s$-wave pairing. The
  bosonic superconducting regime can be readily accessed at intermediate and
  low densities. A $k_0/k_F - g$ version is shown in the inset. The legends
  are the same as in Fig.~\ref{n-g_Phase}.}
\label{n-g-S}
\end{figure}

To make the distinction between $d$- and $s$-wave clearer, we plot the $n-g$
phase diagram for a $s$-wave superconductor on a quasi-2D lattice, as shown
in Fig.~\ref{n-g-S}(a). For a 3D jellium, the density can be roughly
quantified by $(k_F/k_0)^3$. Our results are shown as a a $k_F/k_0$ (and
$k_0/k_F$) vs $g$ plot, in Fig.~\ref{n-g-S}(b) (inset). Figure ~\ref{n-g-S}
demonstrates that, for an $s$-wave superconductor, the bosonic regime can be
readily accessed for intermediate and low densities.  Meanwhile, there is a
narrow range of $n$ (or $k_0/k_F$), for which $T_c$ disappears. It then reappears at larger $g$. 

The threshold coupling for the $s$-wave lattice case is very small, in
comparison with the $d$-wave case.  This can be attributed to the $d$-wave
pairing symmetry factor $\phik = \cos k_x - \cos k_y$ which is very small
near the Fermi surface at low density. Therefore, the net coupling
strength, $|g\phik^2|$, is very weak. This depressed coupling does not occur
for an $s$-wave lattice. The situation with 3D, $s$-wave jellium is slightly
more complicated, since $g_c$ is related to $k_0$ via $g_c = -4\pi /mk_0$.
For small $k_F/k_0$ and fixed $k_F$, $g_c$ decreases with decreasing
$k_F/k_0$, therefore, it takes a larger $g_{th}/g_c$. On the other hand, for
large $k_F/k_0$, the net coupling at Fermi energy, $g\phik^2 \approx
g(k_0/k_F)^2$, becomes small as $k_0$ decreases. This effect dominates the
increase of $g_c$, and, therefore, $g_{th}$ increases gradually with
$k_F/k_0$, as shown in Fig.~\ref{n-g-S}(b) and the inset.

For both $s$-wave cases, the high density part of the $T_c=0$ curve closely
follows the $\mu=0$ curve. For the lattice $s$-wave case, these two
essentially merge, as shown in Fig.~\ref{n-g-S}(a), so $T_c$ vanishes as the
system begins to enter the bosonic regime. This is very different from the
$d$-wave case shown in Fig.~\ref{n-g_Phase}.

\section{Effective pair mass}

The $T_c=0$ curves in the pseudogap regime in the above $n-g$ phase diagrams
are calculated via the divergence of the effective pair mass, $M^*$, at
$T=0$. Additional differences between $s$- and $d$-wave cases come from the
$g$ dependence of $M^*$.  $M^*$ can be calculated at $T=0$ via the expansion
of the inverse $T$ matrix, Eq.~(\ref{Omegaq}), assuming BCS is valid at
$T=0$.  In this way, we will obtain a negative pair mass in the
non-superconducting region IV and V in Fig.~\ref{n-g_Phase}.

\begin{figure}
\centerline{
\includegraphics[width=3in]{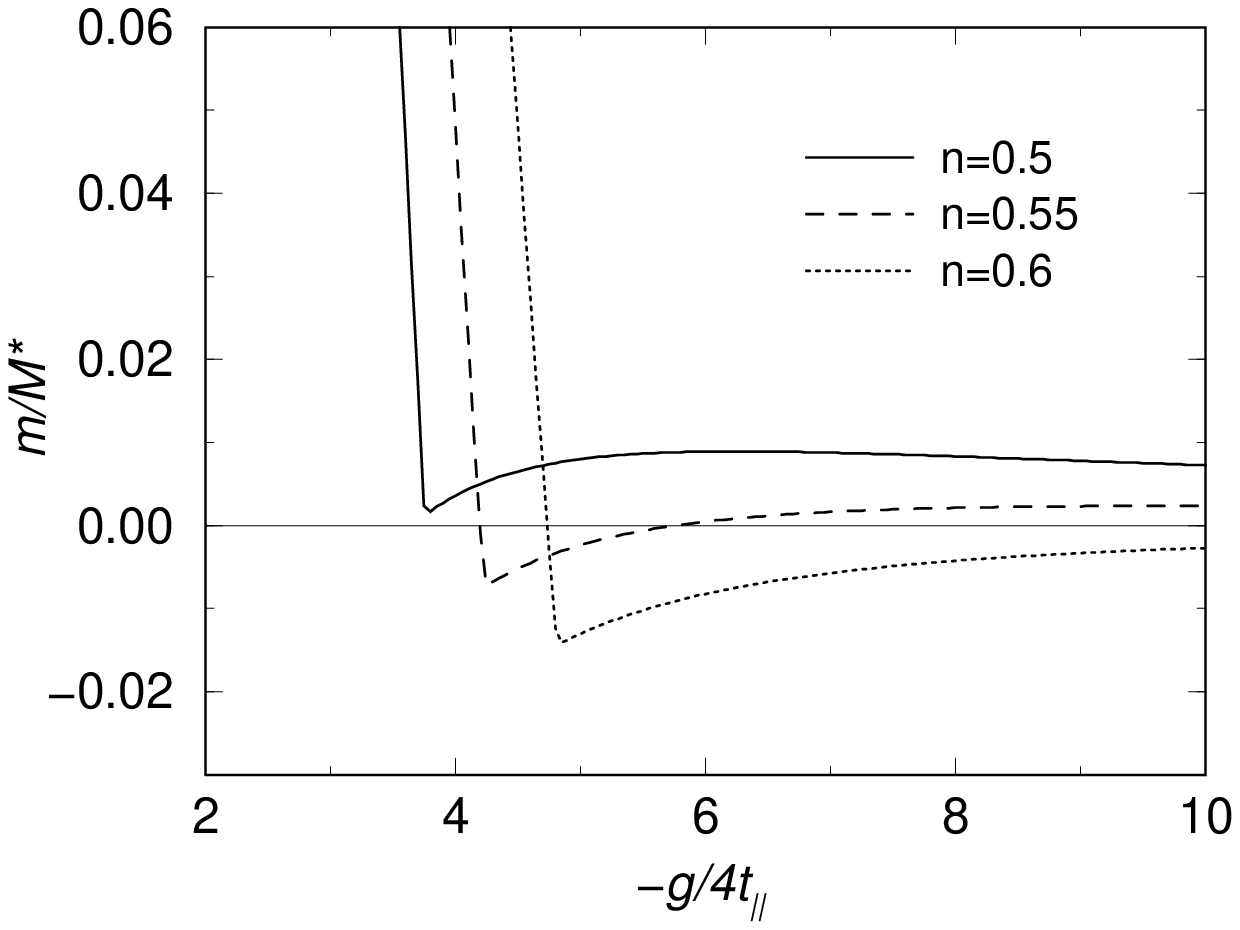}
\includegraphics[width=3in]{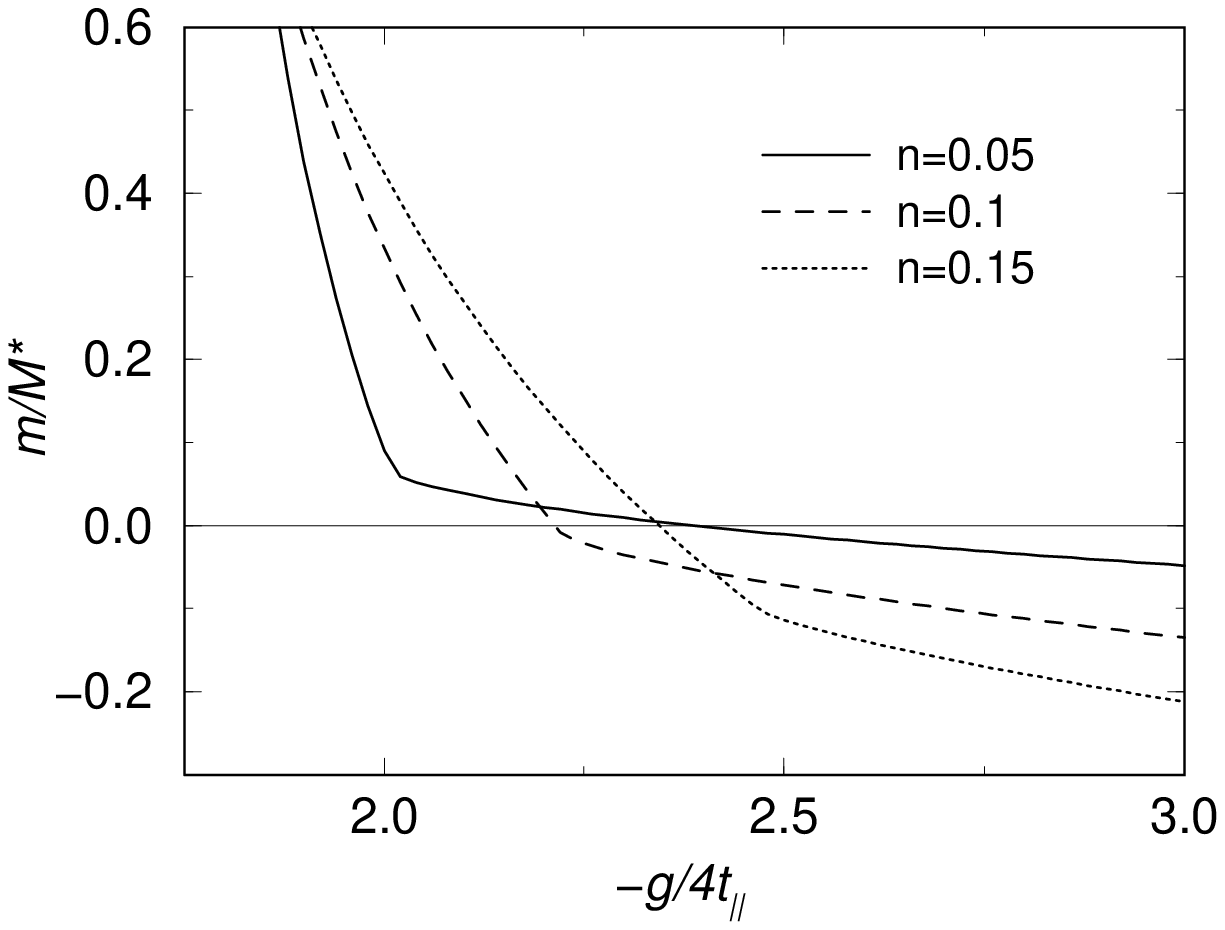}
}
\vskip -0.8in
\hskip 0.7in {\textbf{(a)}} \hskip 2.75in {\textbf{(b)}}
\vskip 0.5in
\caption[Evolution of the effective in-plane pair mass with respect to the
coupling constant in quasi-2D with $s$- and $d$-wave pairing. ]{In-plane
  effective pair mass $M^*$ as a function of $g$ in quasi-2D with (a)
  $s$-wave and (b) $d$-wave pairing, calculated at densities close to the
  critical value for a bosonic superconductor to exist. The parameter
  $t_\perp/t_\parallel$ takes $10^{-2}$ and $10^{-3}$, respectively.}
\label{PairMass}
\end{figure}

Plotted in Fig.~\ref{PairMass}(a) and (b) are the in-plane $M^*$ vs $g$ for
quasi-2D $s$- and $d$-wave pairing, respectively. They are calculated for
densities close to the critical value at which the bosonic superconducting
regime exists. In both Fig.~\ref{PairMass}(a) and Fig.~\ref{PairMass}(b),
the curve with the smallest $n$ remains positive into the bosonic regime
(i.e., $\mu < 0$), whereas the one with the highest $n$ already becomes
negative before entering the bosonic regime. For the $s$-wave $n=0.55$ case,
$M^*$ becomes positive again later, signaling a re-entrant  $T_c$.

It is interesting to note that in both cases, there is always a kink in the
$m/M^*$ curve. This kink corresponds to $\mu=0$, that is, the boundary
between fermionic and bosonic regimes.  However, there is one obvious
difference between the $s$- and $d$-wave cases. For the former, $m/M^*$
increases slightly upon entering the bosonic regime, whereas for the latter,
$m/M^*$ continues to decrease. Usually, a high density will severely limit
the mobility of the pairs, and thus make the pairs heavy. This is due to the
Pauli exclusion between the component fermions of the pairs. For the
$d$-wave case, the interaction will become repulsive if the two electrons
within a pair come too close. Therefore, a $d$-wave superconductor is
effectively always in the large density limit. Consequently, $m/M^*$ always
monotonically decreases with $g$.

What happens when $M^*$ diverges? When $M^*$ diverges, the leading order
term in the pair dispersion becomes $\Omegaq \propto q^4$, and the critical
dimension becomes 4. Therefore, $T_c$ vanishes in 3D. The calculations
beyond this point may not be meaningful. A negative pair mass means that the
pair chemical potential is not zero, and thus the BCS gap equation should
not be satisfied. We speculate that the Leggett ground state (BCS mean field
theory + self-consistently varying $\mu$) may not be applicable to the
negative pair mass regime.

\section{Pair size or correlation length}
\label{Sec_PairSize}

In this subsection, we study the pair size as a function of $g$, for the two
pairing symmetries. This is presumed to correspond to the correlation length
$\xi$.

Following Leggett \cite{Leggett,Leggett3}, the pair size at $T=0$ can be
calculated in momentum space by
\begin{equation}
\xi_0^2 = \frac{\langle \psi_{\mathbf{k}}| -\vec\nabla_{\mathbf{k}}^2
  | \psi_{\mathbf{k}} \rangle} {\langle \psi_{\mathbf{k}}
  | \psi_{\mathbf{k}} \rangle} \:,
\end{equation}
where 
\begin{equation}
\psi_{\mathbf{k}} = \uk^* \vk = \frac{\Delta\phik}{2\Ek}
\end{equation}
is the pair wavefunction in momentum space.
Since
%
%\begin{equation}
$
\langle \psi_{\mathbf{k}}| -\vec\nabla_{\mathbf{k}}^2
  | \psi_{\mathbf{k}} \rangle = \langle
  \vec\nabla_{\mathbf{k}}\psi_{\mathbf{k}}
  | \vec\nabla_{\mathbf{k}}\psi_{\mathbf{k}} \rangle
$
%\end{equation}
%
and
\begin{equation}
\vec\nabla_{\mathbf{k}} \psi_\mb{k} = \frac{\Delta\ek}{2\Ek^3} \left(\ek
  \vec\nabla_{\mathbf{k}} \phik - \phik \vec\nabla_{\mathbf{k}} \ek \right) 
\:,
\end{equation}
we integrate by parts, and obtain
\begin{equation}
\xi_0^2 = \frac{\displaystyle \sumk  \frac{\ek^2}{\Ek^6}\left[ \ek^2 \,
      ( \vec\nabla_{\mathbf{k}} \phik)^2 + \phik^2
      ( \vec\nabla_{\mathbf{k}} \ek)^2 -2\ek\phik 
      ( \vec\nabla_{\mathbf{k}} \ek)\cdot 
      ( \vec\nabla_{\mathbf{k}} \phik) \right] }
{\displaystyle \sumk \frac{\phik^2}{\Ek^2}} \:.
\label{CorrLength_Eq}
\end{equation}

\begin{figure}
\centerline{
\includegraphics[width=3in]{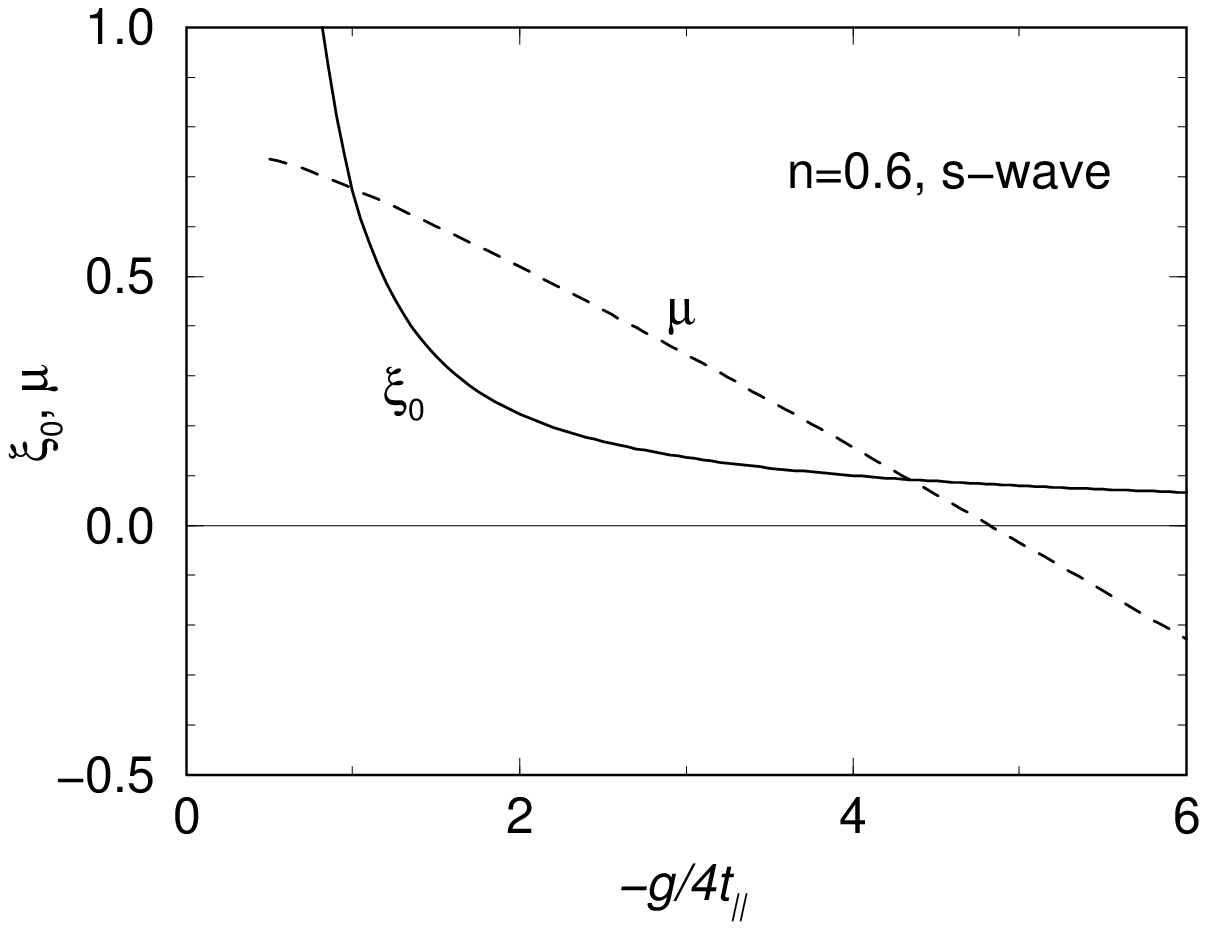} \hfill
\includegraphics[width=2.9in]{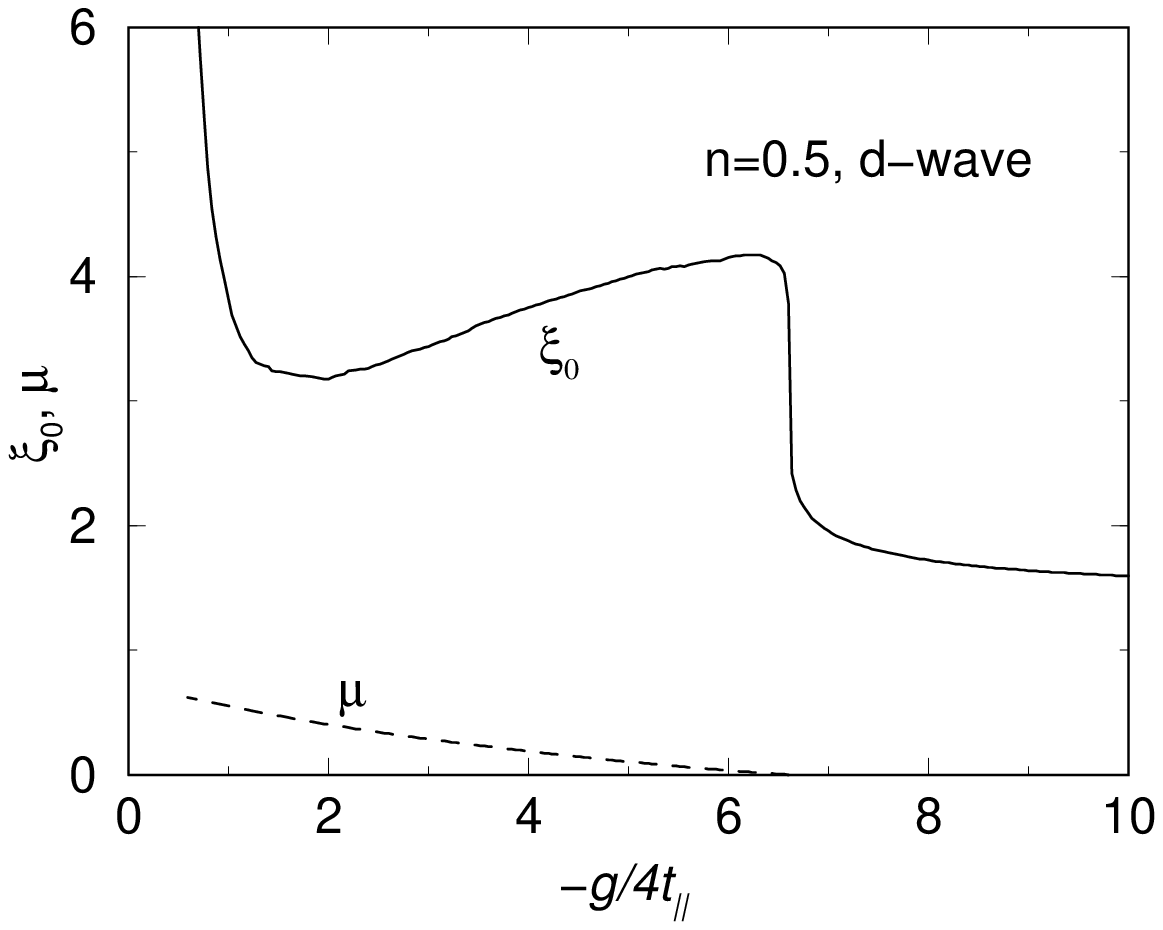}
}
\vskip -0.8in
\hskip 0.7in {\textbf{(a)}} \hskip 4.55in {\textbf{(b)}}
\vskip 0.5in
\caption[Different behavior of the pair size as a function of $g$ between
$s$- and $d$-wave pairing on a quasi-2D lattice.]{Pair size $\xi_0$ as a
  function of $g$ with (a) $s$- and (b) $d$-wave pairing on a quasi-2D
  lattice. Also plotted is $\mu$.  There is a jump in $\xi_0$ with $d$-wave
  pairing across the fermionic-bosonic boundary. We set a lower bound
  $10^{-12}$ for $\phik^2$ in (b). The unit for $\xi_0$ is $a$, the lattice
  constant. }
\label{PairSize}
\end{figure}

Shown in Fig.~\ref{PairSize}(a) is the pair size $\xi_0$ as a function of
$g$ for a quasi-2D lattice with $s$-wave pairing.  Also plotted here is the
chemical potential, $\mu$. As expected, for $s$-wave pairing, as $g$
increases, the pair size $\xi_0$ shrinks, and $\mu$ decreases.

In contrast, for the $d$-wave case, the situation is very different. In the
fermionic regime ($\mu > 0$), the integrand of the numerator in
Eq.~(\ref{CorrLength_Eq}) contains four \textit{essential} singularities
(one in each quadrant) where the Brillouin zone diagonals cross the Fermi
surface (so that $\Ek=0$). In contrast to a simple pole, these
singularities have a different limiting behavior when approached from
different directions.  When approached along the Fermi surface, the
integrand vanishes identically, whereas when approached along the diagonals,
as well as any other direction, the integrand diverges as $1/\delta k^2$,
where $\delta k$ is the deviation of the momentum away from these
singularities.  These singularities lead to a logarithmic divergence of the
integral ($\propto \ln \delta k$). This reflects the non-local effect that the
correlation length in the nodal directions diverges (as $1/\Delta_{\bf k}$).
The logarithmic divergence may be regularized by a small scattering rate
between quasiparticles so that $\Ek$ never vanishes identically.
Equivalently, one may add a tiny lower bound to $\phik^2$. However, in the
bosonic regime, the $d$-wave gap nodes disappear, and the minimum excitation
gap is given by $|\mu|$ at the original nodal points.  Plotted in
Fig.~\ref{PairSize}(b) is the pair size $\xi_0$ for the $d$-wave case. As
$g$ increases, $\xi_0$ first decreases rapidly, and then increases slightly
before $\mu=0$ is reached. As is evident from the figure, there is a jump
when $\mu$ changes sign.  This jump is associated with the full gapping of
the excitation spectrum. In the bosonic regime, the pair size shrinks
gradually as $g$ further increases. In this respect, the system behaves like
an anisotropic $s$-wave superconductor.  On the other hand, the pairing
interaction still has a $d$-wave symmetry, and therefore, the pair size
cannot be smaller than the lattice spacing $a$. As can be seen from the
figure, $\xi_0$ is much larger in the $d$-wave case than in the $s$-wave
case. The former is always larger than $a$, whereas the latter is much
smaller than $a$.

\vskip 1ex

In summary, in this Appendix, we studied in detail the difference between
$s$- and $d$-wave superconductors as a function of $g$ and $n$. We
constructed a phase diagram for a generic quasi-2D superconductor as an
$n-g$ plot, and compared with that for an $s$-wave superconductor,
associated with either a quasi-2D lattice or 3D jellium. We found that
$d$-wave symmetry leads to a very different behavior.  The cuprate
superconductors are far from the bosonic regime in the $n-g$ phase diagram.
Finally, we showed that $d$-wave symmetry also leads to different behavior
of the effective pair mass and pair size, as a function of $g$.
Due to the large pair size, the inverse pair mass $m/M^*$ monotonically
decreases with $g$ in the $d$-wave case, whereas in the $s$-wave case it
starts to increase again at intermediate and low densities upon entering the
bosonic regime. For the very dilute limit, we find that, unlike the $s$-wave
case, there are kinks in the $T_c$, $\mu$, and $\Delta$ curves as the
coupling $g$ crosses the fermionic-bosonic boundary (where $\mu=0$).
Associated with these kinks, there is a jump in the pair size across this
boundary.

\chapter{Extrapolation for the coupled equations above \textit{T}$_{\bf \lowercase{c}}$}
\label{App_AboveTc}

Above $T_c$, the effective chemical potential, $\mu_{pair}$ is no longer
zero, so that the gap equation no longer holds. Close to $T_c$, we can still
use the $T$ matrix expansion, Eq.~(\ref{t_Expansion}), which implies
\beq
t^{-1}_{pg}({\bf 0},0) = a_0 \mu_{pair} \:.
\eeq

Since now $t_{pg}(Q)$ no longer diverges at $Q=0$, in principle,
Eq.~(\ref{Sigma_PG_Approx}) may no longer be a good approximation. However,
for a \textit{crude} estimate, it may still be reasonably good when $T$
is close to $T_c$, where $\mu_{pair}$ is small.

Associated with this non-vanishing $\mu_{pair}$, the pair dispersion
$\Omegaq$ must be replaced by $\Omegaq-\mu_{pair}$. Neglecting the effects
of a finite life time for both the fermions and the pairs, the number
equation remains unchanged. Finally, we arrive at (with $\Delta=\Delta_{pg}$)
\begin{mathletapp}
\label{AboveTc_Eq}
\bea
a_0 g \mu_{pair} & = & 1 + g \sumk \frac{1-2f(\Ek)}{2\Ek}\phik^2 \:, \\
n &= &  2\sumk \bigg[ \vk^2 + \frac{\ek}{\Ek} f(\Ek)\bigg] \:, \\
a_0 \Delta^2 &=& \sum_{\mathbf{q}} b(\Omegaq-\mu_{pair}) \:.
\eea
This set of equations should be supplemented by 
\beq
a_0 \Delta^2 = \frac{1}{2} \bigg[ n-2\sumk f(\ek)\bigg] \:.
\eeq
\end{mathletapp}

Equations (\ref{AboveTc_Eq}) can be used to obtain an \textit{approximate}
solution for $(\Delta, \mu, \mu_{pair})$, as well as $a_0$, for given $g$ and
$T \ge T_c$.

\chapter{Derivation for the pairon contribution to the specific heat}
\label{App_Cv}

In this appendix, we derive Eq.~(\ref{S_Pair_Eq}) for the pair excitation
contribution to the entropy, starting from Eq.~(\ref{ThermoPotential_Eq}).
We will use the analytically continued real frequency representation for the
$T$ matrix and the pair susceptibility $\chi$. Since $t_{pg}(Q)=t(Q)$ for
$Q\ne 0$, we will drop the subscript ``pg'' of the $T$ matrix for
brevity. 

\begin{figure}
\centerline{\hskip 1cm \includegraphics[width=2.3in]{3djelEe-g1k4T_5-3} 
\hfill \includegraphics[width=2.8in]{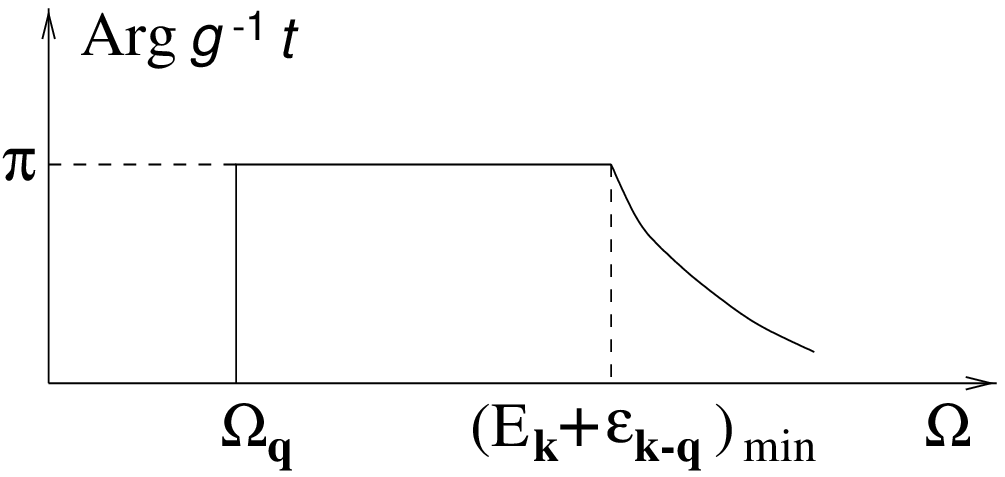}}
\vskip -1.25in 
\hskip 2.3in {\bf (a)} \hskip 2.6in {\bf (b)}
\vskip 1.0in
\caption[Typical behavior of the pair dispersion and of the 
scattering phase shift in the $T$ matrix with $s$-wave pairing.]{Typical
  behavior of (a) the pair dispersion $\Omegaq$ and of (b) the scattering
  phase shift in the $T$ matrix with $s$-wave pairing. $\Omegaq$ in (a) is
  calculated at $T=0.5T_c$, $g=g_c$, and $k=4k_F$ in 3D jellium.}
\label{PhaseShift}
\end{figure}

Equation (\ref{ThermoPotential_Eq}) can be rewritten as 
\begin{equation}
  \Omega_{pair} = \sum_{\mb{q}\ne\mb{0}} \int_C \frac{\mathrm{d}\Omega}{2\pi
    i} \: b(\Omega) \ln\left[ 1+ g\chi(\mb{q},\Omega)\right] \:,
\end{equation}
where the contour $c$ encircles the real axis clockwise but excludes the
origin. The entropy is given by
\begin{eqnarray}
  S_{pair} &=& -\frac{\partial \Omega}{\partial T} \nonumber\\ 
& =& -\sumq
  \int_C \frac{\mathrm{d}\Omega}{2\pi i} \, \left[ \frac{\partial
      b(\Omega)}{\partial T} \ln \left[ 1+g\chi(\mb{q},\Omega)\right] -
    b(\Omega) t(\mb{q},\Omega) \frac{\partial \chi(\mb{q},\Omega)}{\partial
      T} \right] \nonumber\\ &\equiv & (\mbox{I}) + (\mbox{II}) \:.
\label{Cv_S_Eq}
\end{eqnarray}
The first term is associated with the scattering phase shift: 
\begin{eqnarray}
  (\mbox{I}) &=& \sumq \int_{-\infty}^{\infty} \frac{\mathrm{d}\Omega}{\pi}
  \: \frac{\partial b(\Omega)}{\partial T}
  \mbox{Arg}\frac{t(\mb{q},\Omega+i0)}{g} \nonumber\\ 
& = & \sumq
  \int_{-\infty}^{\infty} \frac{\mathrm{d}\Omega}{\pi} \: \frac{\partial
    b(\Omega)}{\partial T} \theta(\Omega-\Omegaq)\left[ \pi -
  \tan^{-1} \frac{\Gammaq}{\Omega-\Omegaq}\right] \:,%\nonumber\\
%&\approx& \sumq \int_{\Omegaq}^{(\Ek+\ekq)_{min}} \mathrm{d} \Omega \:
%  \frac{\partial b(\Omega)}{\partial T} \nonumber\\
\label{Cv_Term1_Eq}
\end{eqnarray}
where in the second step, we have used the $T$ matrix expansion,
Eq.~(\ref{t_Expansion}) (with $\mu_{pair}=0$), and the fact that $g< 0$ as
well as $a_0>0$. Here $\theta(x)$ is the usual step function. We have
replaced $\Gamma_{\mb{q},\Omega}$ by $\Gamma_{\mb{q},\Omegaq}\equiv\Gammaq$.
As can be seen from Eq.~(\ref{Omega_q:Gamma_q}), $\Gammaq=0$ when
$-(\Ek-\ekq)_{min} < \Omegaq < (\Ek+\ekq)_{min}$. When $\Omegaq >
(\Ek+\ekq)_{min}$, the pair dispersion intersects the particle-particle
continuum, and therefore becomes damped, with $\Gammaq>0$.  Shown in
Fig.~\ref{PhaseShift}(a) is an example, calculated at intermediate coupling
in 3D jellium with $s$-wave pairing. In this context, the typical behavior
of the phase shift is $\delta = \mbox{Arg}\: g^{-1}t(\mb{q},\Omega)$ for
$s$-wave pairing is plotted in Fig.~\ref{PhaseShift}(b). $\delta=\pi$ before
$\Omegaq$ intersects the particle-particle continuum. For $d$-wave pairing,
$\min(\Ek+\ekq)=0$, nevertheless, $\Gammaq$ is small and $\delta \approx
\pi$ for small {\bf q}, since the phase space for low energy (nodal)
quasiparticle excitations is small. Furthermore, the main contribution in
Eq.~(\ref{Cv_Term1_Eq}) comes from the small {\bf q} region, therefore we
may neglect the contribution of $\Gammaq$ so that
\begin{equation} 
(\mbox{I}) \approx \sumq \int_{\Omegaq}^\infty \mathrm{d} \Omega \:
\frac{\partial b(\Omega)}{\partial T} \:.  
\end{equation}
This approximation amounts to treating the pair excitations as free bosons.

The second term in Eq.~(\ref{Cv_S_Eq}) can be further split into two terms:
\begin{eqnarray}
  (\mbox{II}) &=& -\sumq \int_{-\infty}^\infty \frac{\mathrm{d}\Omega}{\pi}
  b(\Omega) \left[ \mbox{Im}\, t(\mb{q},\Omega+i0) \mbox{Re}
    \frac{\partial \chi(\mb{q},\Omega)}{\partial T} + \mbox{Re}\,
    t(\mb{q},\Omega+i0) \mbox{Im} \frac{\partial
      \chi(\mb{q},\Omega)}{\partial T} \right] \nonumber\\ 
&\equiv&  (\mbox{II}a) + (\mbox{II}b) \:,
\label{Cv_Term2_Eq}
\end{eqnarray}
where 
\begin{equation}
  \mbox{Im}\, t(\mb{q},\Omega+i0) = -\frac{1}{a_0}
  \frac{\Gammaq}{(\Omega-\Omegaq)^2 + \Gammaq^2} \approx
  -\frac{\pi}{a_0}\delta(\Omega-\Omegaq) \:,
\end{equation}
\begin{equation}
\mbox{Re} \frac{\partial \chi(\mb{q},\Omega)}{\partial T} =
\frac{\partial}{\partial T} \left[ a_0 (\Omega-\Omegaq)\right] \approx -a_0
\frac{\partial \Omegaq}{\partial T\;} \:,
\end{equation}
and
\begin{equation}
\mbox{Im}\frac{\partial \chi(\mb{q},\Omega)}{\partial T} =
\frac{\partial}{\partial T} \left( a_0 \Gammaq\right) \approx 0 \:.
\end{equation}
Here use has been made of the fact that $a_0$ is essentially temperature
independent. (See, e.g., Figs.~\ref{3D_BT_Gaps} and \ref{Cuprate_Gaps}).
In this way, we have
\begin{equation}
  (\mbox{II}a) \approx -\sumq b(\Omegaq) \frac{\partial \Omegaq}{\partial
    T\;}\:,
\qquad
\mbox{and}
\qquad
  (\mbox{II}b) 
%= - \sumq \int_{-\infty}^\infty \frac{\mathrm{d}\Omega}{\pi} b(\Omega)
% \mbox{Re} t(\mb{q},\Omega) \mbox{Im} \frac{\partial 
%\chi(\mb{q},\Omega)}{\partial T}
\approx 0 \:.
\label{Cv_Term2ab_Eq}
\end{equation}
Substituting Eqs.~(\ref{Cv_Term1_Eq}), (\ref{Cv_Term2_Eq}),
and (\ref{Cv_Term2ab_Eq}) into Eq.~(\ref{Cv_S_Eq}), we
finally obtain
\begin{equation}
S_{pair} = \sumq \int_{\Omegaq}^\infty \mathrm{d}\Omega \: \frac{\partial
  b(\Omega)} {\partial T} - \sumq b(\Omegaq) \frac{\partial
  \Omegaq}{\partial T\;} \:.
\label{Cv_S_Eq2}
\end{equation}
The first term corresponds to a free boson contribution, and the second term
reflects the effect of the $T$ dependence of the pair dispersion. In
general, $\Omegaq$ does not change much with temperature, especially in the
low $T$ regime, and may thereby be neglected. In this way, we obtain the
free boson-like expression, Eq.~(\ref{S_Pair_Eq}), for the pairon
contribution to the entropy.
The frequency integration in Eq.~(\ref{Cv_S_Eq2}) can be easily carried out.
We then obtain a simple expression for $\Omega_{pair}$:
\begin{equation}
\Omega_{pair} = -\sumq \int^\infty_{\Omegaq} b(\Omega)\,\mathrm{d}\Omega \:.
\end{equation}
This describes free bosons with dispersion $\Omegaq$.

To see the equivalence between the first term of Eq.~(\ref{Cv_S_Eq2}) and
Eq.~(\ref{S_Pair_Eq}), it suffices to note that 
\begin{equation}
\frac{\partial b(\Omega)}{\partial T} = - \frac{\partial b(\Omega)}{\partial
  \Omega} \frac{\Omega}{T} \:,
\qquad\mbox{and}\qquad
\frac{b(\Omega)}{T} = -\frac{\partial \ln [1+b(\Omega)]} {\partial \Omega} \:,
\end{equation}
so that 
\begin{eqnarray}
\int_{\Omegaq}^\infty \mathrm{d} \Omega \frac{\partial b(\Omega)}{\partial T} 
&=& -\int_{\Omegaq}^\infty \mathrm{d} \Omega \: \frac{\Omega}{T} \frac{\partial
  b(\Omega)}{\partial \Omega} \nonumber\\
& = & b({\Omegaq}) \frac{{\Omegaq}}{T\;} - \int_{\Omegaq}^\infty \mathrm{d} 
\Omega \: \frac{\partial \ln \left[1+b(\Omegaq)\right]} {\partial \Omega} \nonumber\\
& = & \left[1+b(\Omegaq)\right]\ln \left[1+b(\Omegaq)\right] - b(\Omegaq)\ln b(\Omegaq) \:.
\end{eqnarray}

\chapter{Evaluation of the Vertex Corrections}
\label{App_Vertex}

In this appendix we demonstrate an explicit cancellation between the
Maki-Thompson (MT) and Aslamazov-Larkin (AL) diagrams of Fig.~3.  In this
way we prove that the contribution to the vertex correction $\delta \Lambda$
from the superconducting order parameter is given by the Maki-Thompson
diagram, and the pseudogap contribution $\delta \Lambda_{pg}$ comes from MT
and AL diagrams.  It is easy to demonstrate a cancellation between the MT
diagram and the AL diagrams, which will greatly simplify the calculations.
In general, we have

\begin{equation}
  \delta \Lambda_{pg}^\mu (K,K-Q)q_\mu = -(MT)_{pg} + \sum_P
  t_{pg}(P)G_0 (P-K) \frac{\partial
    \varphi^2_{{\mathbf{k}}-{\mathbf{p}}/2-{\mathbf{q}}/2}} 
   {\partial {\bf k}}\cdot {\bf q}\,,
\label{cancellation}
\end{equation}
where $(MT)_{pg}$ refers to the MT diagram contribution, and
$t_{pg}(Q\neq 0)$ is the $T$ matrix or pair propagator.

%---------------------------------

To prove this cancellation, we notice that the vertex corrections in
the (four-)current--current correlation functions can be
obtained from proper vertex insertions in the single particle Green's
functions in the self-energy diagram.  In the pairing
approximation ( $G_0G$ scheme) we have
\begin{equation}
\Sigma_{pg}(K)=\sum_L t_{pg}(K+L)
G_0 (L) \varphi^2_{(K-L)/2},
\label{Eq:self-energy}
\end{equation}
where $L$ is the four-momentum of the fermion loop, this procedure leads
to one Maki-Thompson diagram and two Aslamazov-Larkin diagrams. 

Obviously, $L$ in the Eq.~(\ref{Eq:self-energy}) is a dummy variable so
that its variation does not change $\Sigma(K)$, namely,
\begin{eqnarray}
  0&=&\sum_L [ t_{pg}(K+L+\Delta L)G_0(L+\Delta L)
  \varphi^2_{(K-L-\Delta L)/2} - t_{pg}(K+L)G_0(L)
  \varphi_{(K-L)/2}^2]\nonumber\\ 
&=&\sum_L \Big\{\left[t_{pg}(K+L+\Delta
    L)-t_{pg}(K+L)\right] G_0(L+\Delta L) \varphi^2_{(K-L-\Delta
    L)/2}\nonumber\\ 
&&{} +t_{pg}(K+L) [G_0(L+\Delta L) -G_0(L)]
  \varphi^2_{(K-L-\Delta L)/2}\nonumber\\ 
&&{} + t_{pg}(K+L) G_0(L) [
  \varphi^2_{(K-L-\Delta L)/2} - \varphi^2_{(K-L)/2} ]\Big\}
\label{Eq:diff_Sigma}
\end{eqnarray}
Using $G(K)G^{-1}(K)=1$, we obtain 
\begin{mathletapp}
\begin{eqnarray}
\label{Eq:vertex-insertion}
G(K+\Delta K)-G(K) &=& -G(K)[G^{-1}(K+\Delta K) - G^{-1} (K)] G(K+\Delta
K)\nonumber\\
&=&-G(K)\Lambda_\mu (K+\Delta K, K) G(K) \Delta K^\mu \;,
\end{eqnarray}
where $G^{-1}(K+\Delta K) - G^{-1} (K)\approx \Lambda_\mu (K+\Delta K, K)
\Delta K^\mu$ is the full vertex. Similarly, we have
\begin{eqnarray}
  G_0(K+\Delta K)-G_0(K) &=& -G_0(K)[G_0^{-1}(K+\Delta K) - G_0^{-1}
  (K)] G_0(K+\Delta K)\nonumber\\ 
  &=&-G_0(K)\lambda_\mu (K+\Delta K, K)G_0(K) \Delta K^\mu \;,
\end{eqnarray}
\end{mathletapp}
where $G_0^{-1}(K+\Delta K) - G_0^{-1} (K)\approx \lambda_\mu (K+\Delta
K, K) \Delta K^\mu$ is the bare vertex, and $\lambda^\mu (K+\Delta K, K)=(1,
\vec \nabla_{\bf k} \epsilon_{{\bf k}+\Delta {\bf k}/2})$.
Equations (\ref{Eq:vertex-insertion}) correspond to the vertex insertions
diagrammatically along the full and bare Green's functions, respectively.

Using ${\displaystyle t_{pg}(K+L)=\frac{g}{1+g\chi (K+L)}}\:$, we obtain
\begin{equation}
t_{pg}(K+L+\Delta L)-t_{pg}(K+L)= -t_{pg}(K+L+\Delta L)[\chi(K+L+\Delta L)-\chi(K+L)]t_{pg}(K+L)
\label{Eq:diff_t}
\end{equation}
Writing ${\displaystyle \chi(K+L)=\sum_{L^\prime} G(L^\prime)G_0(K+L-L^\prime)
\varphi^2_{L^\prime-(K+L)/2}}$, we have
\begin{eqnarray}
\lefteqn{\chi(K+L+\Delta L)-\chi(K+L)} \nonumber\\
 &=&  \sum_{L^\prime} G(L^\prime)
\Big\{\left[G_0(K+L-L^\prime+\Delta L)-G_0(K+L-L^\prime)\right]
  \varphi^2_{L^\prime-(K+L+\Delta L)/2}\nonumber\\
&&{} + G_0(K+L-L^\prime)[
    \varphi^2_{L^\prime-(K+L+\Delta L)/2}
    -\varphi^2_{L^\prime-(K+L)/2}] \Big\}.
\label{Eq:diff_chi1}
\end{eqnarray}
On the other hand, writing ${\displaystyle \chi(K+L)=\sum_{L^\prime}
G(K+L-L^\prime)G_0(L^\prime) \varphi^2_{(K+L)/2-L^\prime}}$, we get
\begin{eqnarray}
\lefteqn{\chi(K+L+\Delta L)-\chi(K+L)} \nonumber\\
& =& \sum_{L^\prime} 
\Big\{\left[G(K+L-L^\prime+\Delta L)-G(K+L-L^\prime)\right]G_0(L^\prime)
  \varphi^2_{(K+L+\Delta L)/2-L^\prime}\nonumber\\
&&{} + G(L^\prime)G_0(K+L-L^\prime)[
    \varphi^2_{L^\prime-(K+L-\Delta L)/2}
    -\varphi^2_{L^\prime-(K+L)/2}] \Big\}.
\label{Eq:diff_chi2}
\end{eqnarray}
Combining Eq.~(\ref{Eq:diff_chi1}) and Eq.~(\ref{Eq:diff_chi2}), we
obtain to the first order of $\Delta L$
\begin{eqnarray}
\lefteqn{\chi(K+L+\Delta L)-\chi(K+L)} \nonumber\\
& =& \frac{1}{2} \sum_{L^\prime} 
\Big\{ G(L^\prime)\left[G_0(K+L-L^\prime+\Delta
    L)-G_0(K+L-L^\prime)\right]\varphi^2_{L^\prime-(K+L+\Delta L)/2}
\nonumber\\ 
&&{} +
\left[G(K+L-L^\prime+\Delta L)-G(K+L-L^\prime)\right]G_0(L^\prime)
  \varphi^2_{L^\prime-(K+L+\Delta L)/2}\Big\}\;,
\label{Eq:diff_chi}
\end{eqnarray}
where we have assumed in general $\varphi^2_K =\varphi^2_{-K}$.
Substituting Eq.~(\ref{Eq:diff_chi}) and  Eq.~(\ref{Eq:diff_t})
into Eq.~(\ref{Eq:diff_Sigma}), we obtain
\begin{eqnarray}
0&=&-\frac{1}{2} \sum_{L L^\prime} t_{pg}(K+L+\Delta L) 
t_{pg}(K+L) \Big\{  G(L^\prime) \left[ G_0(K+L-L^\prime+\Delta
    L)\right.\nonumber\\
&&{} -\left. G_0(K+L-L^\prime)\right]\varphi^2_{L^\prime-(K+L+\Delta
  L)/2} \nonumber\\
&&{} +\left[G(K+L-L^\prime+\Delta L)-G(K+L-L^\prime)\right] \nonumber\\
&&{} \times G_0(L^\prime)
  \varphi^2_{L^\prime-(K+L+\Delta L)/2}\Big\} G_0 (L+\Delta L)
  \varphi^2_{(K-L-\Delta L)/2}\nonumber\\
&&{} + \sum_L  t_{pg}(K+L) \left[ G_0 (L+\Delta L) -G_0(L)\right]
  \varphi^2_{(K-L-\Delta L)/2} \nonumber\\
&&{} + \sum_L t_{pg}(K+L)G_0(L) [\varphi^2_{(K-L-\Delta L)/2} -
  \varphi^2_{(K-L)/2}]
\label{Eq:diff_Sigma2}
 \end{eqnarray}
 Comparing this with the analytical expressions corresponding to the
 diagrams in Fig.~3, it is easy to identify the first two terms
 as the two AL diagrams (which we denote by $AL_1$ and $AL_2$) and the
 third one with the MT diagram for the pseudogap vertex corrections.
 Therefore,
\begin{equation}
  \frac{1}{2}\big[(AL_1) + (AL_2)\big] + (MT)_{pg} + \sum_L
  t_{pg}(K+L)G_0(L)\left[\varphi^2_{(K-L-\Delta L)/2} -
    \varphi^2_{(K-L)/2} \right] =0\;.
\end{equation}
Finally, we have
\begin{eqnarray}
  \delta \Lambda_{pg}^\mu(K, K-\Delta L) \Delta L_\mu &=& (AL_1) +(AL_2) +
  (MT)_{pg}\\ 
&=&-(MT)_{pg}-2\sum_L t_{pg}(K+L)G_0(L)\frac{\partial      
\varphi^2_{(K-L-\Delta L)/2}} {\partial L}\cdot \Delta L \;. \nonumber
\end{eqnarray}
Changing variables $K+L\rightarrow P, \Delta L \rightarrow Q$ leads to
Eq.~(\ref{cancellation}).

%-----------------------------
The two contributions which enter Eq.~(\ref{Vertex}) result from
adding the superconducting gap and pseudogap terms which are given,
respectively, by
\begin{mathletapp}
\begin{equation}
\delta \Lambda_{sc} (K, K-Q) = -\Delta_{sc}^2\varphi_{\mathbf{k}} 
\varphi_{\mathbf{k-q}}G_0(-K)G_0(Q-K)\lambda(Q-K,-K)
\end{equation}
and
\begin{eqnarray}
 \lefteqn{\delta \Lambda_{pg}^\mu (K, K-Q)} \nonumber\\
& =& -\sum_P
  t_{pg}(P)\varphi_{\mathbf{k-p/2}} \varphi_{\mathbf{k-q-p/2}}
  G_0(P-K)G_0(P+Q-K) \lambda^\mu (P+Q-K,P-K) \nonumber\\ 
& & {}+ \sum_P  t_{pg}(P)G_0 (P-K) \frac{\partial
    \varphi^2_{{\mathbf{k}}-{\mathbf{p}}/2-{\mathbf{q}}/2}} 
{\partial  k_\mu}\,,
\end{eqnarray}
\end{mathletapp}

\chapter{Full Expressions for the correlation functions $\tensor{\bf P}$,
  ${\bf P}_0$, and $P_{00}$}

\label{App_CorrFunc}

It is useful here to write down the component contributions to the
different correlation functions in the electromagnetic response.  After
adding the superconducting and pseudogap contributions one finds for the
current-current correlation function
\begin{eqnarray}
  \tensor{\frac{\bf n}{\bf m}}+\tensor{\bf P} & = & 2\sum_{\mathbf{k}}
  \frac{\Delta_{sc}^2}{E_{\mathbf{k}}^2} \left[
    \frac{1-2f(E_{\mathbf{k}})}{2E_{\mathbf{k}}} + f^\prime
    (E_{\mathbf{k}}) \right] \left[\varphi_{\mathbf{k}}^2 
(\vec{\nabla}
    \epsilon_{\mathbf{k}})(\vec{\nabla} \epsilon_{\mathbf{k}}) -
    \frac{1}{4} (\vec{\nabla} \epsilon_{\mathbf{k}}^2)
    (\vec{\nabla}\varphi_{\mathbf{k}}^2)\right] 
\nonumber\\ 
&&-2\sum_{\mathbf{k}} f^\prime (E_{\mathbf{k}}) 
\frac{\Omega^2}{\Omega^2
  - ({\mathbf{q}}\cdot \vec{\nabla} E_{\mathbf{k}})^2} (\vec{\nabla}
\epsilon_{\mathbf{k}}) (\vec{\nabla}
\epsilon_{\mathbf{k}}) \nonumber\\ 
&&+ \sum_{\mathbf{k}} \frac{\Delta_{pg}^2}{E_{\mathbf{k}}^2} 
f^\prime
(E_{\mathbf{k}}) \frac{\Omega^2}{\Omega^2 - ({\mathbf{q}}\cdot 
\vec{\nabla}
  E_{\mathbf{k}})^2} \left[\varphi_{\mathbf{k}}^2 (\vec{\nabla}
  \epsilon_{\mathbf{k}})(\vec{\nabla} \epsilon_{\mathbf{k}}) -
  \frac{1}{4} (\vec{\nabla} \epsilon_{\mathbf{k}}^2)
  (\vec{\nabla}\varphi_{\mathbf{k}}^2)\right] \,,\nonumber\\
\label{Q00}
\end{eqnarray}
and for the current-density correlation function
\begin{equation}
  {\mathbf{P}}_0=-2\Omega \sum_{\mathbf{k}} 
\frac{\epsilon_{\mathbf{k}}}
  {E_{\mathbf{k}}} f^\prime (E_{\mathbf{k}}) \frac{{\mathbf{q}}\cdot
    \vec{\nabla} E_{\mathbf{k}}} {\Omega^2 - ({\mathbf{q}}\cdot 
\vec{\nabla}
    E_{\mathbf{k}})^2} \vec{\nabla} \epsilon_{\mathbf{k}} \,,
\label{Q03}
\end{equation}
and finally for the density-density correlation function
\begin{equation}
  P_{00}=-2\sum_{\mathbf{k}}
  \frac{\Delta_{sc}^2\varphi_{\mathbf{k}}^2}{E_{\mathbf{k}}^2} \left[
    \frac{1-2f(E_{\mathbf{k}})}{2E_{\mathbf{k}}} + f^\prime
    (E_{\mathbf{k}}) \right] + 2\sum_{\mathbf{k}} f^\prime
  (E_{\mathbf{k}}) \frac{\Omega^2 
\Delta_{sc}^2\varphi_{\mathbf{k}}^2 -
    E_{\mathbf{k}}^2 ({\mathbf{q}}\cdot \vec{\nabla} 
E_{\mathbf{k}})^2}
    {E_{\mathbf{k}}^2 \left[ \Omega^2 - ({\mathbf{q}}\cdot 
\vec{\nabla}
        E_{\mathbf{k}})^2\right]} \,.
\label{Q33}
\end{equation}
In deriving the first of these we have integrated by parts
to evaluate
\begin{eqnarray}
\tensor{\frac{\bf n}{\bf m}} &=& 2\sum_K \frac{\partial^2 \epsilon_{\mathbf{k}}
    }{\partial {\mathbf{k}} \partial
  {\mathbf{k}}} G(K) %\nonumber\\
= -2  \sum_{\mathbf{k}} G^2(K) (\vec{\nabla}
\epsilon_{\mathbf{k}}) \left[ \vec{\nabla}
\epsilon_{\mathbf{k}} + \vec{\nabla} \Sigma(K)\right]\nonumber\\
&=& 2\sum_{\mathbf{k}} \frac{\Delta^2}{E_{\mathbf{k}}^2} \left[
  \frac{1-2f(E_{\mathbf{k}})}{2E_{\mathbf{k}}} + f^\prime
  (E_{\mathbf{k}}) \right] \left[\varphi_{\mathbf{k}}^2 (\vec{\nabla}
  \epsilon_{\mathbf{k}}) (\vec{\nabla} \epsilon_{\mathbf{k}}) -
  \frac{1}{4} (\vec{\nabla} \epsilon_{\mathbf{k}}^2)
  (\vec{\nabla}\varphi_{\mathbf{k}}^2)\right] \nonumber\\
&& -2\sum_{\mathbf{k}}
f^\prime (E_{\mathbf{k}})(\vec{\nabla} \epsilon_{\mathbf{k}})
(\vec{\nabla} \epsilon_{\mathbf{k}}) \,.
\label{n/m}
\end{eqnarray}
These expressions are then used to evaluate Eq.~(\ref{q.P.q}) in the 
text.

\bibliographystyle{prsty} 
\addcontentsline{toc}{chapter}{Bibliography}
%\bibliography{Thesis}

\end{document}